\documentclass[preprint,12pt]{elsarticle}
\usepackage{amssymb}
\usepackage{amsmath}
\usepackage{amsthm}
\usepackage{graphicx, subfigure, color}
\usepackage{cases}
\usepackage{exscale}
\usepackage{xcolor}
\usepackage{relsize}
\usepackage[noend]{algpseudocode}
\usepackage{algorithmicx}
\usepackage{algorithm}
\usepackage{setspace}
\usepackage{diagbox}
\usepackage{enumerate}
\usepackage{pythonhighlight}
\usepackage[export]{adjustbox}
\usepackage{multirow}
\usepackage[colorlinks=true, linkcolor=blue, anchorcolor=blue, citecolor=blue,]{hyperref}

\numberwithin{figure}{subsection}
\numberwithin{table}{subsection}
\newtheorem{remark}{Remark}

\allowdisplaybreaks

\journal{Computers and Mathematics with Applications}

\begin{document}
\begin{frontmatter}
\title{Solving Unbalanced Optimal Transport on Point Cloud by Tangent Radial Basis Function Method}

\author[label1]{Jiangong Pan}
\ead{mathpjg@sina.com}
\author[label2]{Wei Wan}
\ead{weiwan@ncepu.edu.cn}
\author[label3,label4]{Chenlong Bao}
\ead{clbao@tsinghua.edu.cn}
\author[label3,label4]{Zuoqiang Shi\corref{CA}}
\ead{zqshi@tsinghua.edu.cn}

\affiliation[label1]{organization={Department of Mathematical Sciences},
            addressline={Tsinghua University},
            city={Beijing},
            postcode={100084},
            country={China}}
\affiliation[label2]{organization={School of Mathematics and Physics},
            addressline={North China Electric Power University},
            city={Beijing},
            postcode={100006},
            country={China}}
\affiliation[label3]{organization={Yau Mathematical Sciences Center},
            addressline={Tsinghua University},
            city={Beijing},
            postcode={100084},
            country={China}}
\affiliation[label4]{organization={YanqiLake Beijing Institute of Mathematical Sciences and Applications},
            city={Beijing},
            postcode={101408},
            country={China}}
\cortext[CA]{Corresponding author.}

\begin{abstract}
     In this paper, we solve unbalanced optimal transport (UOT) problem on surfaces represented by point clouds. 
     Based on alternating direction method of multipliers algorithm, the original UOT problem can be solved by an iteration consists of three steps. 
     The key ingredient is to solve a Poisson equation on point cloud which is solved by tangent radial basis function (TRBF) method. 
     The proposed TRBF method requires only the point cloud and normal vectors to discretize the Poisson equation which simplify the computation significantly. 
     Numerical experiments conducted on point clouds with varying geometry and topology demonstrate the effectiveness of the proposed method. 
\end{abstract}

\begin{highlights}
\item This work provides a novel method to solve unbanlanced optimal transport on point cloud. By combining the dynamic formulation and ADMM, the key ingredient of the computation of UOT problem on point cloud becomes to efficiently solve a Poisson equation.
\item The Poisson equation is solved by a meshless tangent radial basis function method. The method only requires points and normal vectors, which significantly simplify the computation. 
\end{highlights}

\end{frontmatter}

\section{Introduction}\label{Sec-Intro}
In 1781, the French mathematician Gaspard Monge introduced the seminal optimal transport (OT) problem, aimed at minimizing the total cost associated with material transport, originally framed within the context of ``moving sand" in civil engineering \cite{monge1781memoire}. 
Now, OT problem has evolved into a cornerstone of research across diverse disciplines such as mathematics, economics, and computational science \cite{feydy2019fast, vissio2020evaluating, bonneel2023survey}. 
In the 1940s, Kantorovich relaxed the stringent constraints of Monge's formulation, leading to a series of pioneering optimization algorithms \cite{kantorovich1942translocation}. 
Several decades later, Brenier demonstrated the equivalence between Monge's problem and Kantorovich's reformulation in the continuous case \cite{brenier1989least}. 
At the dawn of the 21st century, Benamou and Brenier expanded the scope of OT by introducing the dynamic OT problem from the perspective of fluid mechanics \cite{benamou2000computational}, significantly broadening the methodological landscape of OT research.

The essence of the OT problem lies in achieving a precise mass balance between the source and target distributions. 
However, in many problems, the initial and target mass may not be same. Such imbalances have prompted the development of the unbalanced optimal transport (UOT) problem. 
This nuanced problem has widespread applications, spanning image registration, transformation, and generation, as well as climate modeling, style transfer, and medical imaging \cite{lubeck2022neural, li2022arbitrary, chen2021evaluating}. 

Benamou and Brenier incorporated the time dimension into the OT problem, allowing it to be analyzed from the novel perspective of fluid mechanics. 
This approach reformulates the original problem into an optimal control problem constrained by partial differential equations (PDE). 
Owing to the inclusion of time, this variant is referred to as the dynamic OT problem. 
The key point is on finding $\rho$ that satisfies the specified initial and terminal densities $\rho_{0}$ and $\rho_{T}$, and the pair $(\rho, \boldsymbol{v})$ that adheres to the law of mass conservation. 
Despite the temporal extension, the principal objective remains the minimization of total cost over all feasible pairs $(\rho, \boldsymbol{v}) \in \mathcal{C}(\rho_{0}, \rho_{T})$. 
The problem can be formally expressed as follows:
\begin{equation*}\label{dyOT}
    \begin{aligned}
        \min_{(\rho, \boldsymbol{v}) \in \mathcal{C}\left(\rho_{0}, \rho_{T}\right)} 
        \int_{T} \int_{\Omega} 
        \frac{1}{2} \rho(t, \boldsymbol{x}) \|\boldsymbol{v}(t,\boldsymbol{x})\|^{2} 
        \mathrm{d} \boldsymbol{x} \mathrm{d} t,
    \end{aligned}
\end{equation*}
where
\begin{equation*}\label{dyOTC}
    \begin{aligned}
        \mathcal{C}(\rho_0,\rho_{T}): = 
        \{(\rho, \boldsymbol{v}): &\partial_t \rho + \text{div}(\rho\boldsymbol{v}) = 0, \\
        & \rho(0, \boldsymbol{x}) = \rho_0(\boldsymbol{x}),\ \rho(T, \boldsymbol{x}) = \rho_{T}(\boldsymbol{x}),\ \rho(\boldsymbol{x})\geq0\}.
    \end{aligned}
\end{equation*}
Based on above dynamic formulation, it is easy to
generalized to the UOT problem by introducint proper source term \cite{benamou2015augmented, ruthotto2020machine, fu2023high}. 

Significant advancements have been made in addressing UOT challenges in recent years. 
Pham et al. \cite{pham2020unbalanced} demonstrated that the complexity of the Sinkhorn algorithm for obtaining an $\varepsilon$-approximate solution to the UOT problem is of order $\mathcal{O}(n^2/\varepsilon)$. 
Sato's team \cite{sato2020fast} established that under the strong exponential time hypothesis, neither the Kantorovich-Rubinstein distance nor optimal partial transport in the Euclidean metric can be computed in strongly subquadratic time. 
Li et al. \cite{li2022application} introduced a mixed $L^1$ Wasserstein distance that maintains convex properties with respect to operations such as shift, dilation, and amplitude change. 
Bauer and collaborators \cite{bauer2022square} characterized the square root normal field (SRNF) shape distance as the pullback of the WFR unbalanced optimal transport distance. 
Gallou{\"e}t et al. \cite{gallouet2019unbalanced} employed a constructive approach, utilizing alternating minimization movements for the Wasserstein and Fisher-Rao distances, to establish the existence of weak solutions for general scalar reaction-diffusion-advection equations. 

However, there has been relatively limited research on UOT issues in the context of point clouds. 
Fathi and Figalli \cite{fathi2010optimal} extended results established for the Monge transport problem on compact manifolds to non-compact settings, specifically for costs derived from Tonelli Lagrangians. 
Grange and colleagues \cite{grange2023computational} demonstrated how to learn optimal transport maps from samples that correspond to probability distributions defined on manifolds. 
Berti et al. \cite{berti2023numerical} adapted the DMK model from Euclidean to Riemannian contexts and hypothesized the equivalence with solutions to the Monge-Kantorovich equations, a PDE-based formulation of the $L^1$ optimal transport problem. 
Lavenant et al. \cite{lavenant2018dynamical} proposed a method for interpolating between probability distributions on discrete surfaces using optimal transport theory, yielding a structure-preserving Riemannian metric on the finite-dimensional space of probability distributions over discrete surfaces. 
Dong and his collaborator \cite{dong2024gradient} introduced a gradient-enhanced alternating direction method of multipliers (ADMM) algorithm for optimal transport on general surfaces, integrating gradient recovery techniques to enhance computational accuracy. 
Yu's team \cite{yu2023computational} formulated the mean-field game Nash equilibrium on manifolds and established the equivalence between the corresponding PDE system and the optimality conditions of the associated variational formulation. 
Unfortunately, all these methods depend on grids constructed from point clouds. 
It is well known that constructing a mesh with high quality is a challenging problem. It would be very useful to develop numerical method to solve UOT problem directly on point cloud without requirement of the mesh.  

We build upon Benamou and Brenier's dynamical formulation to model the UOT problem on surfaces. 
Based on ADMM, original UOT problem is transferred to an iteration consists of three subproblems.
The second subproblem, which becomes the focal point, involves solving the space-time Poisson equation which is highly non-trival on point cloud. Classical method, such as surface finite element method, needs regular mesh which is expensive to generate for unstructured point cloud. 
To circumvent the difficulty associated with mesh generation, we employ the radial basis function (RBF) method from meshless techniques. 
Furthermore, the tangent plane method is integrated to obviate the need for computing the gradient of normal vectors. 
To further expedite the solution of algebraic equations, we incorporate a fast computational algorithm by taking advantage of the regular structure in time direction. 
The proposed method is validated across various point cloud scenarios, yielding promising results.

The structure of this paper is organized as follows: Section \ref{Sec-Pre} provides a concise overview of the surface dynamic UOT problem. 
In Section \ref{Sec-Alg}, we present a detailed description of the proposed algorithm along with the solution strategies for the subproblems. 
Section \ref{Sec-RBF} introduces the algorithm for solving the elliptic equations on point clouds. 
In Section \ref{Sec-Numerical}, we showcase various numerical results. 
Finally, the conclusions of this study are summarized in Section \ref{Sec-Con}.

\section{Formulation of Unbanlanced Optimal Transport on Surfaces}\label{Sec-Pre}
In this section, we provide a brief introduction to the surface dynamic unbalanced optimal transport (SUOT) problem. 
First, we give some related notations. 
Let $\Gamma$ denote a closed and oriented surface within the spatial domain $\Omega \subset \mathbb{R}^3$, with an empty boundary. 
$\mathcal{I}$ denotes the identity operator. 
The surface gradient operator $\nabla_{\Gamma}$ for a function $\xi$ defined on $\Gamma$ is given by
\begin{equation*}\label{xi-extension}
    \begin{aligned}
        \nabla_{\Gamma}\xi(\boldsymbol{x}) = \nabla\bar{\xi}(\boldsymbol{x}) - \nabla\bar{\xi}(\boldsymbol{x})\cdot\boldsymbol{n}(\boldsymbol{x})\boldsymbol{n}(\boldsymbol{x}),
    \end{aligned}
\end{equation*}
where $\bar{\xi}$ is a smooth extension of $\xi$ in $\mathcal{N}$, and the tangential divergence operator $\text{div}_{\Gamma}$ is defined as $Trace(\nabla_{\Gamma})$.

We consider a model defined over the time interval $[0, T]$ on the surface $\Gamma \subset \Omega$. 
Let $\rho: [0, T] \times \Gamma \rightarrow \mathbb{R}^{+}$ denote the density of agents as a function of time $t \in [0, T]$, where $\rho$ is constrained to the positive real numbers. 
Additionally, let $\boldsymbol{v} = (v_{1}, \cdots, v_{d}): [0, T] \times \Gamma \rightarrow \mathbb{R}^{d}$ represent the velocity field associated with the density, indicating the movement of the mass. 
Here, $g: [0, T] \times \Gamma \rightarrow \mathbb{R}$ is a scalar field characterizing the local growth or depletion of mass within the system. 
Our main objective is to analyze the behavior of $(\rho, \boldsymbol{v}, g)$ given specified initial ($\rho_{0}$) and final ($\rho_{T}$) densities. 
The dynamic SUOT problem can then be formulated as
\begin{equation}\label{dySUOT}
    \begin{aligned}
        \mathcal{W}_{WFR} = \min _{(\rho, \boldsymbol{v}, g) \in \mathcal{C}_{1}\left(\rho_{0}, \rho_{T}\right)} 
        \int_{0}^{T} \int_{\Gamma} 
         \frac{1}{2} \rho(t, \boldsymbol{x}) \|\boldsymbol{v}(t, \boldsymbol{x})\|^{2}
        + \frac{1}{\eta} \rho(t, \boldsymbol{x}) g(t, \boldsymbol{x})^2
        \mathrm{d} \sigma \mathrm{d} t,
    \end{aligned}
\end{equation}
where
\begin{equation}\label{dySUOTC}
    \begin{aligned}
        \mathcal{C}_{1}(\rho_0,\rho_{T}): =
        \{(\rho, \boldsymbol{v}): & \partial_t \rho +
        \text{div}_{\Gamma}(\rho\boldsymbol{v})  = \rho g, \\
        & \rho(0, \boldsymbol{x}) = \rho_0(\boldsymbol{x}),\ \rho(T, \boldsymbol{x}) = \rho_{T}(\boldsymbol{x}),\ \rho(\boldsymbol{x})\geq0\}, 
    \end{aligned}
\end{equation}
and $\eta$ is a source coefficient that balances transport with the creation and destruction of mass.

We introduce $(\boldsymbol{m}, f) = (\rho\boldsymbol{v}, \rho g)$ and introduce the auxiliary variables $(\bar{\rho}, \bar{\boldsymbol{m}}, \bar{f})$. Then the original formulation changes to the following form.
\begin{equation}\label{dySUOT-new}
    \begin{aligned}
        \mathcal{W}_{WFR} = \min _{(\rho, \boldsymbol{m}, f) \in \mathcal{C}_{2}\left(\rho_{0}, \rho_{T}\right)} 
        \int_{0}^{T} \int_{\Gamma} 
        \mathcal{M}(\rho, \boldsymbol{m}) + \mathcal{F}(\rho, f)
        \mathrm{d} \sigma \mathrm{d} t,
    \end{aligned}
\end{equation}
where
\begin{equation*}
\small
\begin{aligned}
    \mathcal{M}(\rho, \boldsymbol{m}) =& \left\{\begin{array}{cl}
        \frac{\|\boldsymbol{m}\|^{2}}{2\rho}   & if\ \rho>0, \\
        0 & if\ (\rho, \boldsymbol{m}) = (0, 0), \\
        +\infty & otherwise.
    \end{array} \right. \  
    \mathcal{F}(\rho, f) = \left\{\begin{array}{cl}
        \frac{f^2}{\eta\rho}   & if\ \rho>0, \\
        0 & if\ (\rho, f) = (0, 0), \\
        +\infty & otherwise.
    \end{array} \right.  \\
\end{aligned}
\end{equation*}
and 
\begin{equation}\label{dySUOTC-new}
\begin{aligned}
    \mathcal{C}_{2}(\rho_0,\rho_{T}): =&
    \{(\bar{\rho}, \bar{\boldsymbol{m}}): \partial_t \bar{\rho} +
    \text{div}_{\Gamma}\bar{\boldsymbol{m}}  = \bar{f}, 
    \ \bar{\rho}(0, \boldsymbol{x}) = \rho_0(\boldsymbol{x}),\ \\
    & \bar{\rho}(T, \boldsymbol{x}) = \rho_{T}(\boldsymbol{x}),\ \bar{\rho}(t, \boldsymbol{x}) = \rho(t, \boldsymbol{x}),\ \bar{\boldsymbol{m}}(t, \boldsymbol{x}) = \boldsymbol{m}(t, \boldsymbol{x}),\ \\
    & \bar{f}(t, \boldsymbol{x}) = f(t, \boldsymbol{x}),\ \rho(\boldsymbol{x})\geq0 \}.
\end{aligned}
\end{equation}
It is noteworthy that constrained optimization problem \eqref{dySUOT-new}, \eqref{dySUOTC-new} becomes convex. Then, we can leverage convex optimization algorithms to solve it.

\section{Alternating directional multiplier method for Surface UOT}\label{Sec-Alg}
In this section, we design an algorithm to solve the SUOT problem \eqref{dySUOT-new} and \eqref{dySUOTC-new} using the ADMM framework \cite{glowinski1975approximation}.

First, we construct the augmented Lagrangian function for problem \eqref{dySUOT-new}:
\begin{equation*}\label{ALF}
    \begin{aligned}
        \mathcal{L}_{\alpha}(&\rho, \boldsymbol{m}, f, \bar{\rho}, \bar{\boldsymbol{m}}, \bar{f}, p, \boldsymbol{q}, r) := \int_{0}^{T} \int_{\Gamma} \left[\mathcal{M}(\rho, \boldsymbol{m}) + \mathcal{F}(\rho, f)\right] \mathrm{d} \sigma \mathrm{d} t \\
        &+ \int_{0}^{T} \int_{\Gamma}\frac{\alpha}{2}
        \left[ \left| \rho-\bar{\rho}+p \right|^2 +
        \left\| \boldsymbol{m}-\bar{\boldsymbol{m}}+\boldsymbol{q} \right\|^2 + \left| f-\bar{f}+r \right|^2 \right]\mathrm{d} \sigma \mathrm{d} t,
    \end{aligned}
\end{equation*}
where $\alpha$ is Lagrange multiplier.

Next, leveraging the ADMM framework, we decompose the problem into the following three subproblems.
\begin{itemize}
    \item {\bf Step 1}: Find $(\rho, \boldsymbol{m}, f)$ and satisfy the minimization problem and constraint as follows:
    \begin{equation}\label{ADMM1}
        \begin{aligned}
            & \min\limits_{\rho, \boldsymbol{m}, f} \mathcal{L}_{\alpha}(\rho, \boldsymbol{m}, f, \bar{\rho}^s, \bar{\boldsymbol{m}}^s, \bar{f}^s, p^s, \boldsymbol{q}^s, r^s) \\
            s.t. & \ \rho(t, \boldsymbol{x})\geq0.
        \end{aligned}
    \end{equation}
    \item {\bf Step 2}: Find $(\bar{\rho}, \bar{\boldsymbol{m}}, \bar{f})$ and satisfy the following minimization problem, continuity equation and constraint conditions:
    \begin{equation}\label{ADMM2}
        \begin{aligned}
            & \min\limits_{\bar{\rho}, \bar{\boldsymbol{m}}, \bar{f}} \mathcal{L}_{\alpha}(\rho^s, \boldsymbol{m}^s, f^s, \bar{\rho}, \bar{\boldsymbol{m}}, \bar{f}, p^s, \boldsymbol{q}^s, r^s) \\
            s.t.& \ \partial_t \bar{\rho} +
            \text{div}_{\Gamma}\bar{\boldsymbol{m}}  = \bar{f}, \\
            & \ \bar{\rho}(0, \boldsymbol{x}) = \rho_0(\boldsymbol{x}), \ \bar{\rho}(T, \boldsymbol{x}) = \rho_{T}(\boldsymbol{x}).
        \end{aligned}
    \end{equation}
    \item {\bf Step 3}: Update parameters:
    \begin{equation}\label{ADMM3}
        \begin{aligned}
            p^{s+1} =& p^s + \rho^s - \bar{\rho}^s, \\
            \boldsymbol{q}^{s+1} =& \boldsymbol{q}^s + \boldsymbol{m}^s - \bar{\boldsymbol{m}}^s, \\
            r^{s+1} =& r^s + f^s - \bar{f}^s.
        \end{aligned}
    \end{equation}
\end{itemize}

Next, we will provide detailed algorithm for each subproblem.
\begin{itemize}
    \item {\bf Strategy for Step 1}: 
\end{itemize}

Solution of \eqref{ADMM1} in step 1 can be given by solving a quintic equation:
\begin{equation}\label{Quintic-eq}
    \begin{aligned}
        \alpha(\frac{2}{\eta\alpha}+\rho)^2& (\frac{1}{\alpha}+\rho)^2\left( \rho-\bar{\rho}^s+p^s  \right) \\
        &- \frac{\left\|\bar{\boldsymbol{m}}^s-\boldsymbol{q}^s\right\|^2}{2}(\frac{2}{\eta\alpha}+\rho)^2 - \frac{(\bar{f}^s-r^s)^2}{\eta}(\frac{1}{\alpha}+\rho)^2 = 0, \\
    \end{aligned}
\end{equation}
Then the solution of \eqref{ADMM1} can be obtained by
\begin{equation*}\label{sub-ADMM1-solution}
    \begin{aligned}
        \rho^{s+1} =& \max\left\{\rho, 0\right\},  \\
        \boldsymbol{m}^{s+1} =& \frac{\alpha\rho}{1+\alpha\rho}(\bar{\boldsymbol{m}}^s-\boldsymbol{q}^s), \\
        f^{s+1} =& \frac{\eta\alpha\rho}{2+\eta\alpha\rho}(\bar{f}^s-r^s).
    \end{aligned}
\end{equation*}

\begin{itemize}
    \item {\bf Strategy for Step 2}: 
\end{itemize}
To solve problem \eqref{ADMM2}, we use the Lagrange multiplier method. 
\begin{equation*}\label{sub-ADMM2-L}
\begin{aligned}
    \mathcal{L}_{2}(\bar{\rho}, \bar{\boldsymbol{m}}, \bar{f}, \lambda, \lambda_{0}, \lambda_{T}) &:= \int_{0}^{T} \int_{\Gamma}\frac{1}{2}
    \left[\left| \rho^s-\bar{\rho}+p^s \right|^2 +
    \left\| \boldsymbol{m}^s-\bar{\boldsymbol{m}}+\boldsymbol{q}^s \right\|^2 + \left| f^s-\bar{f}+r^s \right|^2\right]\mathrm{d} \sigma \mathrm{d} t \\
    & + \int_{0}^{T} \int_{\Gamma} \lambda(t, \boldsymbol{x})\left( \partial_t \bar{\rho} +
    \text{div}_{\Gamma}\bar{\boldsymbol{m}}  - \bar{f} \right)\mathrm{d} \sigma \mathrm{d} t \\
    & + \int_{\Gamma} \lambda_{0}(\boldsymbol{x}) \left( \bar{\rho}(0, \boldsymbol{x}) - \rho_0(\boldsymbol{x}) \right)\mathrm{d} \sigma + \int_{\Gamma} \lambda_{T}(\boldsymbol{x}) \left( \bar{\rho}(T, \boldsymbol{x}) - \rho_{T}(\boldsymbol{x}) \right)\mathrm{d} \sigma\\
    &= \int_{0}^{T} \int_{\Gamma}\frac{1}{2}
    \left[\left| \rho^s-\bar{\rho}+p^s \right|^2 +
    \left\| \boldsymbol{m}^s-\bar{\boldsymbol{m}}+\boldsymbol{q}^s \right\|^2 + \left| f^s-\bar{f}+r^s \right|^2\right]\mathrm{d} \sigma \mathrm{d} t \\
    & - \int_{0}^{T} \int_{\Gamma} 
    \partial_{t}\lambda(t, \boldsymbol{x})\bar{\rho} + (\nabla_{\Gamma}\lambda(t, \boldsymbol{x}))\cdot\bar{\boldsymbol{m}} + \lambda(t, \boldsymbol{x})\bar{f} \mathrm{d} \sigma \mathrm{d} t \\
    & + \int_{\Gamma} \lambda(T, \boldsymbol{x})\bar{\rho}(T, \boldsymbol{x}) - \lambda(0, \boldsymbol{x})\bar{\rho}(0, \boldsymbol{x})  \mathrm{d} \sigma \\
    & + \int_{\Gamma} \lambda_{0}(\boldsymbol{x}) \left( \bar{\rho}(0, \boldsymbol{x}) - \rho_0(\boldsymbol{x}) \right)\mathrm{d} \sigma + \int_{\Gamma} \lambda_{T}(\boldsymbol{x}) \left( \bar{\rho}(T, \boldsymbol{x}) - \rho_{T}(\boldsymbol{x}) \right)\mathrm{d} \sigma\\
\end{aligned}
\end{equation*}
where $\lambda, \lambda_{0}, \lambda_{T}$ are the Lagrange multiplier. 

Then, we can get the optimality condition by taking first order variations. 
\begin{equation*}\label{sub-ADMM2-Saddle}
    \begin{aligned}
        & \frac{\delta \mathcal{L}_{2}(\bar{\rho}, \bar{\boldsymbol{m}},\bar{f}, \lambda, \lambda_{0}, \lambda_{T})}{\delta \bar{\rho}} = 
        -\left( \rho^s-\bar{\rho}+p^s \right) - \partial_{t}\lambda(t, \boldsymbol{x}) = 0, \\ 
        & \frac{\delta \mathcal{L}_{2}(\bar{\rho}, \bar{\boldsymbol{m}},\bar{f}, \lambda, \lambda_{0}, \lambda_{T})}{\delta \bar{\boldsymbol{m}}} = 
        -\left( \boldsymbol{m}^s-\bar{\boldsymbol{m}}+\boldsymbol{q}^s \right) - \nabla_{\Gamma}\lambda(t, \boldsymbol{x}) = 0, \\
        & \frac{\delta \mathcal{L}_{2}(\bar{\rho}, \bar{\boldsymbol{m}},\bar{f}, \lambda, \lambda_{0}, \lambda_{T})}{\delta \bar{f}} = 
        -\left( f^s-\bar{f}+r^s \right) -\lambda(t, \boldsymbol{x}) = 0,
    \end{aligned}
\end{equation*}
and
\begin{equation*}\label{sub-ADMM2-Saddle-lm}
    \begin{aligned}
        & \frac{\delta \mathcal{L}_{2}(\bar{\rho}, \bar{\boldsymbol{m}},\bar{f}, \lambda, \lambda_{0}, \lambda_{T})}{\delta \lambda} = \partial_t \bar{\rho} +
        \text{div}_{\Gamma}\bar{\boldsymbol{m}}  - \bar{f} = 0, \\ 
        & \frac{\partial \mathcal{L}_{2}(\bar{\rho}, \bar{\boldsymbol{m}},\bar{f}, \lambda, \lambda_{0}, \lambda_{T})}{\partial \lambda_{0}} = \bar{\rho}(0, \boldsymbol{x}) - \rho_0(\boldsymbol{x}) = 0, \\
        & \frac{\partial \mathcal{L}_{2}(\bar{\rho}, \bar{\boldsymbol{m}},\bar{f}, \lambda, \lambda_{0}, \lambda_{T})}{\partial \lambda_{T}} = \bar{\rho}(T, \boldsymbol{x}) - \rho_{T}(\boldsymbol{x}) = 0.
    \end{aligned}
\end{equation*}
From the first three equations, we can get the expression of $(\bar{\rho}, \bar{\boldsymbol{m}}, \bar{f})$ in terms of $\lambda$:
\begin{equation}\label{sub-ADMM2-solution}
    \begin{aligned}
    \bar{\rho} =& (\rho^s+p^s) + \partial_{t}\lambda(t, \boldsymbol{x}), \\
    \bar{\boldsymbol{m}} =& (\boldsymbol{m}^s+\boldsymbol{q}^s) + \nabla_{\Gamma}\lambda(t, \boldsymbol{x}), \\
    \bar{f} =& (f^s+r^s) + \lambda(t, \boldsymbol{x}).
\end{aligned}
\end{equation}

By substituting the \eqref{sub-ADMM2-solution} to the last three equations of the optimality condition, we can get that $\lambda$ satisfies an elliptic equations in $(0,T)\times \Gamma$:
\begin{equation}\label{eq-lambda}
\begin{aligned}
    \partial_{tt}\lambda + \triangle_{\Gamma}\lambda - \lambda = -\partial_{t}(\rho^s+p^s) - \text{div}_{\Gamma}(\boldsymbol{m}^s+\boldsymbol{q}^s) + (f^s+r^s),
\end{aligned}
\end{equation}
with Neumann boundary conditions in time direction:
\begin{equation}\label{eq-lambda-bc}
\begin{aligned}
    \partial_{t}\lambda(0, \boldsymbol{x}) =& \rho_0(\boldsymbol{x}) - (\rho^s(0, \boldsymbol{x})+p^s(0, \boldsymbol{x})), \\
    \partial_{t}\lambda(T, \boldsymbol{x}) =& \rho_{T}(\boldsymbol{x}) - (\rho^s(T, \boldsymbol{x})+p^s(T, \boldsymbol{x})).
\end{aligned}
\end{equation}

Efficiently solving \eqref{eq-lambda} with \eqref{eq-lambda-bc} on the point cloud is the key ingredient to this work. 
This is a Poisson type equation with Neumann boundary condition. In Euclidean space, many numerical methods can solve it efficiently, such as finite difference method, finite element method, spectral method etc. However, on point cloud, it is not easy to develop an efficient numerical solver due to the lack of mesh structure.

\section{Tangent radial basis function method}\label{Sec-RBF}
In this section, we provide a detailed introduction to the tangent radial basis function (TRBF) that circumvents the need for computing the gradient of normal vectors. 
Finally, we present a fast algorithm for solving the resulting algebraic equations.

\subsection{Radial basis function method}\label{Subsec-RBF}
First, we review some key aspects of the RBF method \cite{shankar2015radial}. Suppose $L$ is a surface operator, and $\left.(L u)\right|_{\boldsymbol{x}_c}$ can be expressed as a linear combination of function values $u_i= u(\boldsymbol{x}_i)$ at $n$ neighboring node locations $\left(\boldsymbol{x}_i\right)_{i=1}^n$ to approximate $L u$ at a specific location $\boldsymbol{x}_c$. This can be represented as
\begin{equation}\label{RBF-OP}
    \begin{aligned}
        \left.(L u)\right|_{\boldsymbol{x}=\boldsymbol{x}_c} \approx \sum\limits_{i=1}^n a_i u_i = \sum\limits_{i=1}^n a_i u(\boldsymbol{x}_i).
    \end{aligned}
\end{equation}
To determine the weights $a_i$, we utilize RBF $u_{rbf}(\boldsymbol{x})$ and polynomials to approximate the $u(\boldsymbol{x})$:
\begin{equation}\label{RBF-IF}
    \begin{aligned}
    u_{rbf}\left(\boldsymbol{x}\right)=\sum_{k=1}^n \lambda_k \phi(\left\|\boldsymbol{x}-\boldsymbol{x}_k\right\|_2) +\sum_{l=1}^{m_p} \xi_l p_l(\boldsymbol{x}).
    \end{aligned}
\end{equation}
The coefficients are determined by fitting $u$ on $\boldsymbol{x}_i$, 
\begin{equation*}\label{RBF-IFD}
    \begin{aligned}
    u_i=\sum_{k=1}^n \lambda_k \phi(\left\|\boldsymbol{x}_i-\boldsymbol{x}_k\right\|_2) +\sum_{l=1}^{m_p} \xi_l p_l(\boldsymbol{x}_i),\quad i=1,2,\cdots, n.
    \end{aligned}
\end{equation*}
Notice that above linear system is under-determined. To find a unique solution, we add following constraints
\begin{equation}\label{RBF-IFC}
    \begin{aligned}
    \sum_{k=1}^n \lambda_k p_j\left(\boldsymbol{x}_k\right)=0, \quad j=1,2,3, \cdots, m_p,
    \end{aligned}
\end{equation}
where $m_p$ is the dimension of the polynomial space up to degree $l$, and $\{p_l\}_{l=1}^{m_p}$ is a basis for this space. 
Substituting \eqref{RBF-IF} into \eqref{RBF-OP} and adding \eqref{RBF-IFC}, we can obtain
\begin{equation}\label{RBF-temp}
    \begin{aligned}
        \sum\limits_{i=1}^n a_i \left( \sum_{k=1}^n \lambda_k \phi\right. & \left.(\left\|\boldsymbol{x}_i-\boldsymbol{x}_k\right\|_2) + \sum_{l=1}^{m_p} \xi_l p_l(\boldsymbol{x}_k) \right) + \sum_{j=1}^{m_p} \eta_j \sum_{k=1}^{n}\lambda_k p_j(\boldsymbol{x}_i)\\
        & = L \left( \sum_{k=1}^n \lambda_k \phi(\left\|\boldsymbol{x}_c-\boldsymbol{x}_k\right\|_2) +\sum_{l=1}^{m_p} \xi_l p_l(\boldsymbol{x}_c) \right),
    \end{aligned}
\end{equation}
where $\eta$ are each approximation polynomial coefficient. 

Since \eqref{RBF-temp} holds for any $\lambda_k$ and $\xi_l$, we can get a unnecessary and sufficient condition:
\begin{equation*}\label{RBF-eq}
    \begin{aligned} 
        \sum_{i=1}^n a_i \phi(\left\|\boldsymbol{x}_i-\boldsymbol{x}_k\right\|_2) + \sum_{j=1}^{m_p} \eta_j p_j(\boldsymbol{x}_k) &= L\phi(\left\|\boldsymbol{x}_c-\boldsymbol{x}_k\right\|_2), \\ 
        \sum_{i=1}^{n} a_i p_l(\boldsymbol{x}_k) &= L p_l(\boldsymbol{x}_c).
    \end{aligned}
\end{equation*}
Further, we can get the following matrix form: 
\begin{equation}\label{RBF-algebraic}
    \begin{aligned}
    \left(\begin{array}{cc}
    A & P \\
    P^T & 0
    \end{array}\right)\binom{a}{\eta}=\binom{\left.L \Phi(\boldsymbol{x})\right|_{\boldsymbol{x}=\boldsymbol{x}_c}}{\left.L \mathbf{p}(\boldsymbol{x})\right|_{\boldsymbol{x}=\boldsymbol{x}_c}},
    \end{aligned}
\end{equation}
where $A = \{\phi_{i j}=\phi\left(\left\|\boldsymbol{x}_i-\boldsymbol{x}_j\right\|_2\right)\}$, 
\begin{equation*}\label{sub-matrix}
\begin{aligned}
P = \left[\begin{array}{cccc}
p_{1}(\boldsymbol{x}_{1}) & p_{2}(\boldsymbol{x}_{1}) & \cdots & p_{m_p}(\boldsymbol{x}_{1}) \\
p_{1}(\boldsymbol{x}_{2}) & p_{2}(\boldsymbol{x}_{2}) & \cdots & p_{m_p}(\boldsymbol{x}_{2}) \\
\vdots & \vdots &  & \vdots \\
p_{1}(\boldsymbol{x}_{n}) & p_{2}(\boldsymbol{x}_{n}) & \cdots & p_{m_p}(\boldsymbol{x}_{n}) \\
\end{array}\right],
\end{aligned}
\end{equation*}
$\Phi(\boldsymbol{x})=\left(\phi\left(\left\|\boldsymbol{x}-\boldsymbol{x}_1\right\|_2\right), \phi\left(\left\|\boldsymbol{x}-\boldsymbol{x}_2\right\|_2\right), \cdots, \phi\left(\left\|\boldsymbol{x}-\boldsymbol{x}_n\right\|_2\right)\right)^T$ and  $\mathbf{p}(\boldsymbol{x})=\left(p_1( \boldsymbol{x})\right.$, $\left.p_2( \boldsymbol{x}), \cdots, p_{m_p}( \boldsymbol{x})\right)^T$. 
Many radial basis functions are available in the literature \cite{fornberg2008choosing}. 
In this study, we employ the infinitely smooth Gaussian radial basis function defined as $\phi(r) = e^{-(\varepsilon r)^2}$. 

\subsection{TRBF for elliptic equation}\label{Subsec-RBF-EE}
First, let time steps $t_i=(i-1)\Delta t, i=1, 2, \cdots, N_t$, $\Delta t = \frac{T}{N_t-1}$. 
Next, we first combine the finite difference method to give the time discrete scheme of equation \eqref{eq-lambda}. 
To incorporate the Neumann boundary condition \eqref{eq-lambda-bc}, we adopt the ghost point method (GPM) \cite{leveque2007finite}, and define the discrete difference operator $\partial_{tt}$ and $\partial_{t}$ (without GPM) as
\begin{equation*}\label{time-diff-op}
    \begin{aligned}
        &\partial_{t t} \lambda_i(\boldsymbol{x}) = 
        \begin{cases}
        \frac{-2 \lambda_i(\boldsymbol{x})+2 \lambda_{i+1}(\boldsymbol{x})}{\Delta t^2} & i=0, \\ 
        \frac{\lambda_{i-1}(\boldsymbol{x})-2 \lambda_i(\boldsymbol{x})+\lambda_{i+1}(\boldsymbol{x})}{\Delta t^2} & 1 \leq i \leq N_t-1, \\ 
        \frac{-2 \lambda_i(\boldsymbol{x})+2 \lambda_{i-1}(\boldsymbol{x})}{\Delta t^2} & i=N_t.
        \end{cases} \\
        &\partial_{t} u_i(\boldsymbol{x}) = 
        \begin{cases}
        \frac{-3u_i(\boldsymbol{x}) + 4u_{i+1}(\boldsymbol{x}) - u_{i+2}(\boldsymbol{x})}{2\Delta t} & i=0, \\ 
        \frac{u_{i+1}(\boldsymbol{x})-u_{i-1}(\boldsymbol{x})}{2\Delta t} & 1 \leq i \leq N_t-1, \\ 
        \frac{3u_i(\boldsymbol{x}) - 4u_{i-1}(\boldsymbol{x}) + u_{i-2}(\boldsymbol{x})}{2\Delta t} & i=N_t.
        \end{cases}
    \end{aligned}
\end{equation*}
Additionally, we define the auxiliary right hand side function as
\begin{equation*}\label{time-gp}
    \begin{aligned}
        \hat{\lambda}_i(\boldsymbol{x})= 
        \begin{cases}
        \frac{2 \lambda_0(\boldsymbol{x})}{\Delta t} & i=0, \\ 
        0 & 1 \leq i \leq N_t-1, \\ 
        \frac{2 \lambda_1(\boldsymbol{x})}{\Delta t} & i=N_t.
        \end{cases}
    \end{aligned}
\end{equation*}

Combined with the above notion, we get semi-discretization of equation \eqref{eq-lambda} is
\begin{equation}\label{eq-lambda-semi-d}
    \begin{aligned}
        \partial_{tt}\lambda_{i} + \triangle_{\Gamma}\lambda_{i} - \lambda_{i} = & -\partial_{t}(\rho_{i}+p_{i}) \\
        & - \text{div}_{\Gamma}(\boldsymbol{m}_{i}+\boldsymbol{q}_{i}) + (f_{i}+r_{i}) + \hat{\lambda}_i, \quad \forall\ 0 \leq i \leq N_t.
    \end{aligned}
\end{equation}

Further, we only need the discrete surface $\Gamma$ to be some points $\boldsymbol{x}_j, j=1, 2, \cdots, N_s$ and normal vectors $\boldsymbol{n}(\boldsymbol{x})$, and do not need the grid structure. This effectively controls the cost of computing. 
Then, we use RBF discrete $\text{div}_{\Gamma}$ in \eqref{eq-lambda-semi-d}. 
Using method in Section \ref{Subsec-RBF} with \eqref{RBF-algebraic}, for each center point $\boldsymbol{x}_c$ and neighbor points $\Lambda(\boldsymbol{x}_c)$ construct $3$-dimensional RBF approximation and sovle algebraic equation:
\begin{equation}\label{RBF-div-algebraic}
    \begin{aligned}
    \left(\begin{array}{cc}
    A & P \\
    P^T & 0
    \end{array}\right)\binom{\boldsymbol{b}}{\boldsymbol{\eta}_b}=\binom{\text{div}_{\Gamma} \Tilde{\Phi}(x_c, y_c, z_c)}{\text{div}_{\Gamma} \Tilde{\mathbf{p}}(x_c, y_c, z_c)},
    \end{aligned}
\end{equation}
where $\Tilde{\Phi} = \{\Phi_1, \Phi_2, \Phi_3\}$ and $\Tilde{\mathbf{p}} = \{\mathbf{p}_1, \mathbf{p}_2, \mathbf{p}_3\}$. 
Solving \eqref{RBF-div-algebraic}, we can get coefficient $\boldsymbol{b}$ and approximation formula for divergence:
\begin{equation}\label{operator-d1}
    \begin{aligned}
        \text{div}_{\Gamma}(\boldsymbol{m}_{i, j}+\boldsymbol{q}_{i, j}) = & \sum\limits_{k\in\Lambda(\boldsymbol{x}_j)} \boldsymbol{b}_{i,  k}\cdot(\boldsymbol{m}_{i, k}+\boldsymbol{q}_{i, k}).
    \end{aligned}
\end{equation}
where $\Lambda(\boldsymbol{x}_j)$ contains points near $\boldsymbol{x}_{j}$. 
Then, we have
\begin{equation}\label{eq-lambda-full-d1}
    \begin{aligned}
        \partial_{tt}\lambda_{i, j} + \triangle_{\Gamma}\lambda_{i, j} & - \lambda_{i, j} = -\partial_{t}(\rho_{i, j}+p_{i, j}) \\
        & - \sum\limits_{k\in\Lambda(\boldsymbol{x}_j)}( \boldsymbol{b}_{i, k}\cdot(\boldsymbol{m}_{i, k})+\boldsymbol{q}_{i, k}) + (f_{i, j}+r_{i, j}) + \hat{\lambda}_{i, j}.
    \end{aligned}
\end{equation}
In \eqref{eq-lambda-full-d1}, we need to discretize $\triangle_{\Gamma}\lambda_{i}$ by following:
\begin{equation}\label{lambda-laplace}
    \begin{aligned}
        \triangle_{\Gamma} \lambda_i = \text{div}_{\Gamma}(\nabla_{\Gamma}\lambda_i) = (P\nabla)\cdot(P\nabla\lambda_i),
    \end{aligned}
\end{equation}
where $P = \mathcal{I} - \boldsymbol{n}\boldsymbol{n}^T$. 
As with the divergence calculation \eqref{operator-d1}, build RBF in $\mathbb{R}^3$ for approximate $\triangle_{\Gamma} \lambda_i$. 
However, it is not difficult to find that we also need to calculate the derivative of the normal gradient at the point $\boldsymbol{x}_j$. 
And the calculation of this derivative is generally complicated. 
Therefore, we take another approach TPM \cite{suchde2019meshfree} to solve it. 

\begin{figure}[htbp]
    \begin{center}
    \includegraphics[width=6cm]{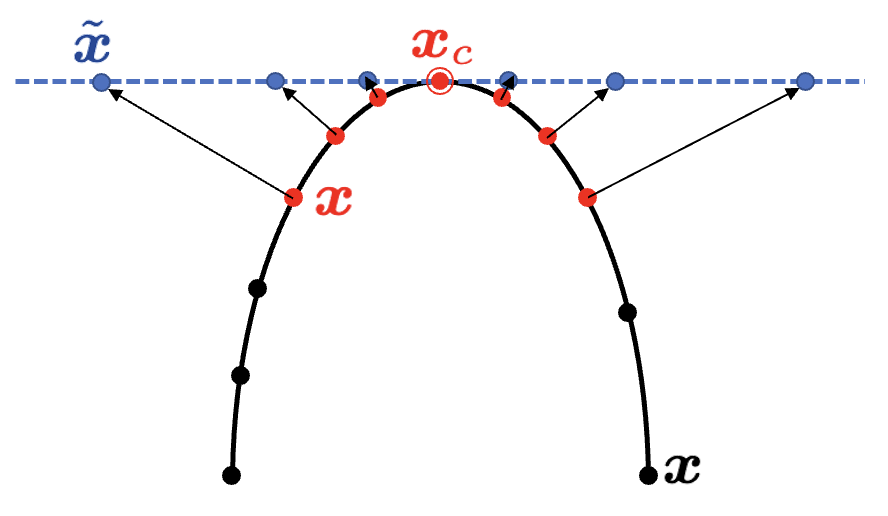}
    \caption{There is an additional circle around the center point $\boldsymbol{x}_c$, and all neighbor points $\boldsymbol{x}\in\Lambda(\boldsymbol{x}_c)$ are marked in red. The remaining black points $\boldsymbol{x}$ on the manifold are shown for reference. The projected position on the tangent space is marked in blue $\Tilde{\boldsymbol{x}}\in\Lambda_P(\boldsymbol{x}_c)$. The manifold is shown in black, while the tangent to the center point is shown as a dashed blue line.}
    \label{pic:pro-normal}
    \end{center}
\end{figure}

Consider a function $\lambda: \Gamma\rightarrow\mathbb{R}$ and a function $\bar{\lambda}: \mathbb{R}^3\rightarrow\mathbb{R}$. $\bar{\lambda}$ is said to be an extension of $\lambda$ if $\bar{\lambda}|_{\Gamma}=\lambda$. Further, $\bar{\lambda}$ is said to be a normal extension of $\lambda$ if $\boldsymbol{n} \cdot \nabla \bar{\lambda}=0$. Thus, in the normal extension of $\lambda$, the surface gradient of a function can be defined as its extended conventional gradient: 
\begin{equation}\label{lambda-grad-extension}
    \begin{aligned}
        \nabla_{\Gamma} \lambda=\nabla \bar{\lambda}-(\boldsymbol{n} \cdot \nabla \bar{\lambda}) \boldsymbol{n}=\nabla \bar{\lambda}.
    \end{aligned}
\end{equation}
Then, finding two orthogonal basis $\boldsymbol{s}_{1}$ and $\boldsymbol{s}_{2}$ of the tangent plane $T_{\boldsymbol{x}_{c}} $ at $\boldsymbol{x}_{c}$, we can construct the local orthonormal coordinate system $\left\{\boldsymbol{s}_{1}, \boldsymbol{s}_{2}, \boldsymbol{n}\right\}$. 
Therefore, 
\begin{equation*}\label{nabla-lambda-bar}
    \begin{aligned}
        \nabla\bar{\lambda} = R^T\nabla_{tn}\bar{\lambda} =R^T 
        \left(\begin{array}{c}
            \boldsymbol{s}_{1}\cdot\nabla\bar{\lambda}  \\
            \boldsymbol{s}_{2}\cdot\nabla\bar{\lambda}  \\
            \boldsymbol{n}\cdot\nabla\bar{\lambda}
        \end{array}
        \right) = R^T
        \left(\begin{array}{c}
            \boldsymbol{s}_{1}\cdot\nabla\bar{\lambda}  \\
            \boldsymbol{s}_{2}\cdot\nabla\bar{\lambda}  \\
            0
        \end{array}
        \right).
    \end{aligned}
\end{equation*}
where rotation matrix $R = (\boldsymbol{s}_{1}, \boldsymbol{s}_{2}, \boldsymbol{n})^T$, and the subscript $tn$ indicates the rotated $\boldsymbol{s}_{1}, \boldsymbol{s}_{2}, \boldsymbol{n}$ coordinate frame at a particular point. 
Next, combining normal direction projection points on this tangent plane, a two-dimensional RBF is constructed to calculate $\Tilde{\nabla}_{T}\bar{\lambda}$. 
Further, consider a vector valued function $\boldsymbol{u} = (u_1, u_2, u_3)$ defined on $\Gamma$. A normal extension of $\bar{\boldsymbol{u}}$ is obtained by a normal extension of each component of $\bar{\boldsymbol{u}}$. i.e. $\bar{\boldsymbol{u}} = (\bar{u}_1, \bar{u}_2, \bar{u}_3)$. Then, the surface divergence of $\boldsymbol{u}$ can be written as
\begin{equation}\label{lambda-div-extension}
    \begin{aligned}
        \text{div}_{\Gamma} \boldsymbol{u} = \text{div} \bar{\boldsymbol{u}} - \boldsymbol{n}\cdot[(\boldsymbol{n} \cdot \nabla) \bar{\boldsymbol{u}}] =\text{div} \bar{\boldsymbol{u}}.
    \end{aligned}
\end{equation}
By combining \eqref{lambda-grad-extension} and \eqref{lambda-div-extension}, we can calculate surface Laplacian
\begin{equation*}\label{lambda-laplace-extension}
    \begin{aligned}
        \triangle_{\Gamma} \lambda &= \text{div}_{\Gamma}\nabla_{\Gamma} \lambda = \triangle\bar{\lambda} = \text{div}\nabla \bar{\lambda} = \text{div}(R^T\nabla_{tn}\bar{\lambda}) \\
        & = (R\nabla)\cdot(RR^T\nabla_{tn}\bar{\lambda}) = \text{div}_{tn}(RR^T\nabla_{tn}\bar{\lambda}) = \triangle_{tn}\bar{\lambda} = \triangle_{T}\bar{\lambda},
    \end{aligned}
\end{equation*}
where $\triangle_{T}$ is the $2$ dimensional Laplacian on the tangent plane, and $RR^T = R^T R = I$. 
In order to compute $\triangle_{T}\bar{\lambda}$, we project each point $\boldsymbol{x}_j$ to the tangential plane $T_{\boldsymbol{x}_c}$ with $\{\boldsymbol{s}_{1}, \boldsymbol{s}_{2}\}$ by the normal direction $\boldsymbol{n}_j$ (A $1$-dimensional manifold for reference, as shown in Figure \ref{pic:pro-normal}). 
$\boldsymbol{x}_{j-T_{\boldsymbol{x}_c}}$ is defined as projection of $\boldsymbol{x}_j$ onto the plane $T_{\boldsymbol{x}_c}$. 
In local coordinates system $\left\{\boldsymbol{s}_{1}, \boldsymbol{s}_{2}, \boldsymbol{n}\right\}$, the point $\Tilde{\boldsymbol{x}}_{j-T_{\boldsymbol{x}_c}}$ is the same point as the point $\boldsymbol{x}_{j-T_{\boldsymbol{x}_c}}$ in $\mathbb{R}^3$.

Using method in Section \ref{Subsec-RBF} with \eqref{RBF-algebraic}, for each center point $\boldsymbol{x}_c$ and projection neighbor points $\Lambda_{P}(\boldsymbol{x}_c)$ in tangent plane construct $2$-dimensional RBF approximation and sovle algebraic equation:
\begin{equation}\label{RBF-laplace-algebraic}
    \begin{aligned}
    \left(\begin{array}{cc}
    A & P \\
    P^T & 0
    \end{array}\right)\binom{a}{\eta_a} = 
    \binom{\triangle_{T} \Tilde{\Phi}(\Tilde{x}_{c}, \Tilde{y}_{c})}
    {\triangle_{T} \Tilde{\mathbf{p}}(\Tilde{x}_{c}, \Tilde{y}_{c})}. 
    \end{aligned}
\end{equation} 
Solving \eqref{RBF-laplace-algebraic}, we can get coefficient $a$ and approximation formula for $\triangle_{T}$($=\triangle_{\Gamma}$) at center node $\boldsymbol{x}_{j}$: 
\begin{equation}\label{operator-d2}
    \begin{aligned}
        \triangle_{\Gamma}\lambda_{i,j} =& \sum\limits_{k\in\Lambda_P(\boldsymbol{x}_j)} a_{i, k} \lambda_{i, k}.
    \end{aligned}
\end{equation}
Since we use the same stencil at each time step, we can simplify $a_{i, k}=a_{k}, \boldsymbol{b}_{i, k}=\boldsymbol{b}_{k}$ in \eqref{operator-d1} and \eqref{operator-d2}. 
Therefore, we can get the following simplification
\begin{equation}\label{operator-d-new}
    \begin{aligned}
    \triangle_{\Gamma}\lambda_{i,j} = & \sum\limits_{k\in\Lambda_P(\boldsymbol{x}_j)} a_{k} \lambda_{i, k}, \\
    \text{div}_{\Gamma}(\boldsymbol{m}_{i, j}+\boldsymbol{q}_{i, j}) = & \sum\limits_{k\in\Lambda(\boldsymbol{x}_j)} \boldsymbol{b}_{k}\cdot(\boldsymbol{m}_{i, k}+\boldsymbol{q}_{i, k}).
    \end{aligned}
\end{equation}

Using \eqref{operator-d-new}, we will get a fully discrete format with $\forall\ 0 \leq i \leq N_{t}$ and $\forall\ 0 \leq j \leq N_{n}$ from \eqref{eq-lambda-full-d1},
\begin{equation}\label{eq-lambda-full-d}
\begin{aligned}
\partial_{tt}\lambda_{i, j} + \sum\limits_{k\in\Lambda_P(\boldsymbol{x}_j)} a_{k} \lambda_{i, k} & - \lambda_{i, j} =  -\partial_{t}(\rho_{i, j}+p_{i, j}) \\
& - \sum\limits_{k\in\Lambda(\boldsymbol{x}_j)}( \boldsymbol{b}_{k}\cdot(\boldsymbol{m}_{i, k})+\boldsymbol{q}_{i, k}) + (f_{i, j}+r_{i, j}) + \hat{\lambda}_{i, j}.
\end{aligned}
\end{equation}
The Step 2 can be completed by solving the algebraic equation \eqref{eq-lambda-full-d}.

\begin{remark} 
    In the RBF kernel function $\varepsilon$ is called the shape parameter, and for spheres $\varepsilon=1$ is sufficient. 
    But for general surfaces, we need to select a specific value. 
    The idea is to choose a shape parameter for each stencil that give a particular target condition number $\kappa_T$ for the RBF interpolation matrix on that stencil. 
    We denote it by $A_{x_c}(\varepsilon)$ since the entries of the matrix depend continuously on the shape parameter. The desired condition number $\kappa_T$ is given as the solution
    \begin{equation}\label{eq:epsilon}
    \begin{aligned}
    F\left(\varepsilon, \kappa_T\right)=\log \left(\kappa\left(A_{x_c}(\varepsilon)\right) / \kappa_T\right)=0,
    \end{aligned}
    \end{equation}
    where $\kappa\left(A_{x_c}\right)$ is the condition number of $A_{x_c}(\varepsilon)$ with respect to $l_2$ norm. 
    Since $A_{x_c}(\varepsilon)$ is symmetric, the condition number is ratio between largest and smallest eigenvalue. 

    Since the condition number of RBF interpolation matrices increase monotonically as the shape parameter decreases to zero \cite{shankar2015radial}, solution of \eqref{eq:epsilon} is unique and can be solved by bisection method efficiently. 
    Since we are using the algorithm on each stencil and each template contains about $7$ nodes, the local matrix is faster to compute. 
    In subsequent numerical experiments, we set the bisection method for 100 iterations or stop when the accuracy reaches 1e-10.
\end{remark}

\subsection{Fast solver for time direction}\label{Subsec-FS}
In this subsection, we present a fast solver for the large linear system \eqref{eq-lambda-full-d}. 
The key idea is to decouple the time discretization using spectral decomposition, akin to the pre-computation approach outlined in \cite{lavenant2018dynamical}. 
The semi-discretization \eqref{eq-lambda-semi-d} can be reformulated in the following vector form:
\begin{equation}\label{eq-lambda-semi-d-f}
    \begin{aligned}
    \boldsymbol{E}\boldsymbol{\lambda}+\mathbf{I} \Delta_{\Gamma}\boldsymbol{\lambda} + \mathbf{I} \boldsymbol{\lambda}=\mathbf{r},
    \end{aligned}
\end{equation}
where $\boldsymbol{E}$ is the matrix of $\partial_{tt}$,  $\boldsymbol{\lambda} = (\lambda_{1}, \ldots, \lambda_{N})^T$, $\boldsymbol{r} = (r_{1}, \ldots, r_{N})^T$ and $r_i= -\partial_{t}(\rho_{i}+p_{i}) - \text{div}_{\Gamma}(\boldsymbol{m}_{i}+\boldsymbol{q}_{i}) + (f_{i}+r_{i}) + \hat{\lambda}_i$. 
Denote the eigen pair of the matrix $\boldsymbol{E}$ by $\left(\gamma_i, \boldsymbol{e}_i\right)$. 
It is not hard to see that $\gamma_i=\frac{-2+2 \cos (i \pi \Delta t)}{\Delta t^2}$ and $\boldsymbol{e}_i=\cos (i \pi \mathbf{t})$ for $i=0, \ldots, N_t$. Notice that $\boldsymbol{e}_i$ forms a basis of $\mathbb{R}^{N}$. 
Then, we can write $\boldsymbol{\lambda}$ and $\boldsymbol{r}$ as
\begin{equation*}\label{basic-lambda}
\begin{aligned}
\boldsymbol{\lambda} = \sum_{i=0}^{N_t} w_i \boldsymbol{e}_i, \quad \boldsymbol{r} = \sum_{i=0}^{N_t} r_i \boldsymbol{e}_i.
\end{aligned}
\end{equation*}
Plugging into the matrix equation \eqref{eq-lambda-semi-d-f} gives
\begin{equation}\label{eq-lambda-eigen}
\begin{aligned}
\sum_{i=0}^{N_t} \gamma_i w_i \boldsymbol{e}_i + \Delta_{\Gamma}\left(\sum_{i=0}^{N_t} w_i \boldsymbol{e}_i\right) + \sum_{i=0}^{N_t} w_i \boldsymbol{e}_i = \sum_{i=0}^{N_t} r_i \boldsymbol{e}_i.
\end{aligned}
\end{equation}
The linear independence of $\left\{\boldsymbol{e}_i\right\}$ suggests that we can decouple the equation \eqref{eq-lambda-eigen} into $N_t+1$ equations
\begin{equation}\label{eq-lambda-eigen-sub}
    \begin{aligned}
    (\gamma_i + 1) w_i+\Delta_{\Gamma} w_i=r_i,
    \end{aligned}
\end{equation}
for $i=0, \ldots, N_t$. Then, the decoupled equations \eqref{eq-lambda-eigen-sub} can be solved using TRBF.

\section{Numerical Experiments}\label{Sec-Numerical}
In this section, we present a series of numerical experiments to demonstrate the efficacy of the proposed algorithm.  
First, we test our method in some toy examples to verify the accuracy and computational time. 
In the second part, we solve the optimal transport and unbalanced optimal transport problem on smooth surfaces which are given by level-set functions. 
In the last part, numerical examples on general point cloud are conducted. 
Further details and the code can be found at https://github.com/Poker-Pan/ADMM-TRBF-SUOT.

\subsection{Numerical examples with exact solutions}\label{Subsec-ES}
First, we solve a classical Poisson equation using the TRBF method to test its accuracy.  
Let the computational domain be $(t, x) \in [0, 1]\times[0, 1]$. 
The equation is 
\begin{equation*}\label{Laplace-eq}
    \begin{aligned}
        u_{tt} + \triangle u =& -8\pi^2\cos(2\pi t)\cos(2\pi x), & & (t, \boldsymbol{x})\in(0,1)\times(0,1),\\
        u_t =& 0, & & (t, \boldsymbol{x})\in\{0,1\}\times[0,1],\\
        u_x =& 0, & & (t, \boldsymbol{x})\in[0,1]\times\{0,1\}.
    \end{aligned}
\end{equation*}
with the exact solution 
$$u_e = \cos(2\pi t)\cos(2\pi x).$$

\begin{table}[htbp]
    \centering
    \caption{The spatiotemporal convergence order of the TRBF method error.}
    \label{tab:L-eq-error-con-time-space}
    \begin{tabular}{c|c|cccc}
        \hline 
        $1/\Delta t$ & $1/h$ & $\|u_{e}-u\|_{l_1}$ & odr & $\|u_{e}-u\|_{l_2}$ & odr \\
        \hline
        8 & 8 & 2.167e-2 & - & 2.846e-2 & - \\
        16 & 16 & 4.042e-3 & 2.20 & 5.577e-3 & 2.13 \\
        32 & 32 & 9.088e-4 & 2.05 & 1.253e-3 & 2.05 \\
        64 & 64 & 2.143e-4 & 2.03 & 2.981e-4 & 2.02 \\
        128 & 128 & 5.209e-5 & 2.01 & 7.276e-5 & 2.01 \\
        256 & 256 & 1.278e-5 & 2.01 & 1.790e-5 & 2.01 \\
        \hline
    \end{tabular}
\end{table}
In Table \ref{tab:L-eq-error-con-time-space}, We can see that the convergence rate is second order as spatial and time step go to zero. 
The similar result was also reported in \cite{adil2022numerical}. 

Regarding the convergence of ADMM method, we compute an example of optimal transport between 1D Gaussian distributions.  
The Gaussian distribution used are given as follows:
\begin{equation*}\label{Gaussian}
    \begin{aligned}
    \rho_{G}(x, \mu, \sigma)=\frac{1}{(2\pi\sigma)^{1/2}}e^{-\frac{1}{2\sigma}(x-\mu)^2}. 
    \end{aligned}
\end{equation*}
Then we choose $\rho_{0}(x)=\rho_{G}(x, 0.45, 0.05)$ as the initial state and $\rho_{1}(x)=\rho_{G}(x, 0.55, 0.05)$ as the termination state. 

\begin{table}[htbp]
    \centering
    \caption{Errors of costs for 1D OT example.}
    \label{tab:admm-rbf-1D-Gau1}
    \begin{tabular}{c|c|c}
        \hline 
        $1/\Delta t$ & $1/h$ & $|W_{WFR-E}-W_{WFR-C}|$ \\
        \hline
        8 & 8 & $1.41\times 10^{-2}$ \\
        16 & 16 & $1.87\times 10^{-3}$ \\
        32 & 32 & $3.81\times 10^{-4}$ \\
        64 & 64 & $1.57\times 10^{-4}$ \\
        128 & 128 & $7.75\times 10^{-5}$ \\
        \hline
    \end{tabular}
\end{table}

As shown in Table \ref{tab:admm-rbf-1D-Gau1}, the error of cost also converges to zero, however the rate is reduced to first order.

\subsection{Unbalanced optimal transport on smooth surfaces}\label{Subsec-OTLS}
In this section, we test our method on smooth surfaces, and the surfaces are given by level set functions. 
\textit{Sphere}, \textit{Ellipsoid}, \textit{Peanut}, \textit{Torus}, and \textit{Opener} are used in the tests. 
The level set functions are given in \ref{App-A}. 
The point cloud is generated by sampling the surfaces through the grid generator \cite{persson2004simple}.
The initial and terminal distributions are given in
Table \ref{tab:SOT-LS} of \ref{App-B}. In the computations, the distributions are normalized on the point cloud. $\beta>0$ is the ratio between the terminal and initial mass. When $\beta=1$, it is classical optimal transport problem. When $\beta\neq 1$, it becomes unbalanced optimal transport problem. Here we set $\beta=1,1.5$ respectively to test the performance of the proposed method for balanced and unbalanced optimal transport problem. 
\begin{figure}[htbp]
    \begin{center}
    \includegraphics[width=2.4cm]{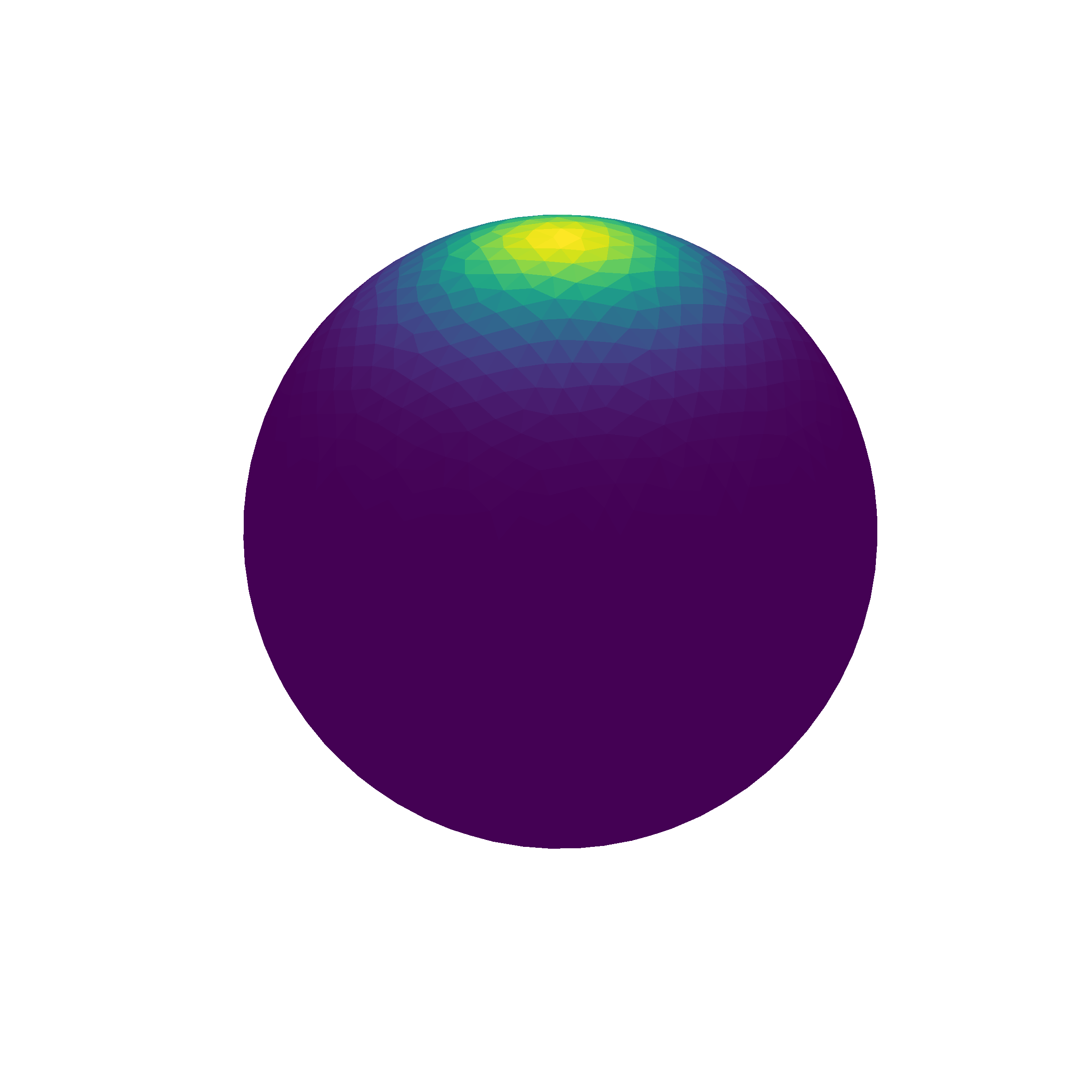}
    \includegraphics[width=2.4cm]{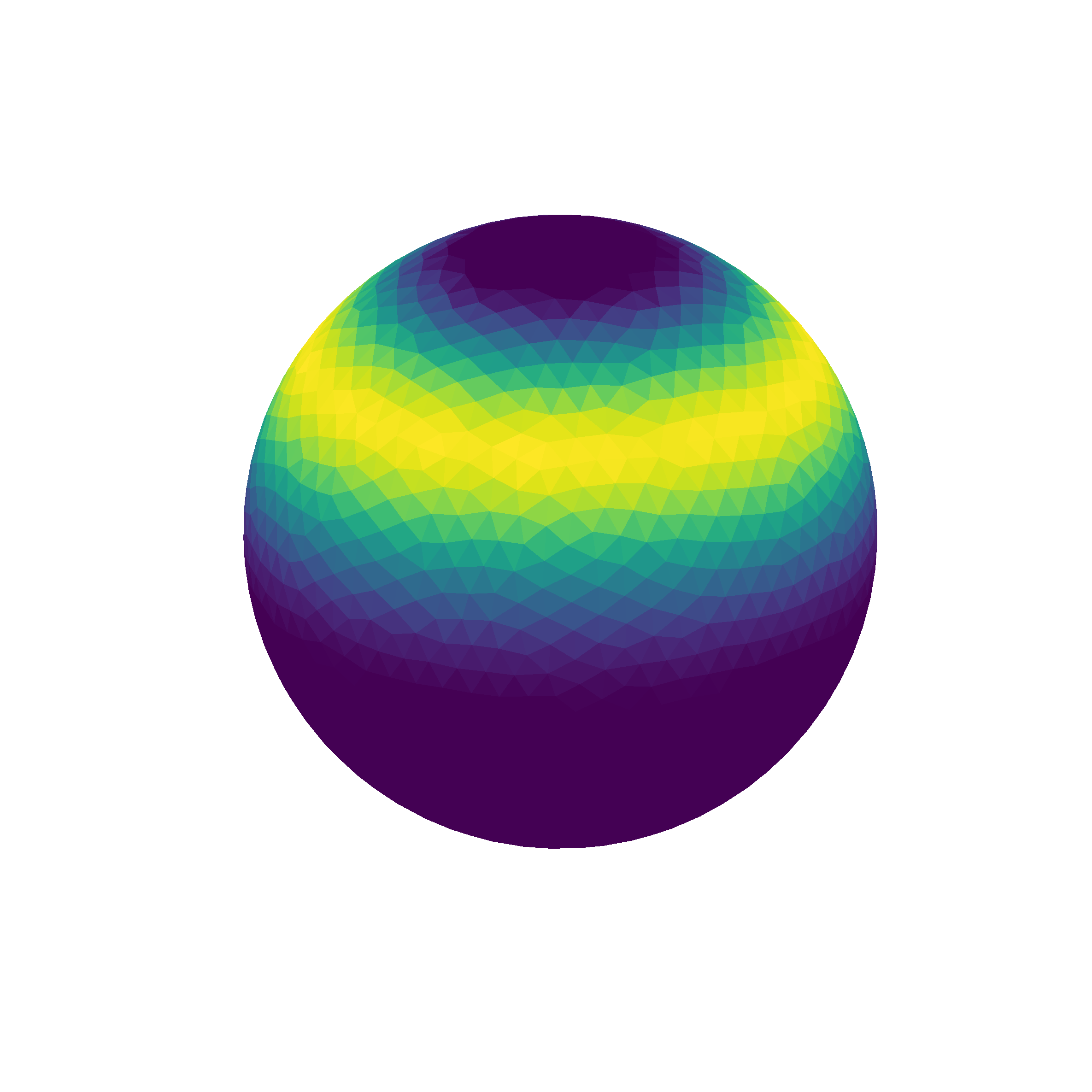}
    \includegraphics[width=2.4cm]{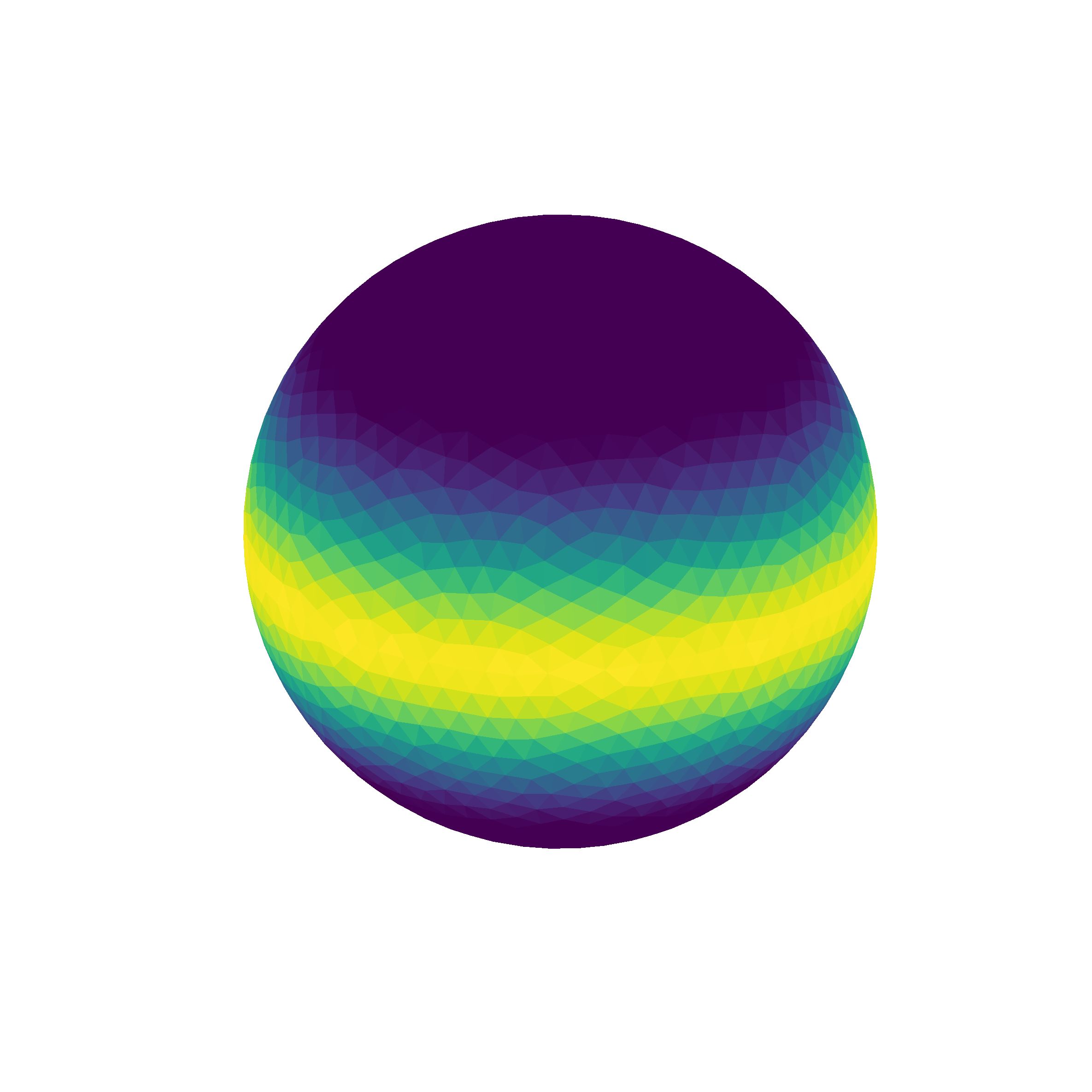}
    \includegraphics[width=2.4cm]{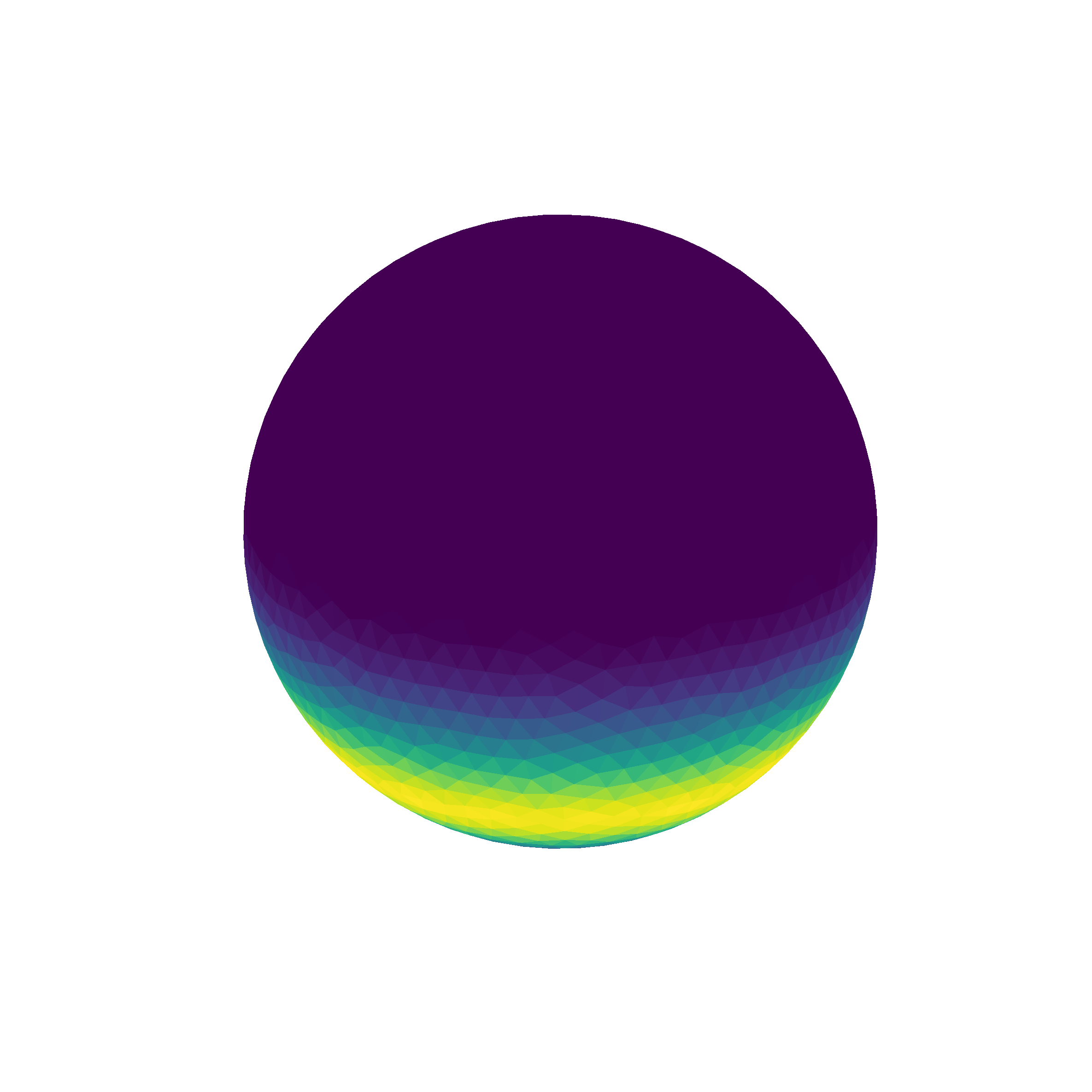}
    \includegraphics[width=2.9cm]{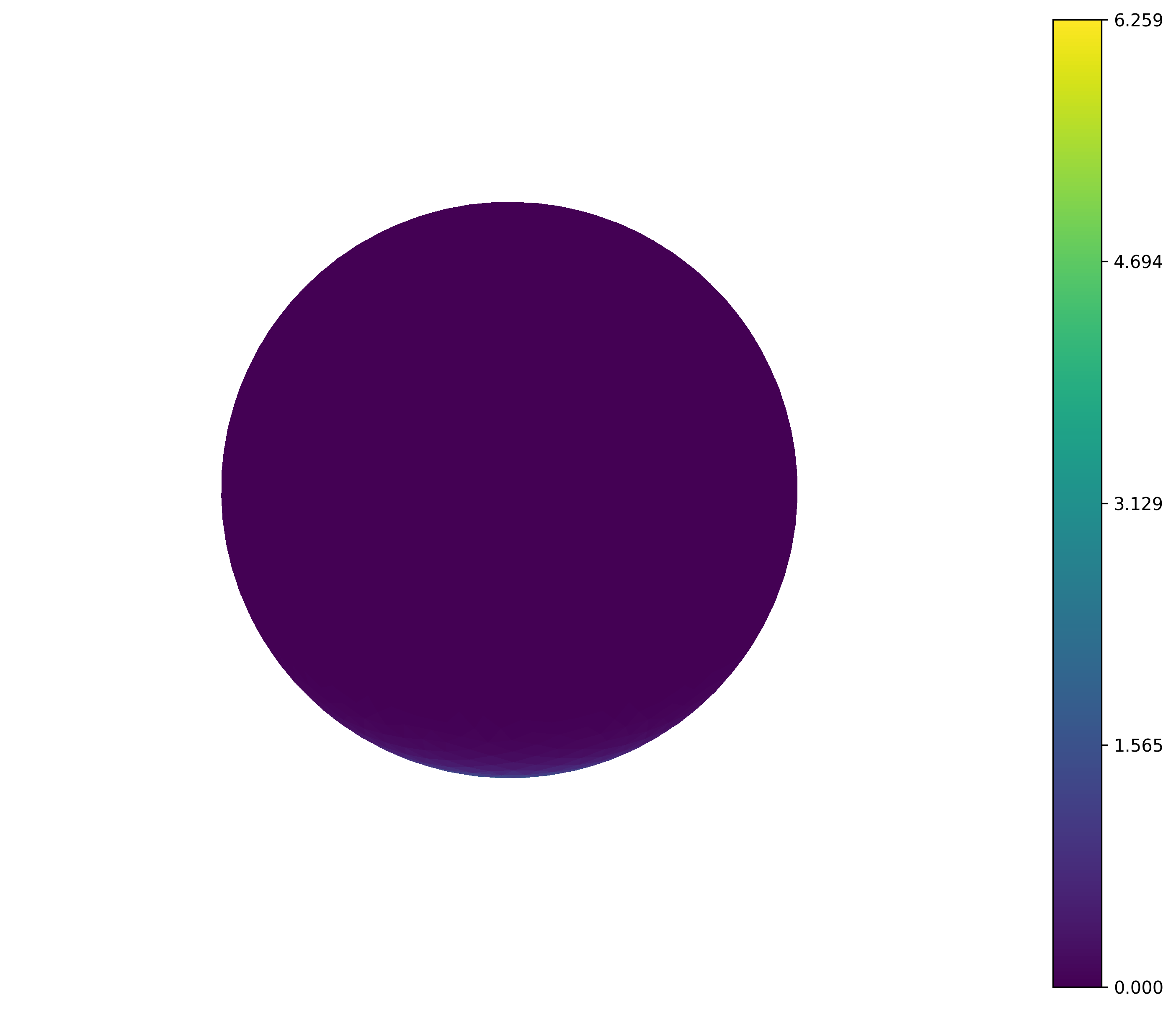}\\
    \vspace{5pt}

    \includegraphics[width=2.4cm]{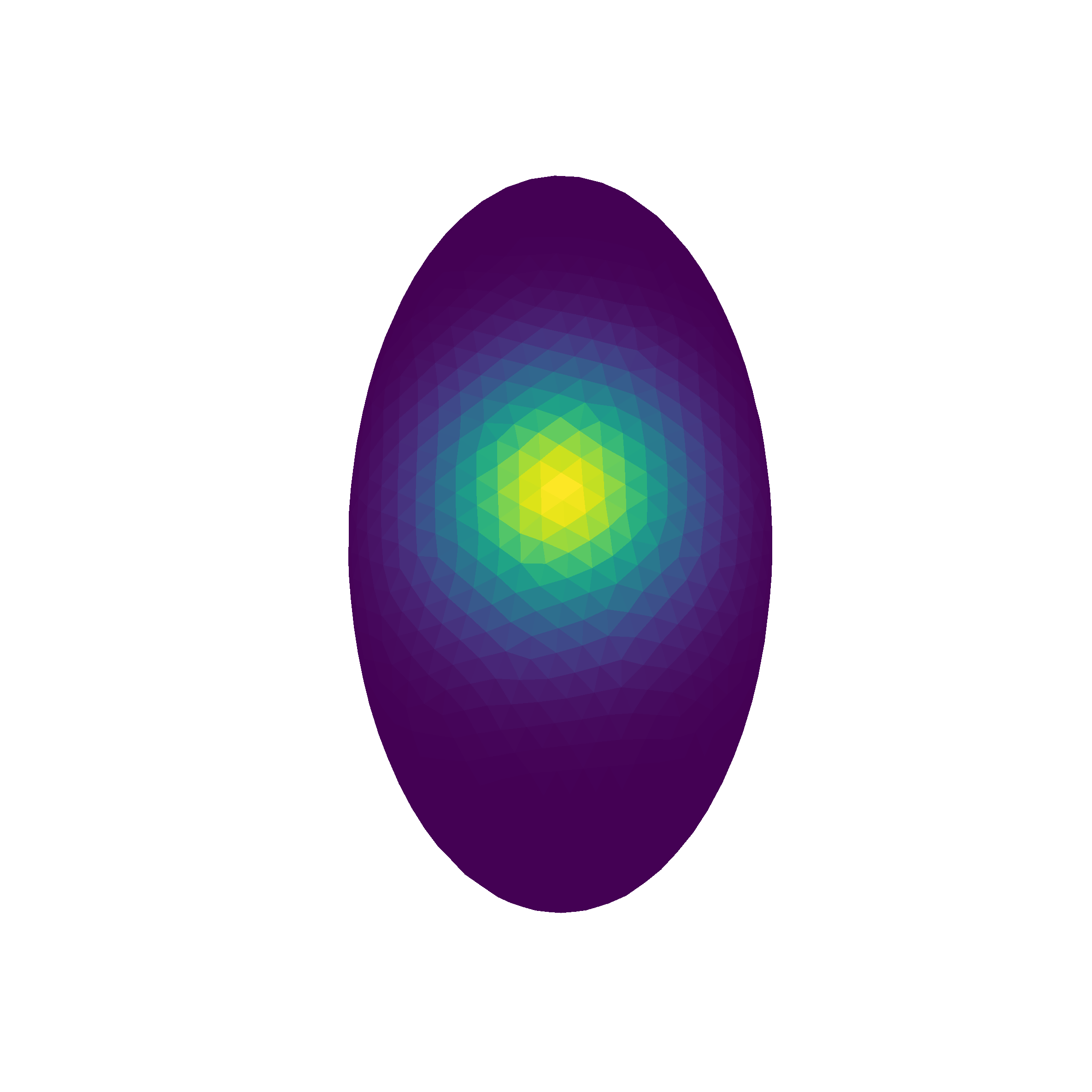}
    \includegraphics[width=2.4cm]{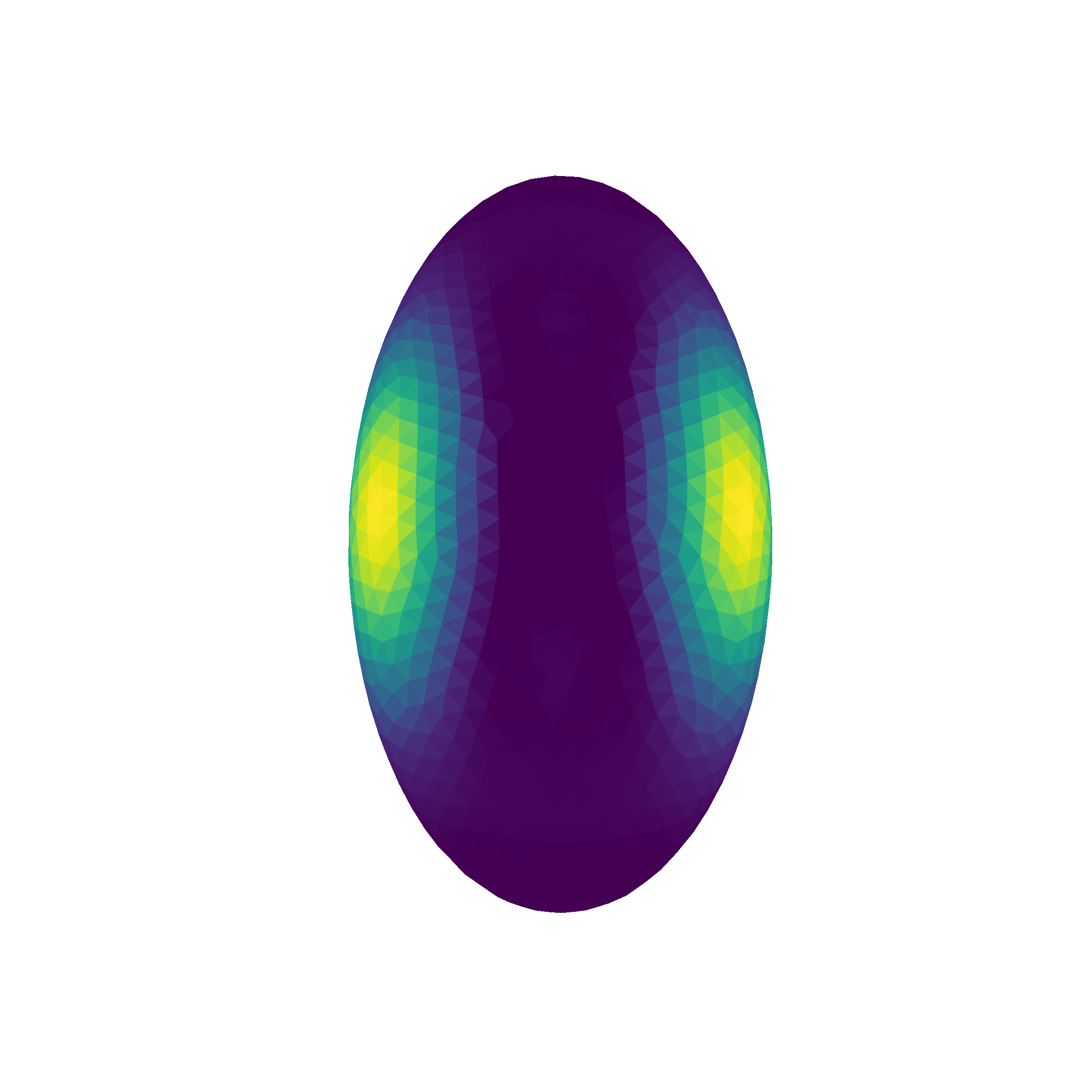}
    \includegraphics[width=2.4cm]{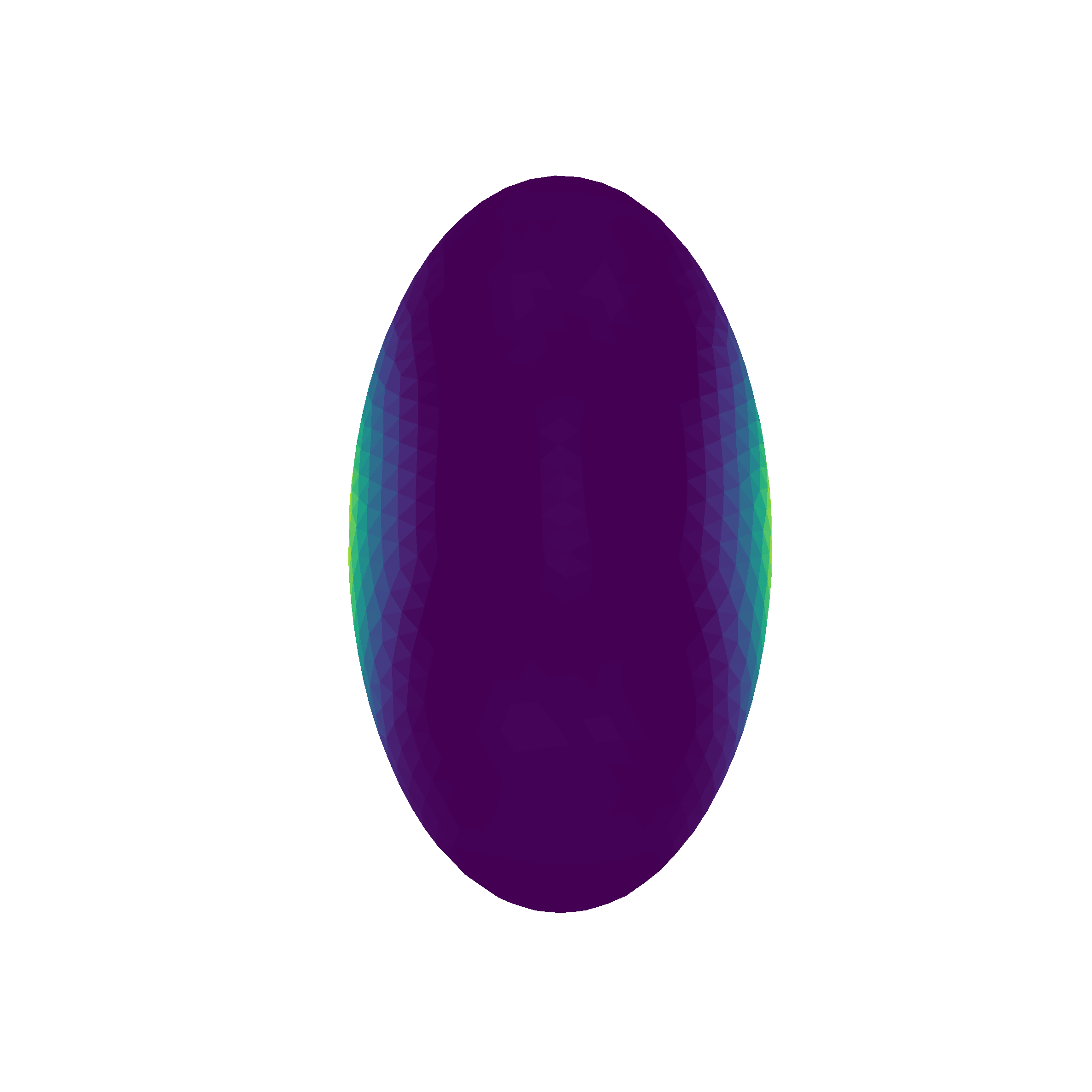}
    \includegraphics[width=2.4cm]{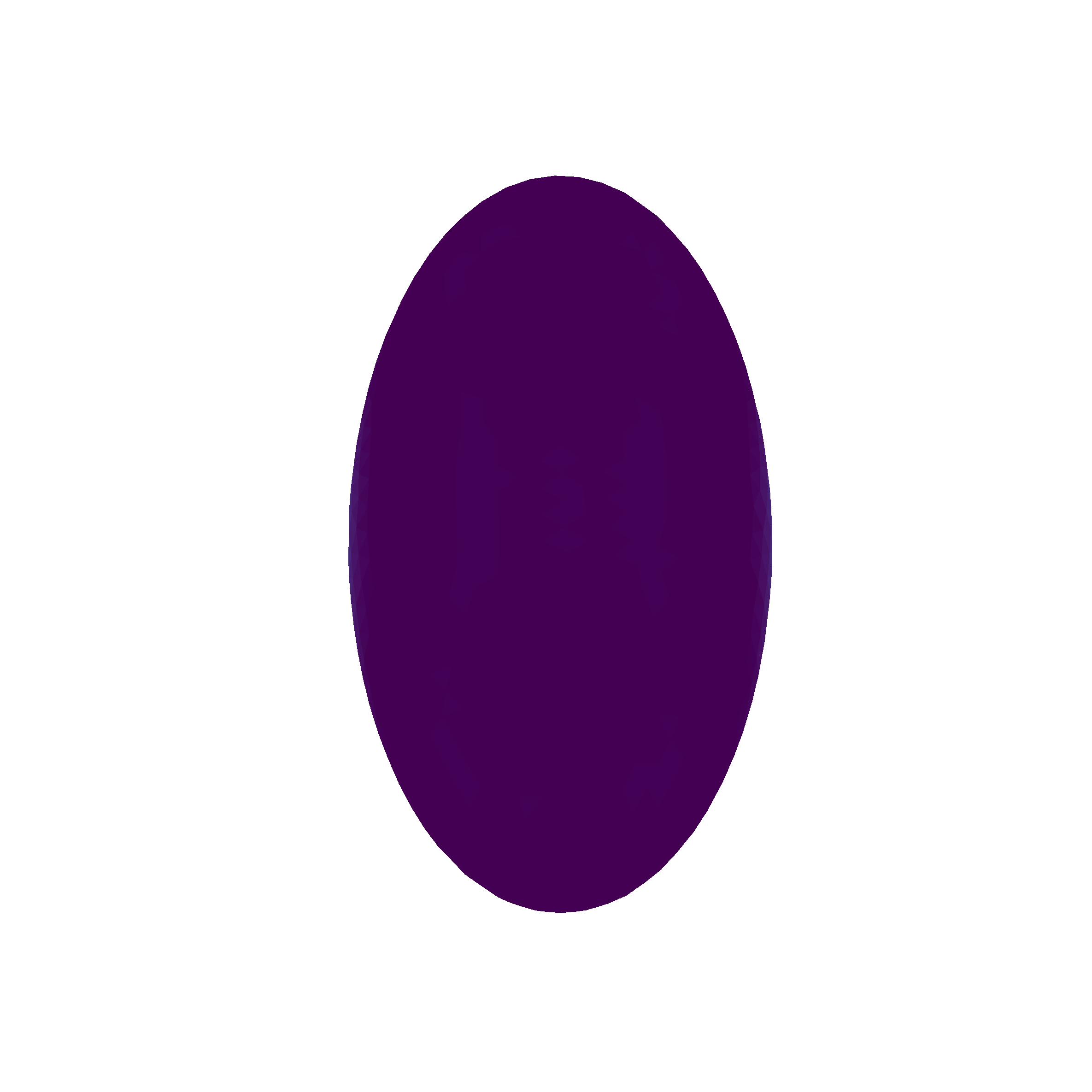}
    \includegraphics[width=2.9cm]{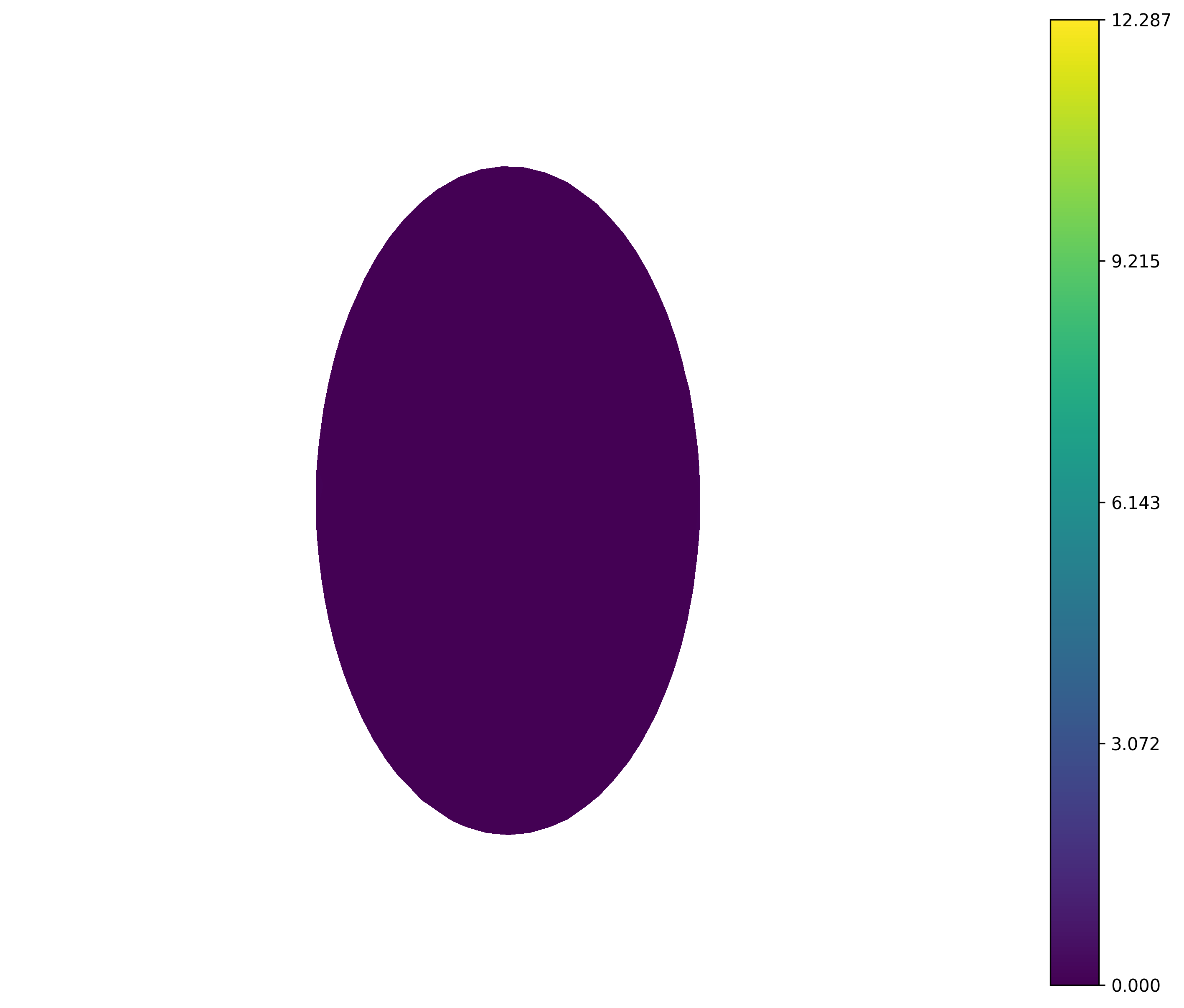}\\
    \vspace{5pt}
    
    \includegraphics[width=2.4cm]{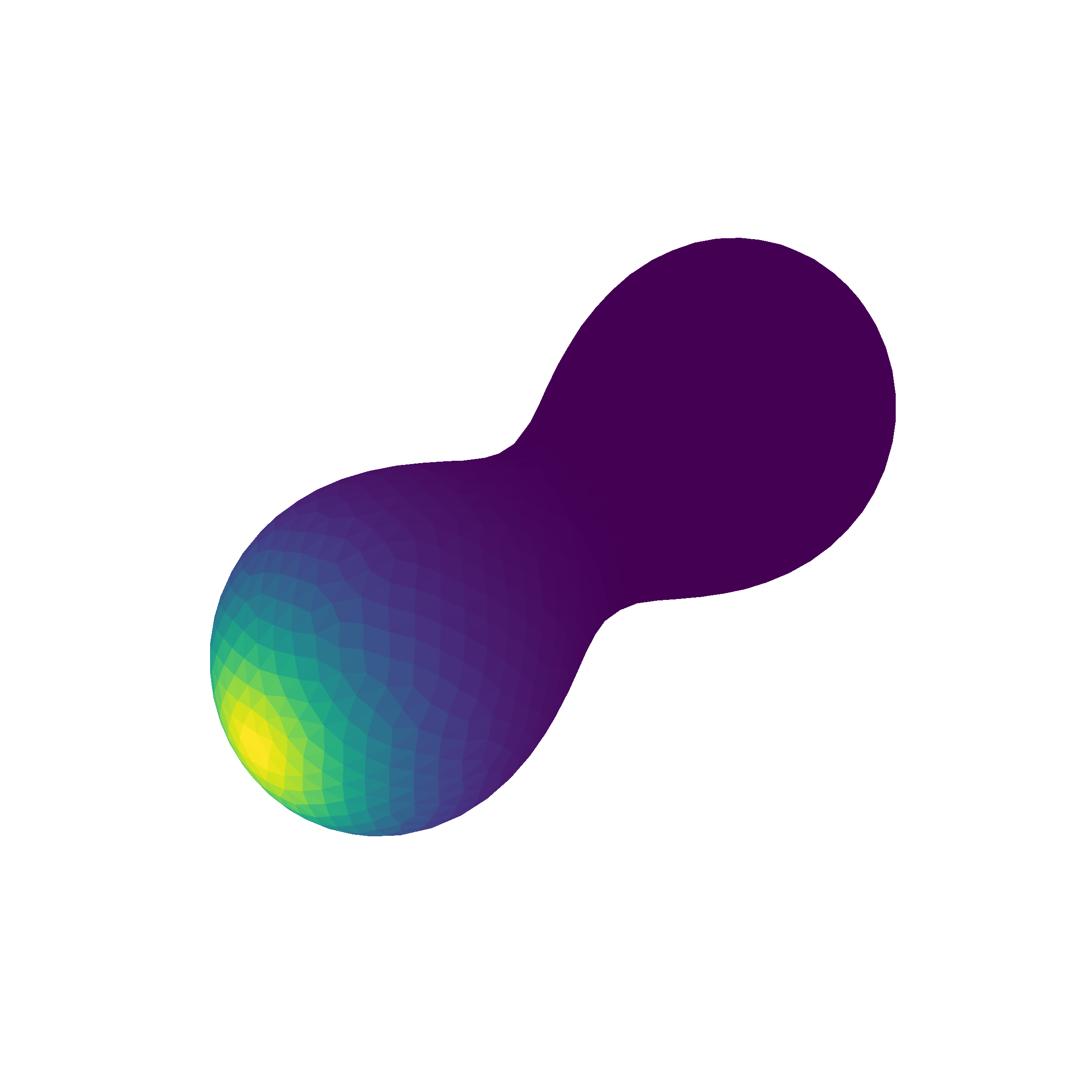}
    \includegraphics[width=2.4cm]{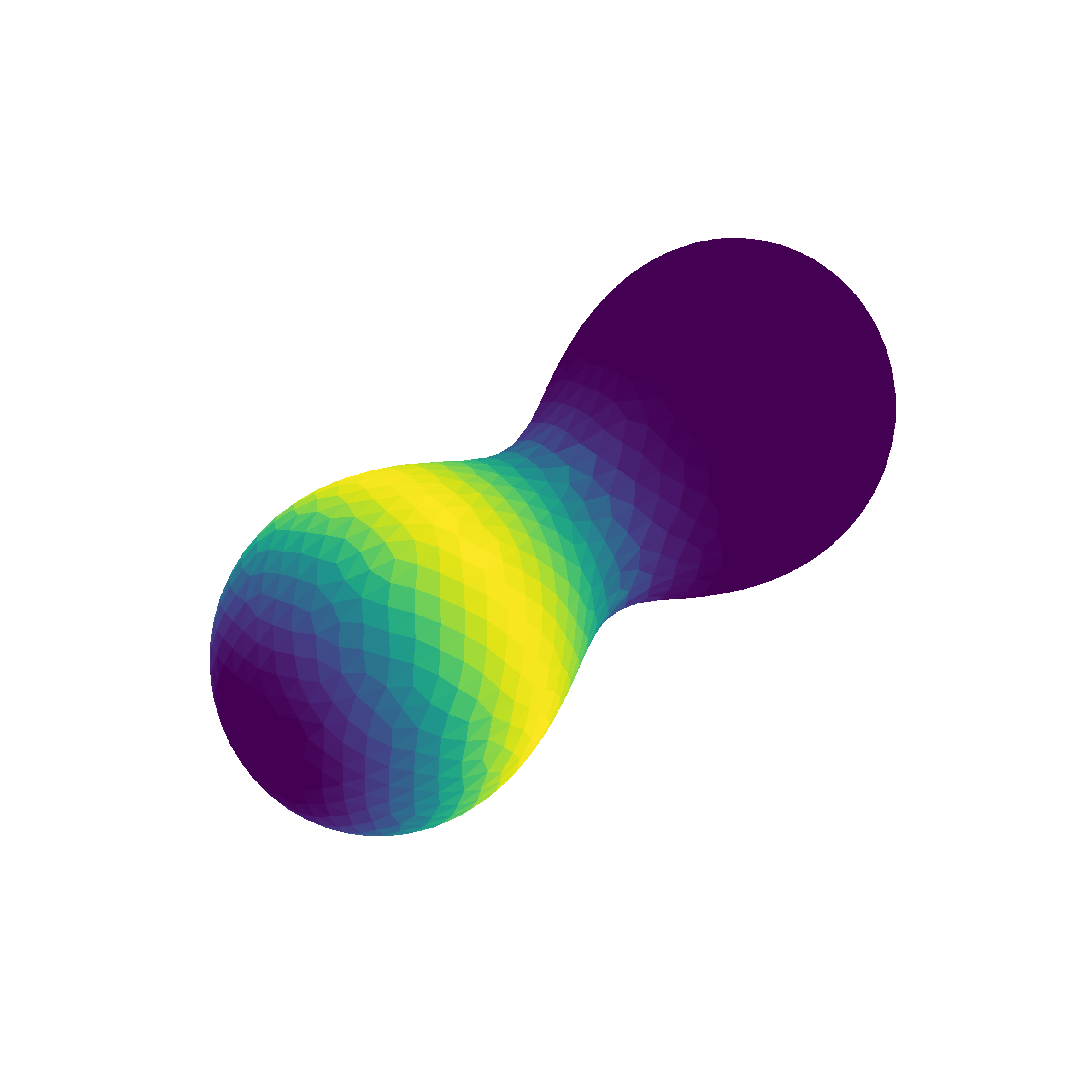}
    \includegraphics[width=2.4cm]{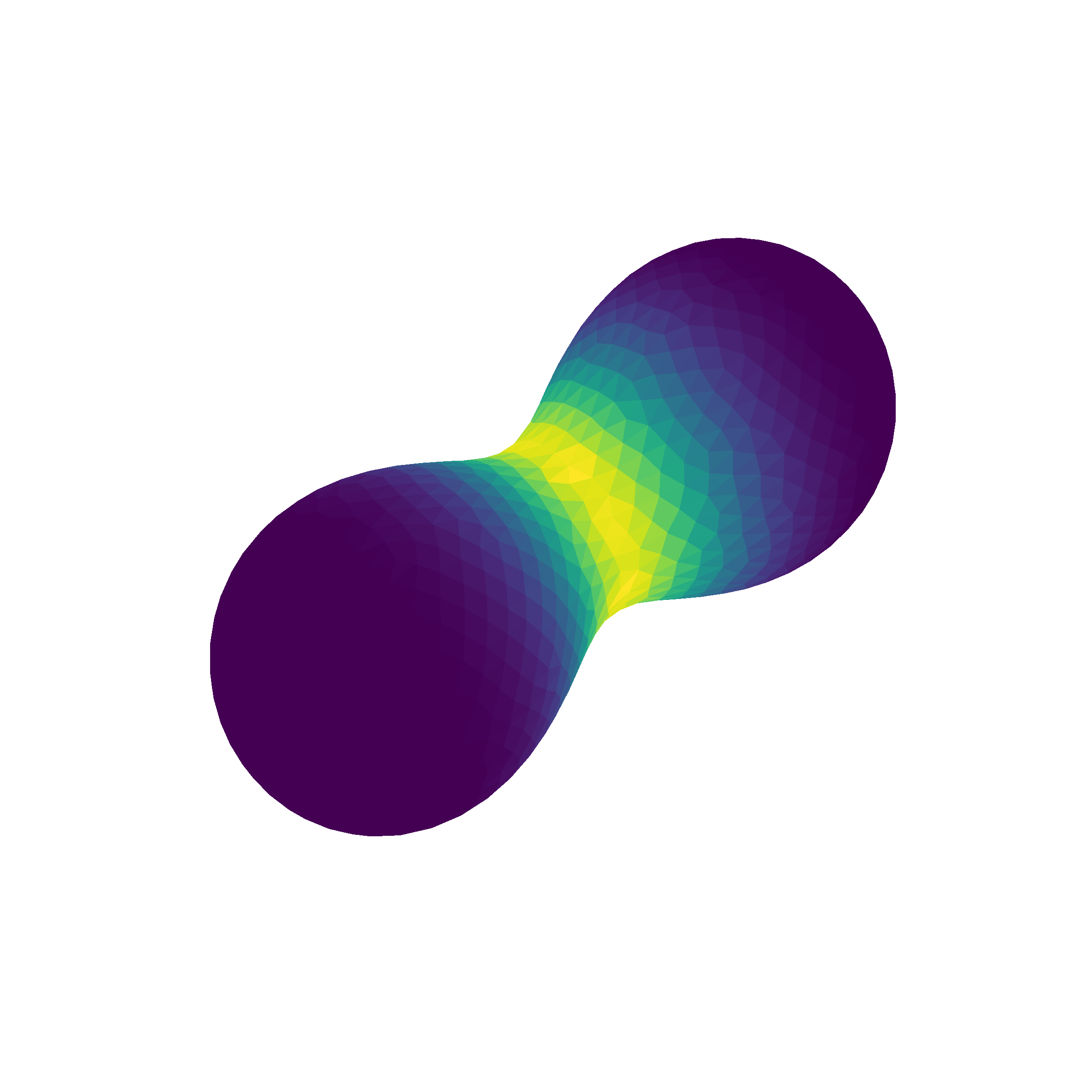}
    \includegraphics[width=2.4cm]{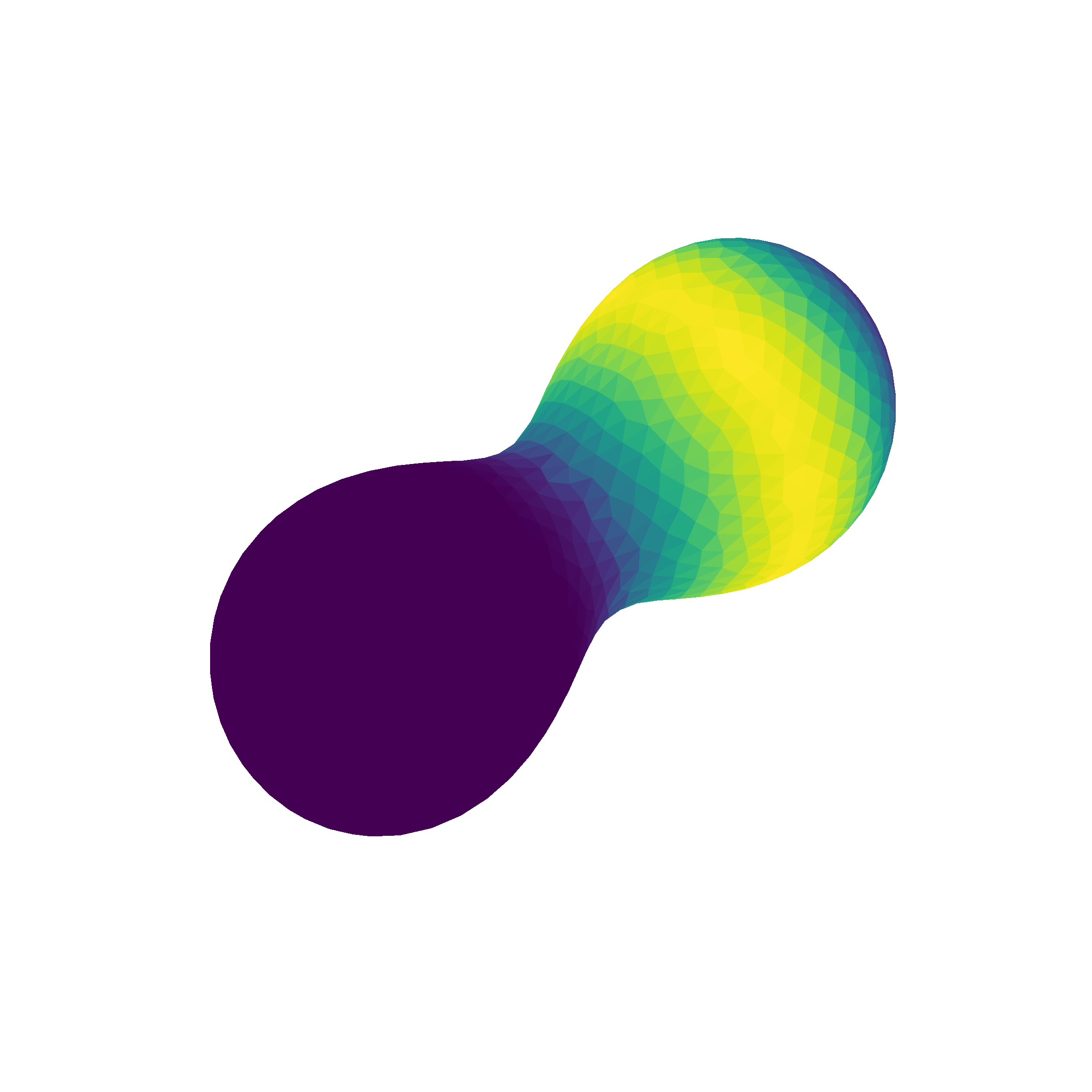}
    \includegraphics[width=2.9cm]{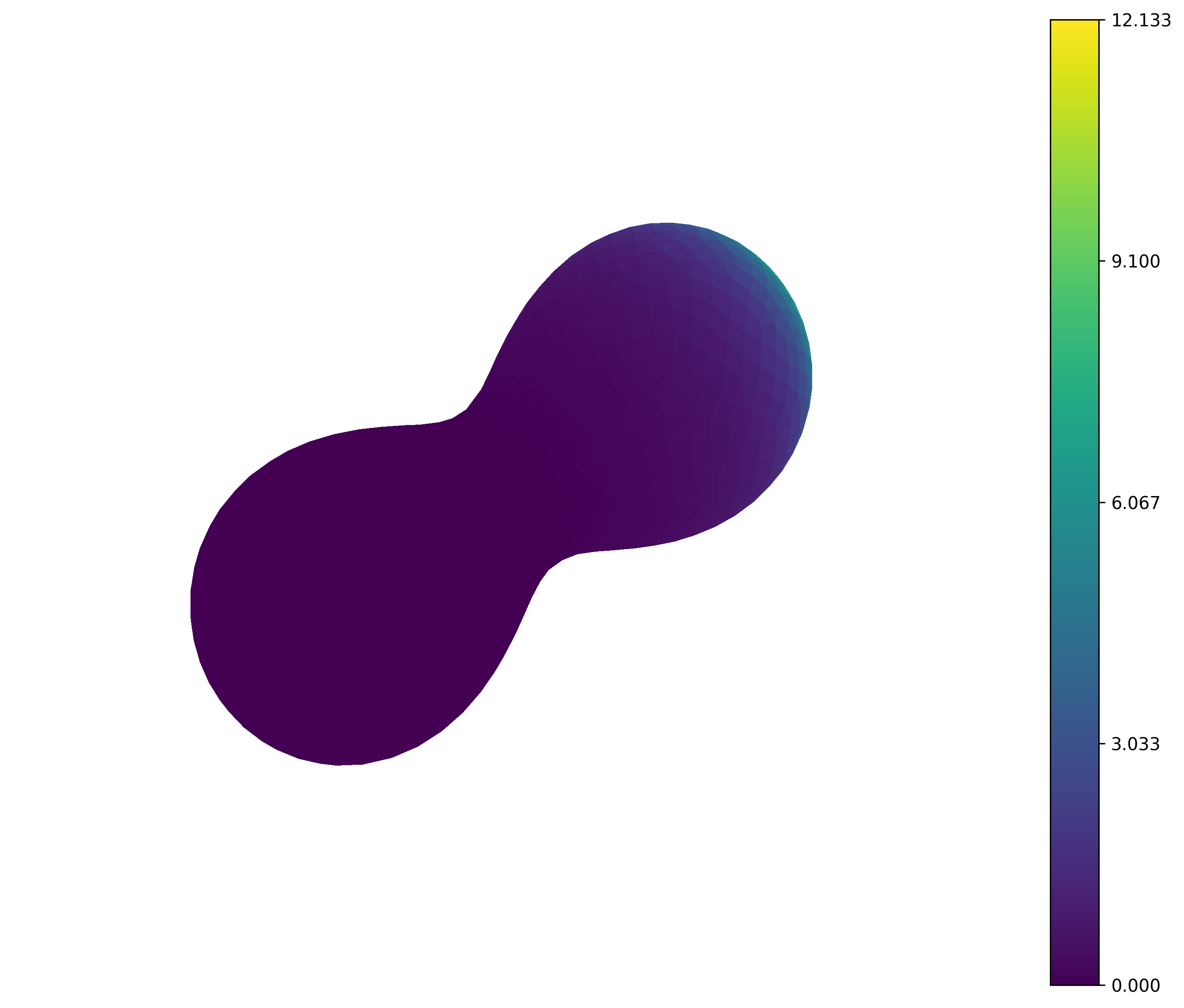}\\
    \vspace{5pt}

    \includegraphics[width=2.4cm]{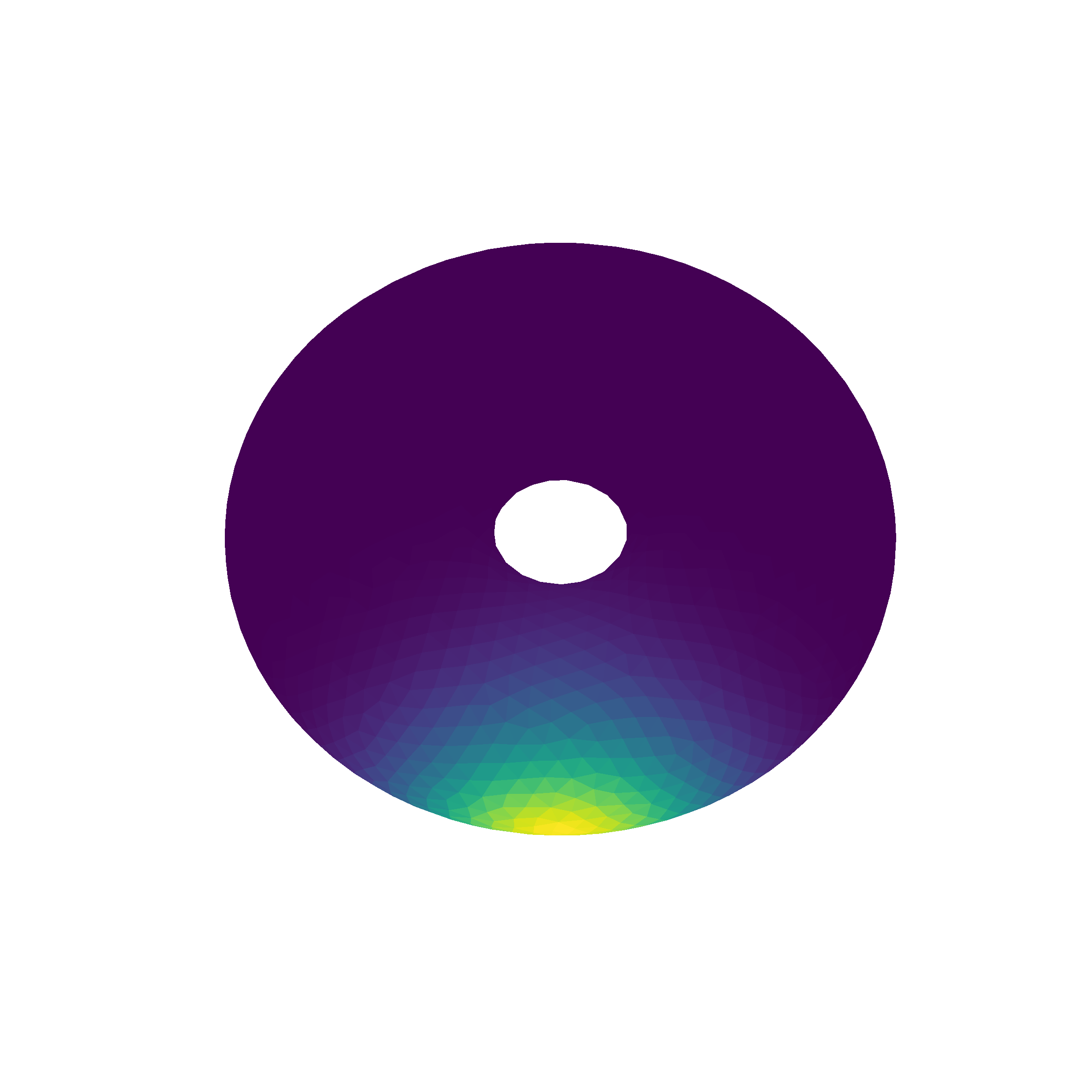}
    \includegraphics[width=2.4cm]{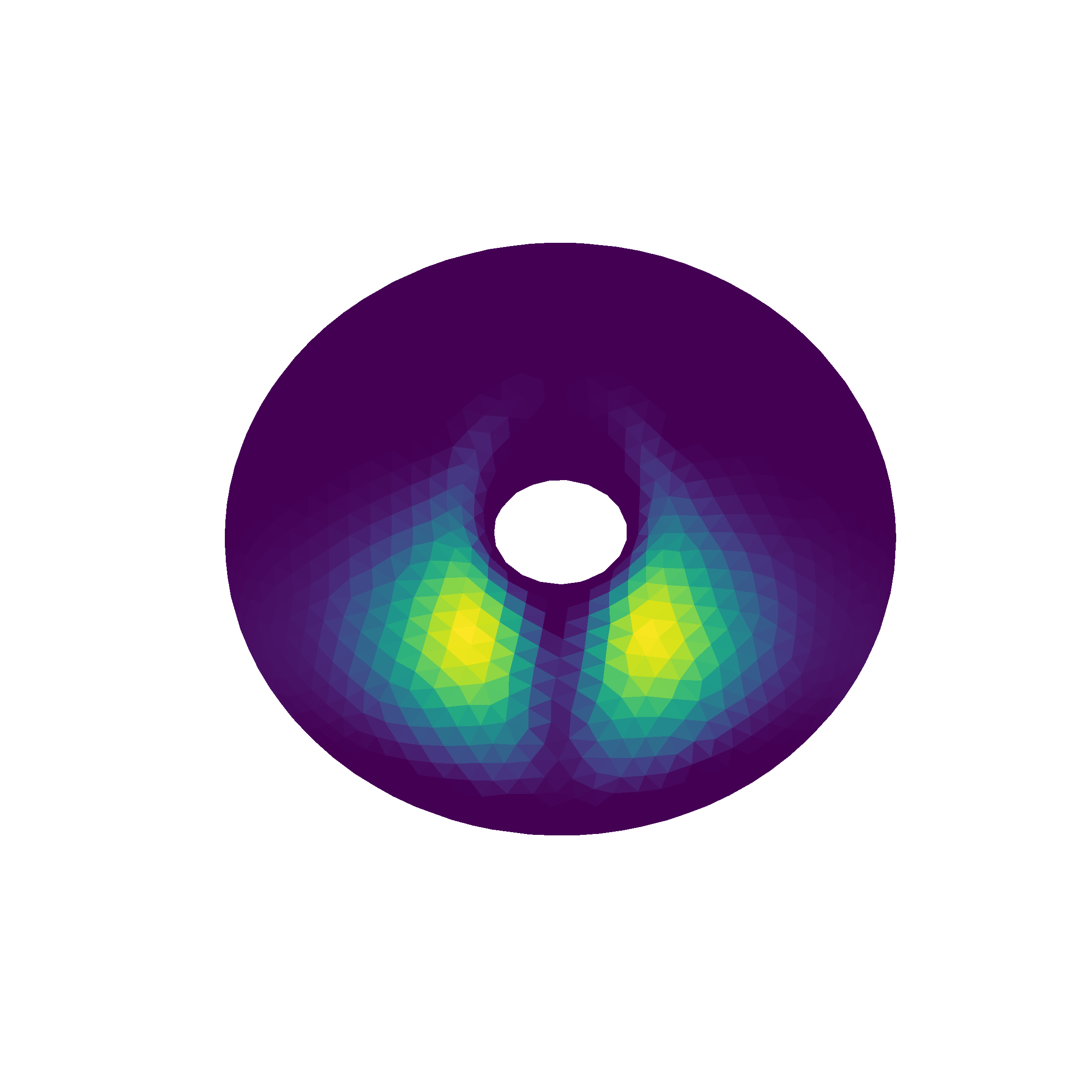}
    \includegraphics[width=2.4cm]{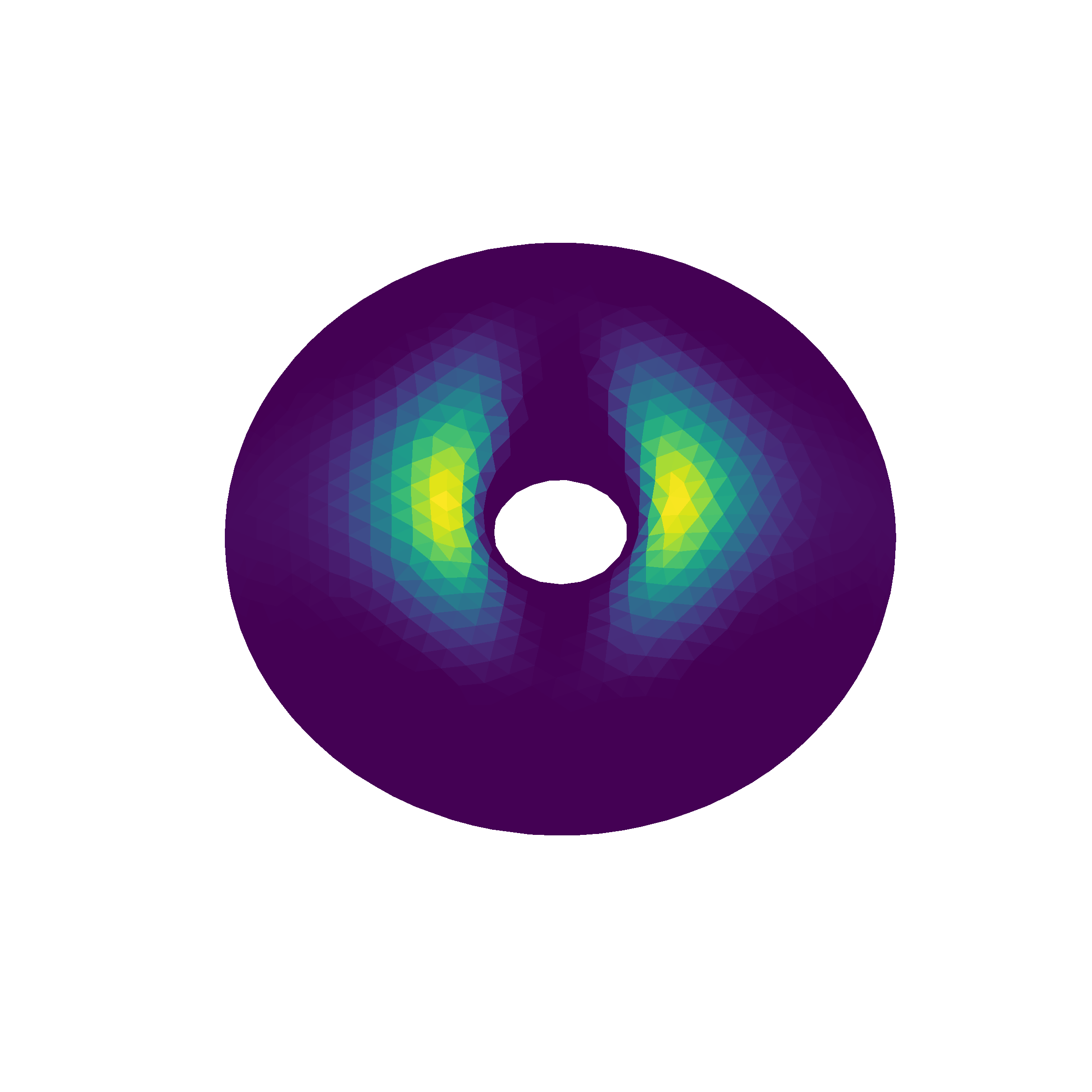}
    \includegraphics[width=2.4cm]{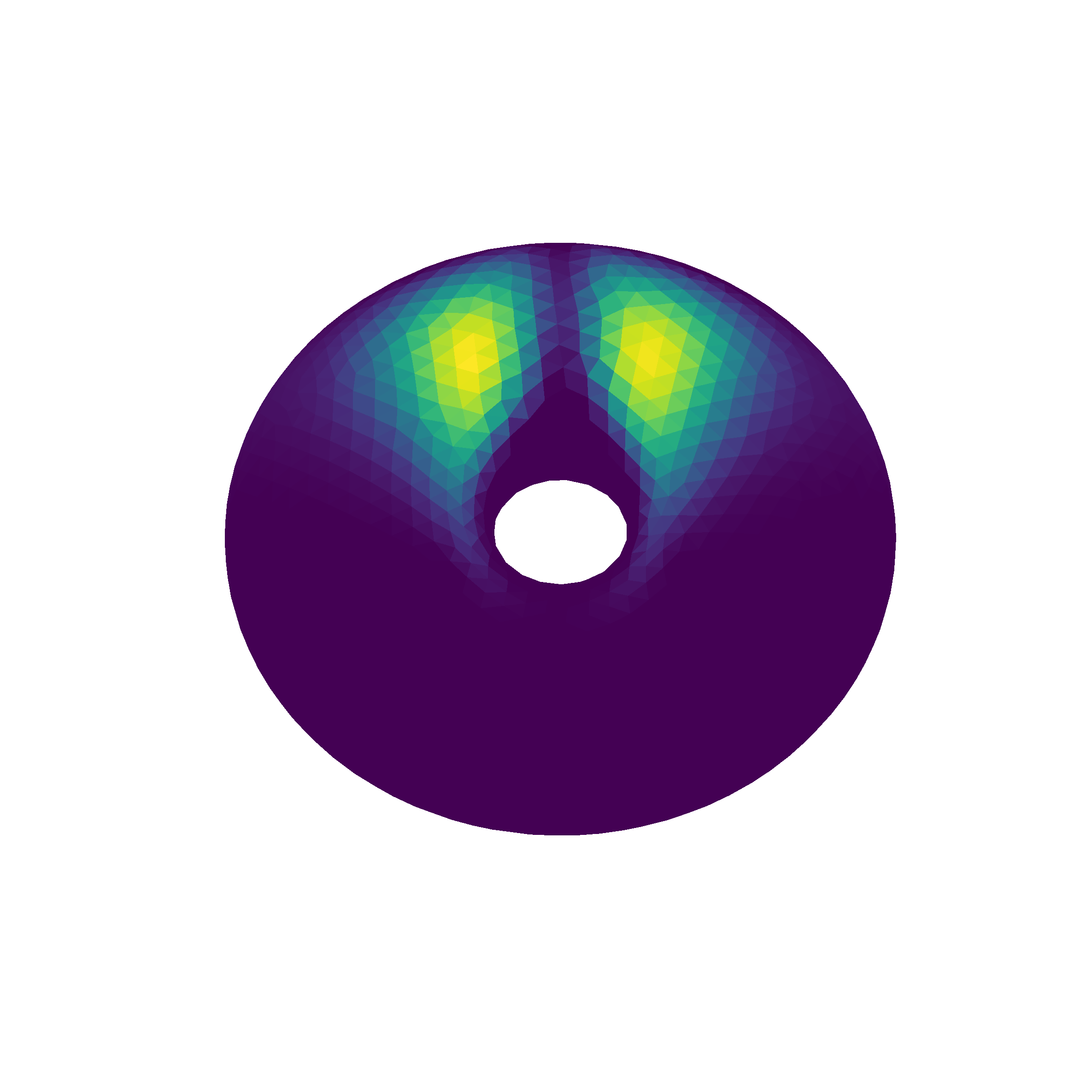}
    \includegraphics[width=2.9cm]{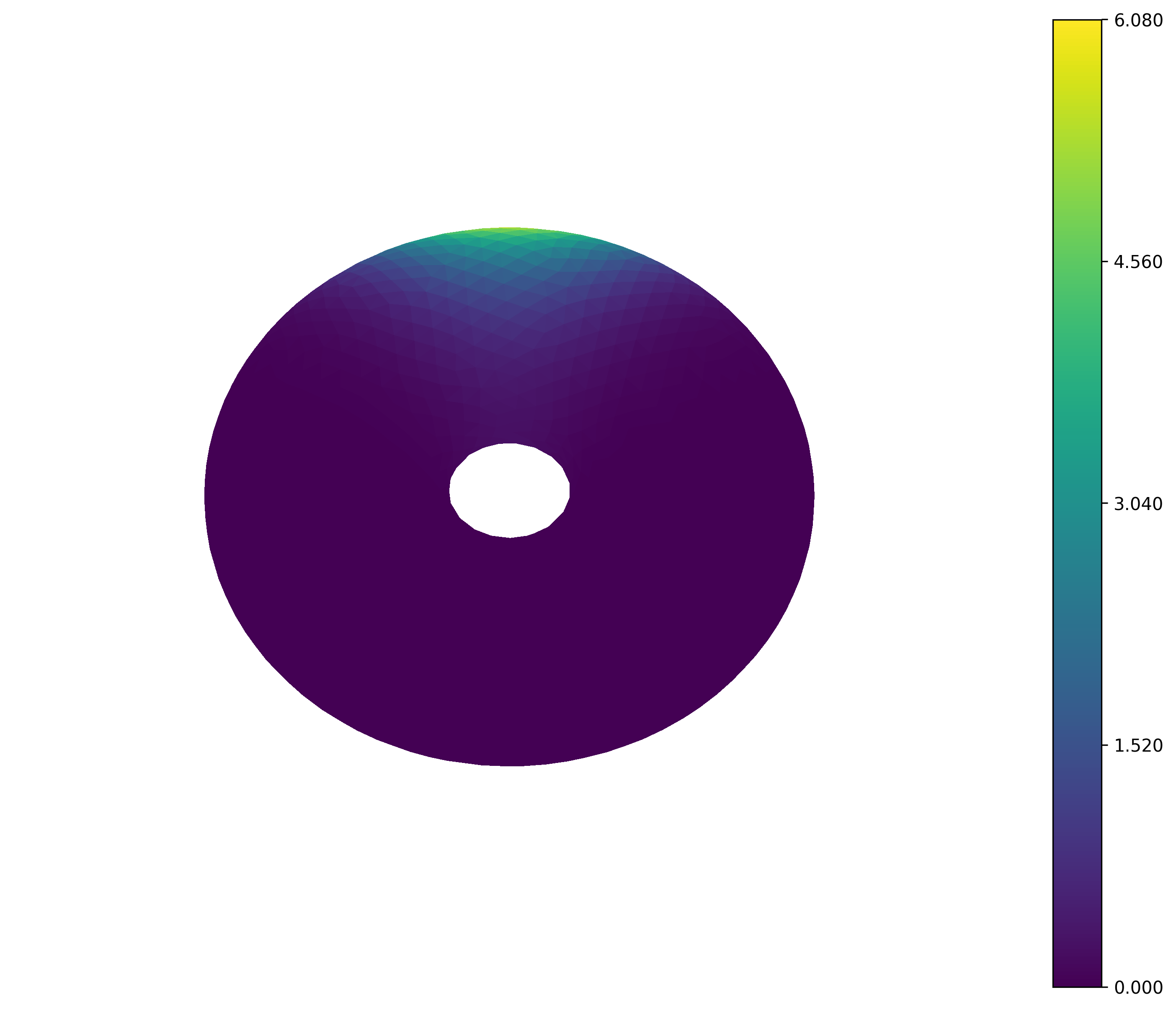}\\
    \vspace{5pt}

    \subfigure[$\rho(0, \boldsymbol{x})$]{
    \includegraphics[width=2.4cm]{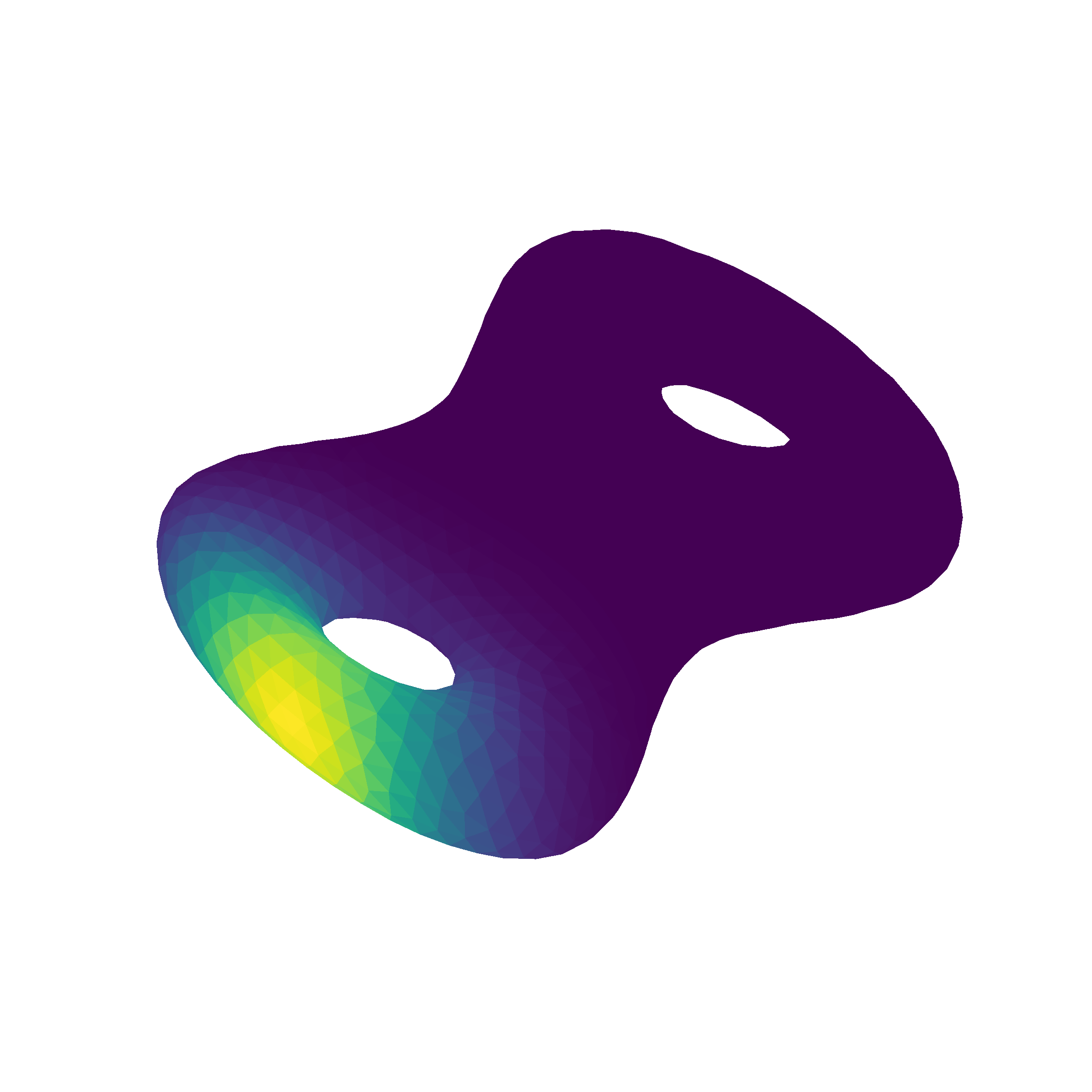}}
    \subfigure[$\rho(0.25, \boldsymbol{x})$]{
    \includegraphics[width=2.4cm]{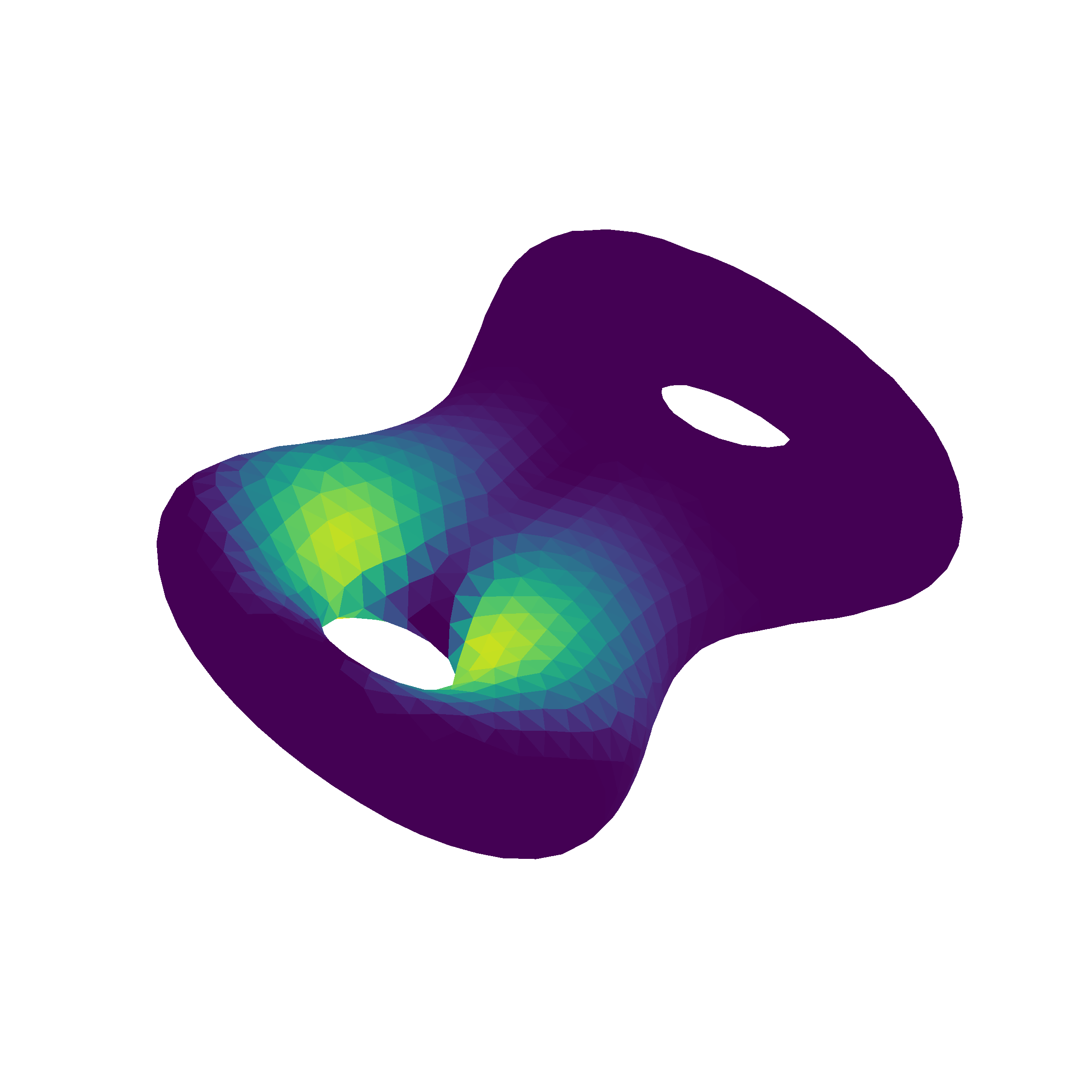}}
    \subfigure[$\rho(0.5, \boldsymbol{x})$]{
    \includegraphics[width=2.4cm]{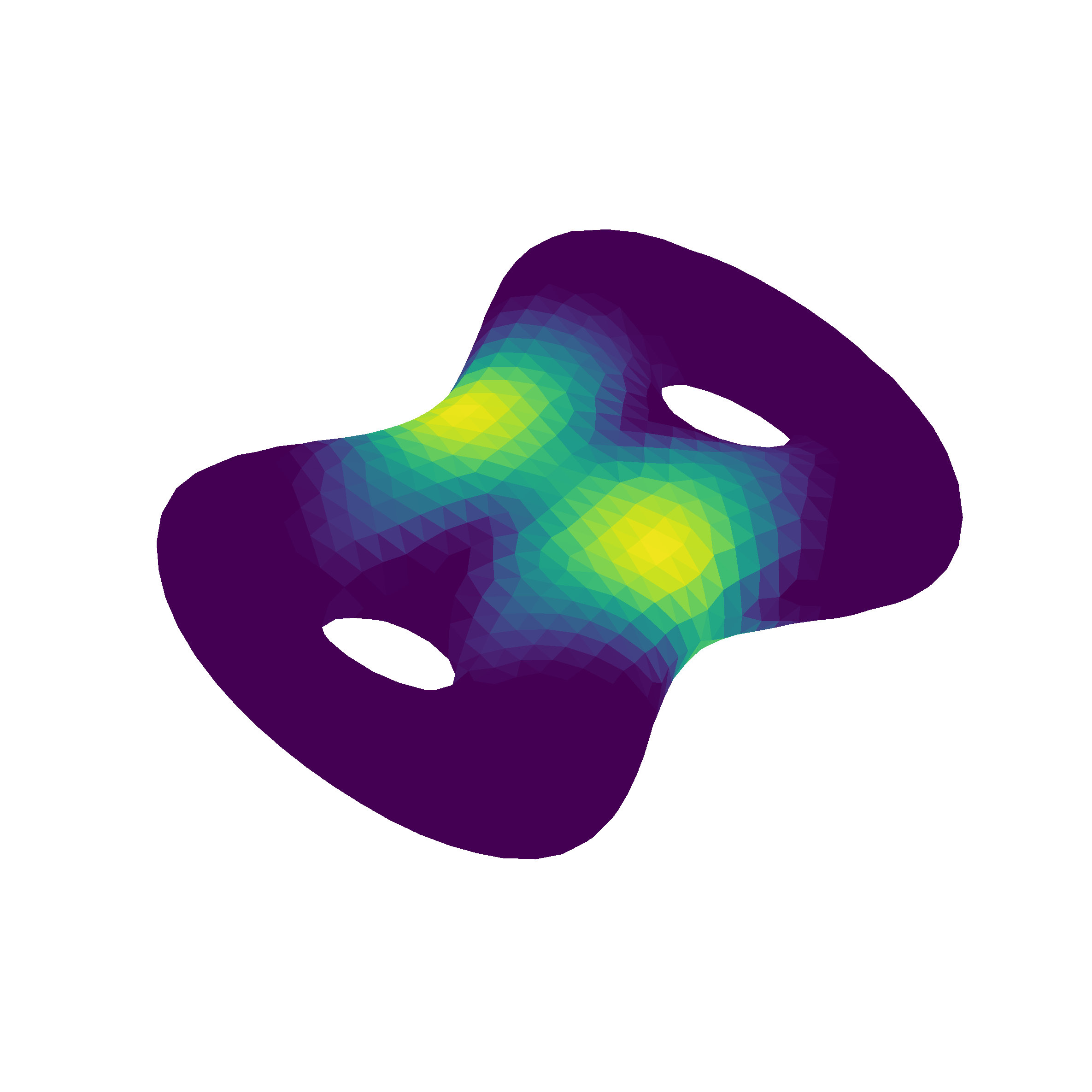}}
    \subfigure[$\rho(0.75, \boldsymbol{x})$]{
    \includegraphics[width=2.4cm]{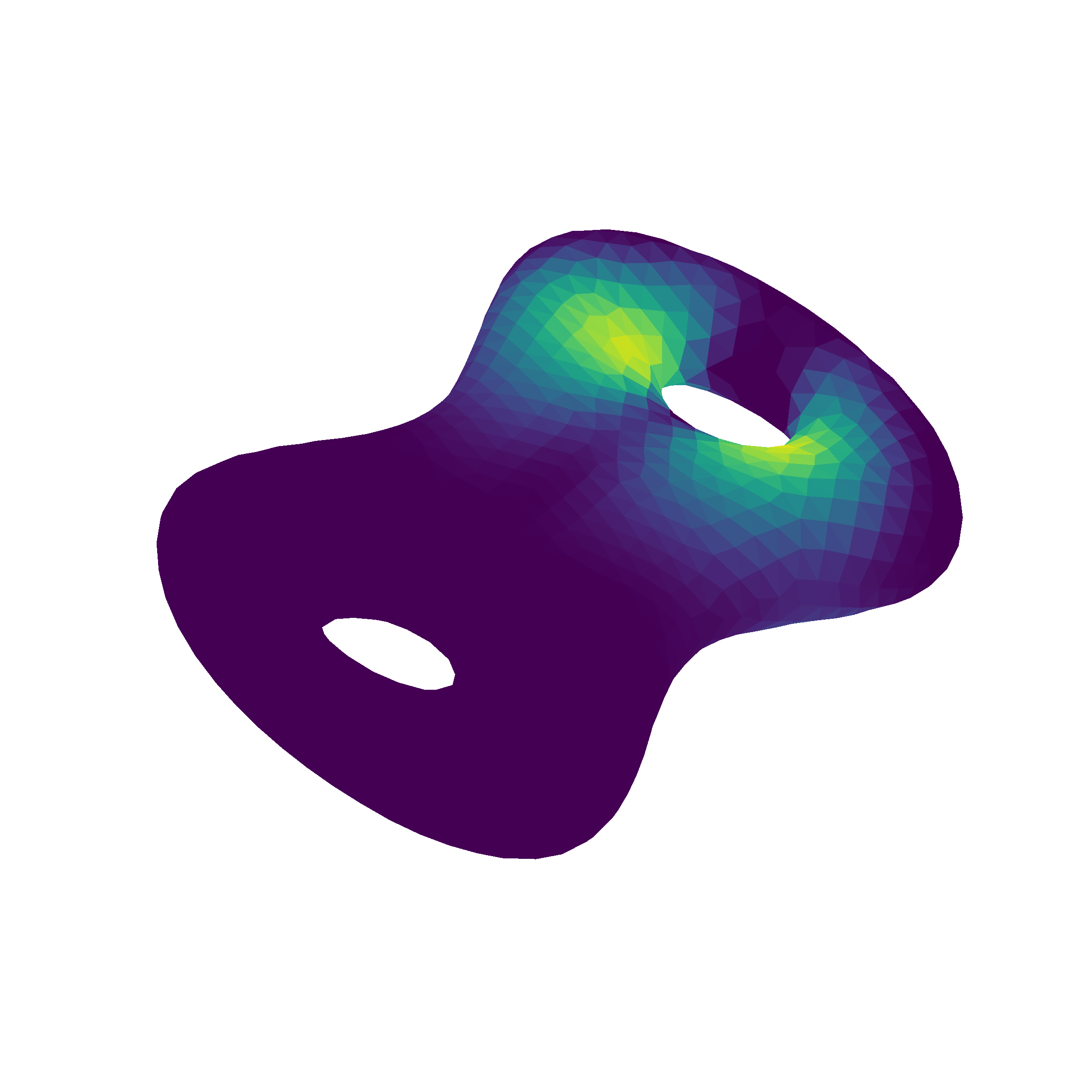}}
    \subfigure[$\rho(1, \boldsymbol{x})$]{
    \includegraphics[width=2.9cm]{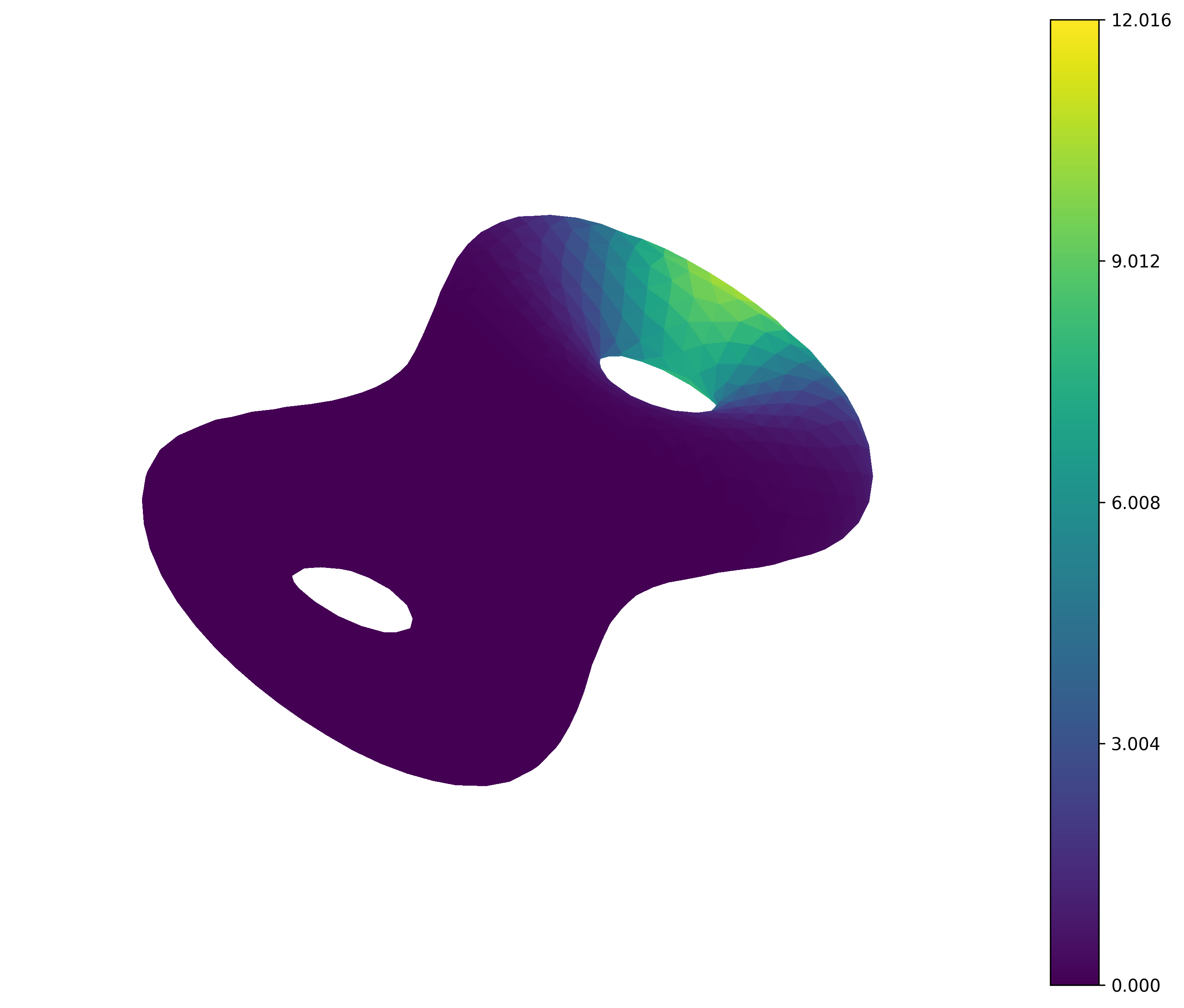}}\\
    \caption{SOT examples ($\beta=1$) and the calculation times are 442s, 156s, 227s, 278s and 186s respectively.}
    \label{SOT-LS}
    \end{center}
\end{figure}

\begin{figure}[htbp]
    \begin{center}
    \includegraphics[width=2.4cm]{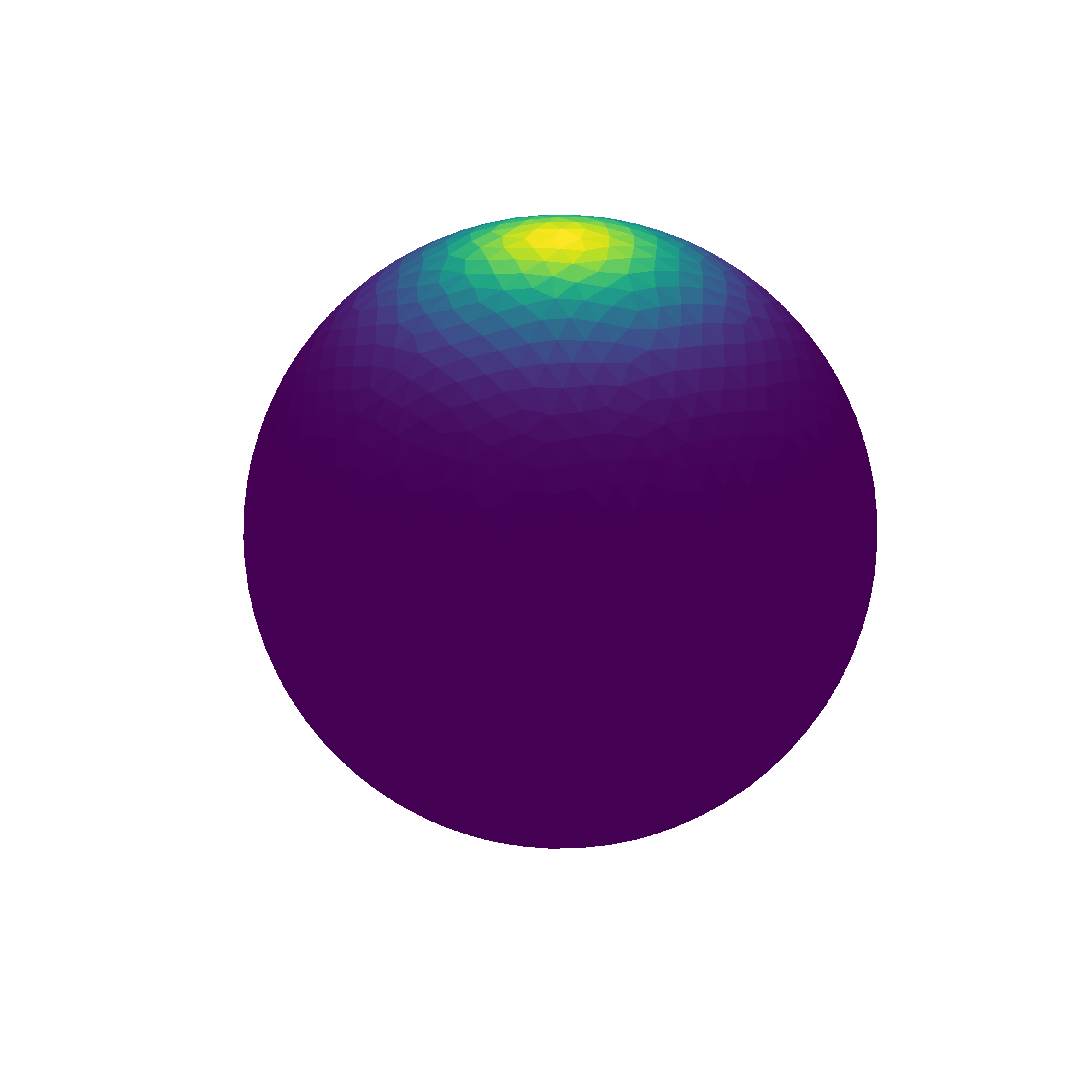}
    \includegraphics[width=2.4cm]{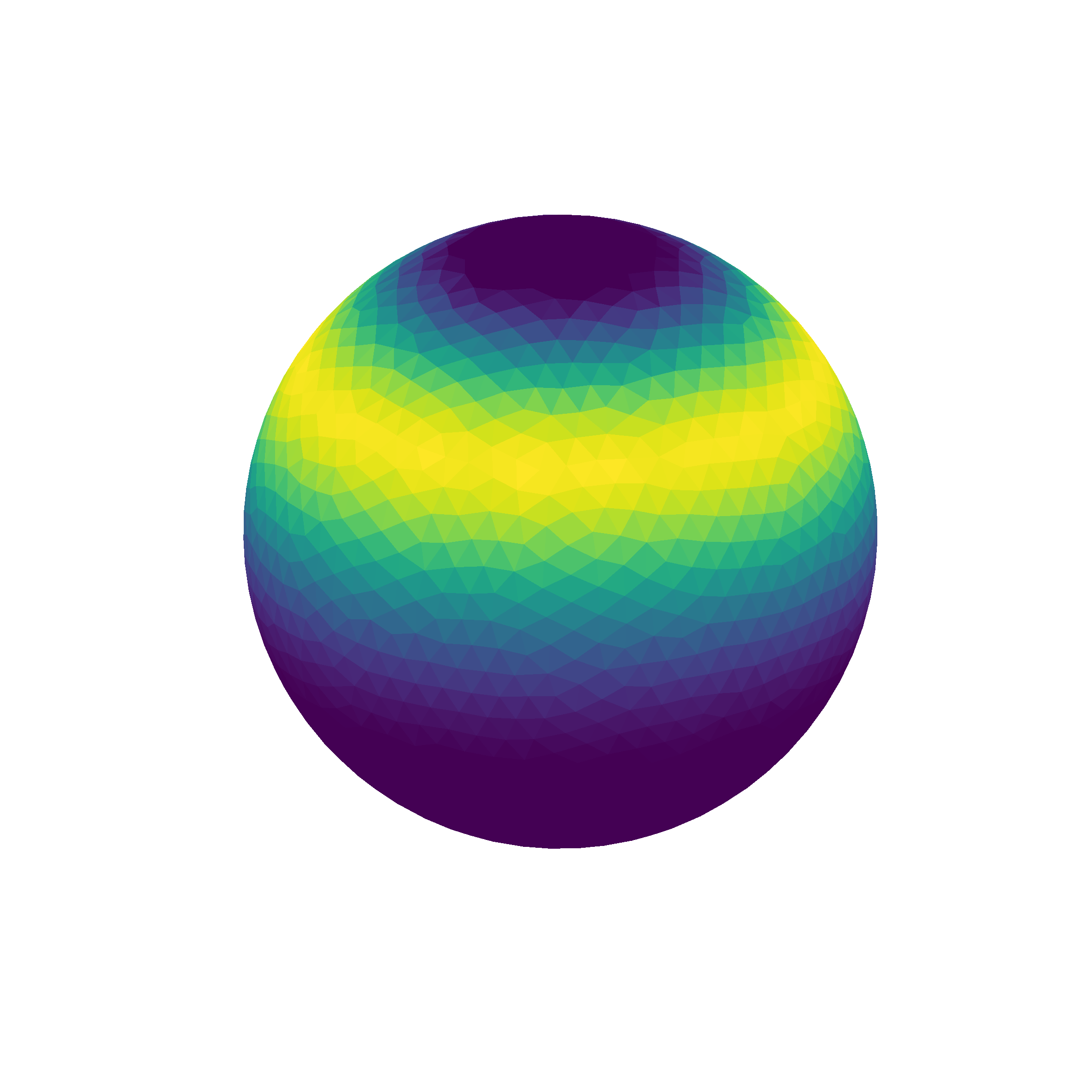}
    \includegraphics[width=2.4cm]{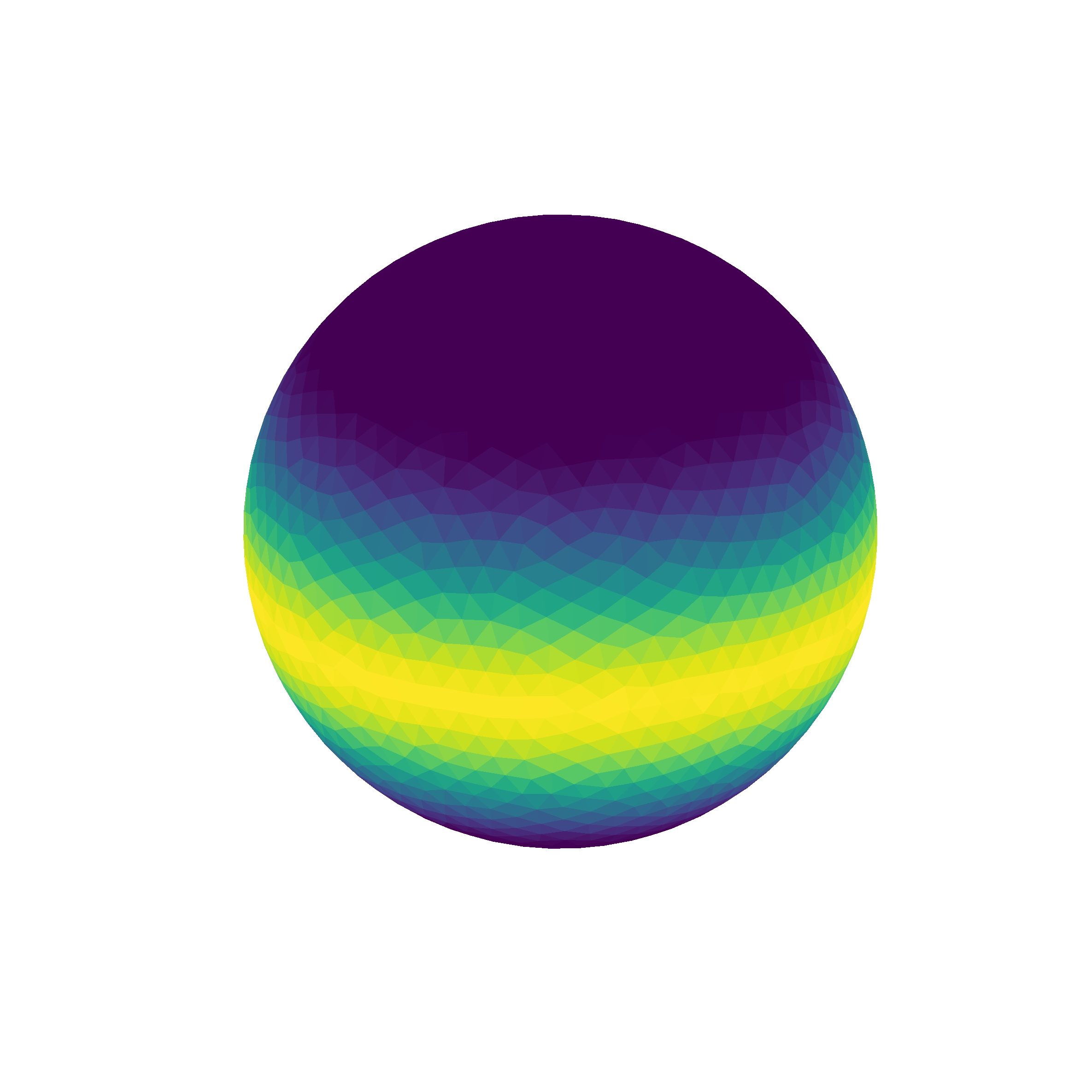}
    \includegraphics[width=2.4cm]{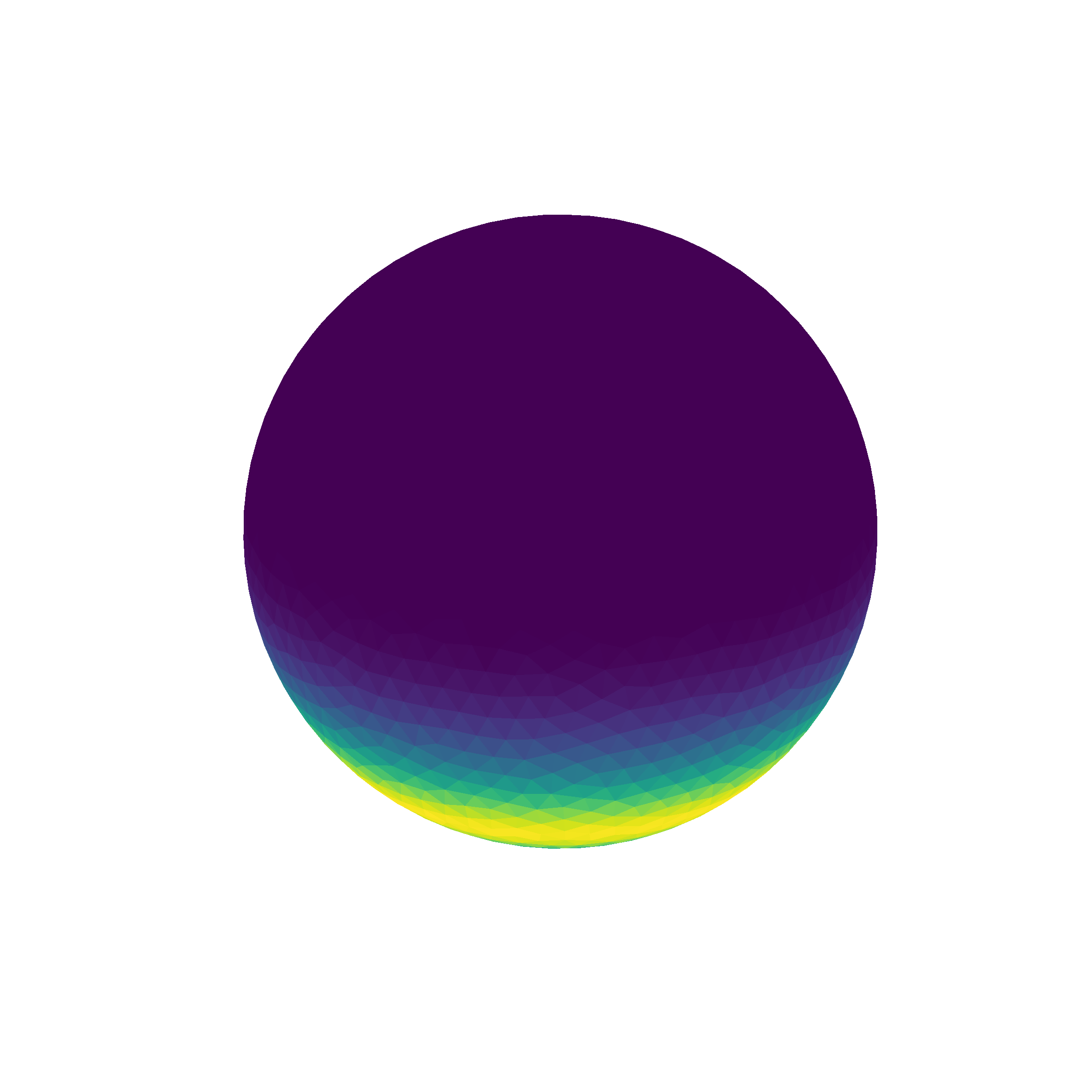}
    \includegraphics[width=2.9cm]{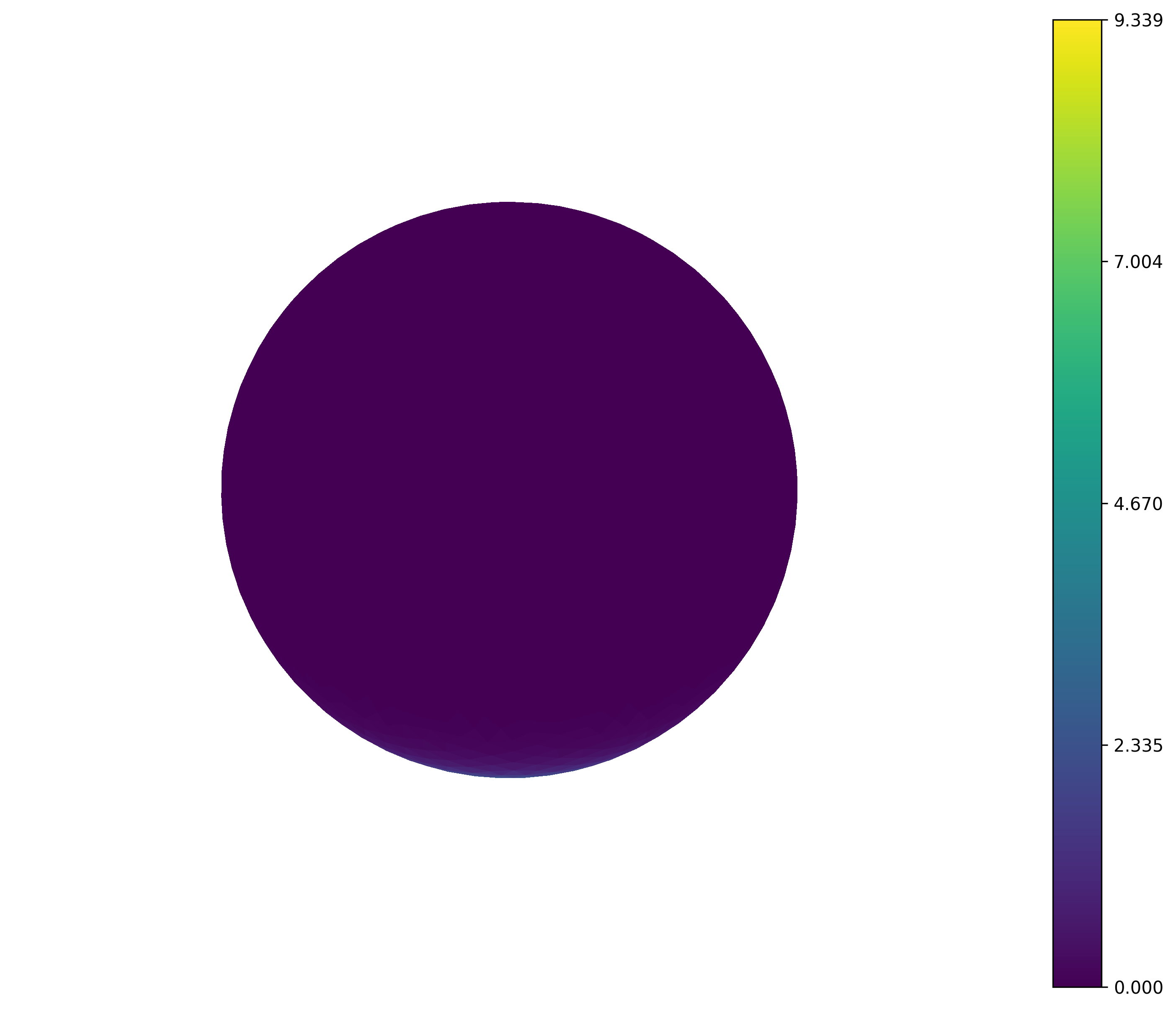}\\
    \vspace{5pt}

    \includegraphics[width=2.4cm]{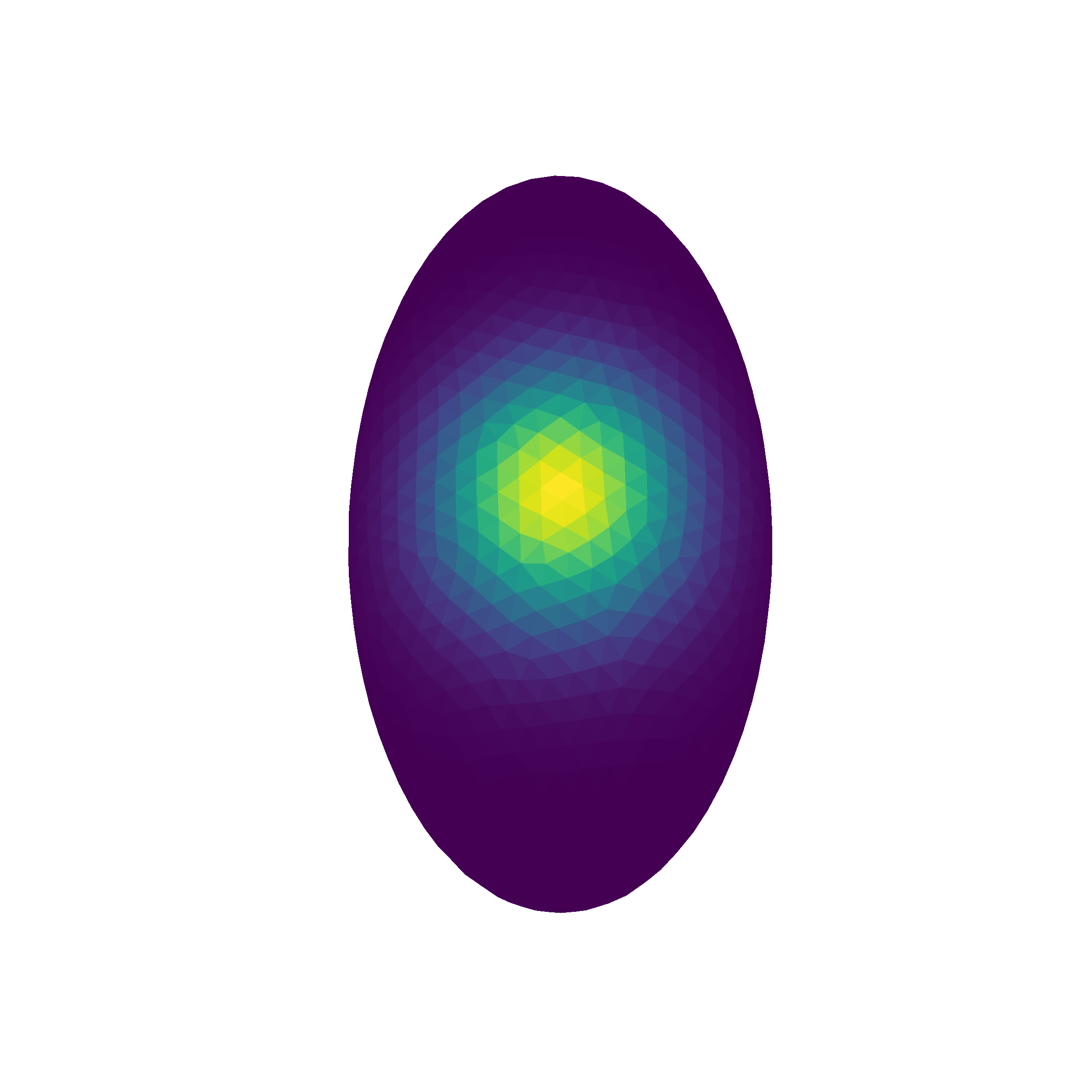}
    \includegraphics[width=2.4cm]{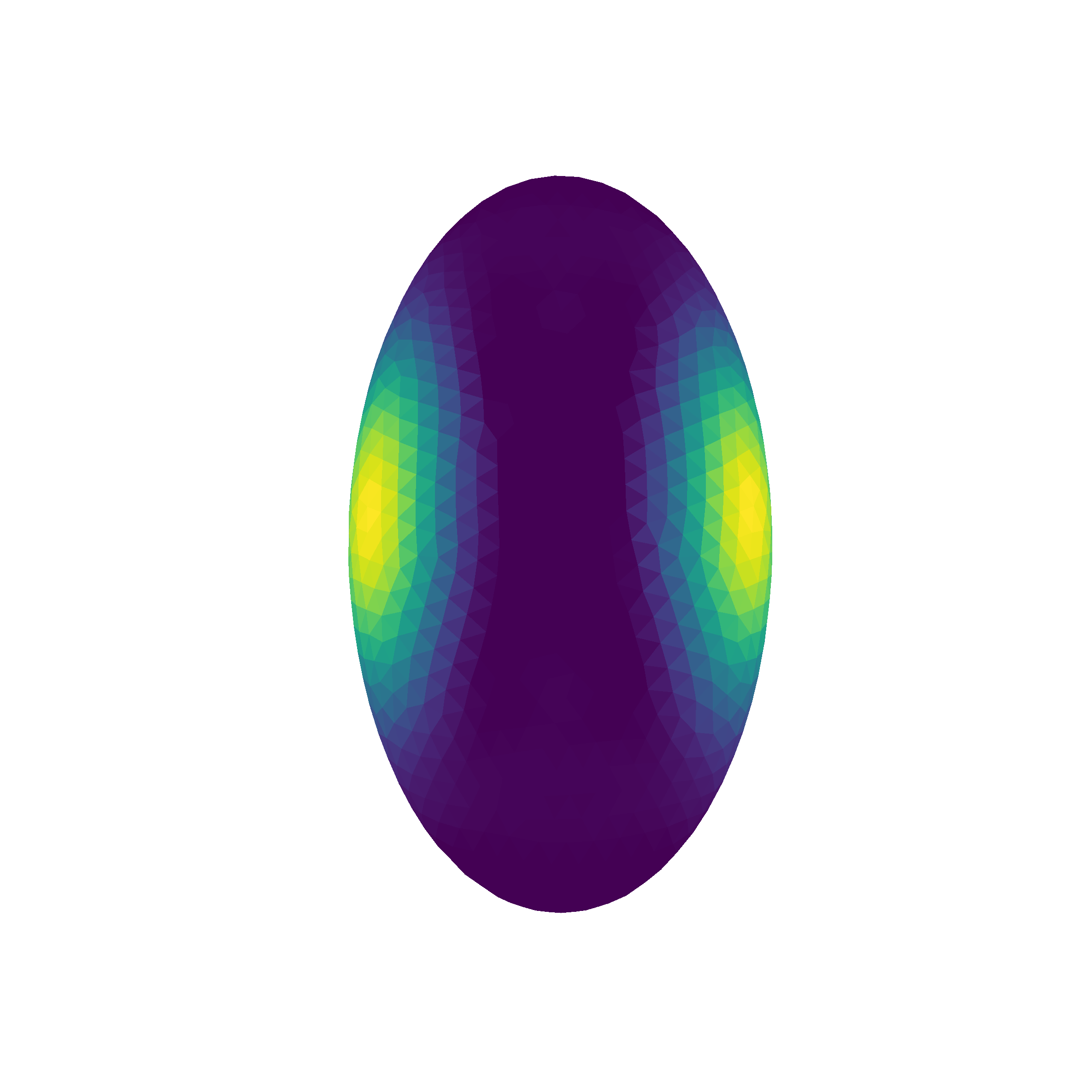}
    \includegraphics[width=2.4cm]{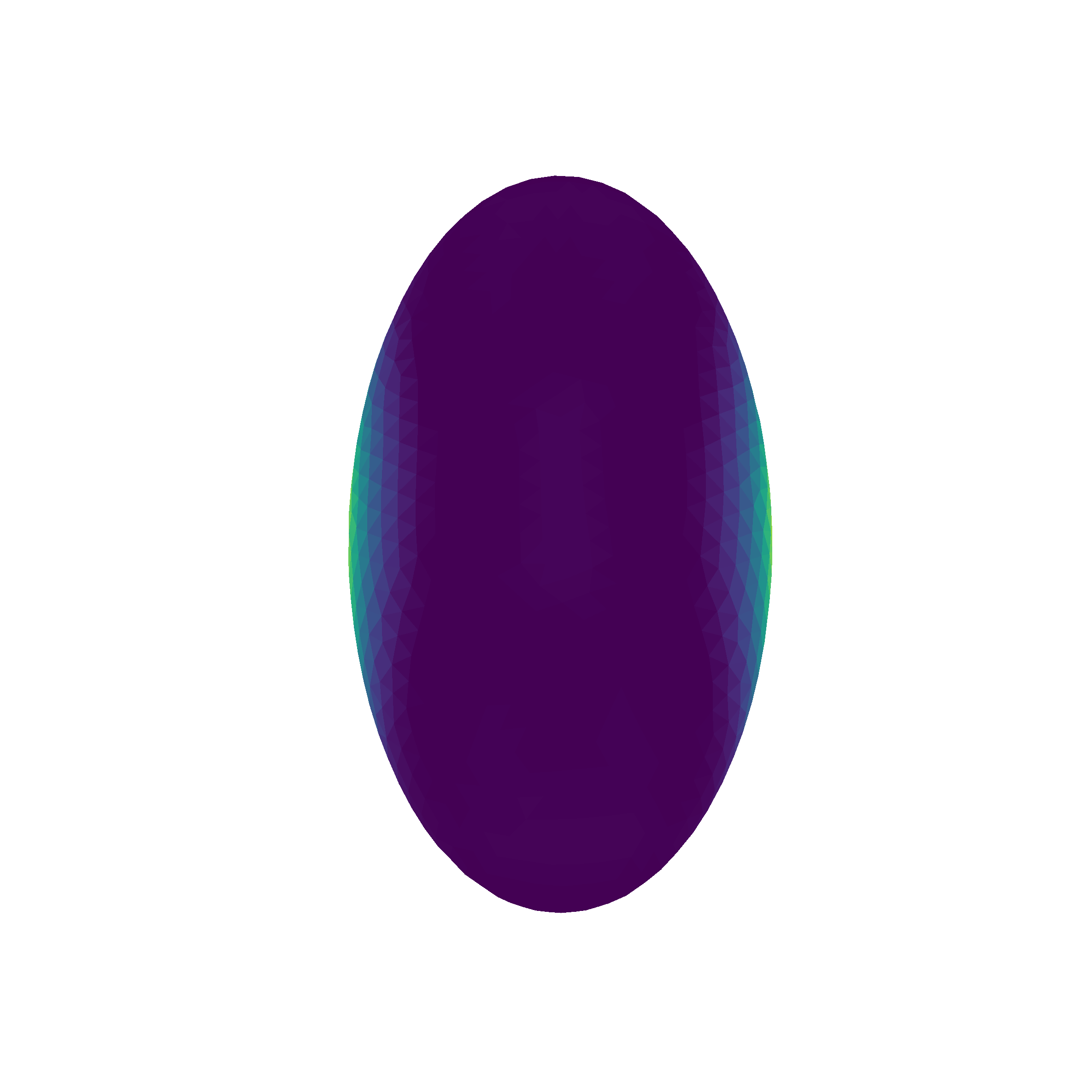}
    \includegraphics[width=2.4cm]{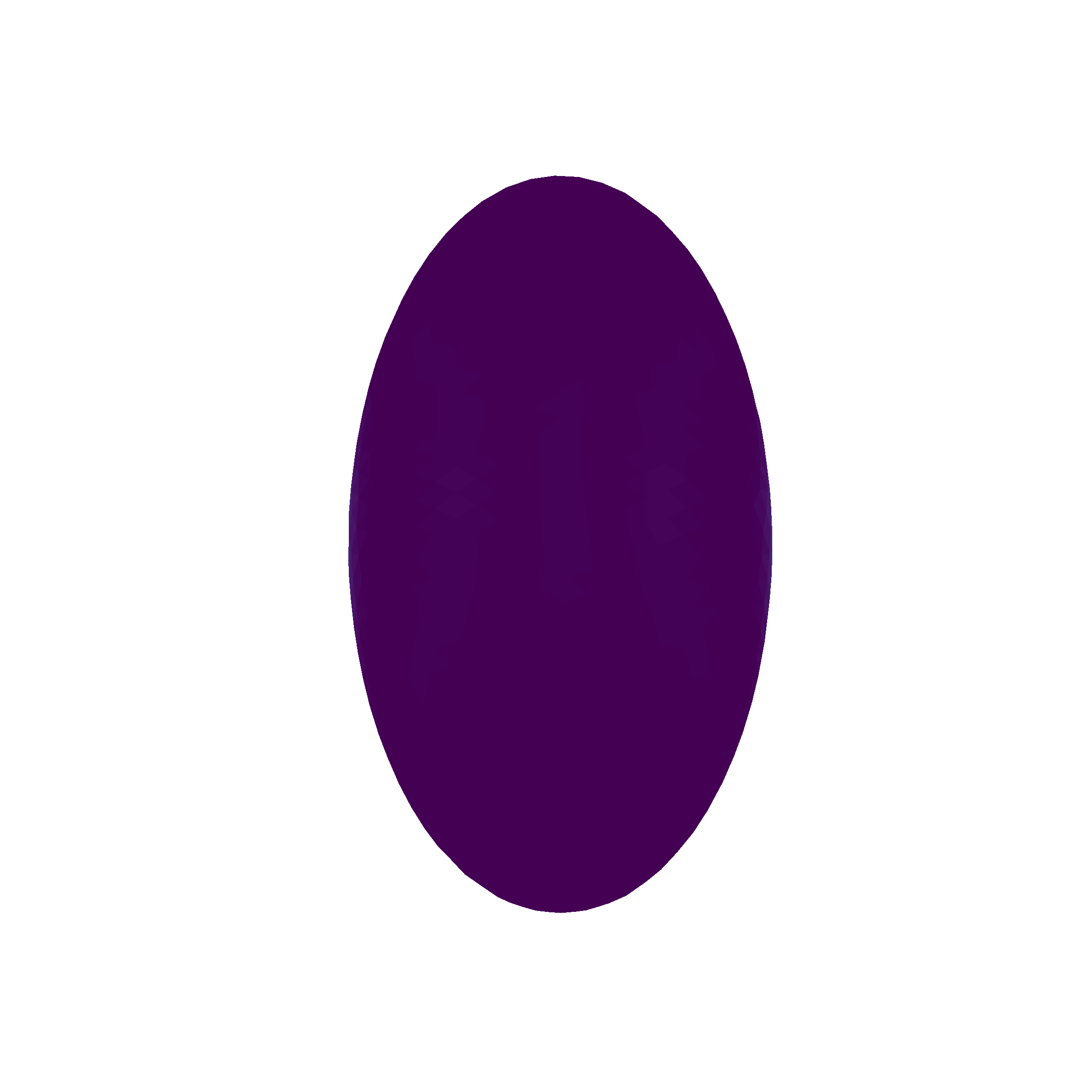}
    \includegraphics[width=2.9cm]{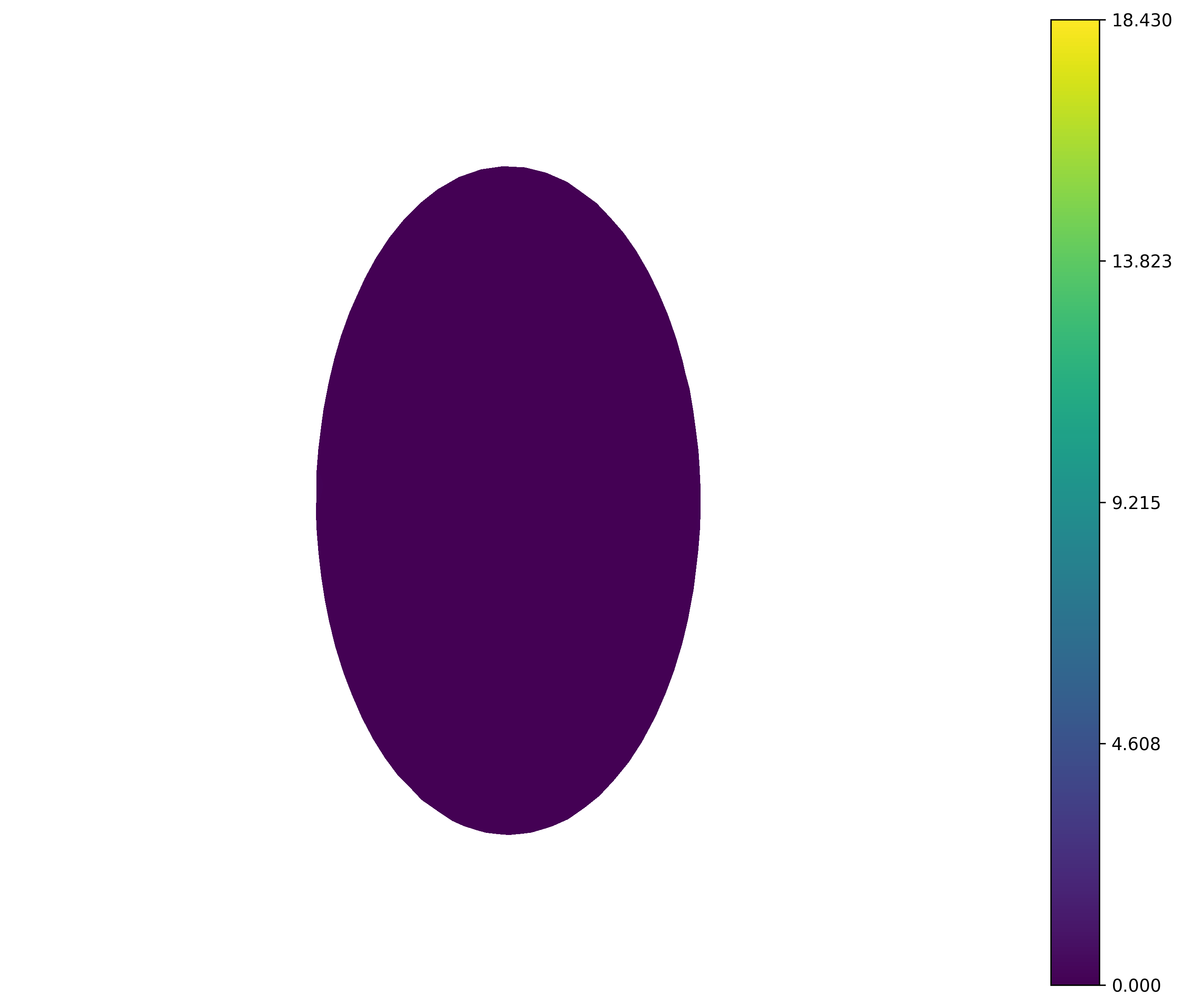}\\
    \vspace{5pt}
    
    \includegraphics[width=2.4cm]{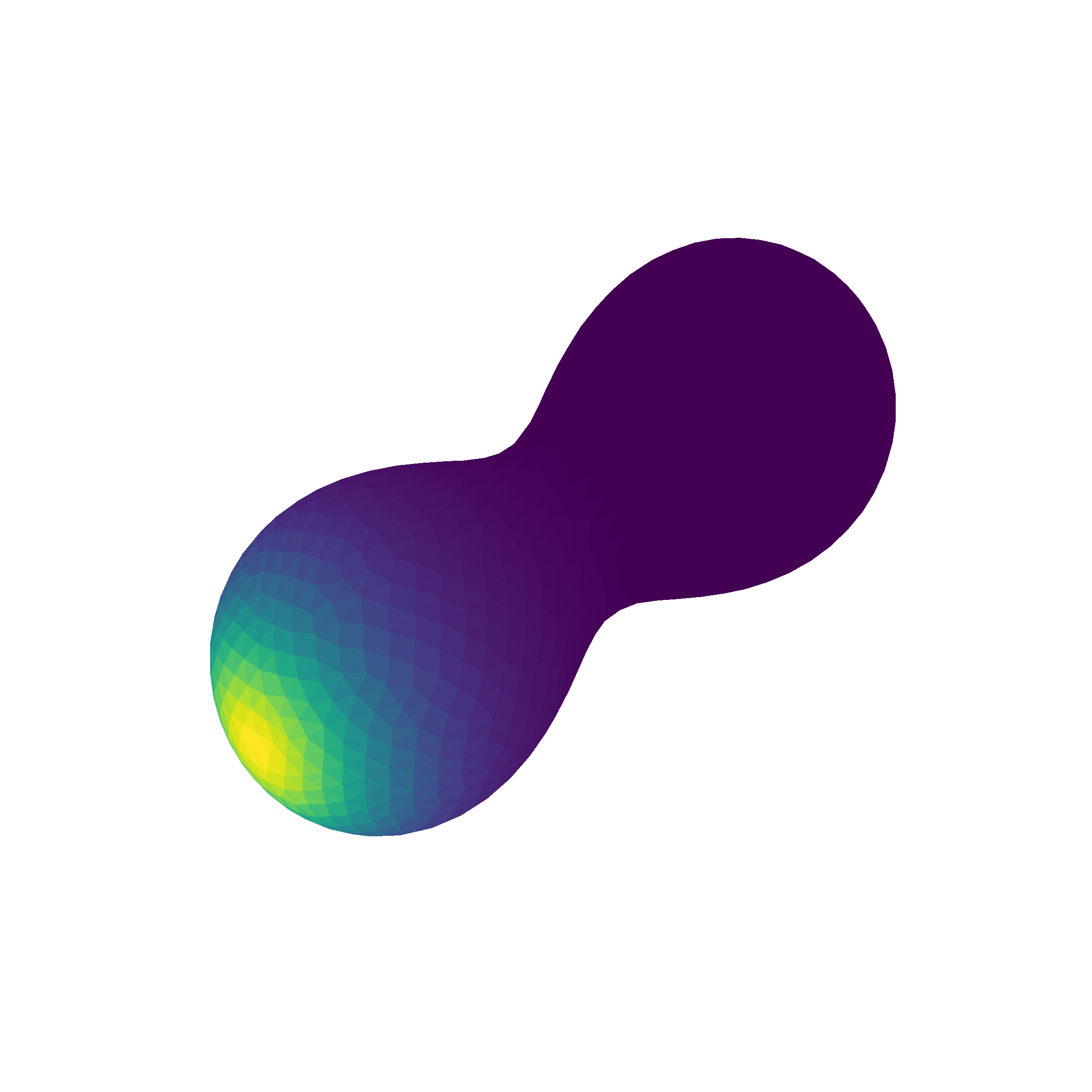}
    \includegraphics[width=2.4cm]{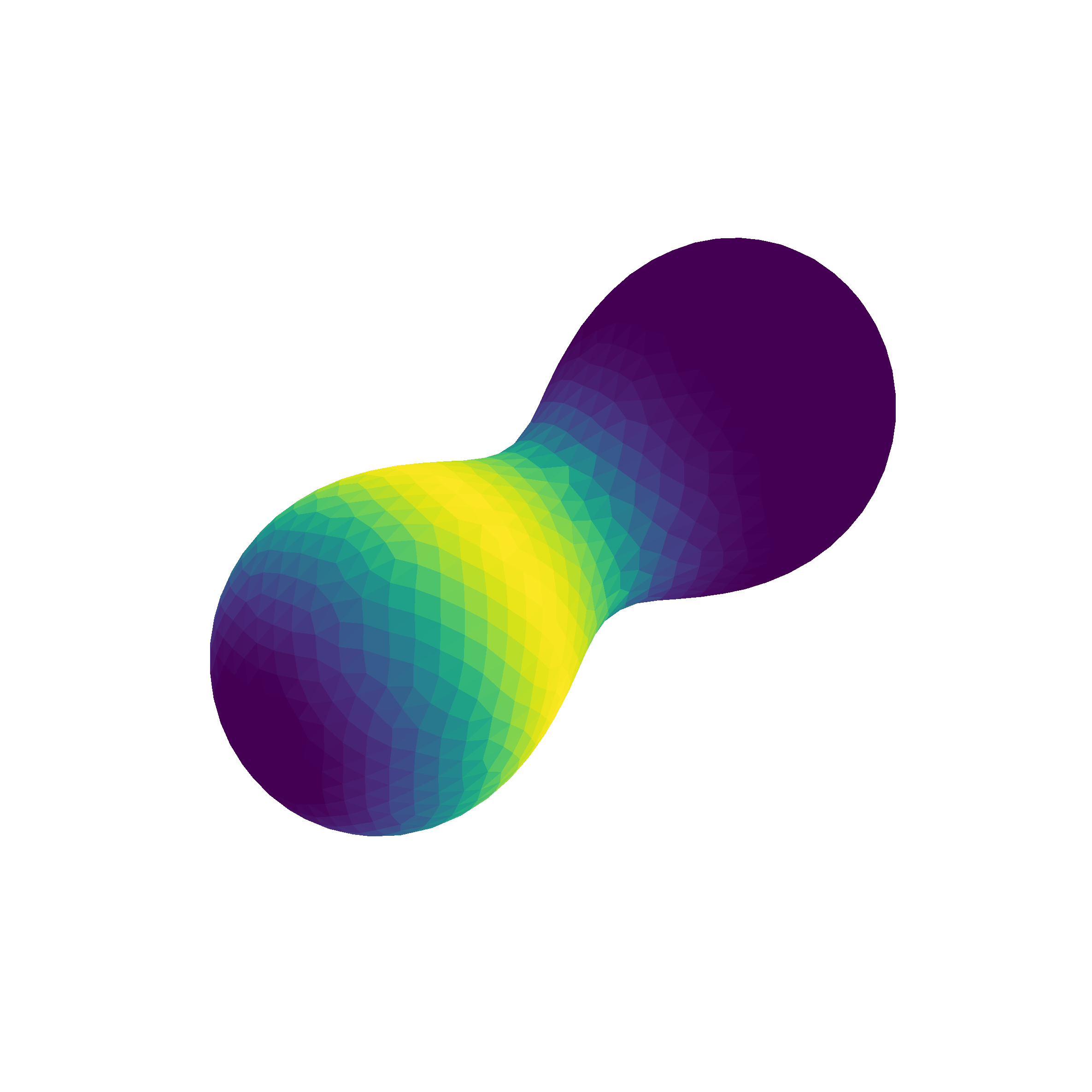}
    \includegraphics[width=2.4cm]{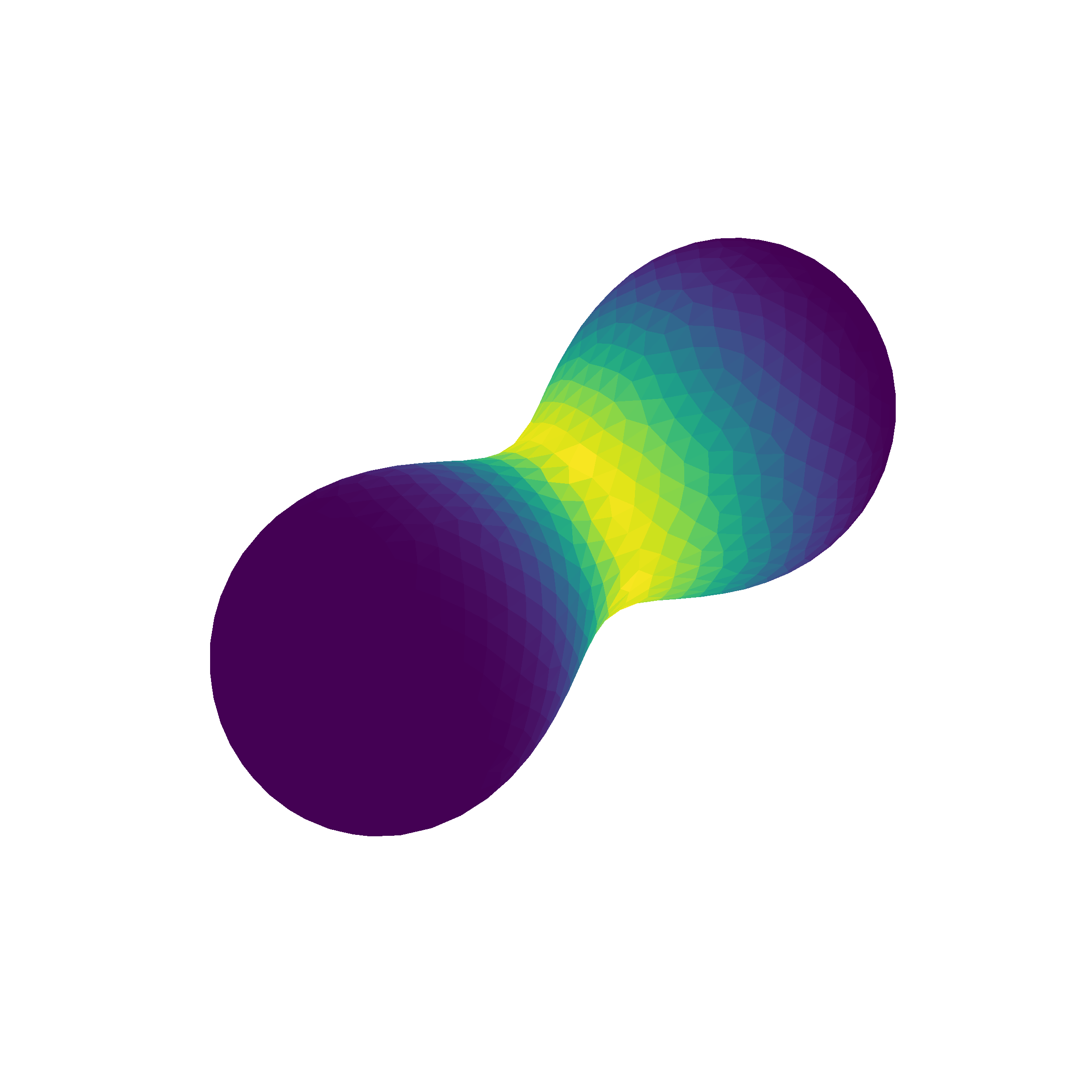}
    \includegraphics[width=2.4cm]{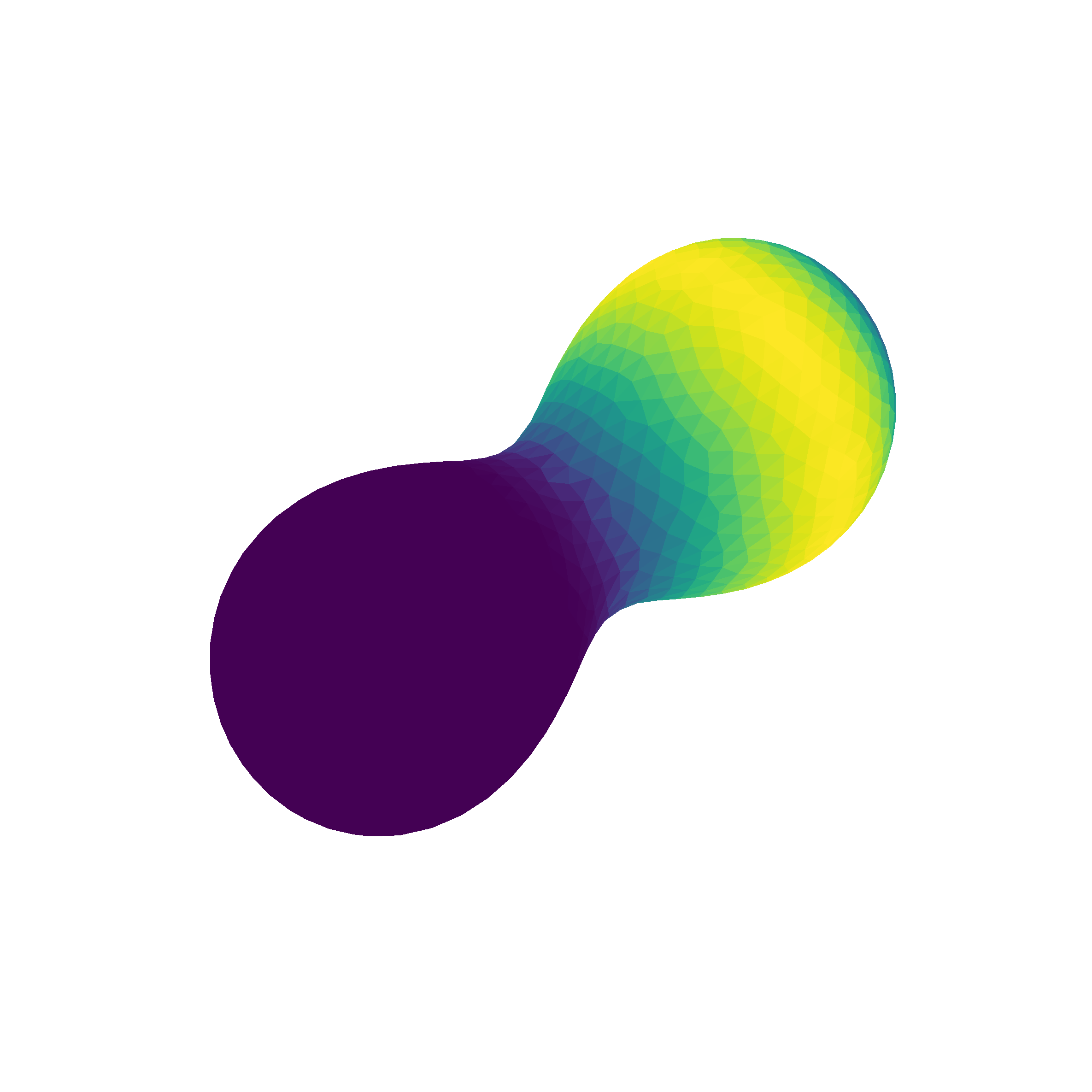}
    \includegraphics[width=2.9cm]{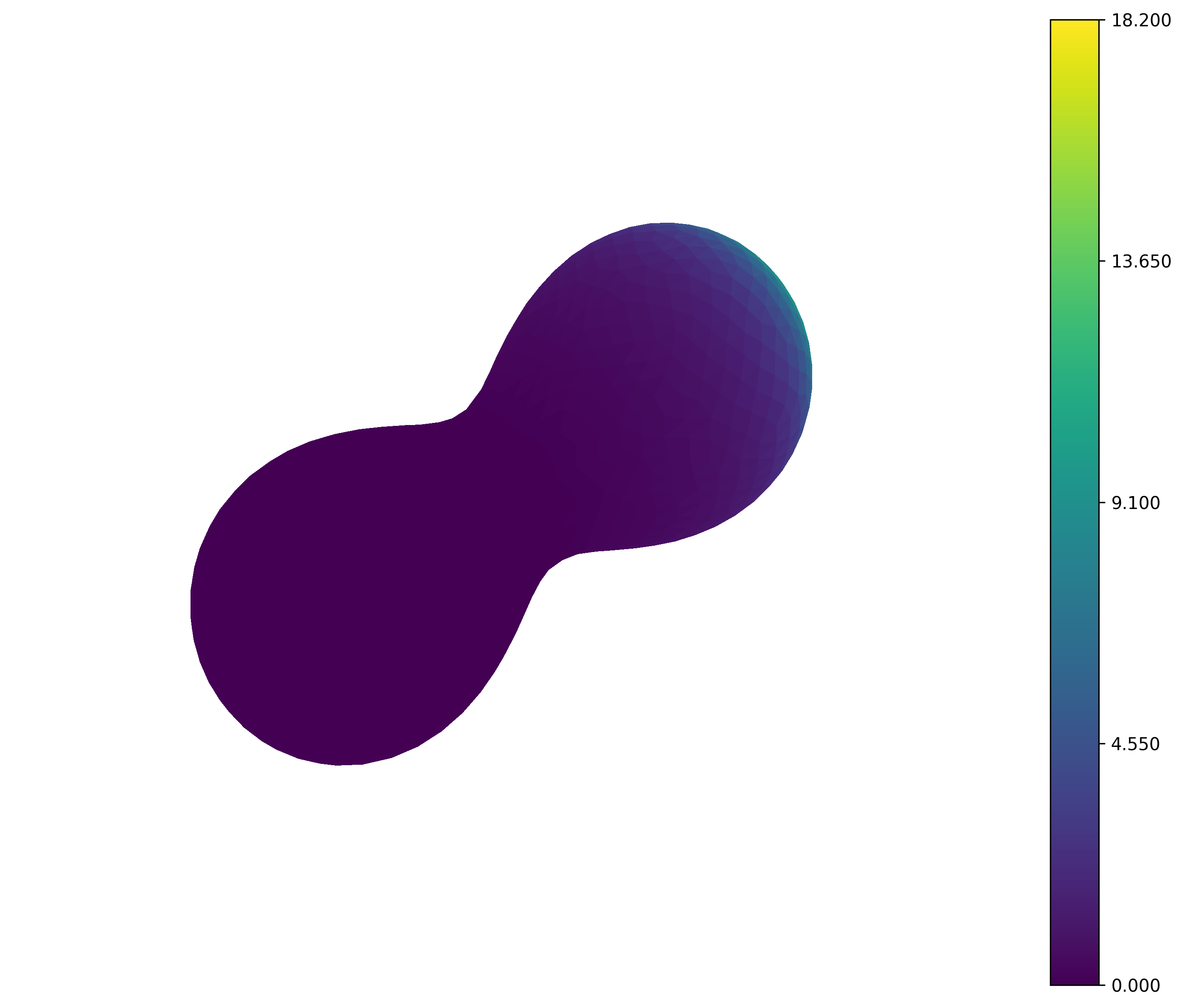}\\
    \vspace{5pt}

    \includegraphics[width=2.4cm]{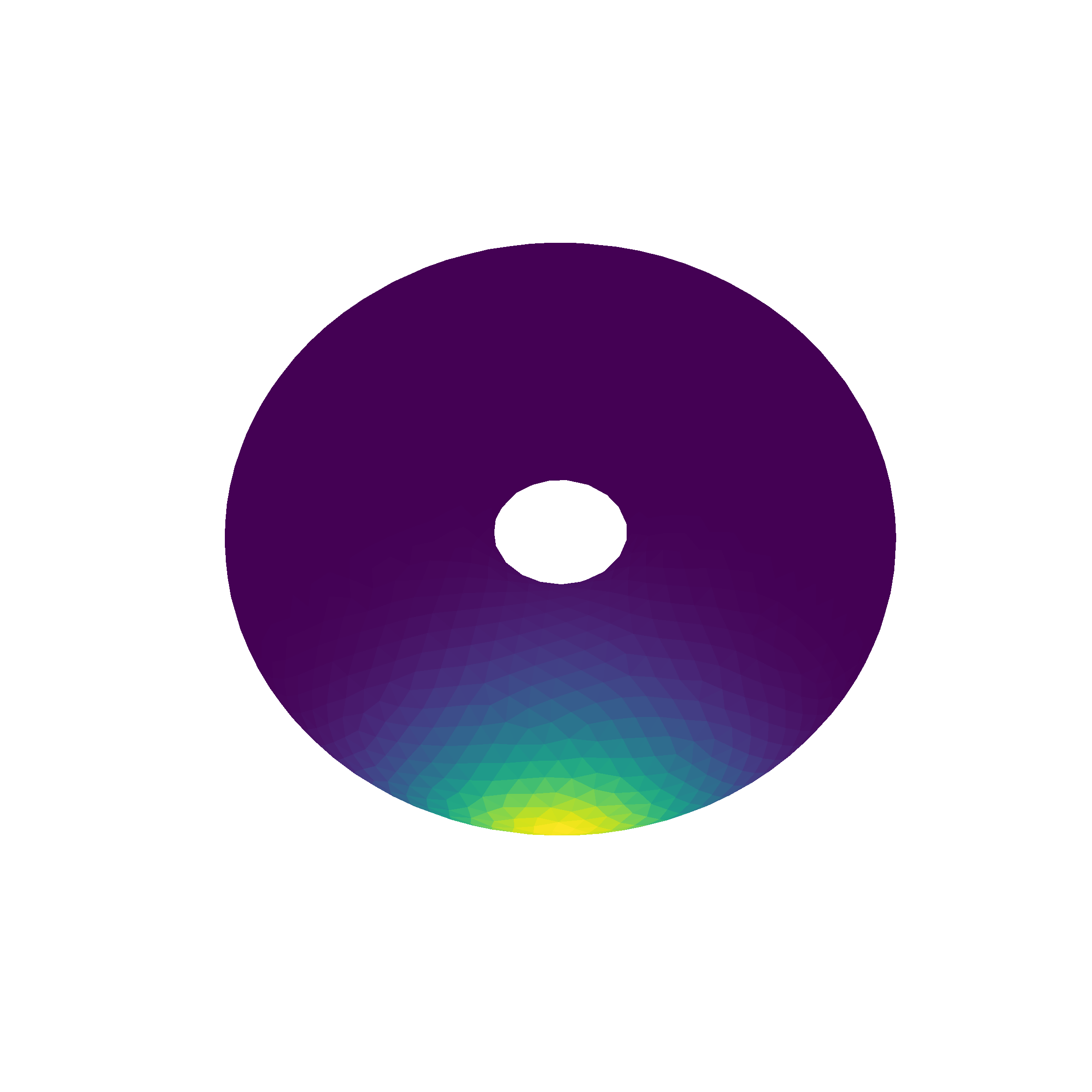}
    \includegraphics[width=2.4cm]{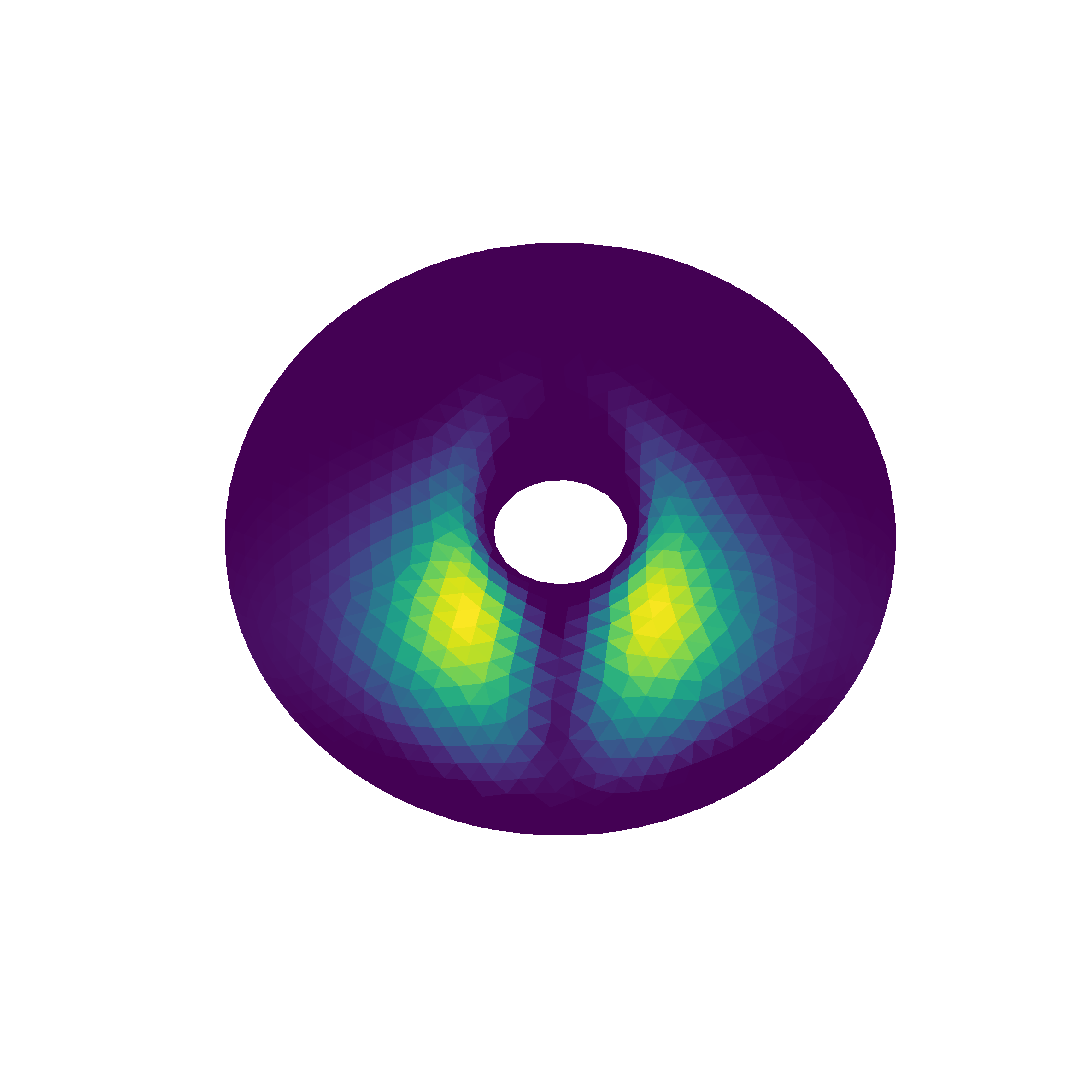}
    \includegraphics[width=2.4cm]{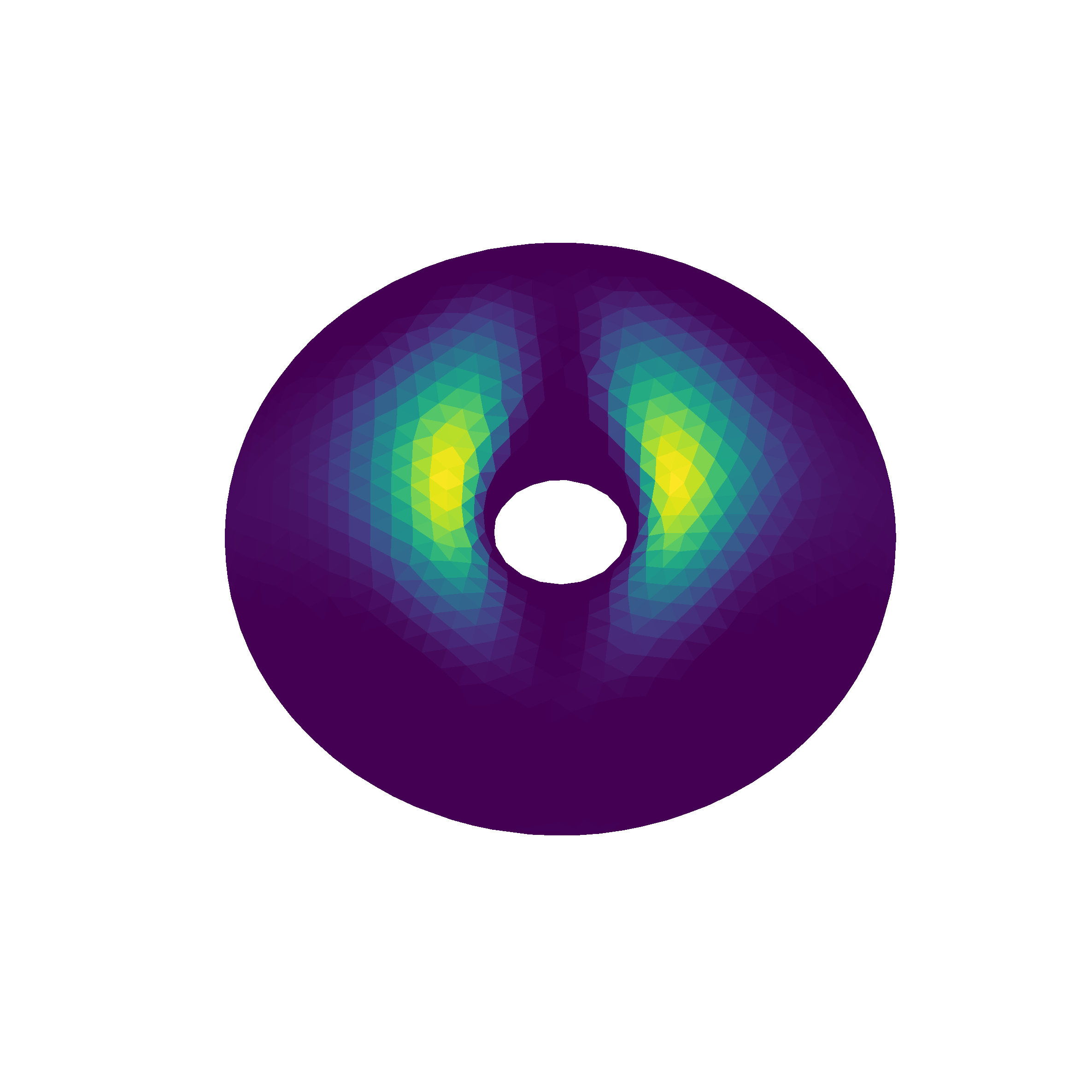}
    \includegraphics[width=2.4cm]{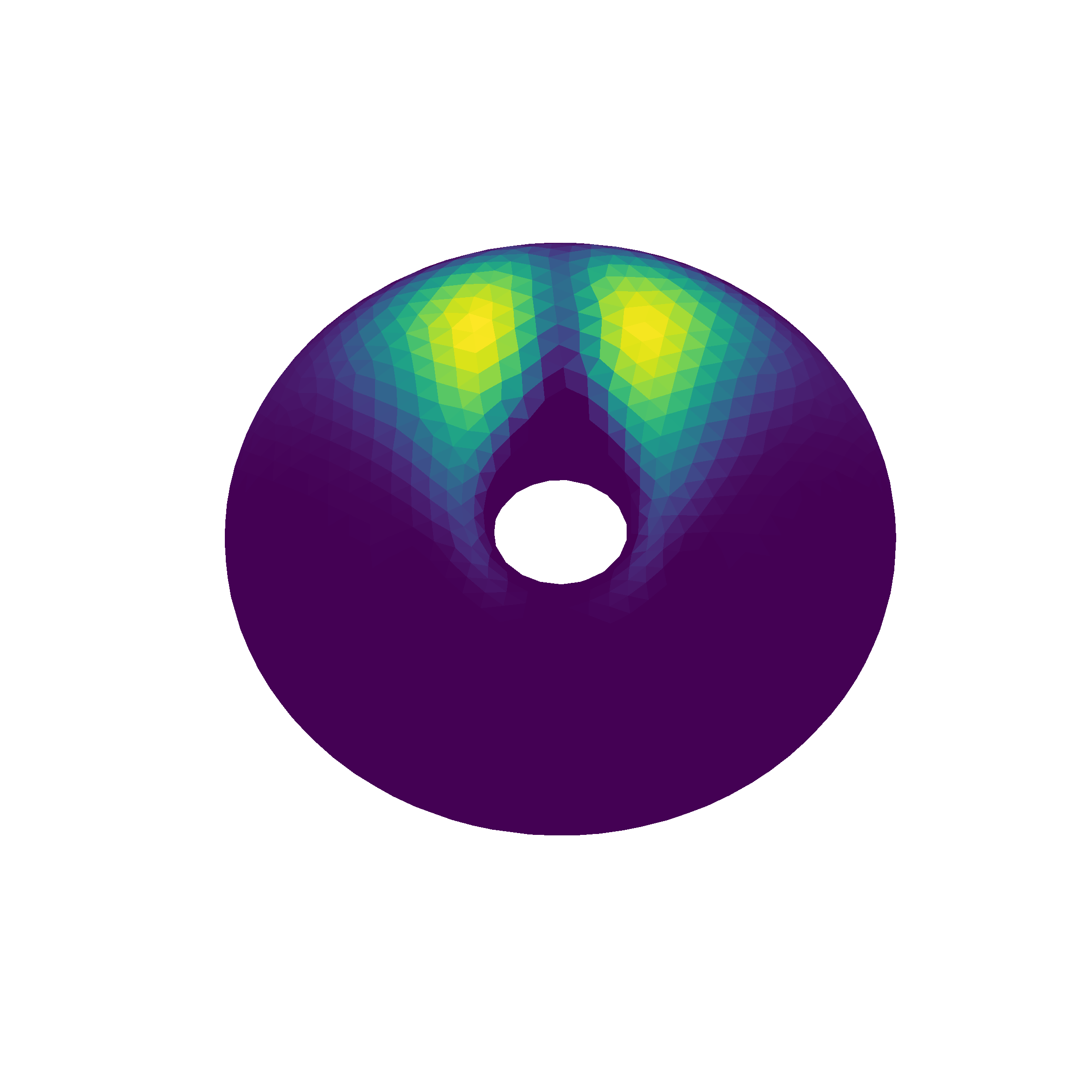}
    \includegraphics[width=2.9cm]{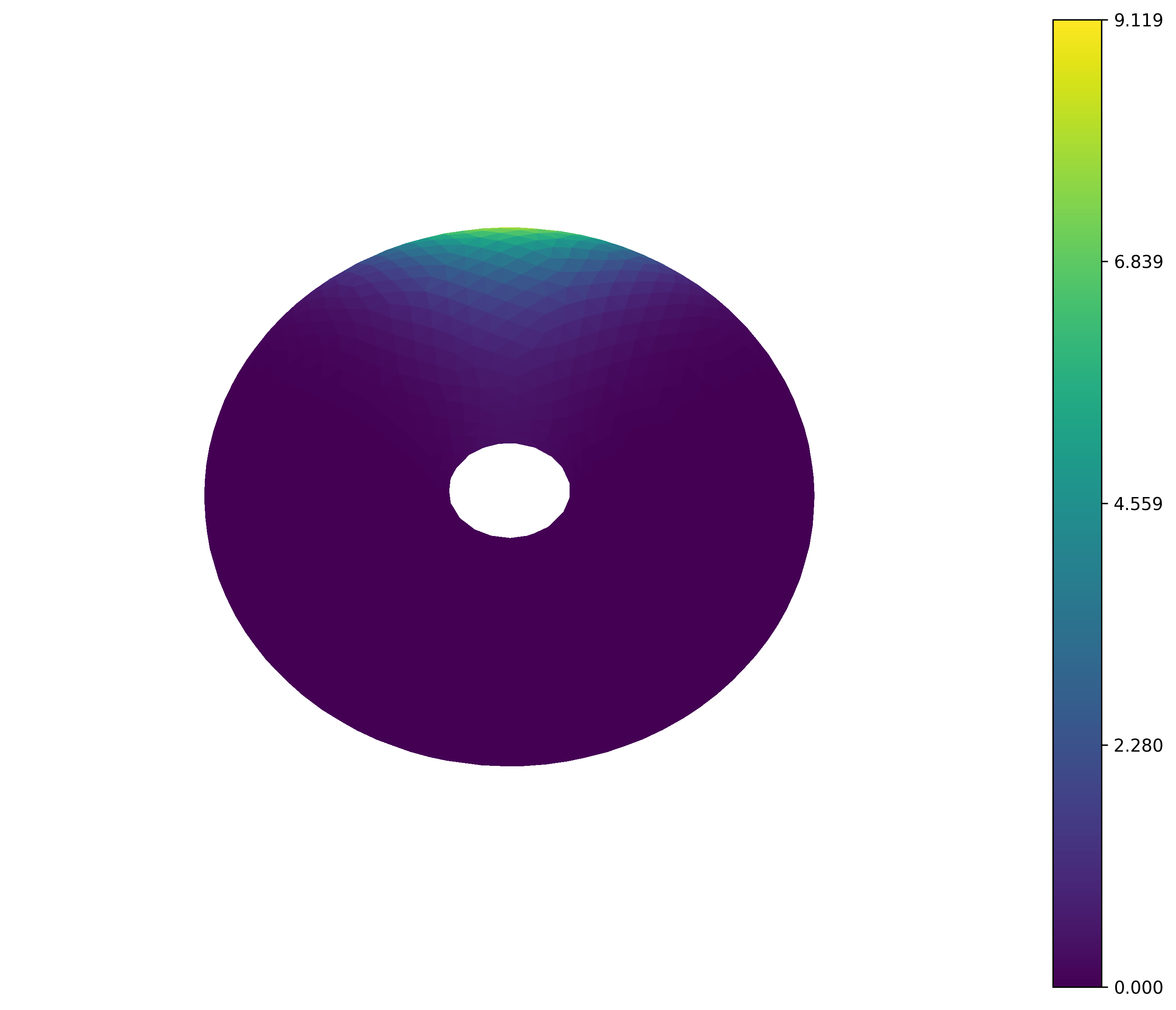}\\
    \vspace{5pt}

    \subfigure[$\rho(0, \boldsymbol{x})$]{
    \includegraphics[width=2.4cm]{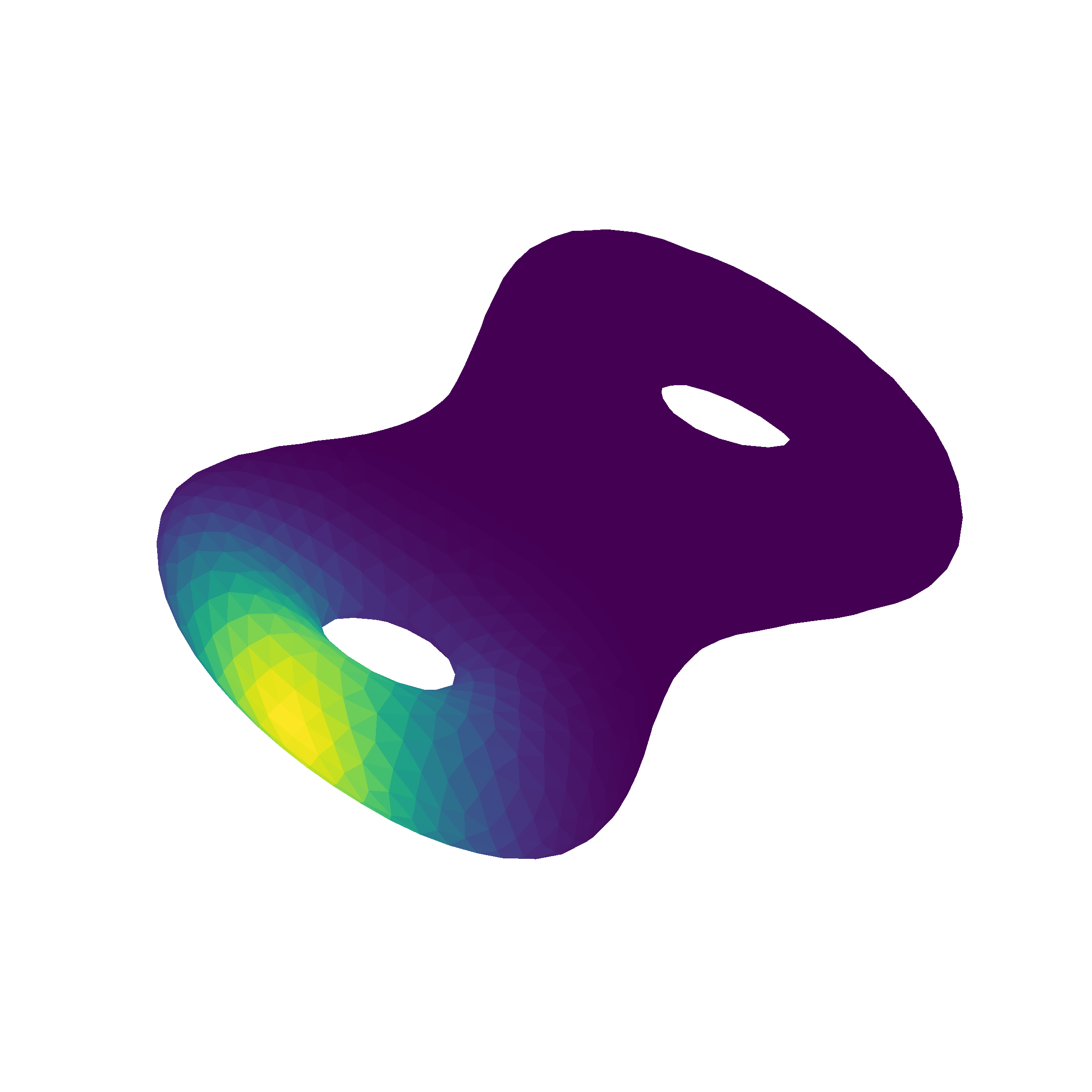}}
    \subfigure[$\rho(0.25, \boldsymbol{x})$]{
    \includegraphics[width=2.4cm]{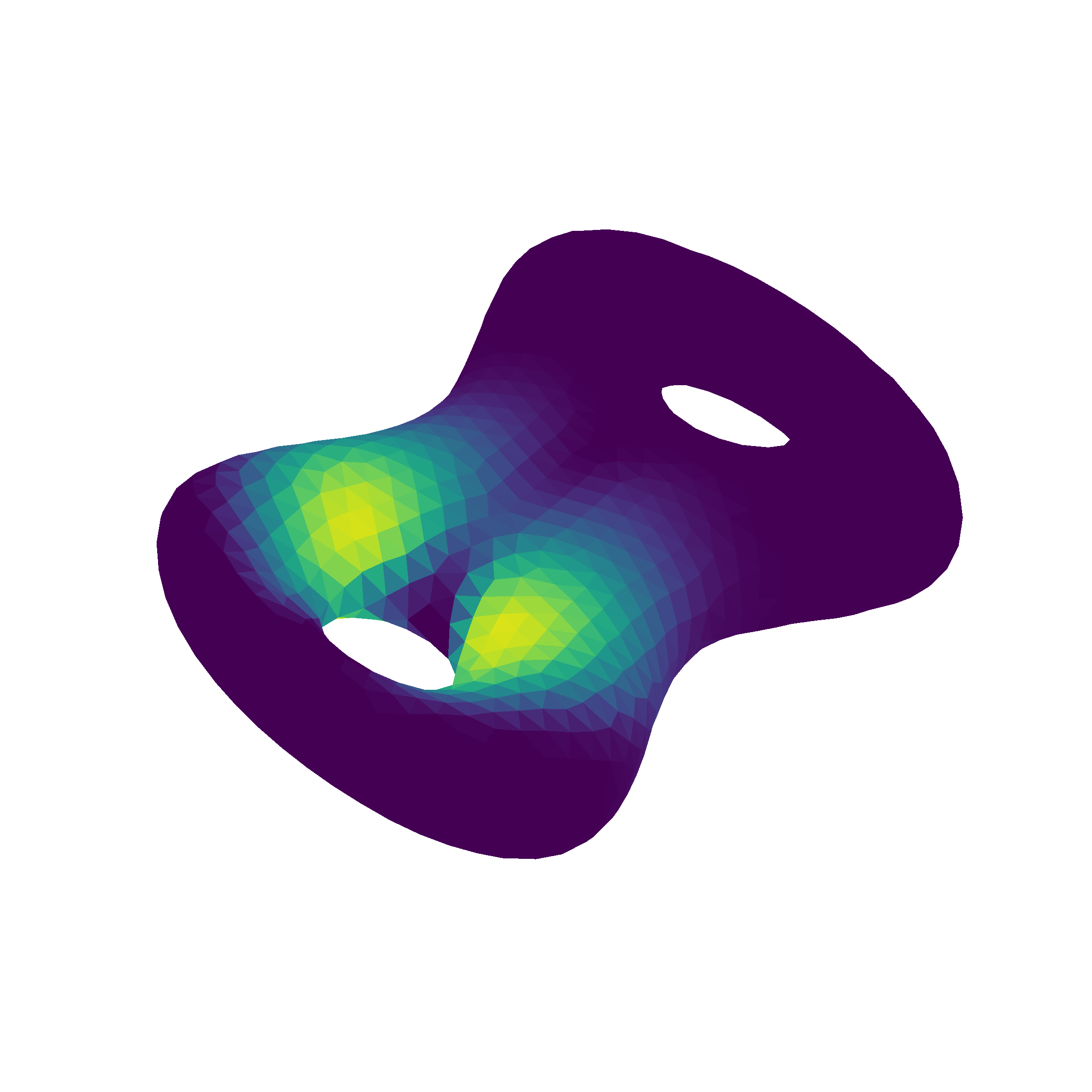}}
    \subfigure[$\rho(0.5, \boldsymbol{x})$]{
    \includegraphics[width=2.4cm]{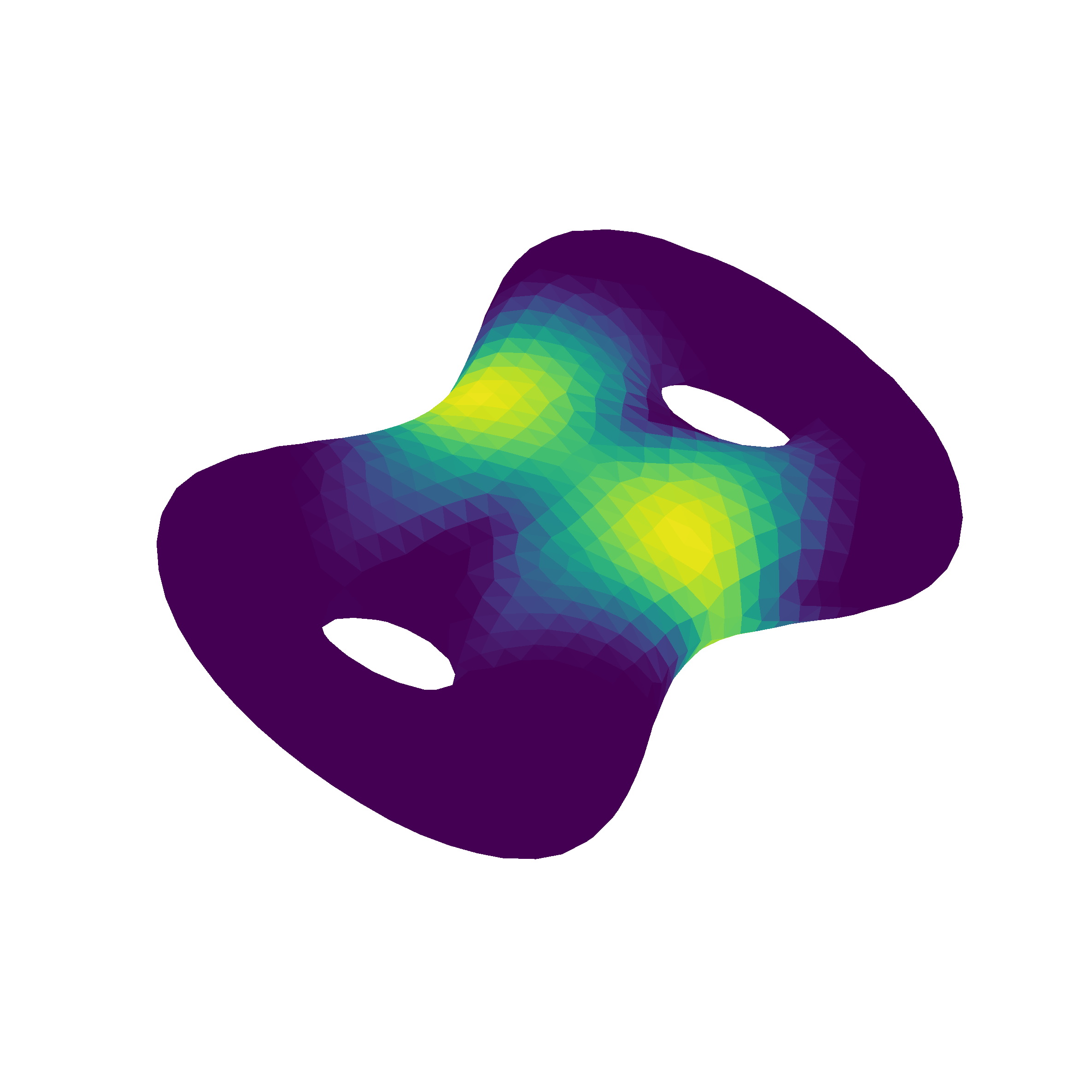}}
    \subfigure[$\rho(0.75, \boldsymbol{x})$]{
    \includegraphics[width=2.4cm]{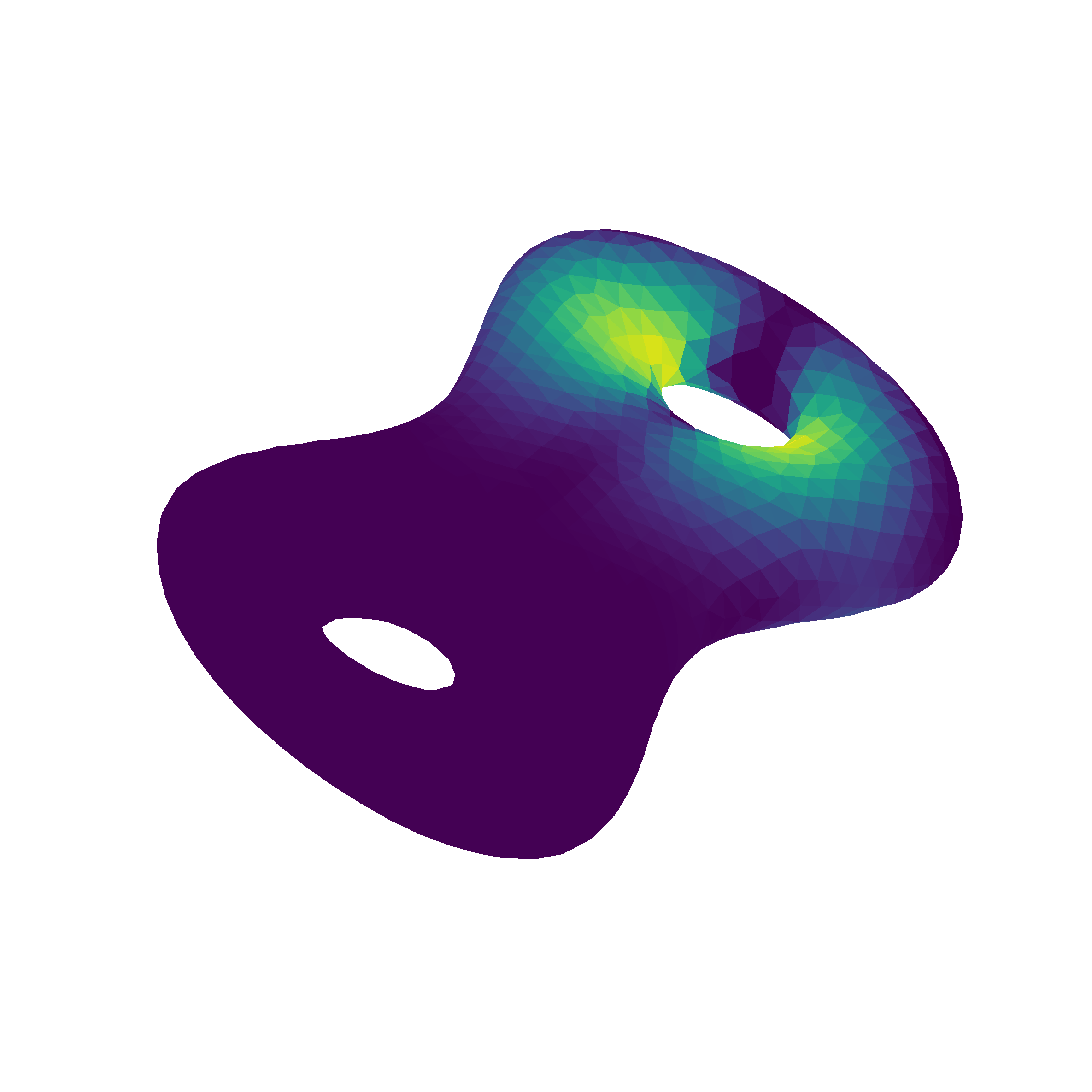}}
    \subfigure[$\rho(1, \boldsymbol{x})$]{
    \includegraphics[width=2.9cm]{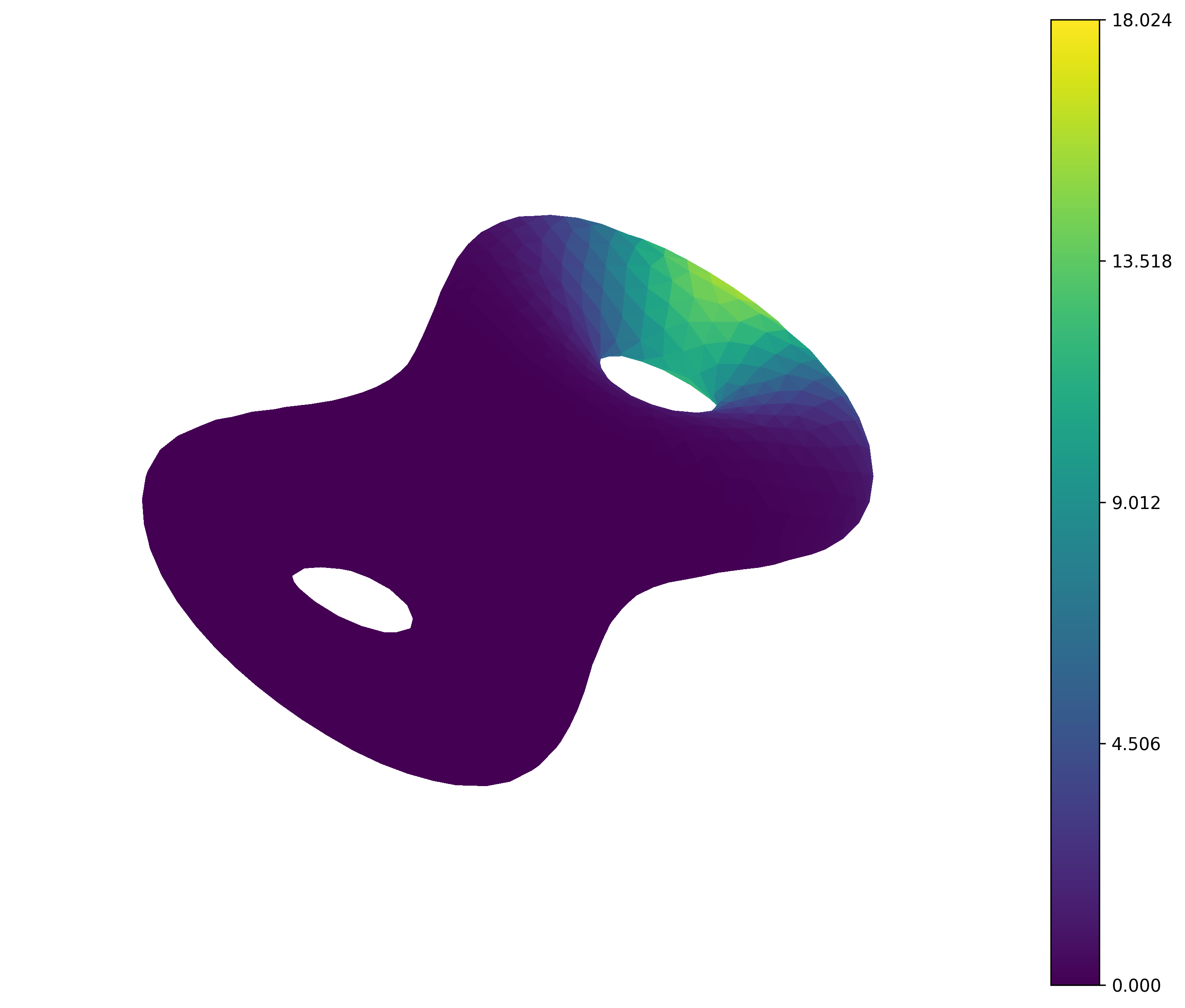}}\\
    \caption{SUOT examples ($\beta=1.5$) and the calculation times are 450s, 160s, 236s, 287s and 192s respectively.}
    \label{SUOT-LS}
    \end{center}
\end{figure}

\begin{figure}[htbp]
    \begin{center}
    \includegraphics[width=2.4cm]{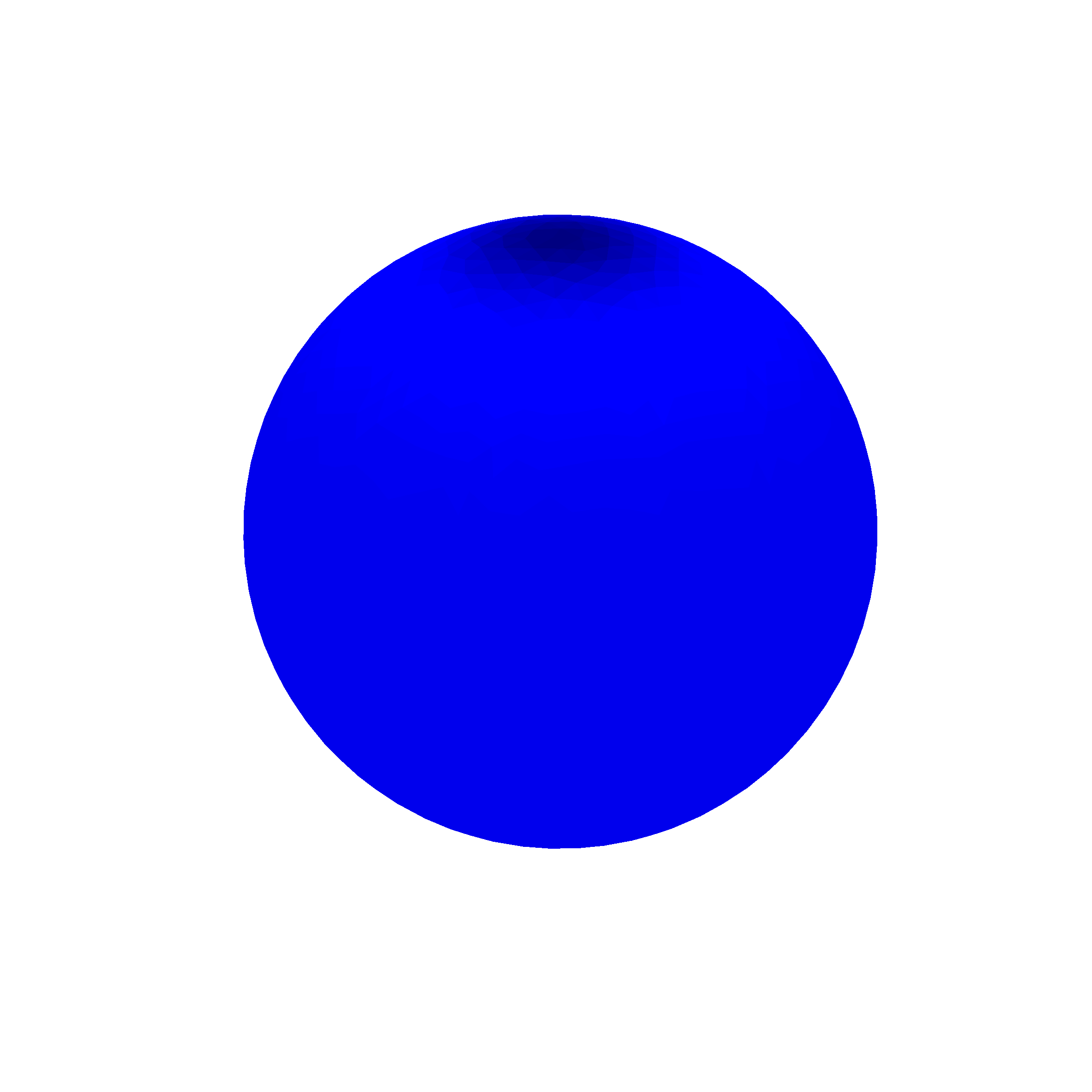}
    \includegraphics[width=2.4cm]{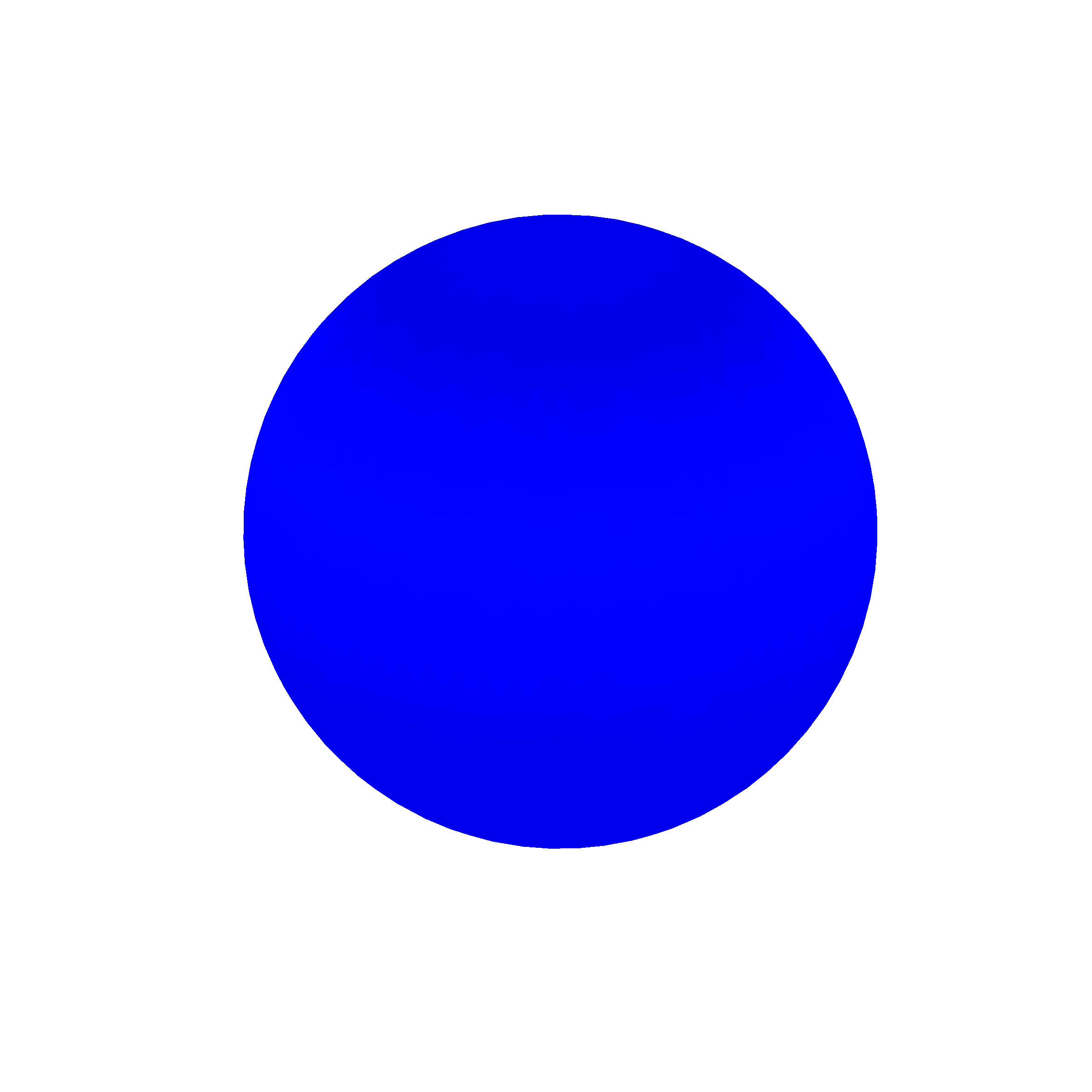}
    \includegraphics[width=2.4cm]{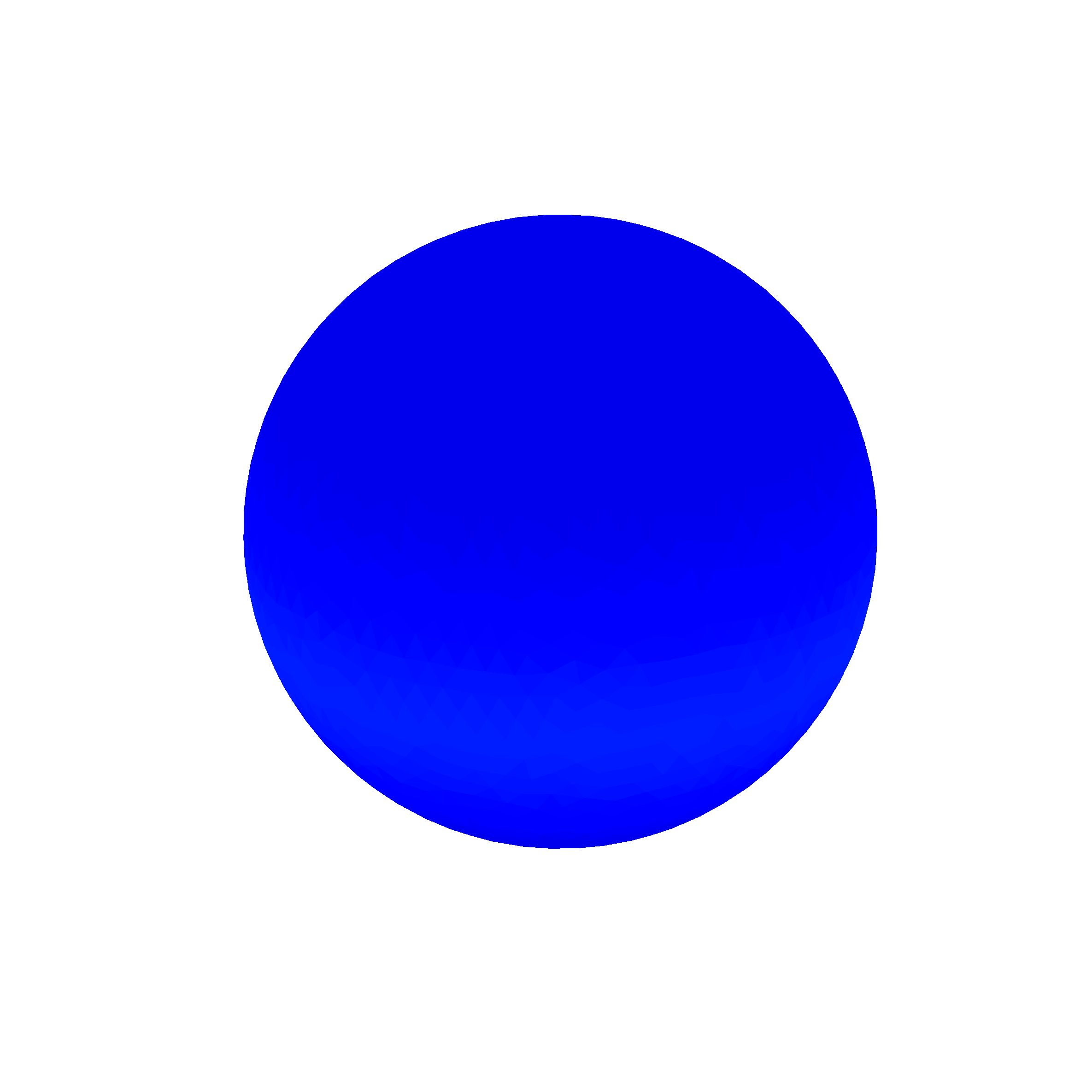}
    \includegraphics[width=2.4cm]{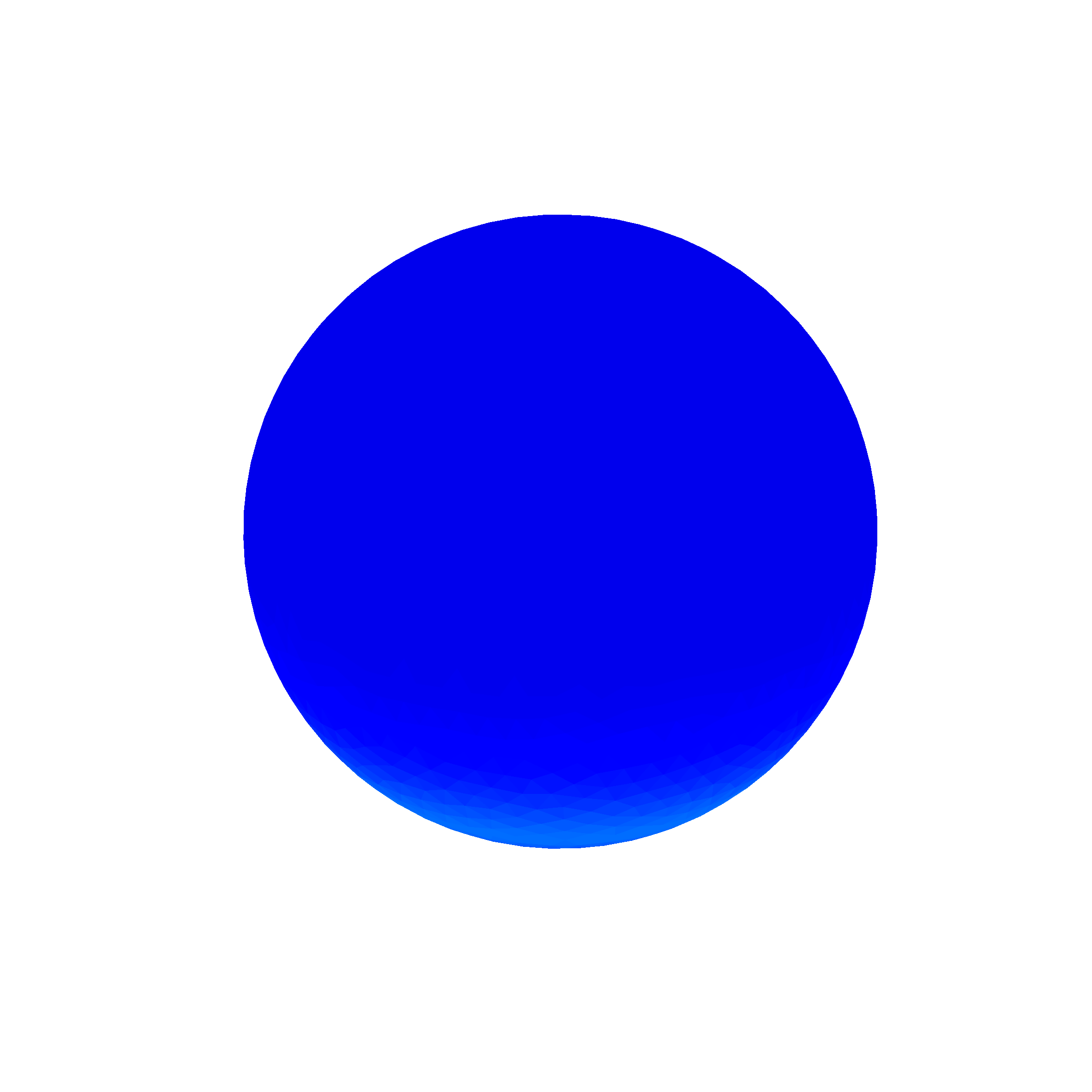}
    \includegraphics[width=2.9cm]{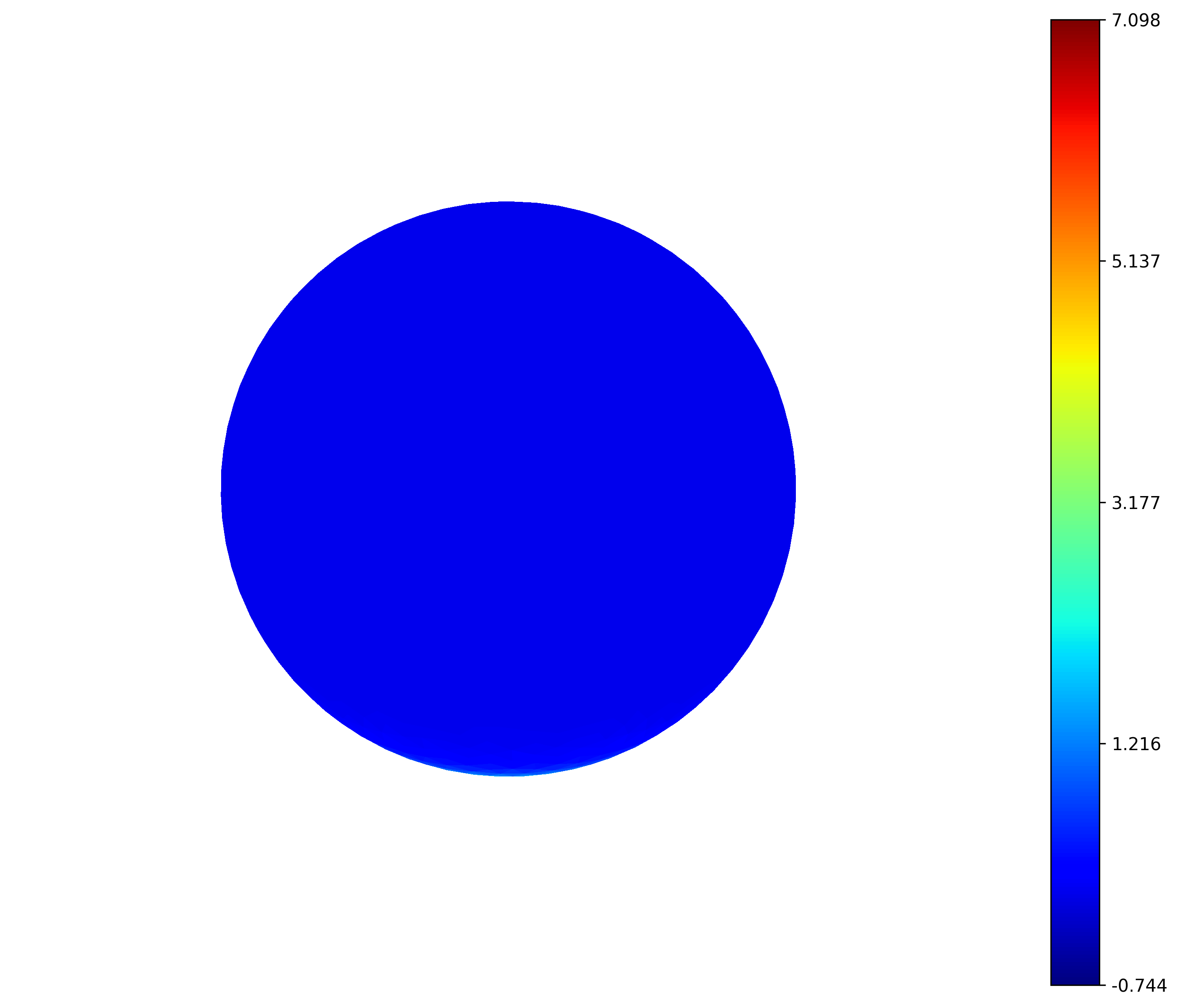}\\
    \vspace{5pt}

    \includegraphics[width=2.4cm]{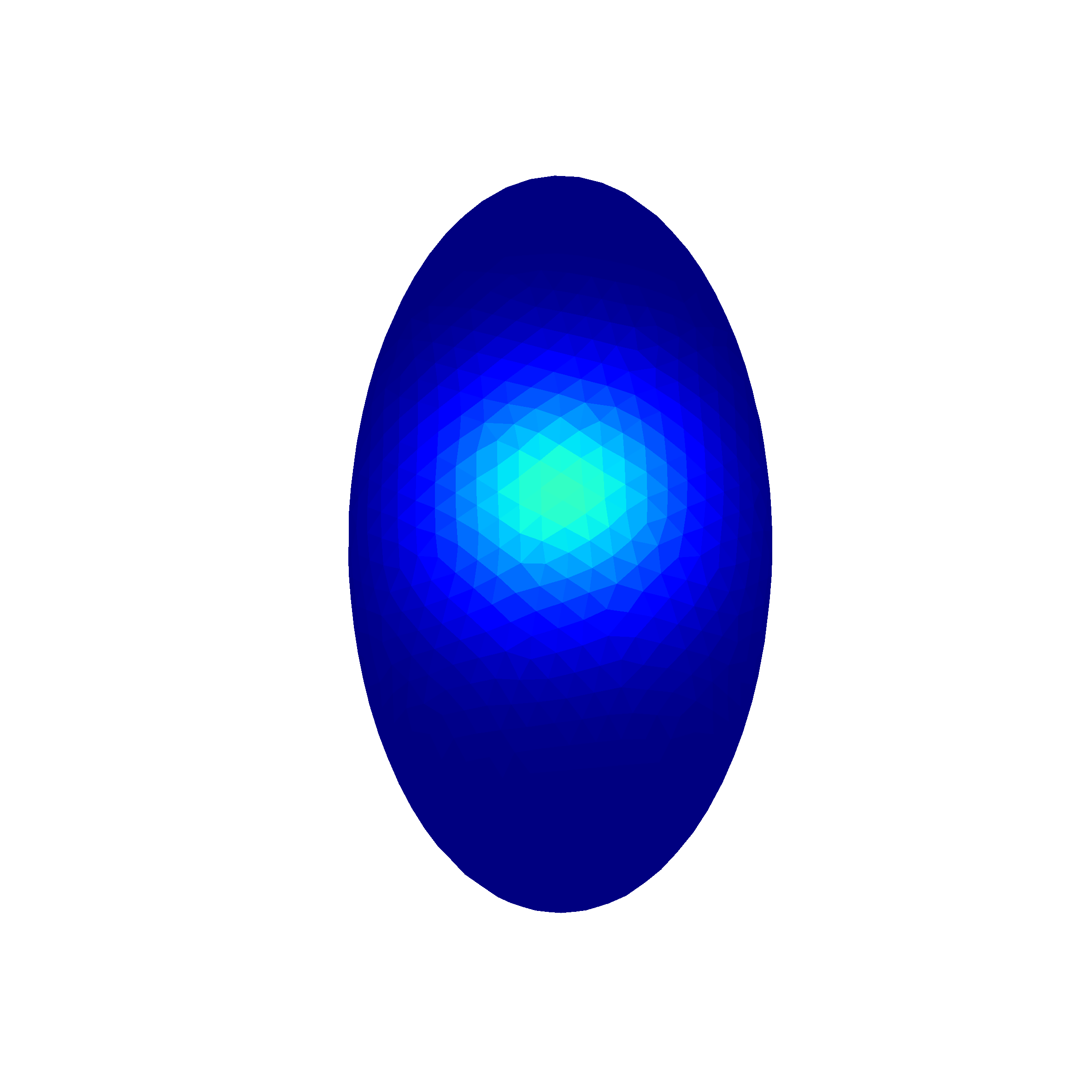}
    \includegraphics[width=2.4cm]{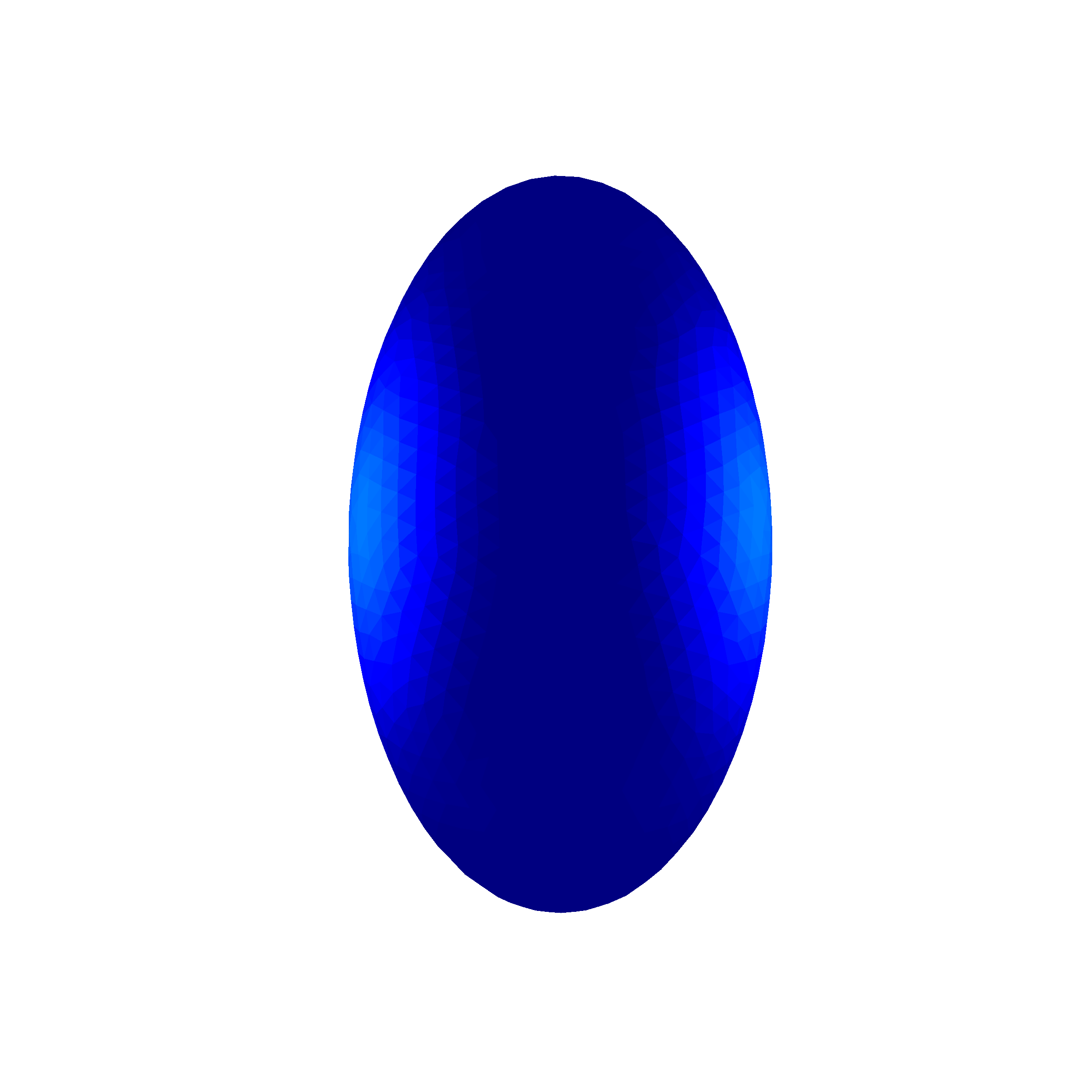}
    \includegraphics[width=2.4cm]{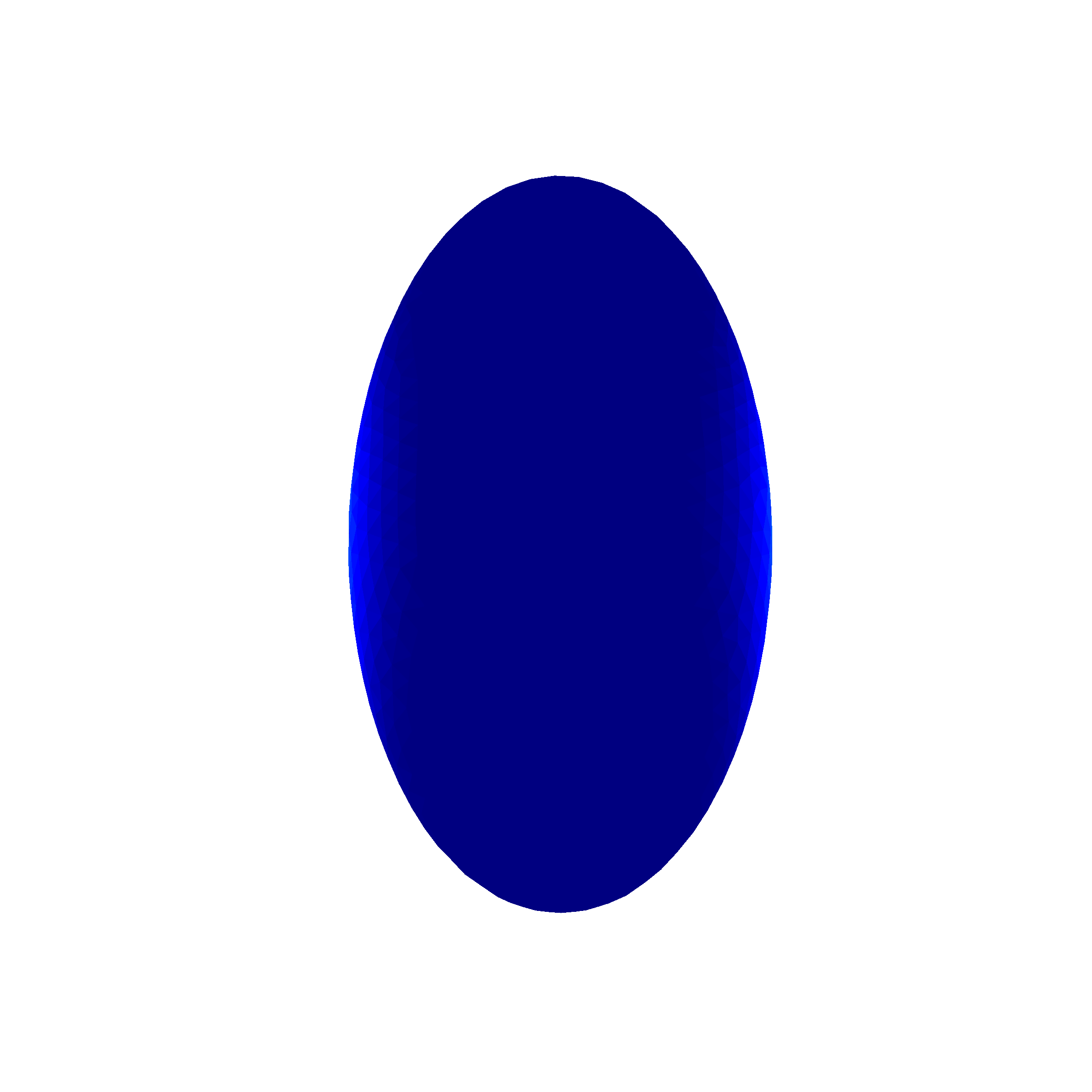}
    \includegraphics[width=2.4cm]{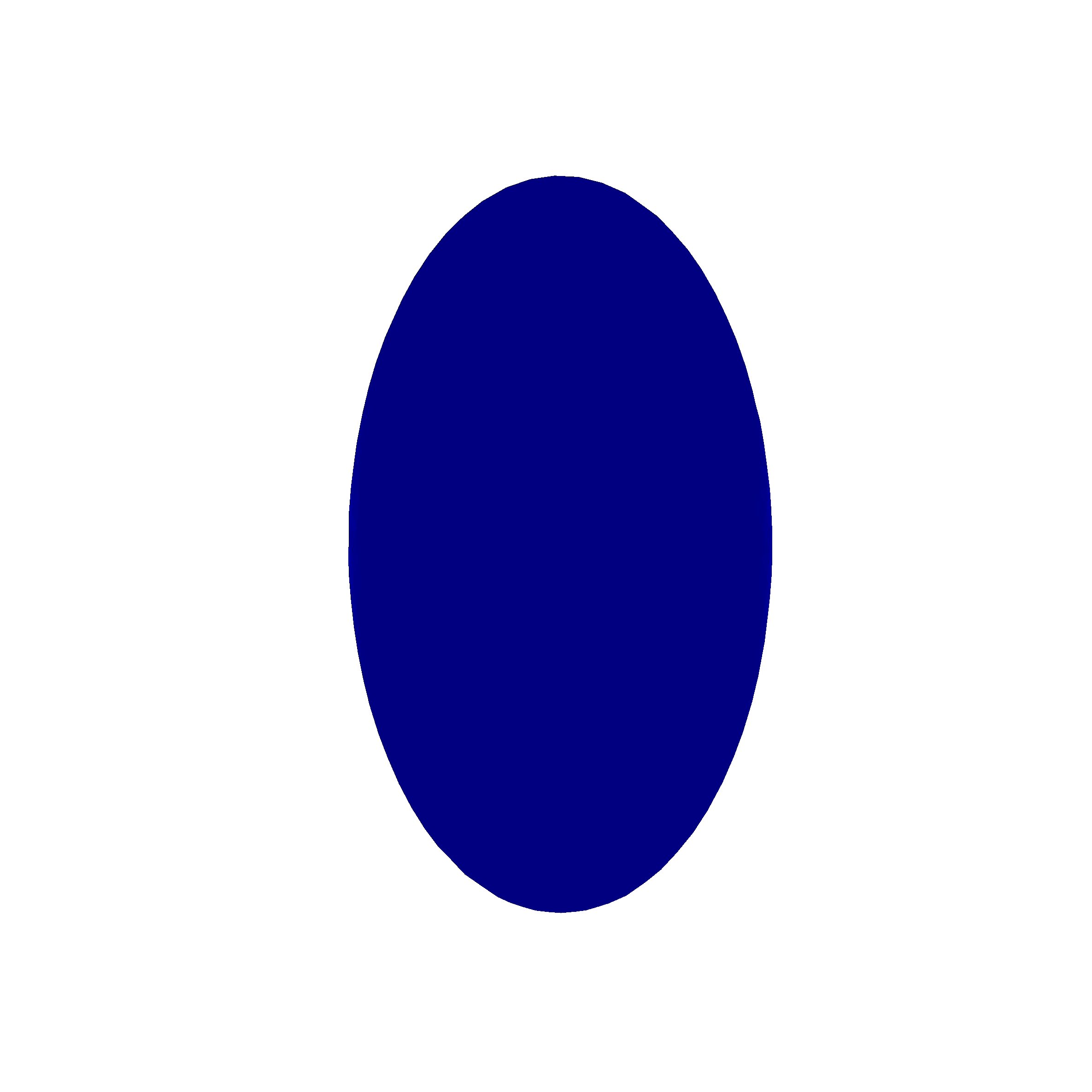}
    \includegraphics[width=2.9cm]{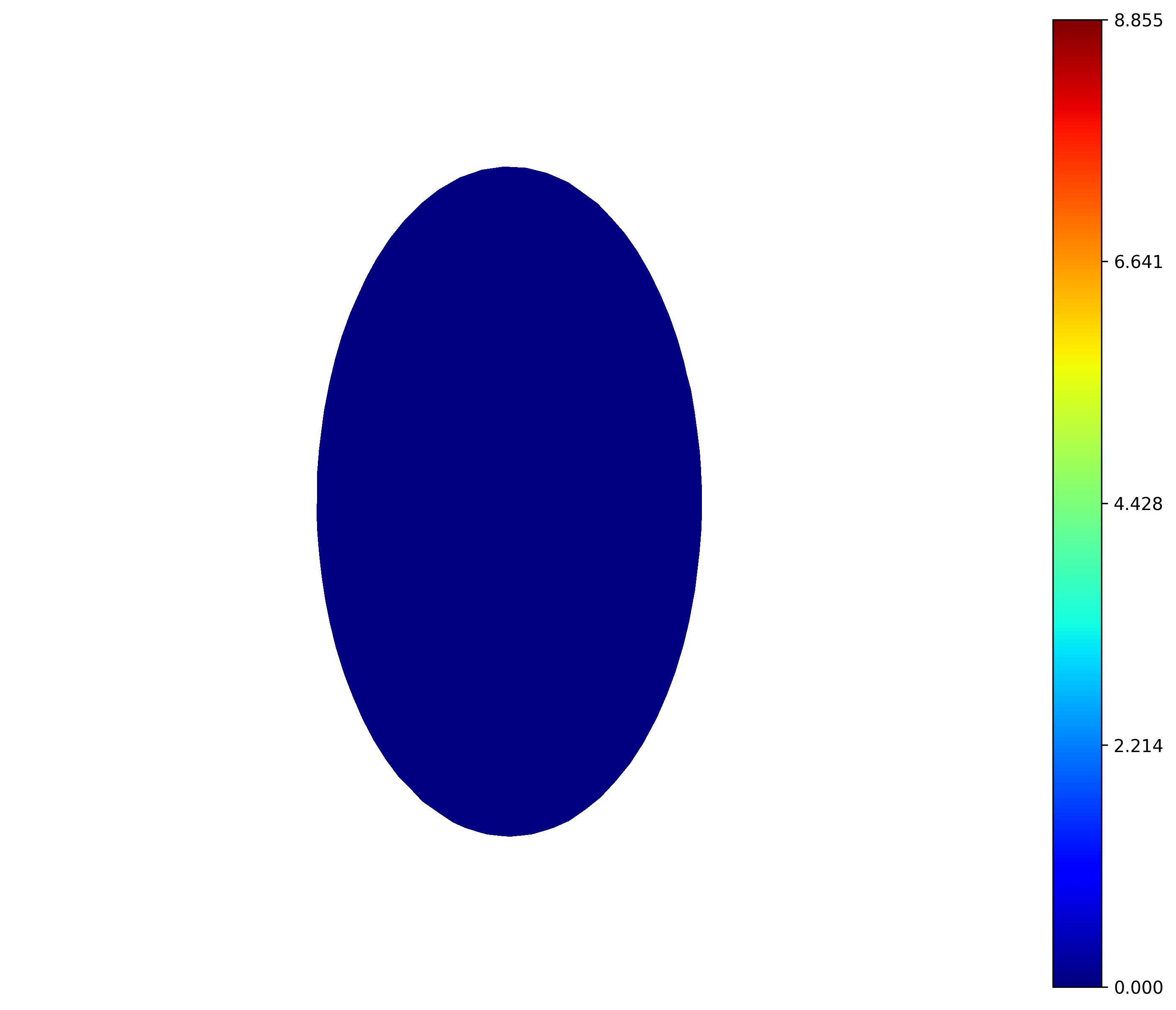}\\
    \vspace{5pt}
    
    \includegraphics[width=2.4cm]{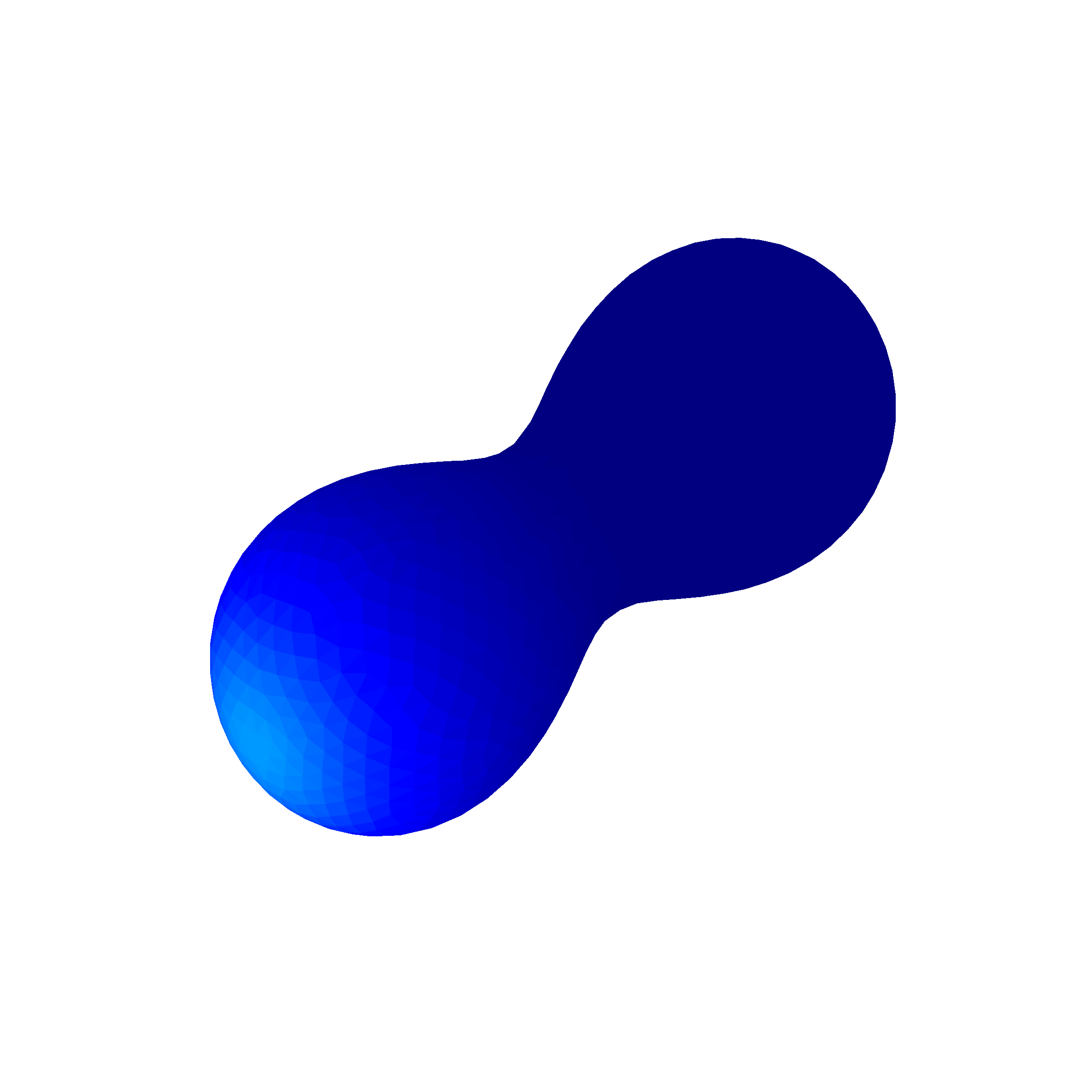}
    \includegraphics[width=2.4cm]{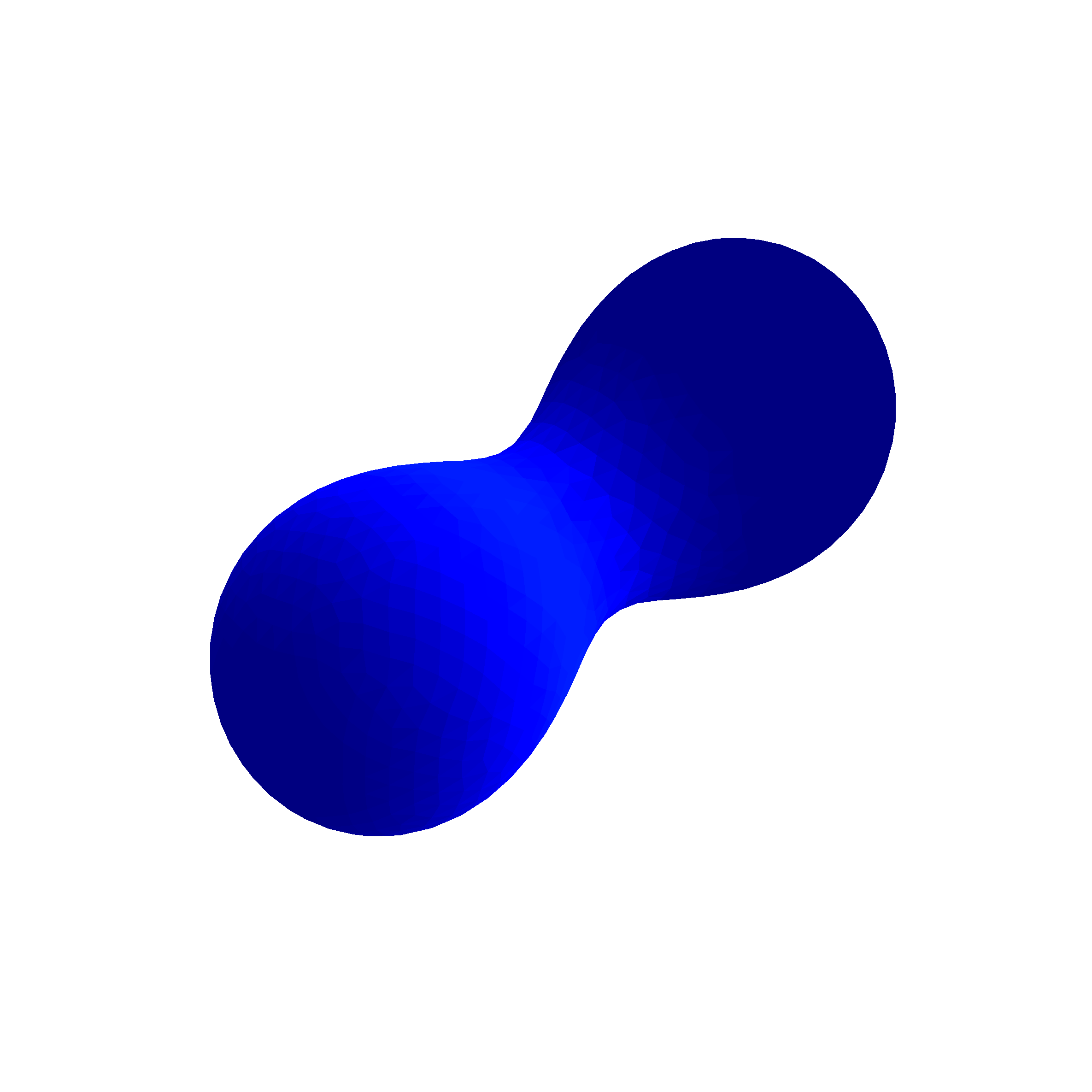}
    \includegraphics[width=2.4cm]{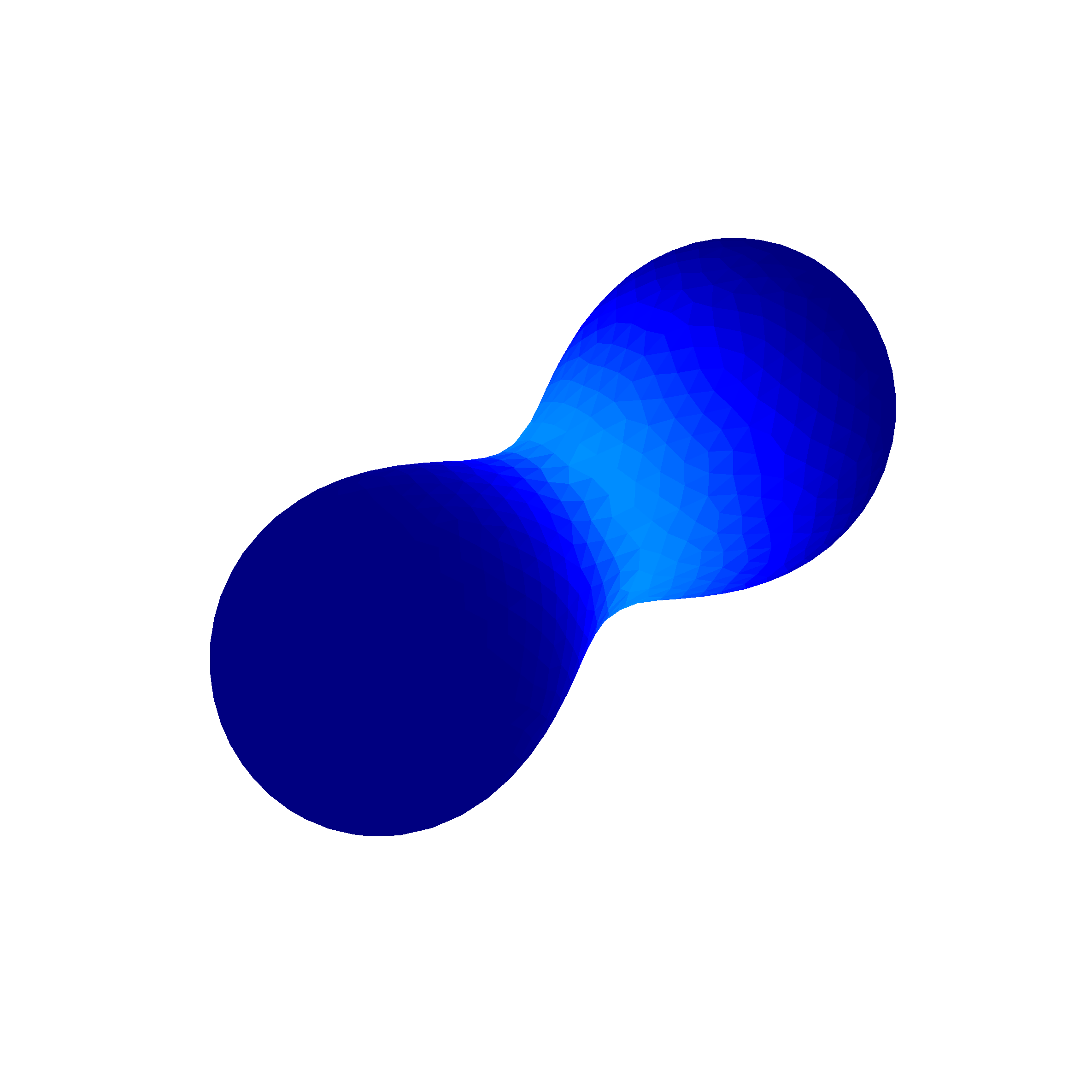}
    \includegraphics[width=2.4cm]{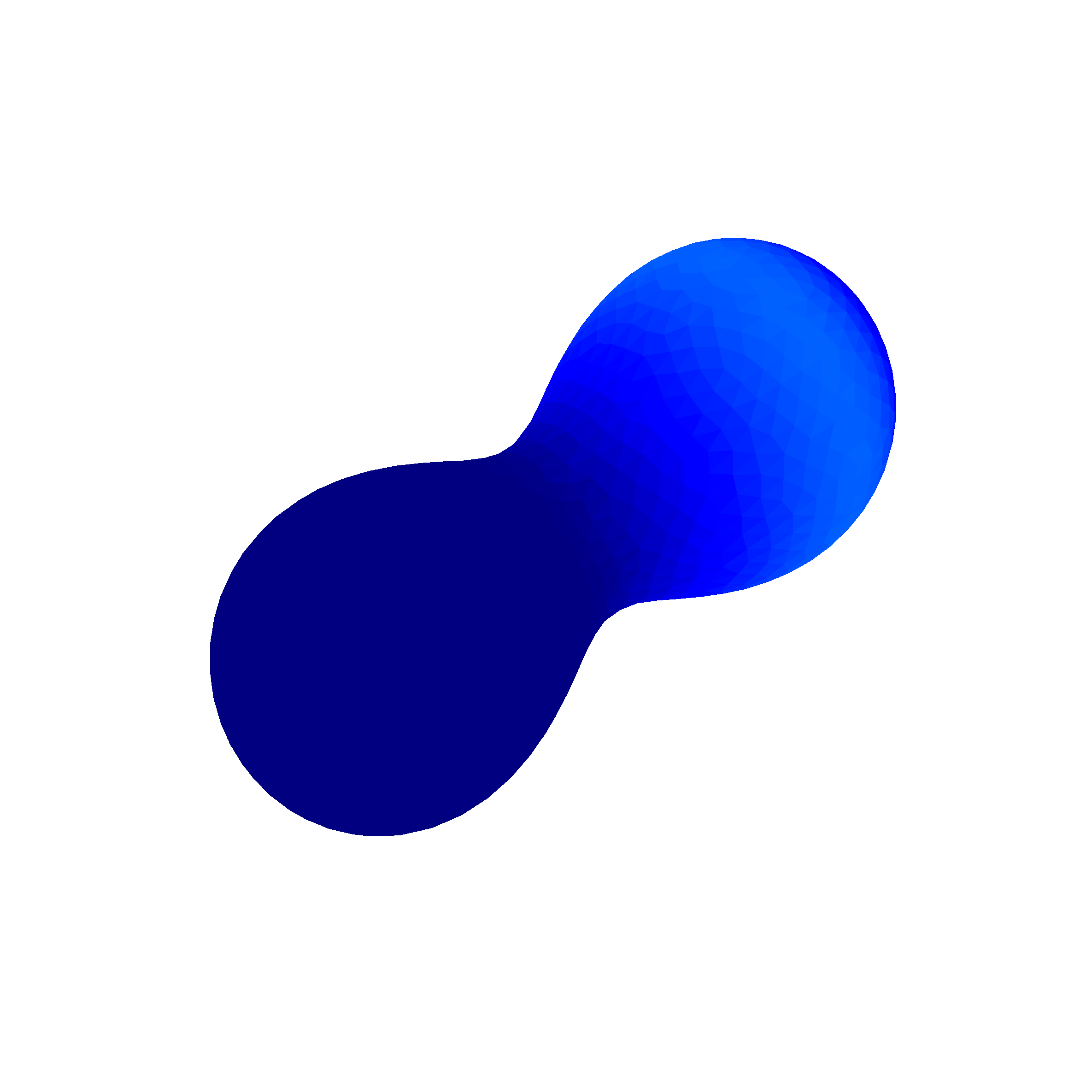}
    \includegraphics[width=2.9cm]{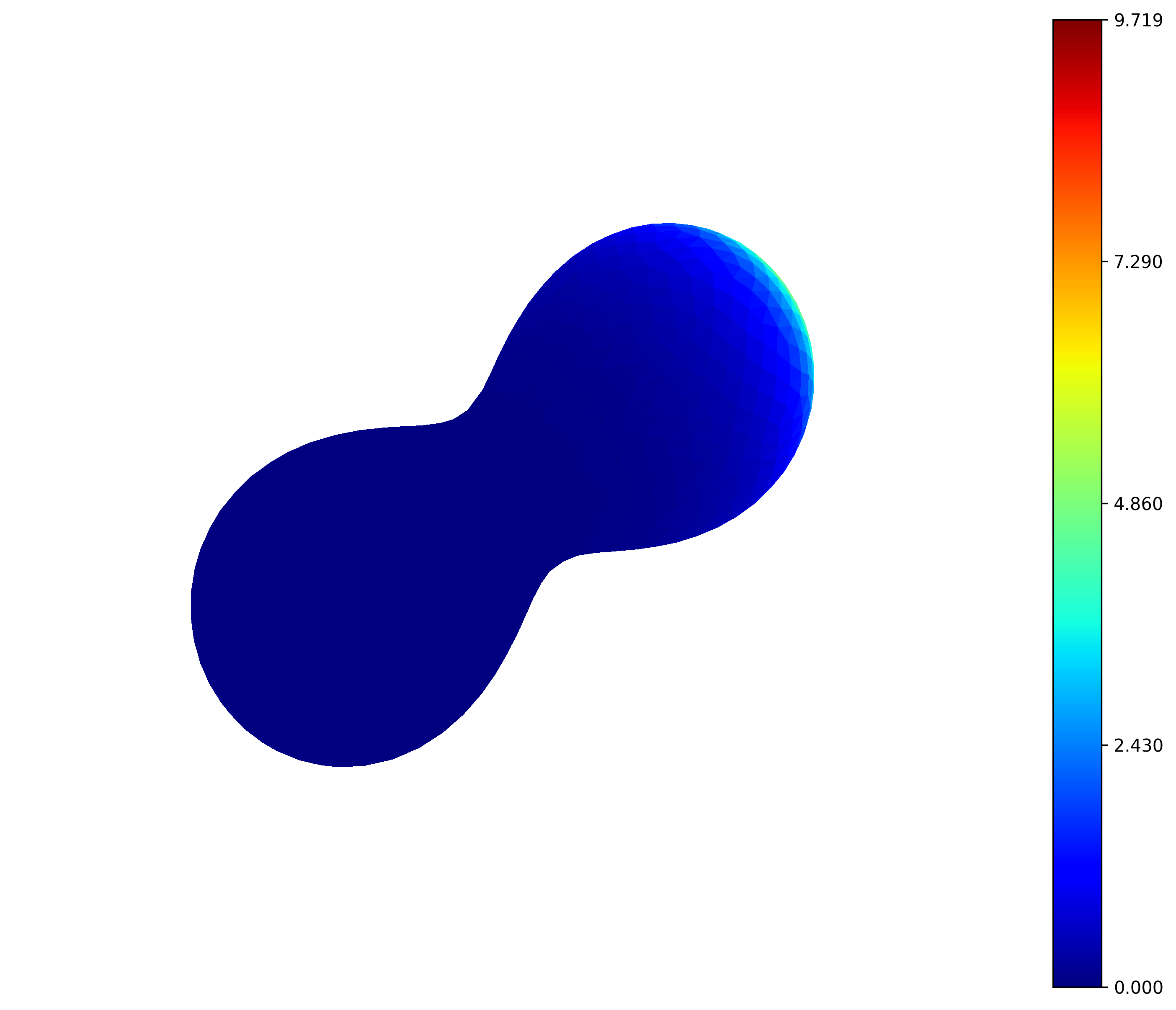}\\
    \vspace{5pt}

    \includegraphics[width=2.4cm]{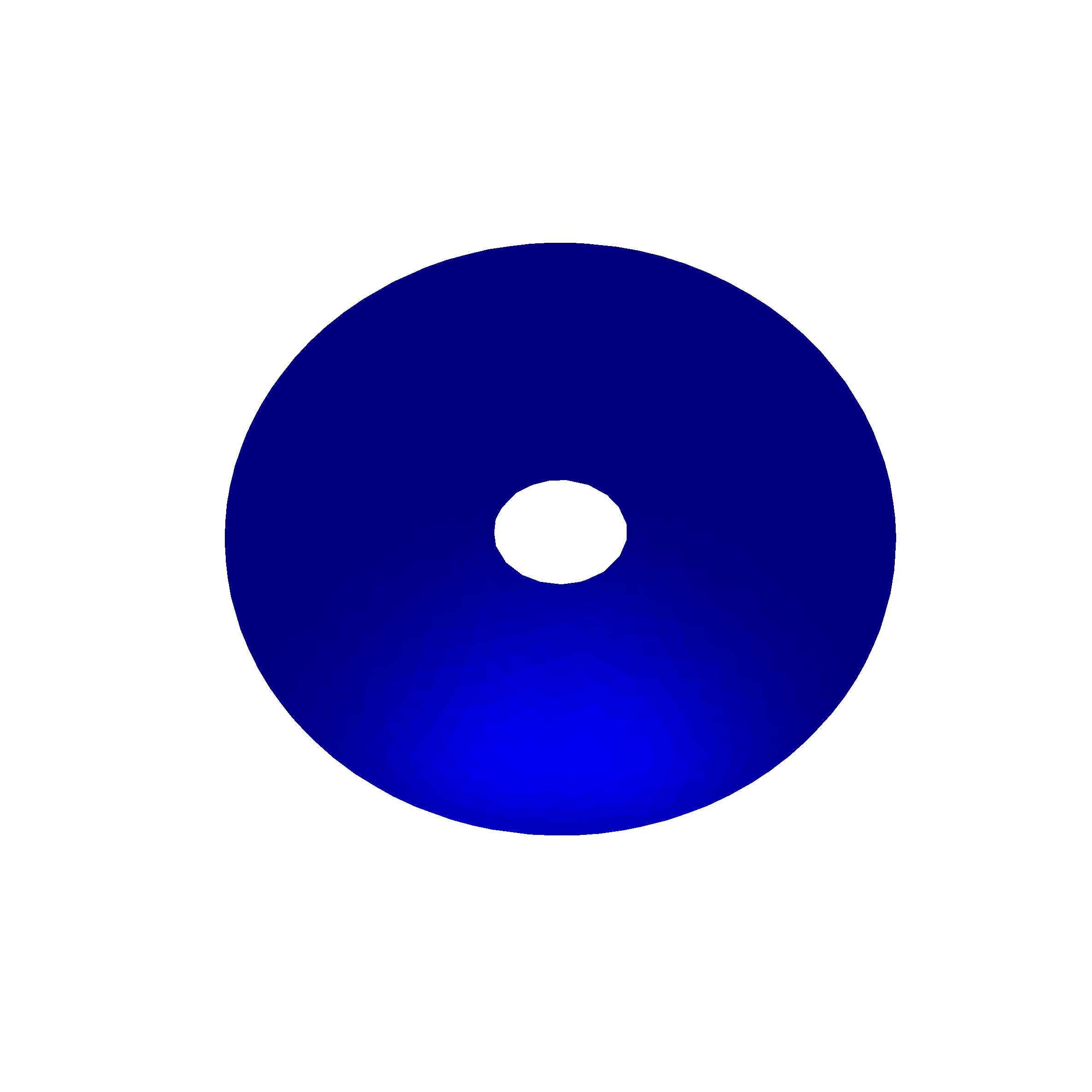}
    \includegraphics[width=2.4cm]{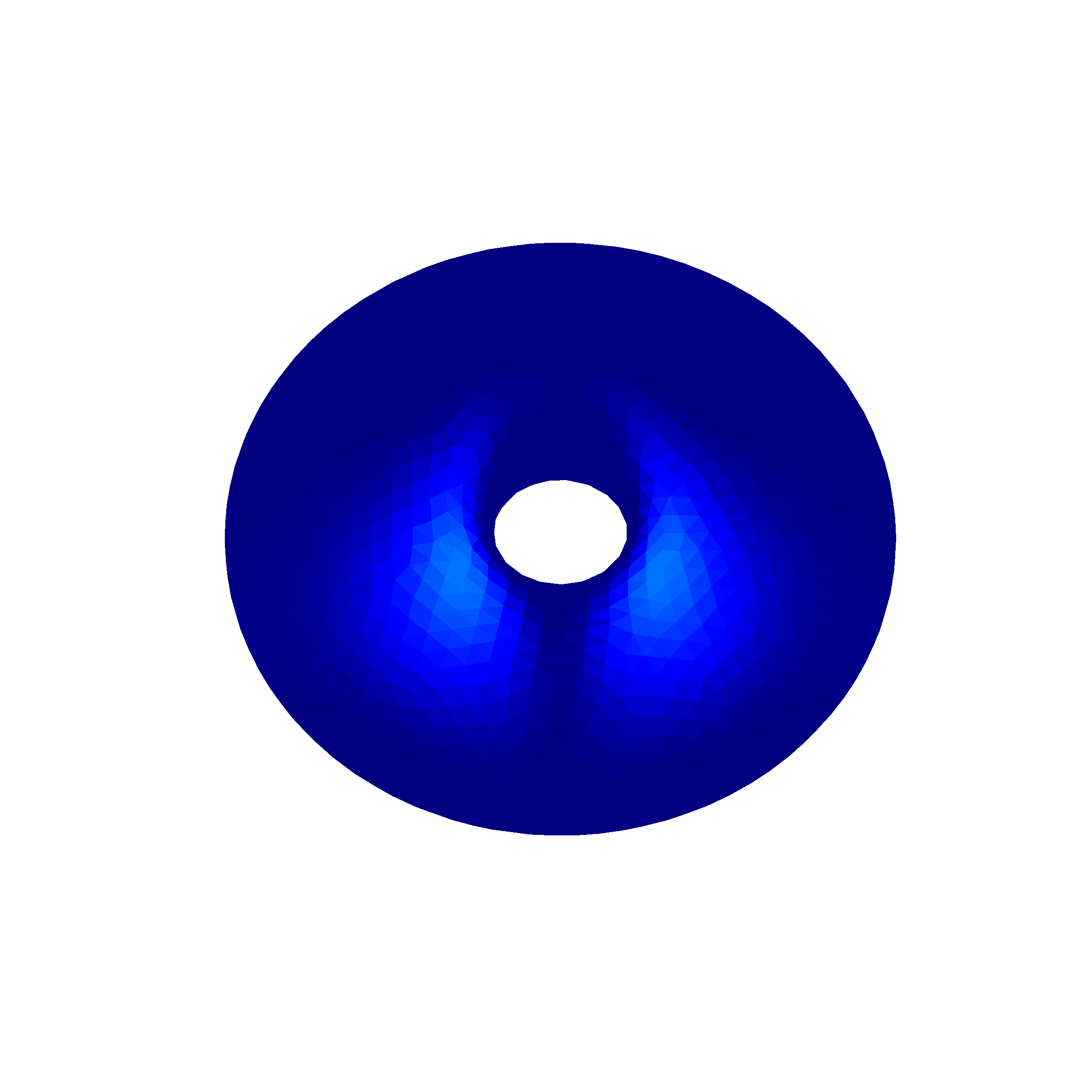}
    \includegraphics[width=2.4cm]{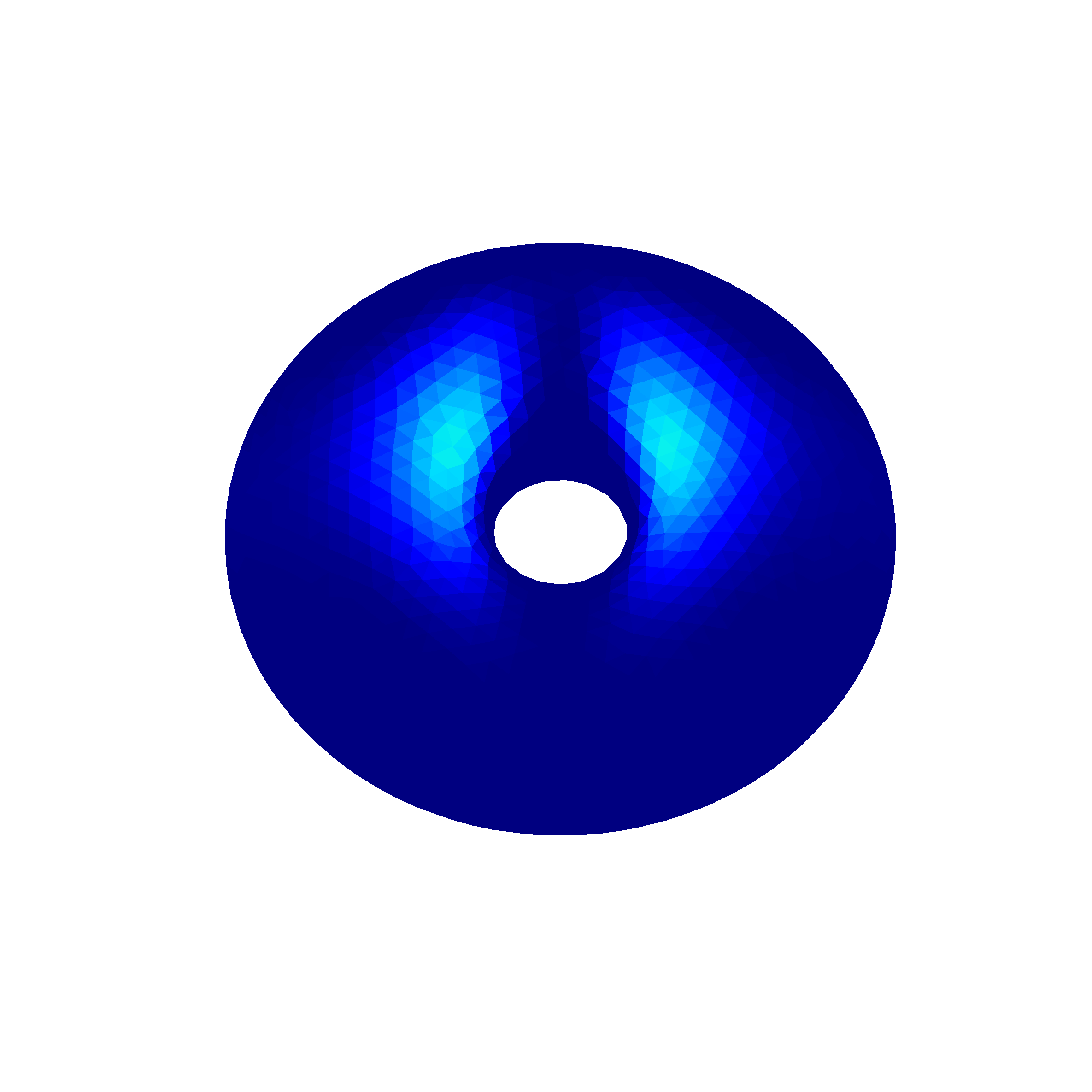}
    \includegraphics[width=2.4cm]{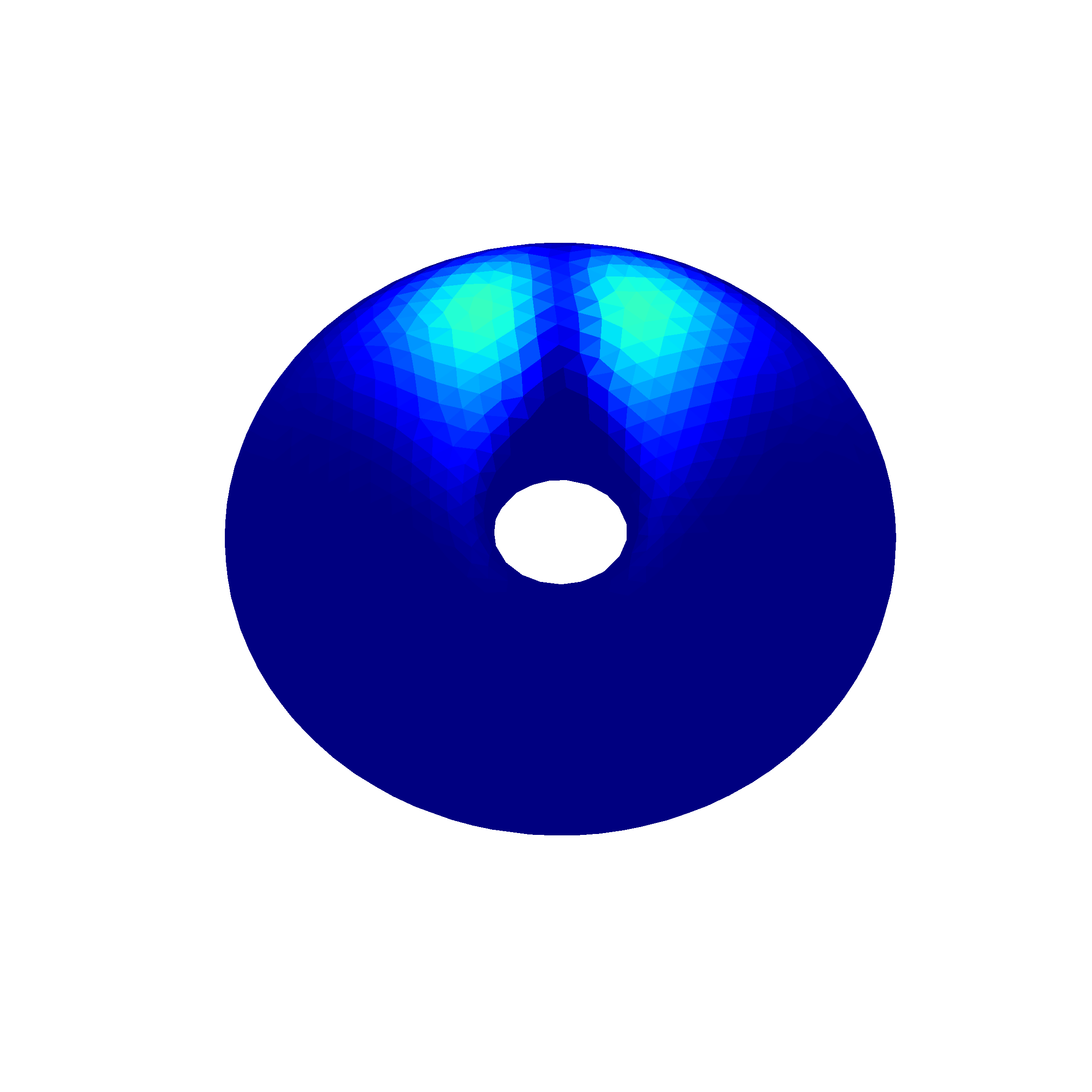}
    \includegraphics[width=2.9cm]{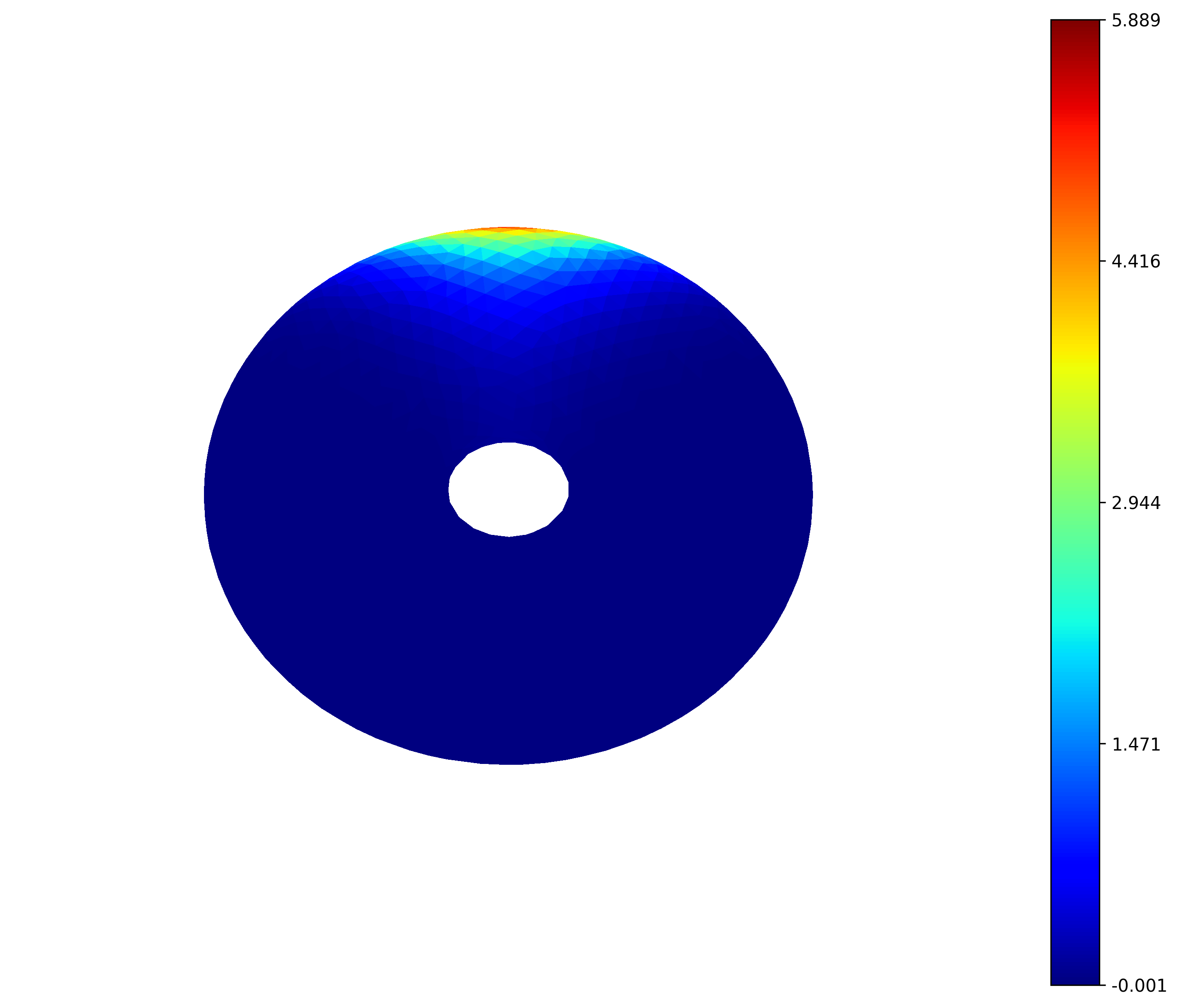}\\
    \vspace{5pt}

    \subfigure[$f(0, \boldsymbol{x})$]{
    \includegraphics[width=2.4cm]{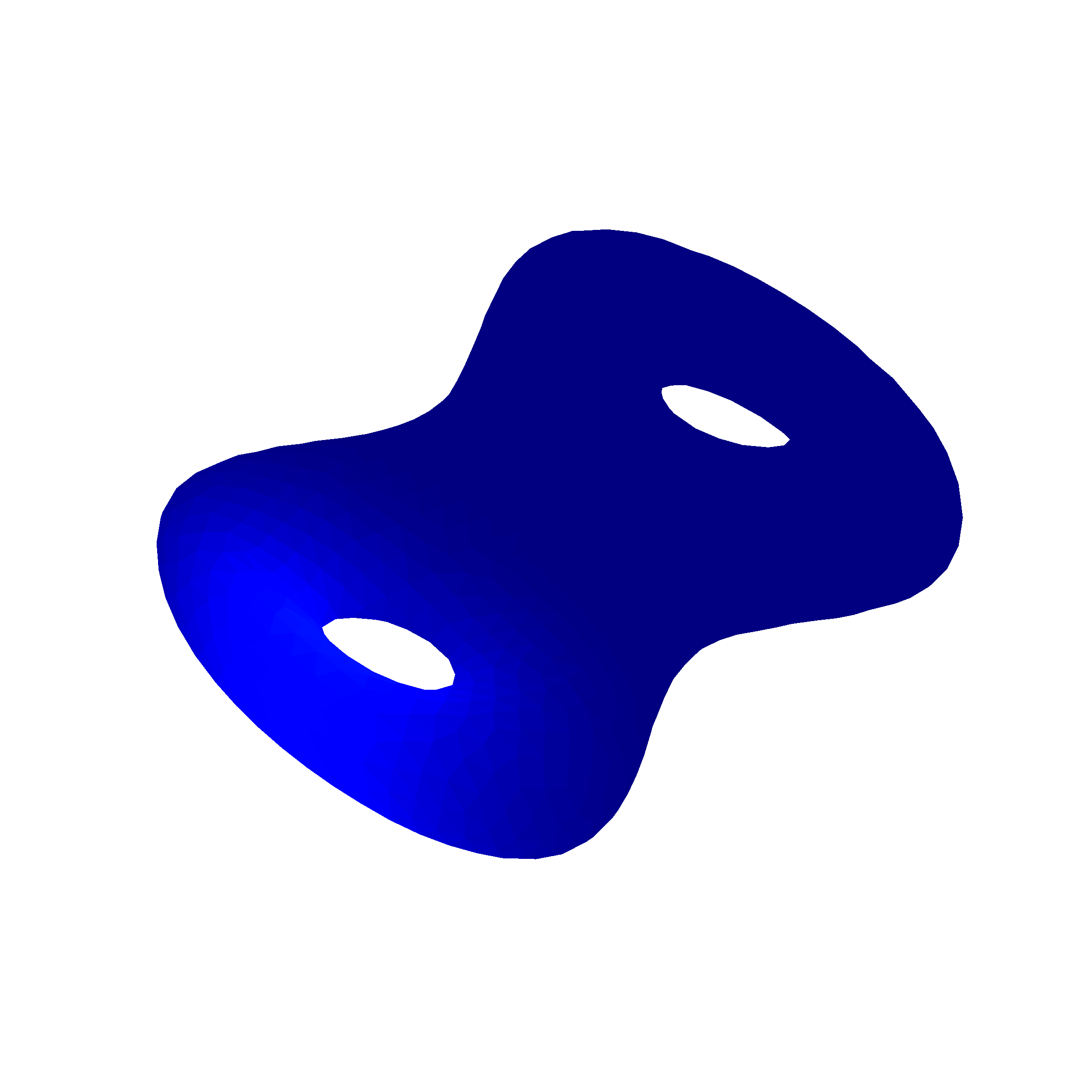}}
    \subfigure[$f(0.25, \boldsymbol{x})$]{
    \includegraphics[width=2.4cm]{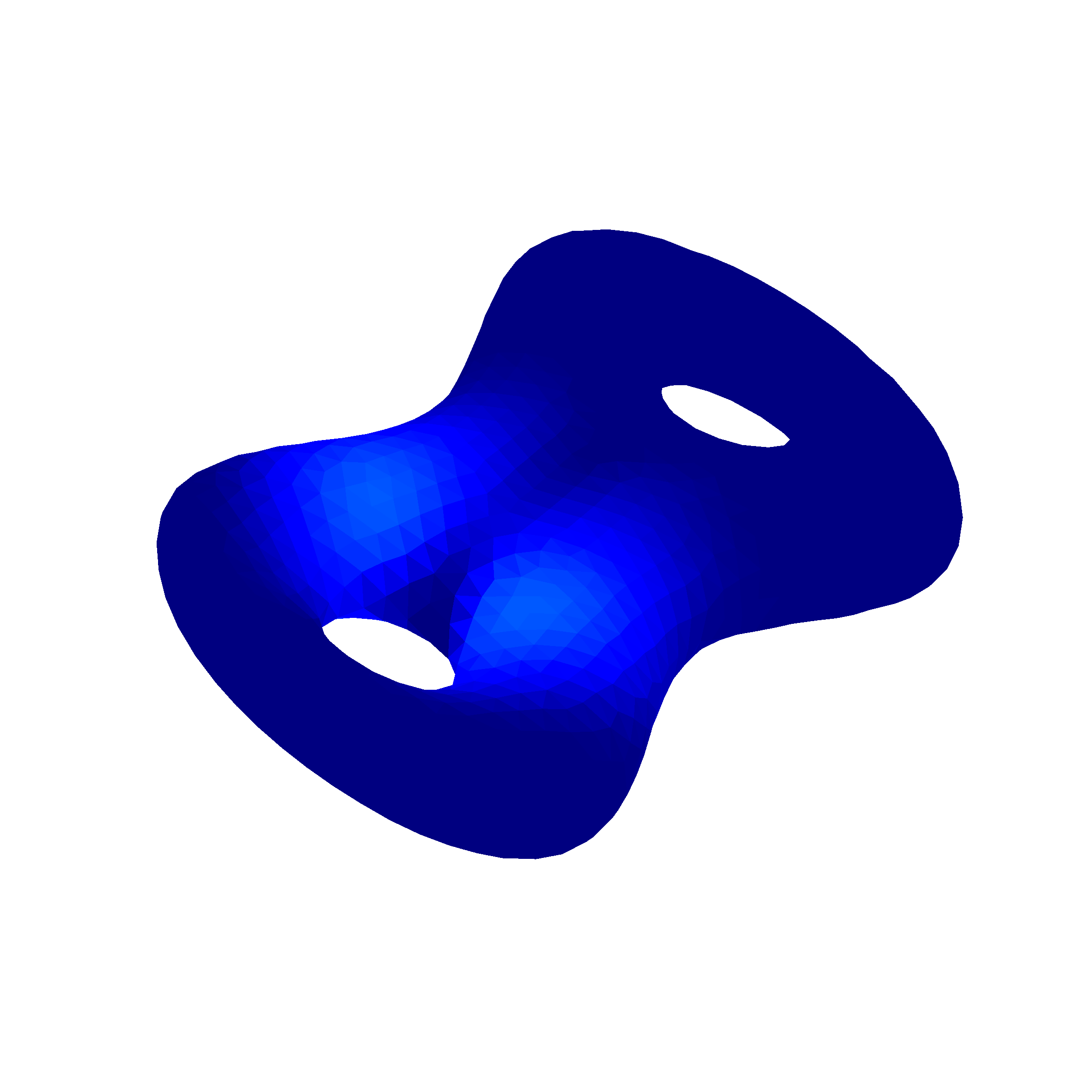}}
    \subfigure[$f(0.5, \boldsymbol{x})$]{
    \includegraphics[width=2.4cm]{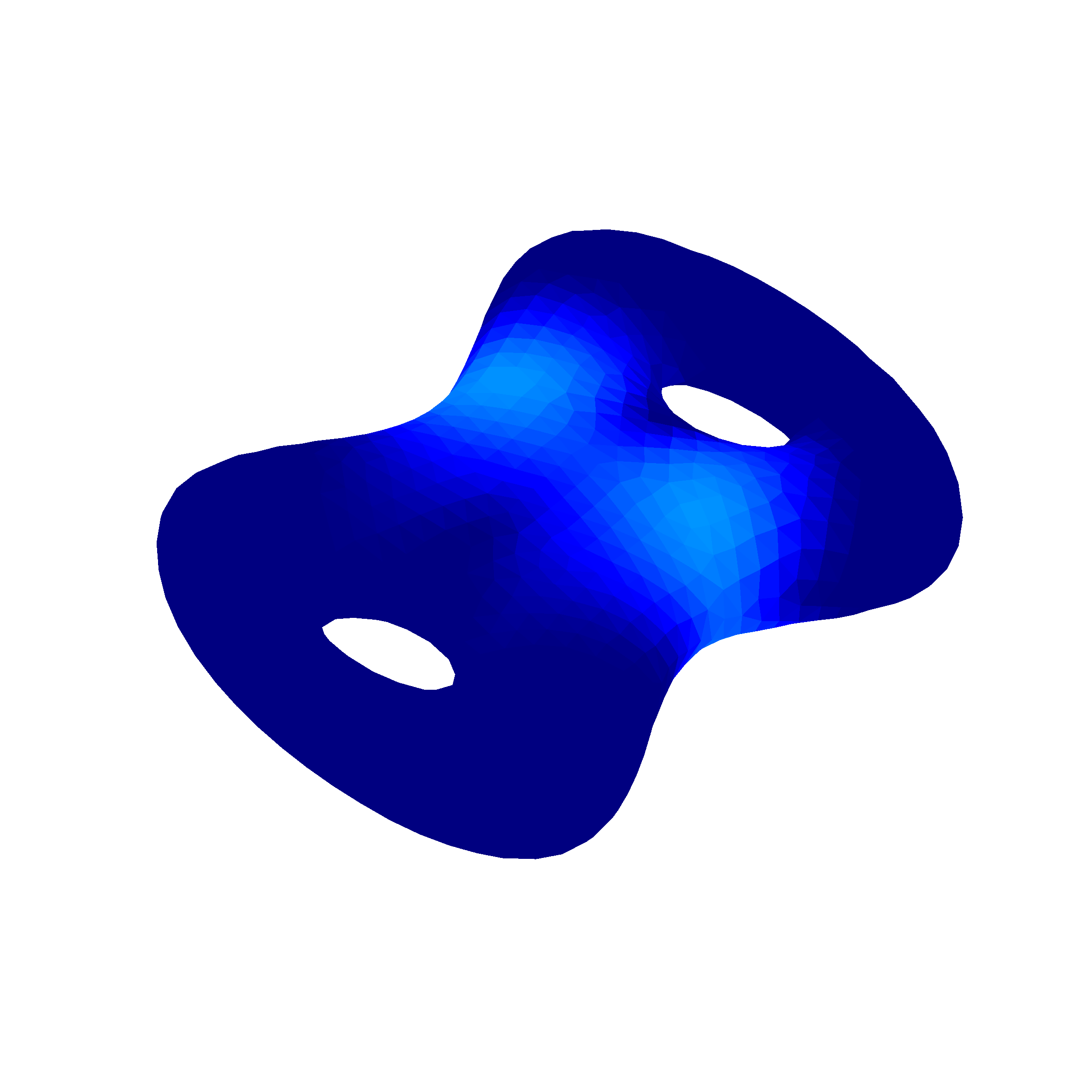}}
    \subfigure[$f(0.75, \boldsymbol{x})$]{
    \includegraphics[width=2.4cm]{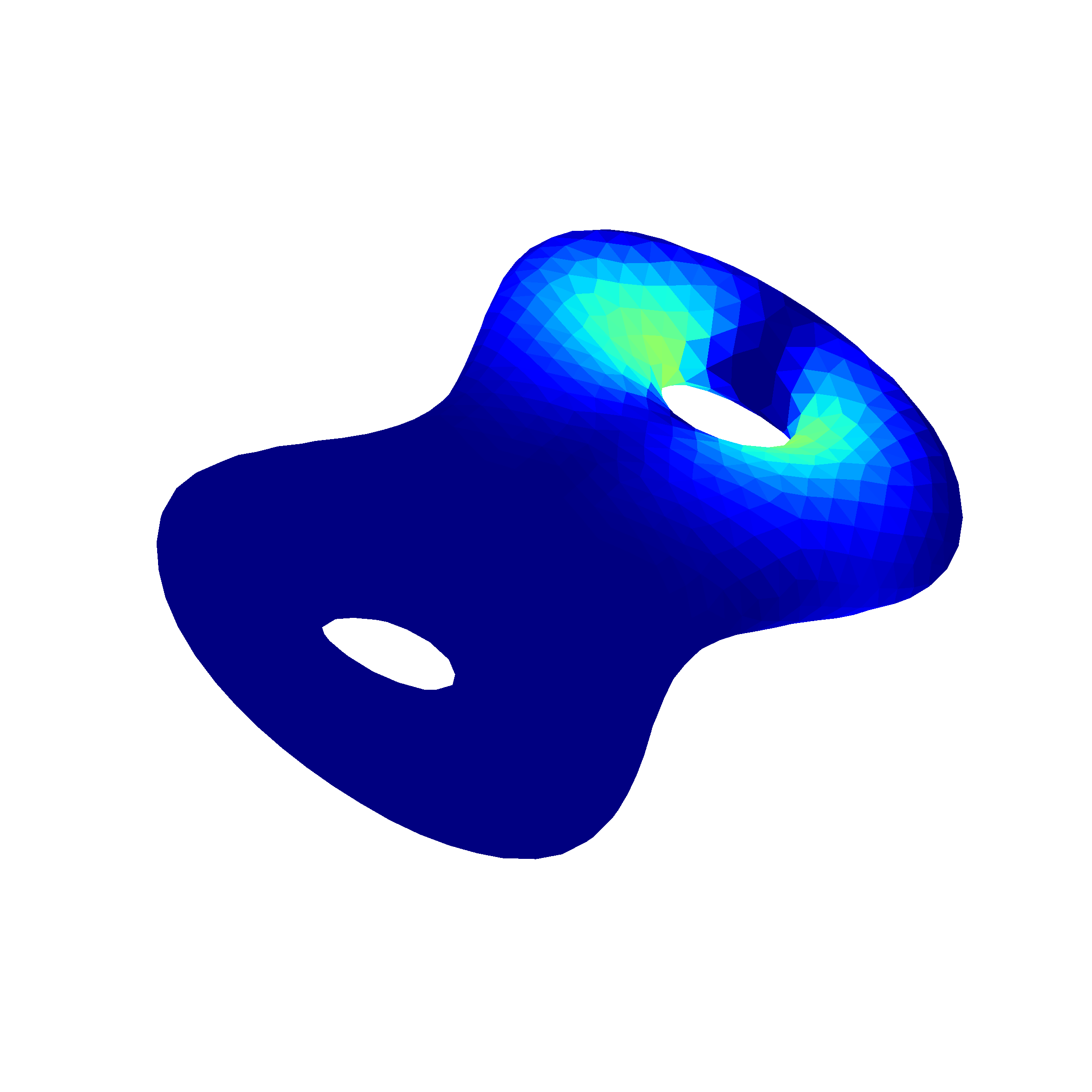}}
    \subfigure[$f(1, \boldsymbol{x})$]{
    \includegraphics[width=2.9cm]{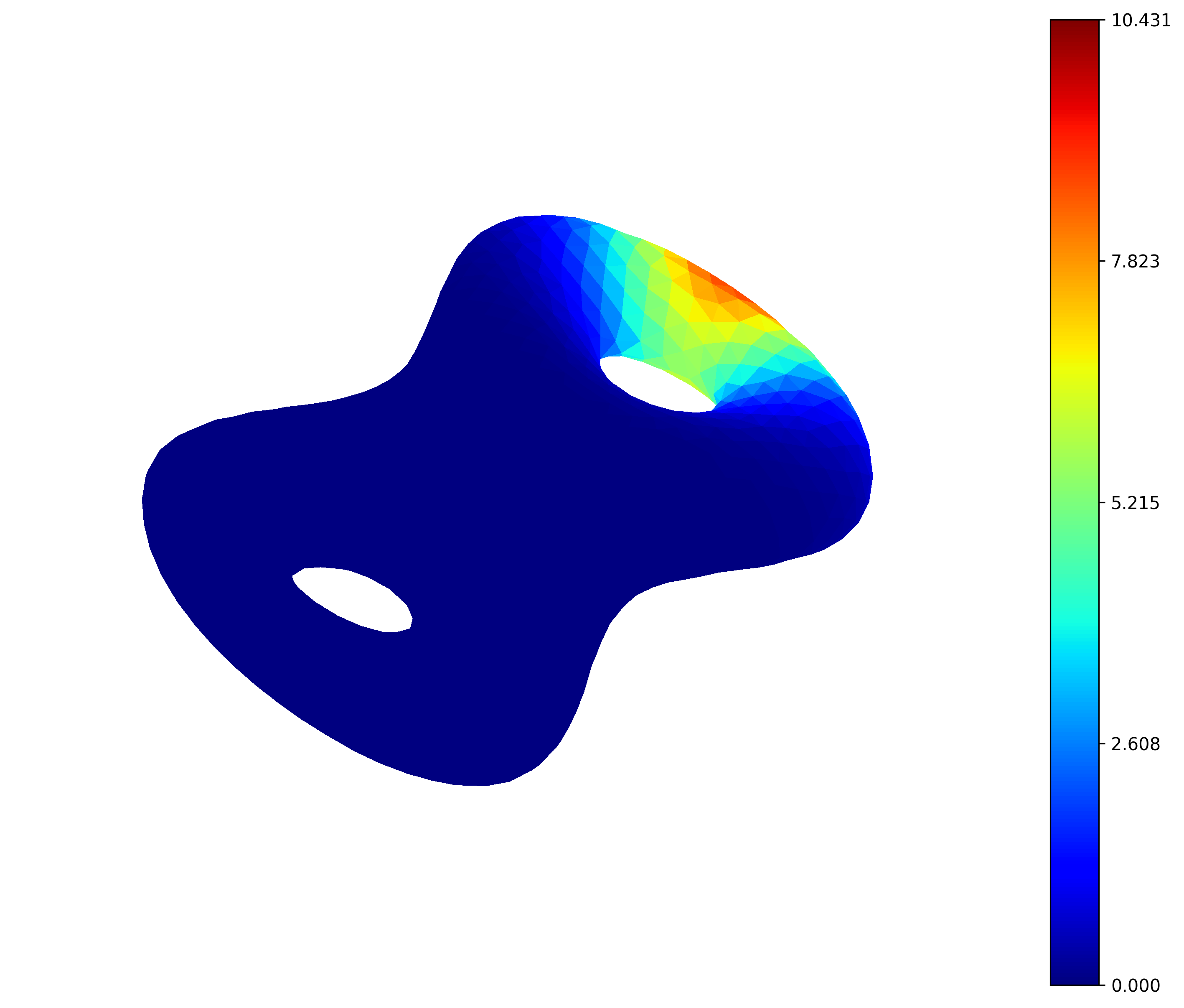}}\\
    \caption{SUOT test on point cloud ($\beta=1.5$): source item $f$.}
    \label{SUOT-LS-f}
    \end{center}
\end{figure}

The results are shown in Figure \ref{SOT-LS}, Figure \ref{SUOT-LS}.
Our method gives good results in all cases. For unbalanced cases, see Figure \ref{SUOT-LS}, the results seem to be qualitatively similar to the results in OT cases. However, the mechanism is totally different. In unbalanced cases, the source term also plays important role as shown in Figure \ref{SUOT-LS-f}.  

Next, we test the algorithm with more complicated distributions. The distributions are set to be mixed Gaussian type as given in Table \ref{tab:SUOT} of \ref{App-B}. In these cases, mass should be splitted or merged, such that the velocity field will be much more complicated. 

\begin{figure}[htbp]
    \begin{center}
    \includegraphics[width=2.4cm]{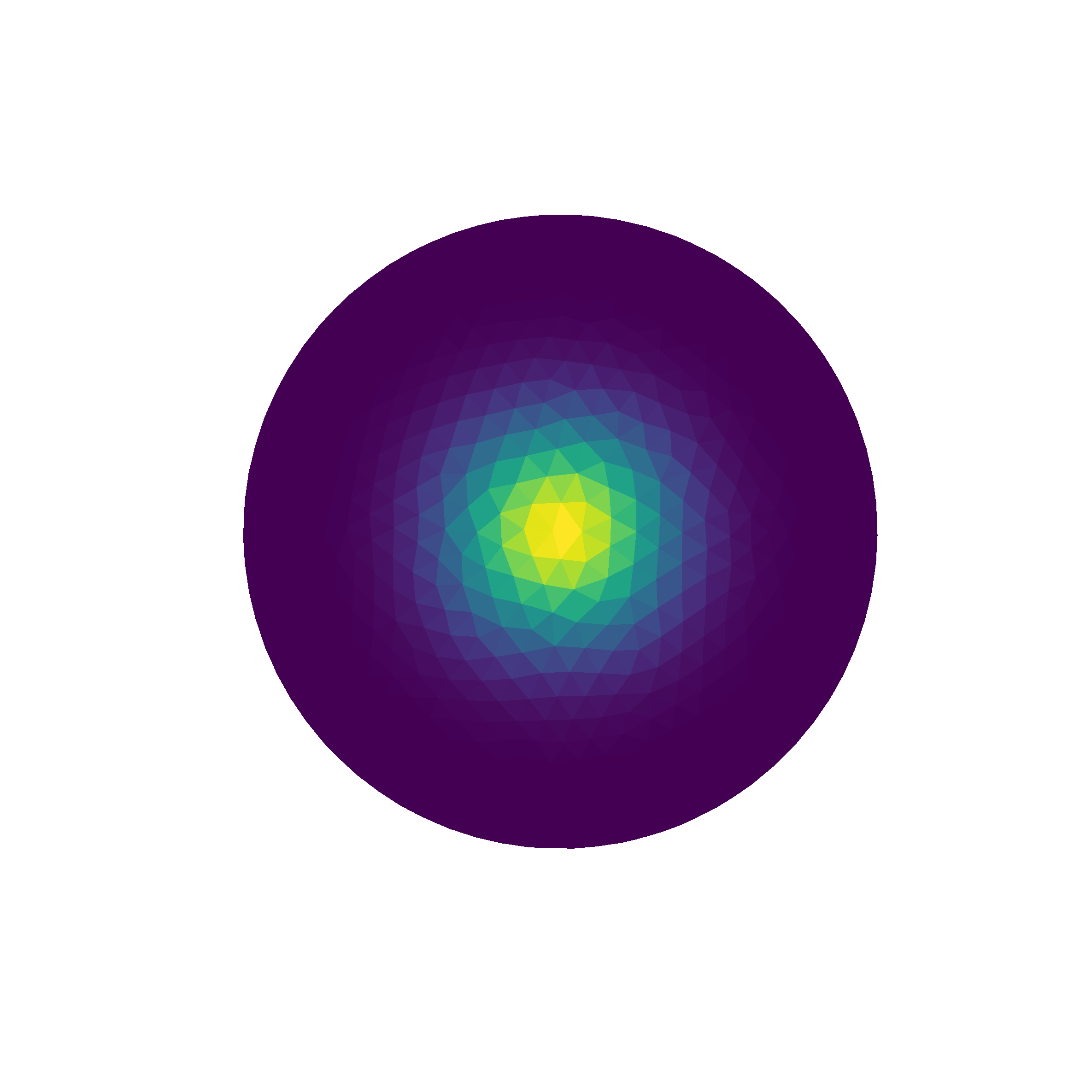}
    \includegraphics[width=2.4cm]{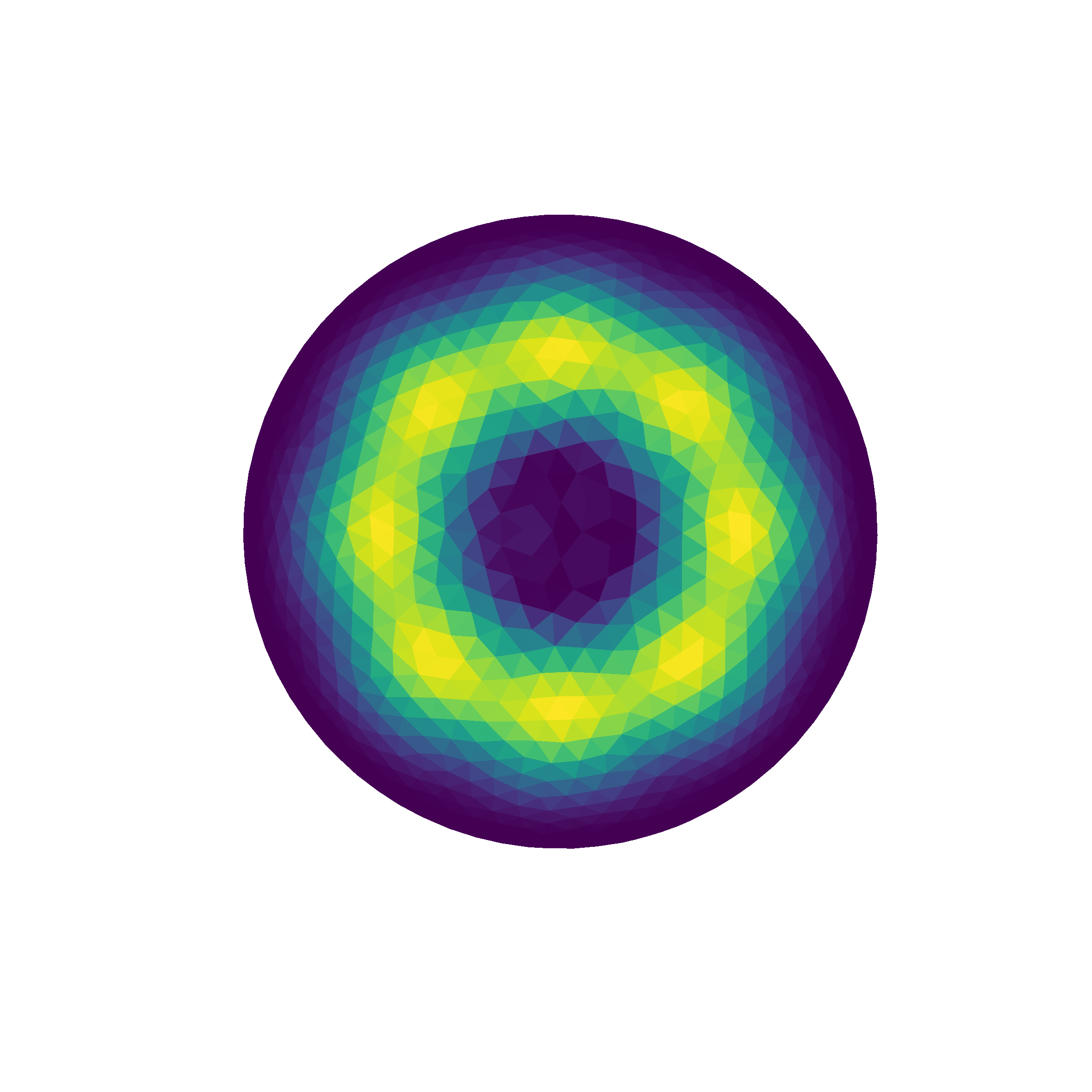}
    \includegraphics[width=2.4cm]{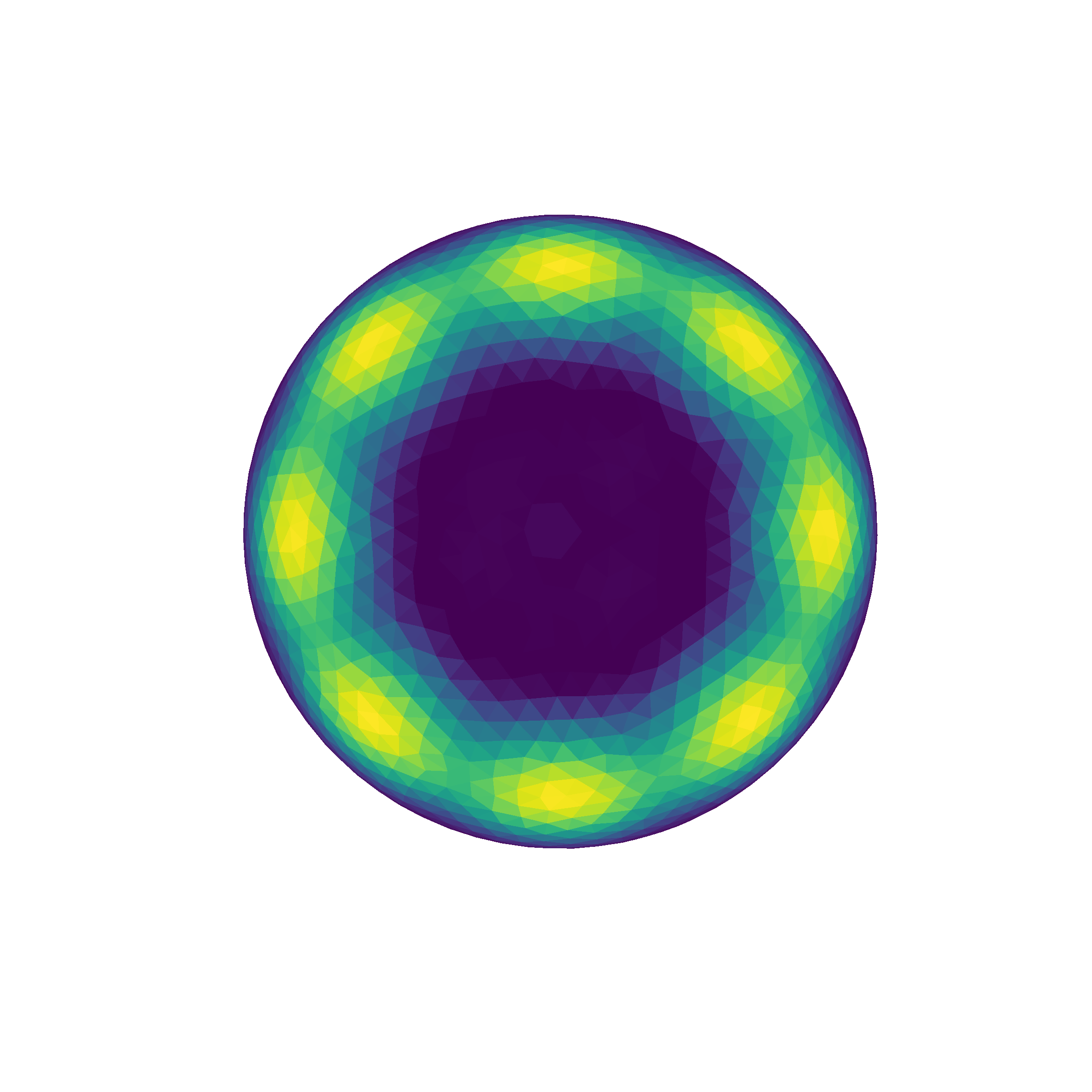}
    \includegraphics[width=2.4cm]{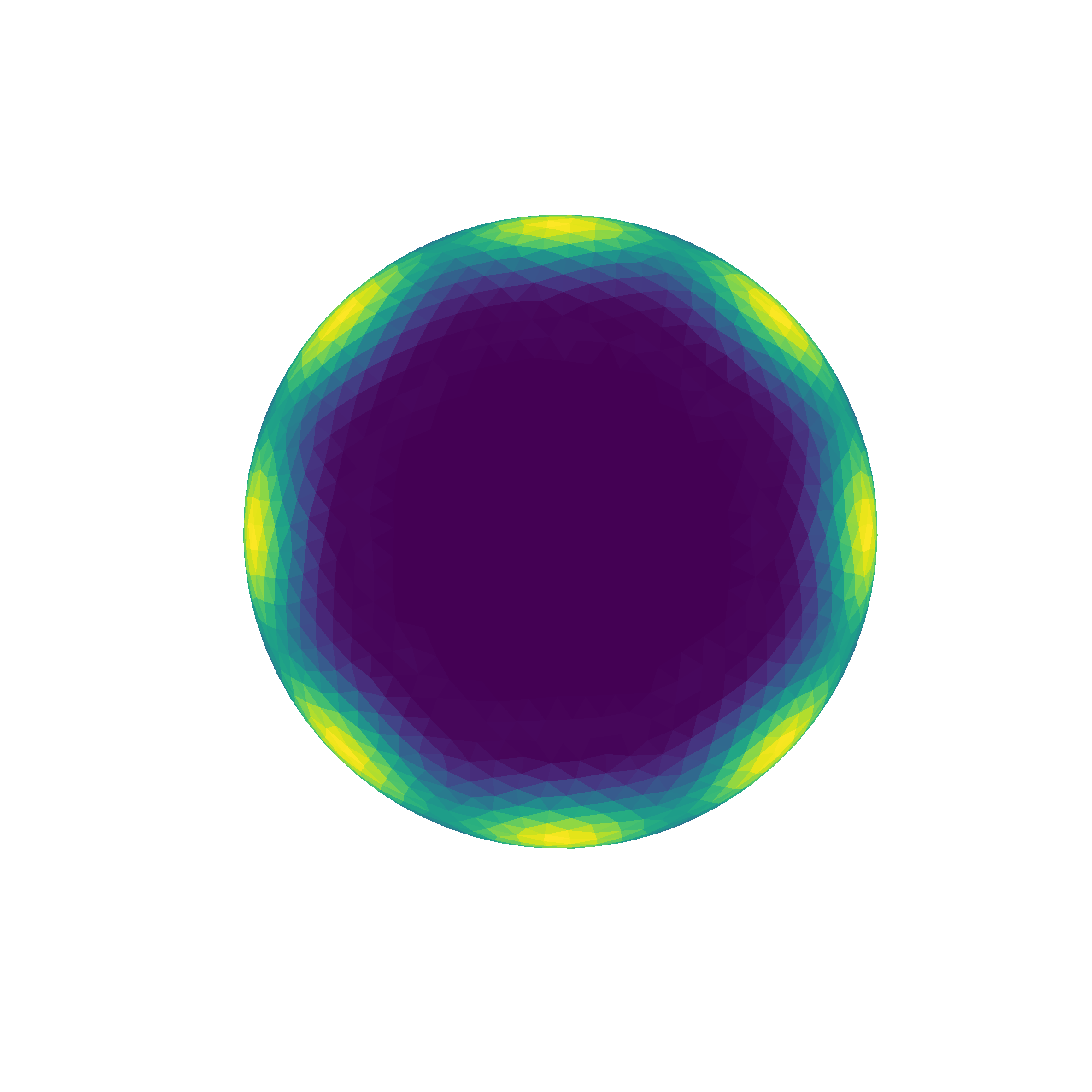}
    \includegraphics[width=2.9cm]{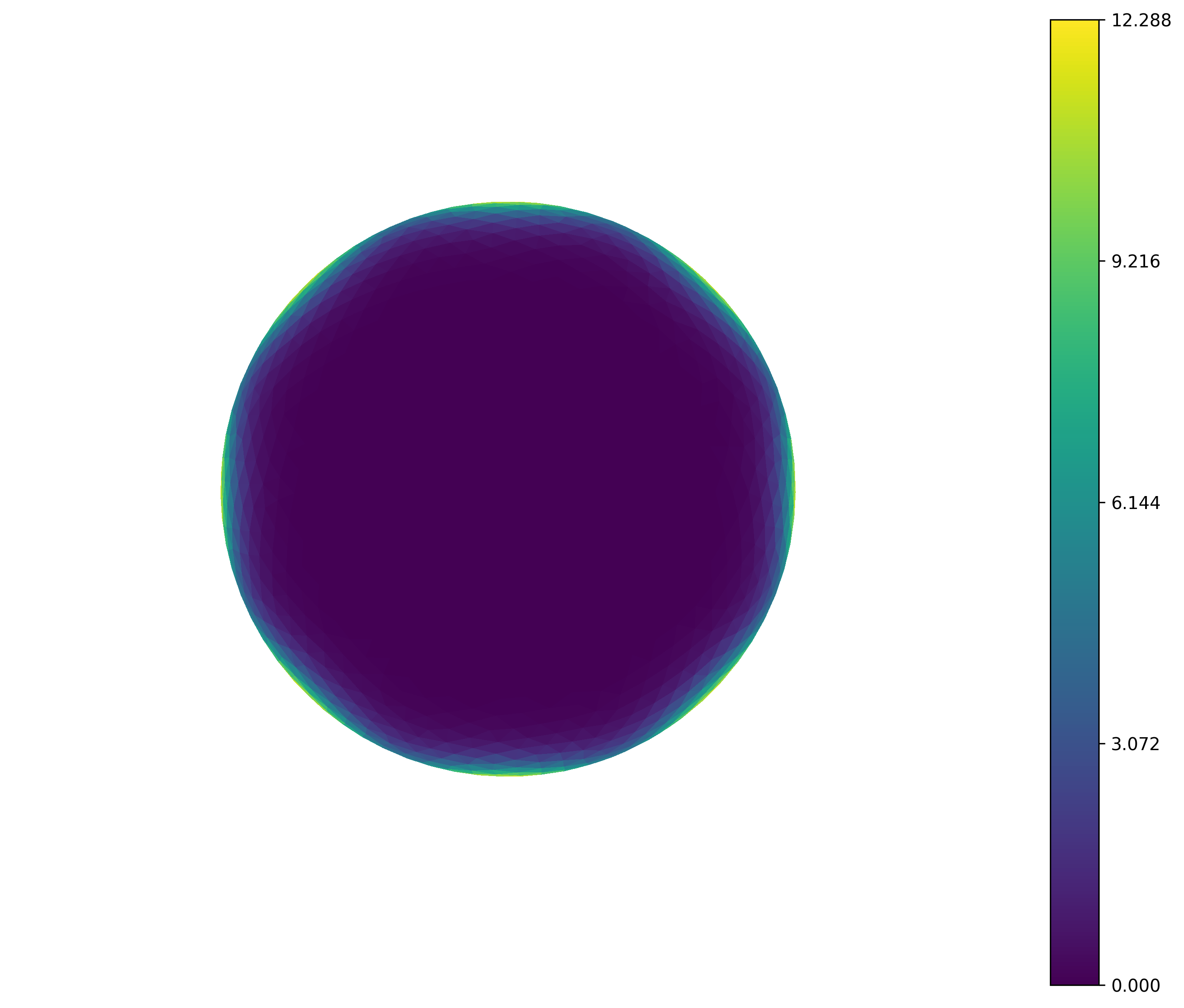}\\
    \vspace{5pt}

    \subfigure[$\rho(0, \boldsymbol{x})$]{
    \includegraphics[width=2.4cm]{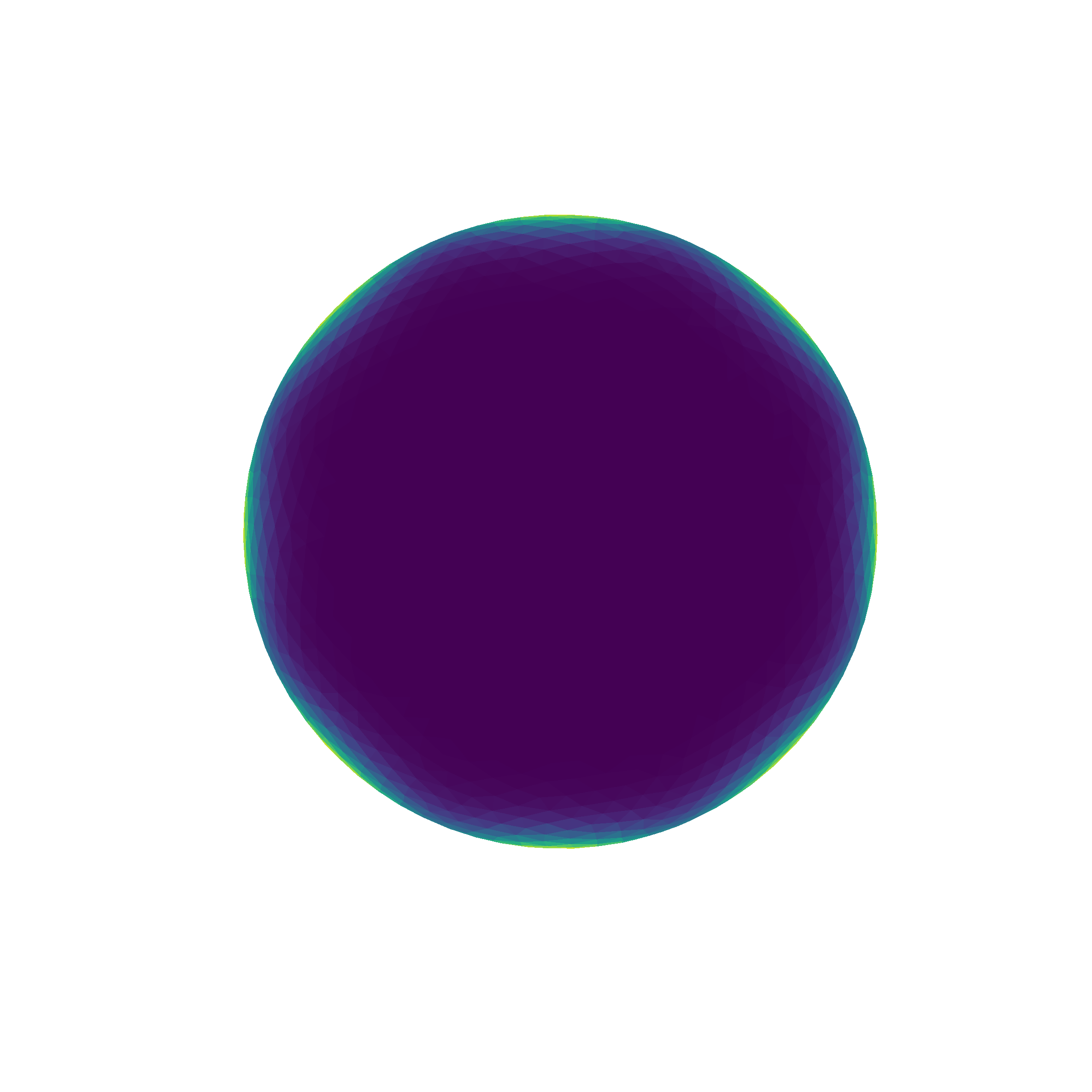}}
    \subfigure[$\rho(0.25, \boldsymbol{x})$]{
    \includegraphics[width=2.4cm]{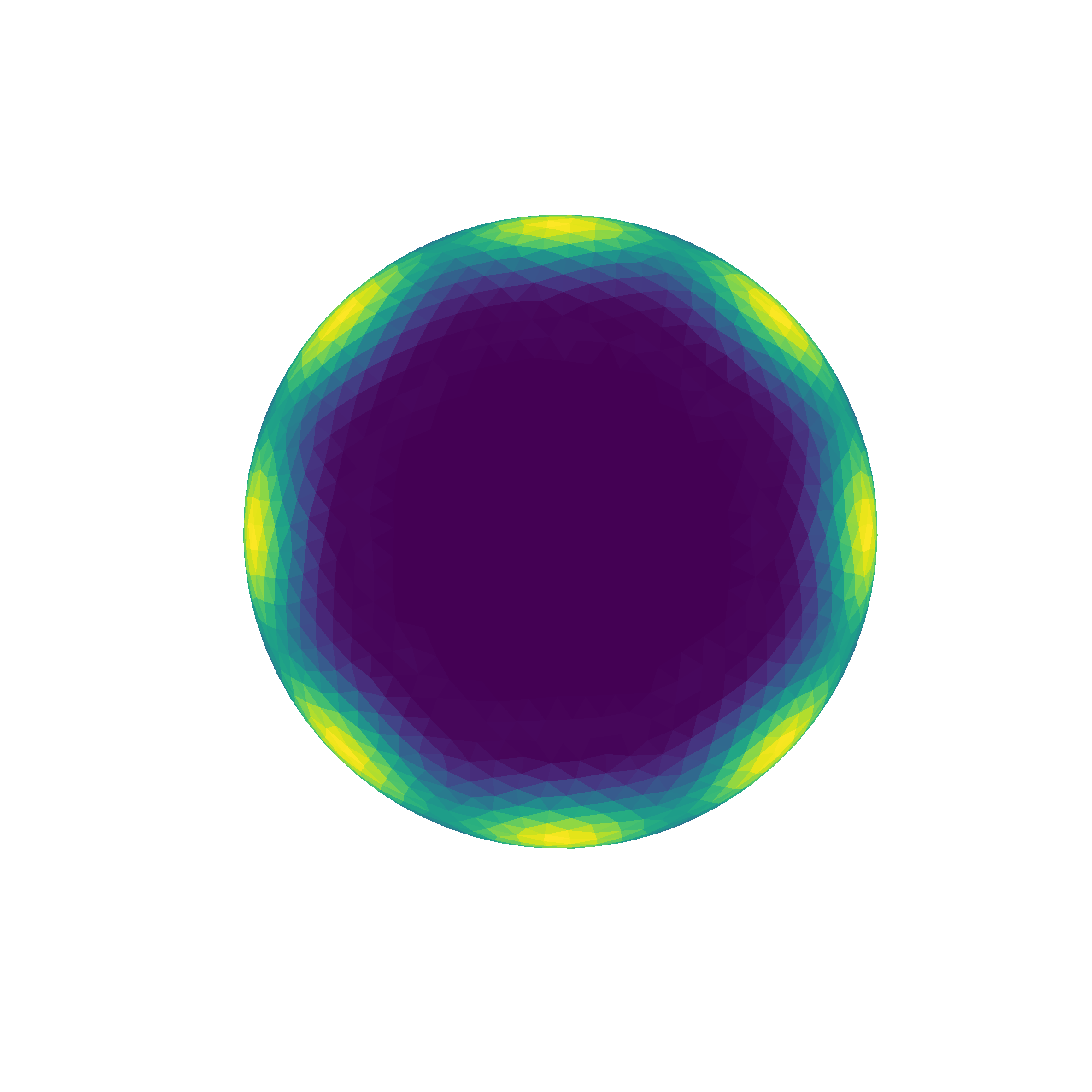}}
    \subfigure[$\rho(0.5, \boldsymbol{x})$]{
    \includegraphics[width=2.4cm]{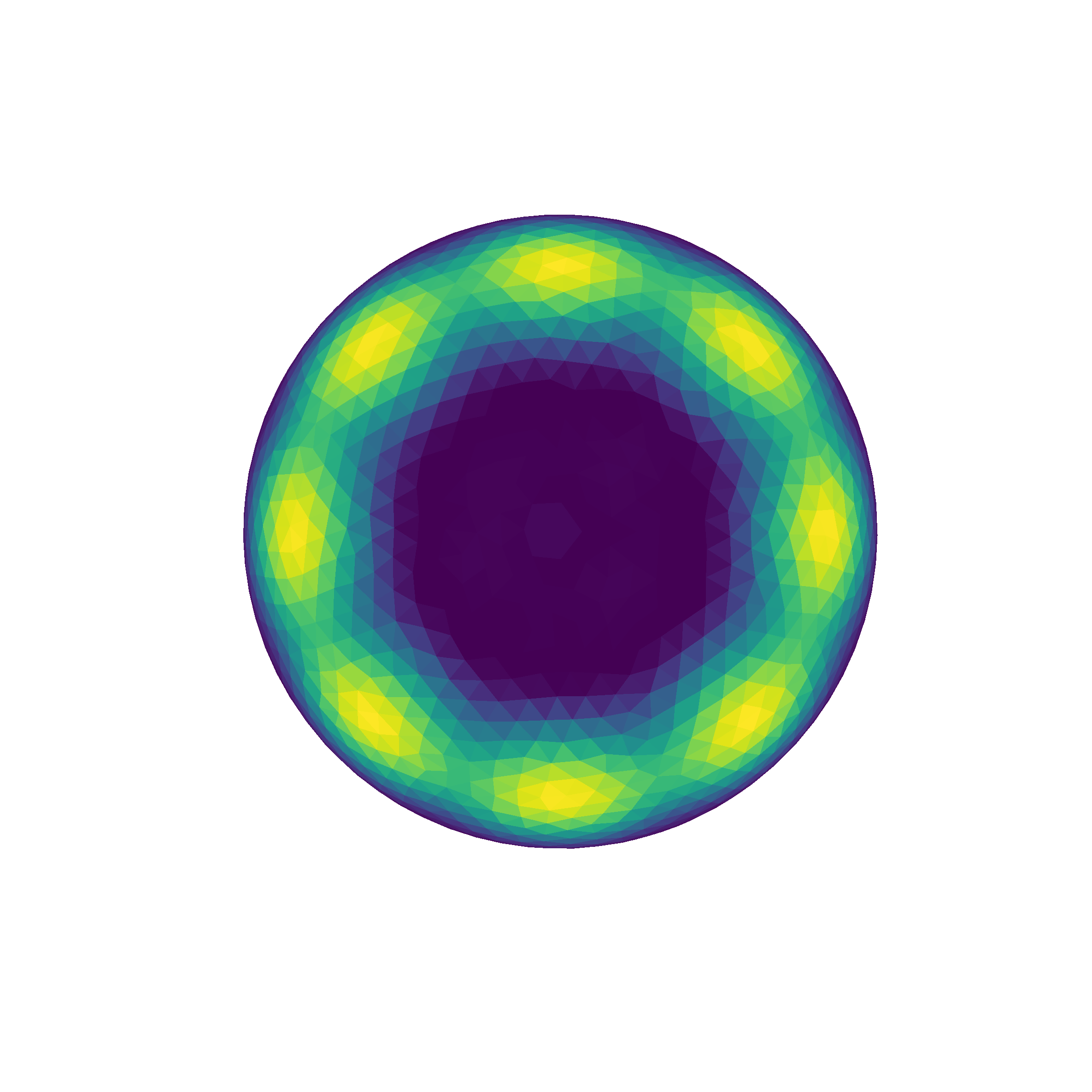}}
    \subfigure[$\rho(0.75, \boldsymbol{x})$]{
    \includegraphics[width=2.4cm]{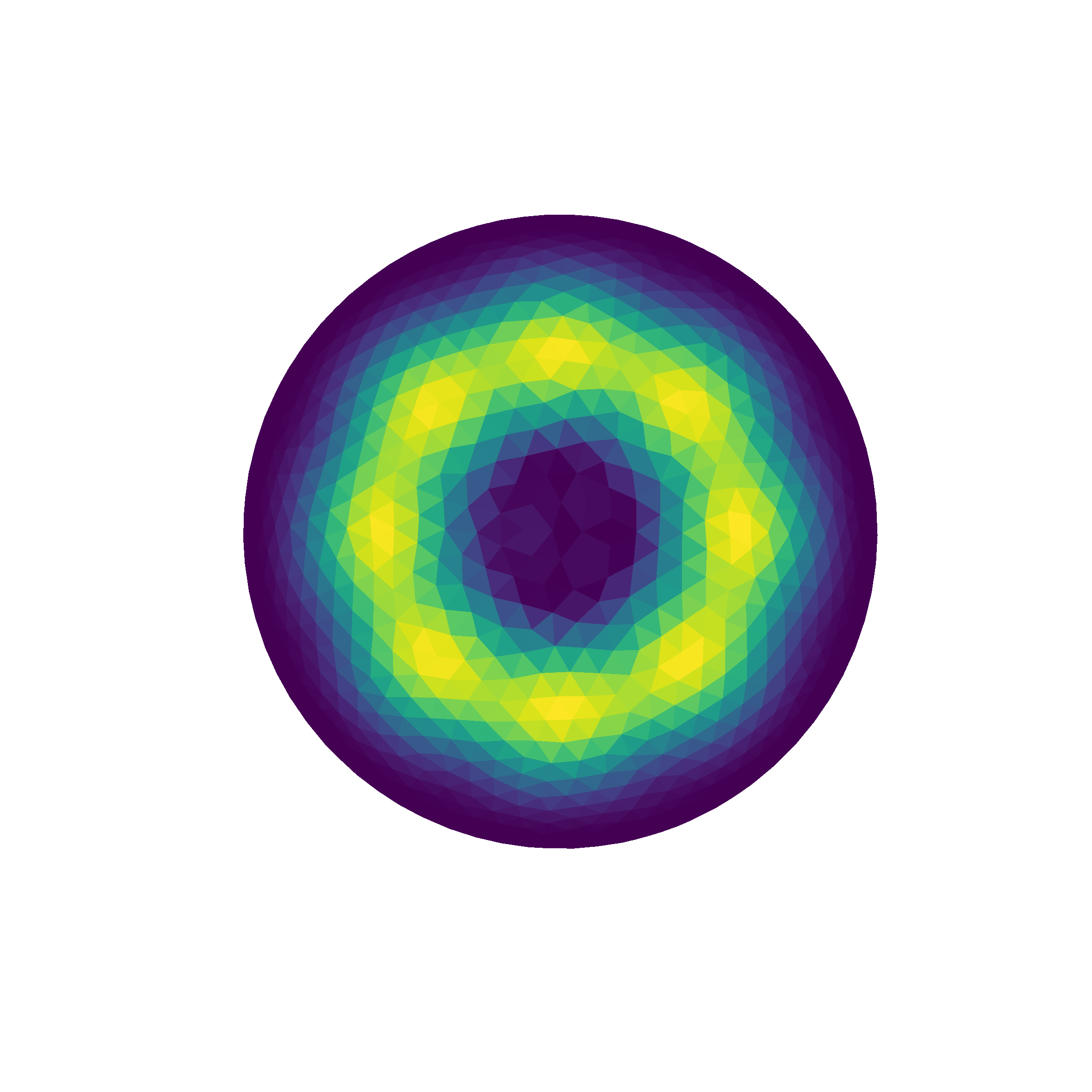}}
    \subfigure[$\rho(1, \boldsymbol{x})$]{
    \includegraphics[width=2.9cm]{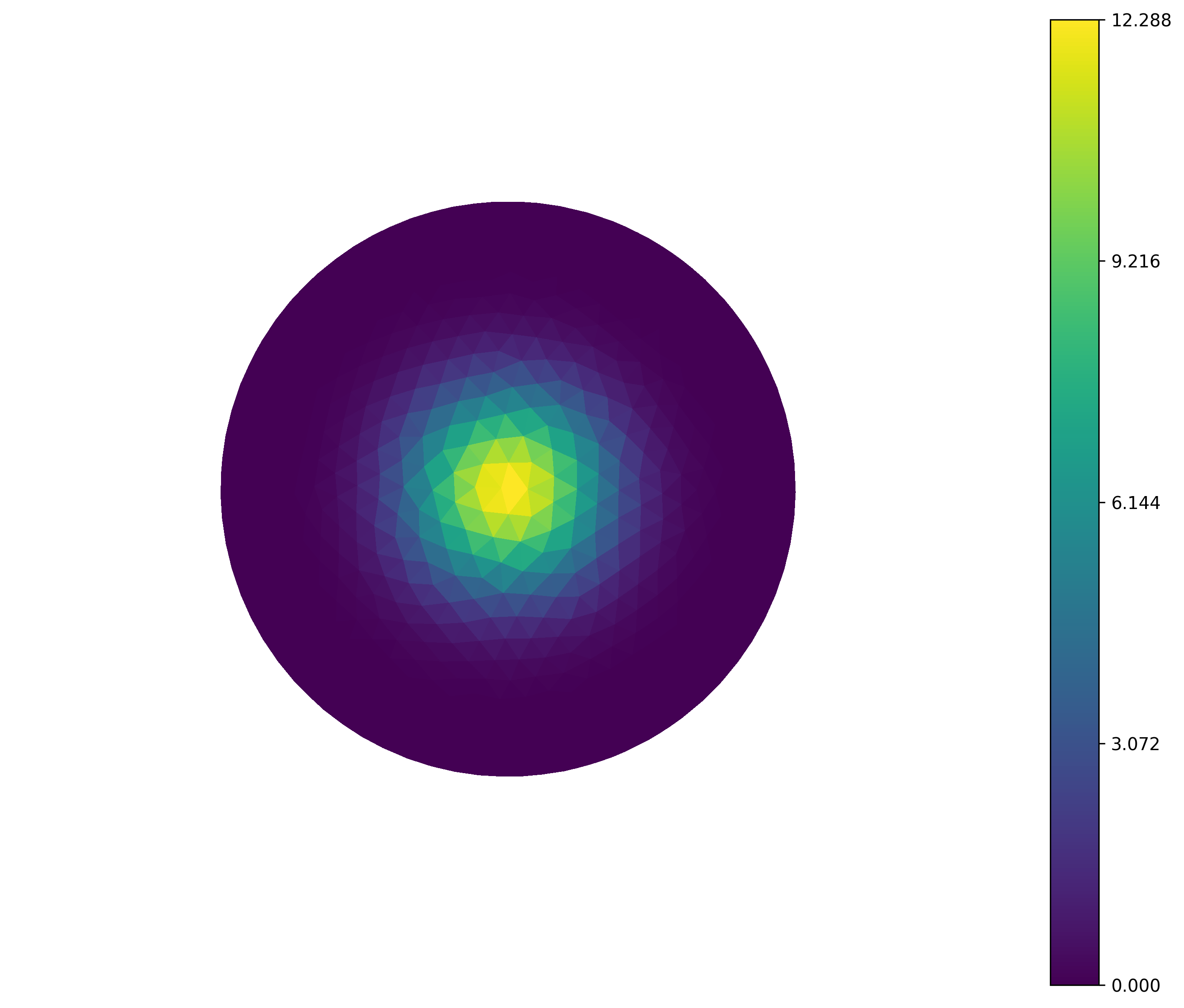}}\\
    \caption{SUOT test on point cloud for $S1$, $S2$ and the calculation times are both 452s.}
    \label{SUOT-2}
    \end{center}
\end{figure}

\begin{figure}[htbp]
    \begin{center}
    \includegraphics[width=2.4cm]{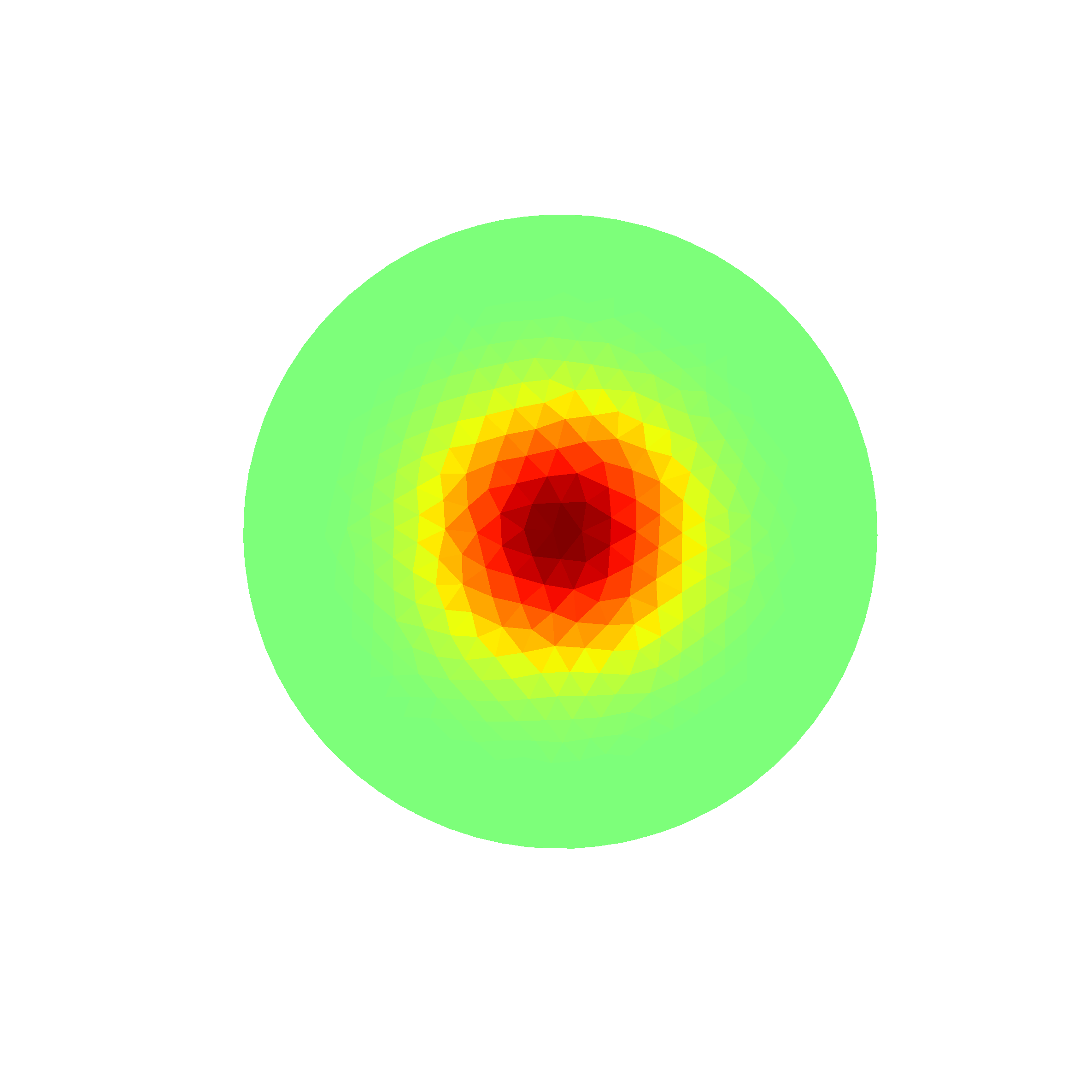}
    \includegraphics[width=2.4cm]{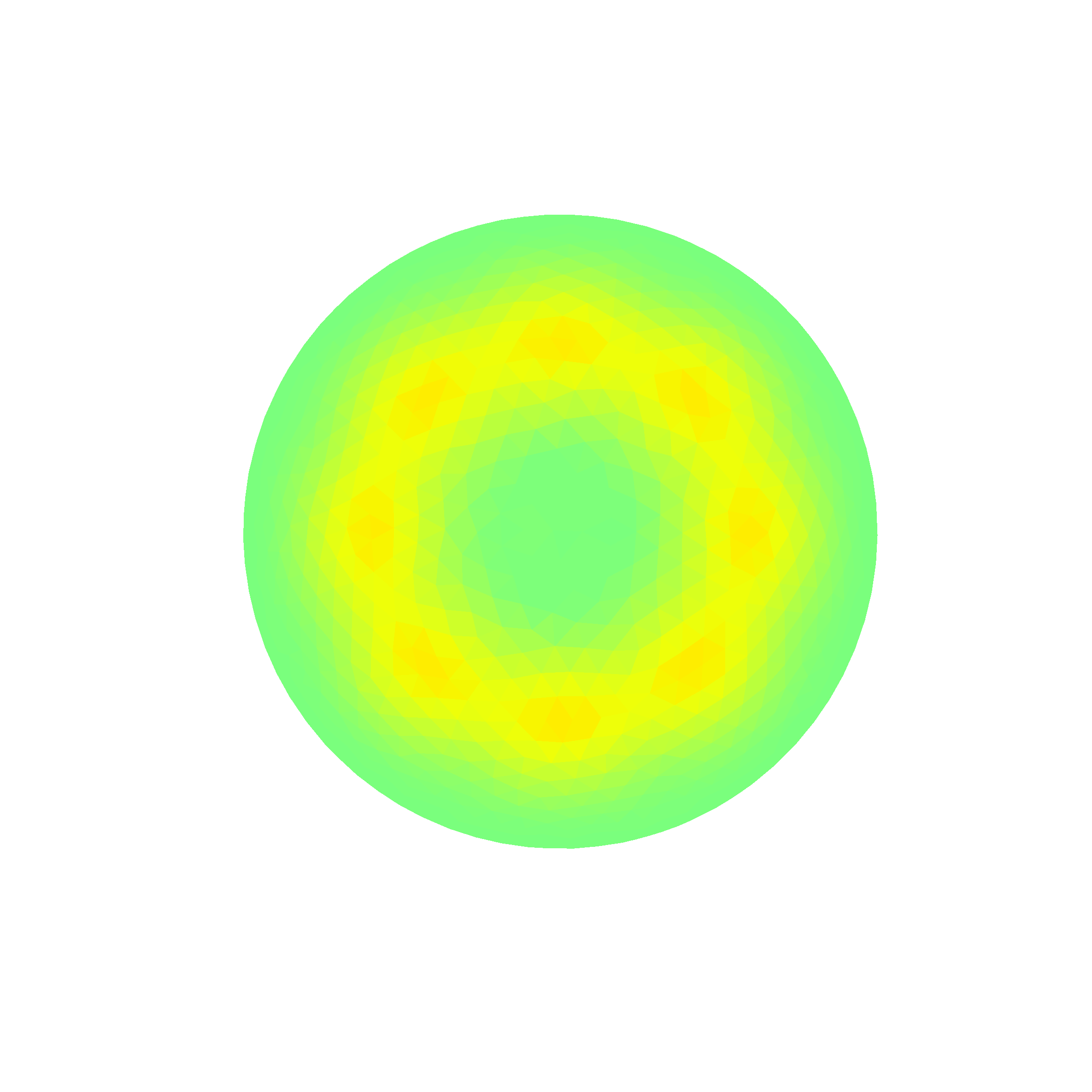}
    \includegraphics[width=2.4cm]{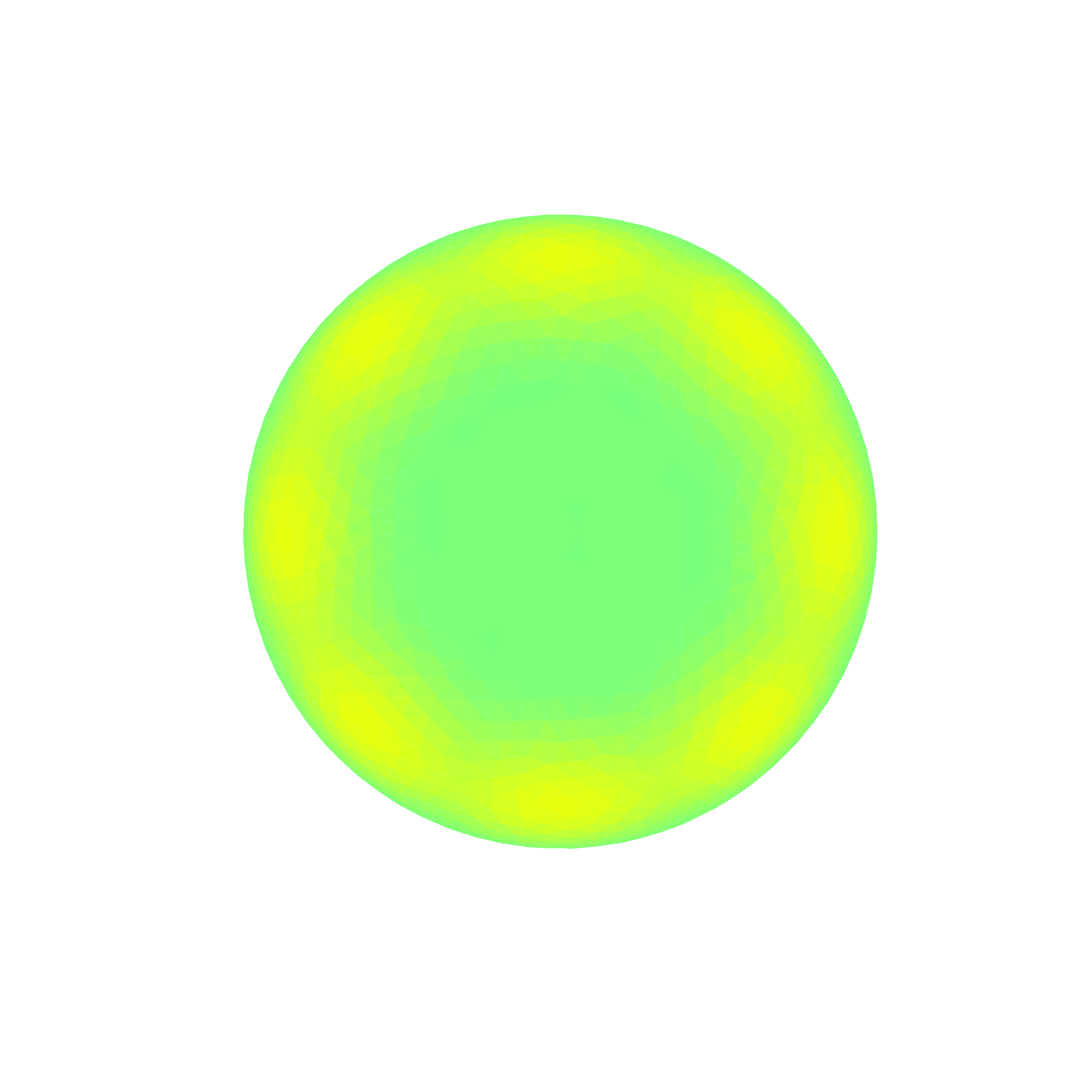}
    \includegraphics[width=2.4cm]{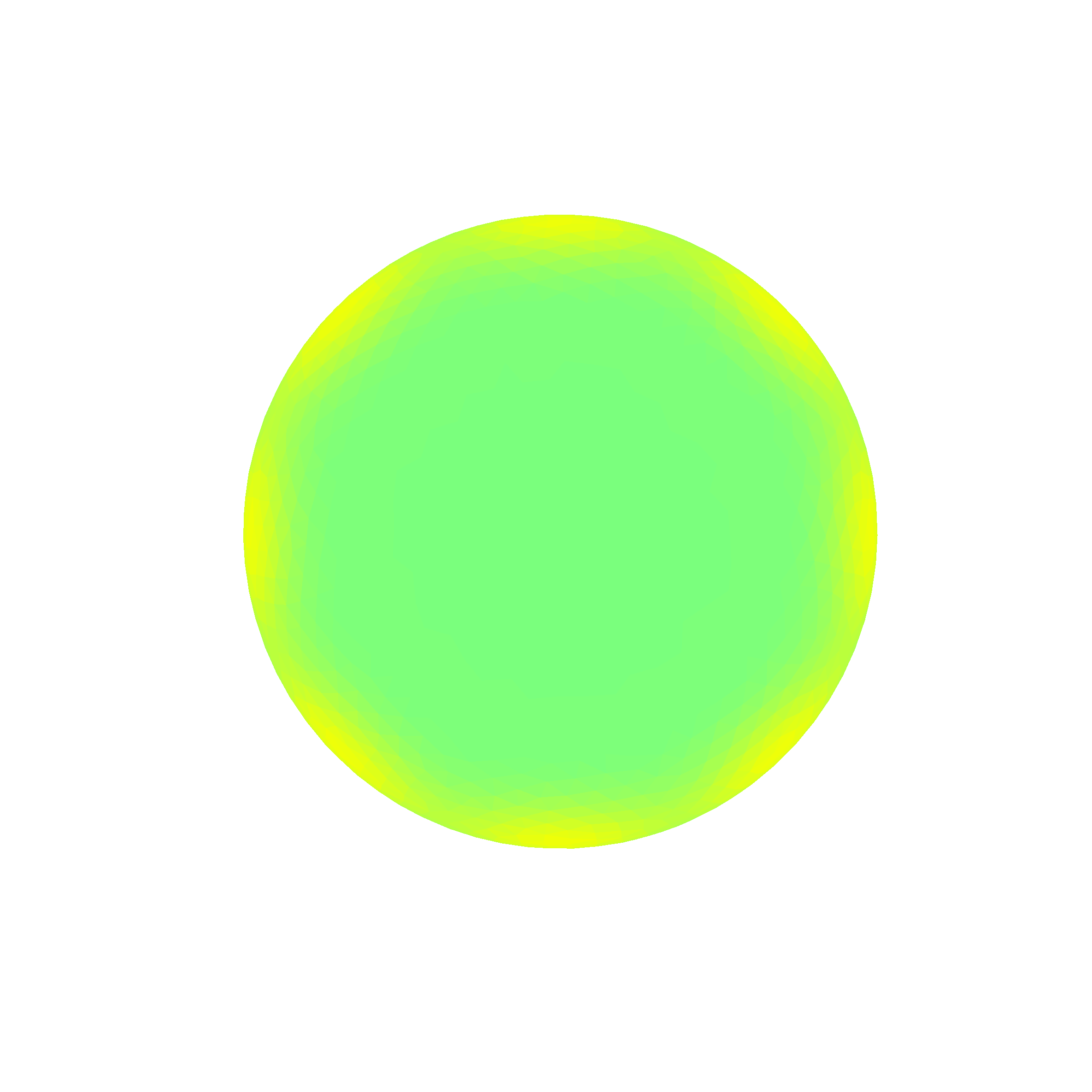}
    \includegraphics[width=2.9cm]{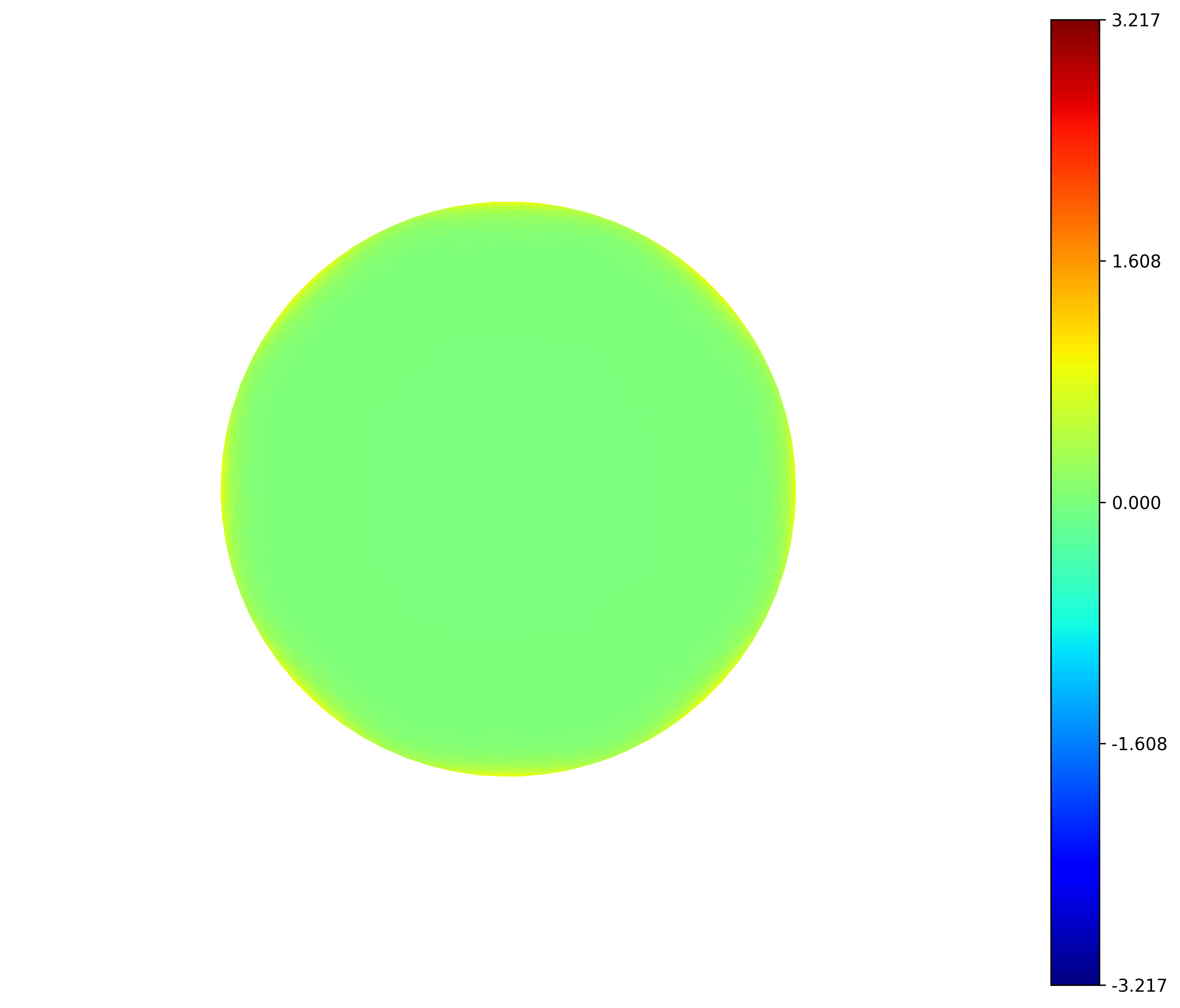}\\
    \vspace{5pt}

    \subfigure[$f(0, \boldsymbol{x})$]{
    \includegraphics[width=2.4cm]{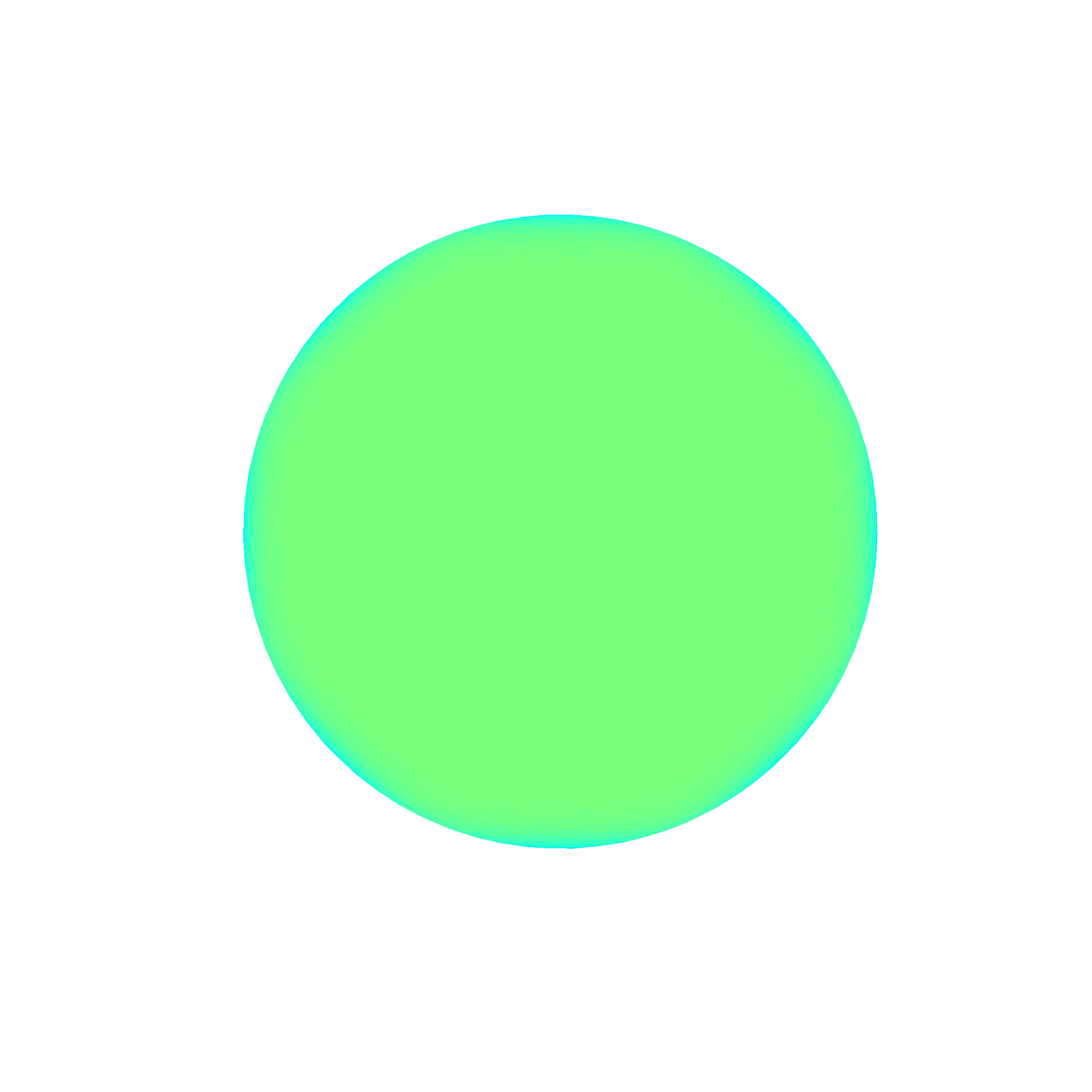}}
    \subfigure[$f(0.25, \boldsymbol{x})$]{
    \includegraphics[width=2.4cm]{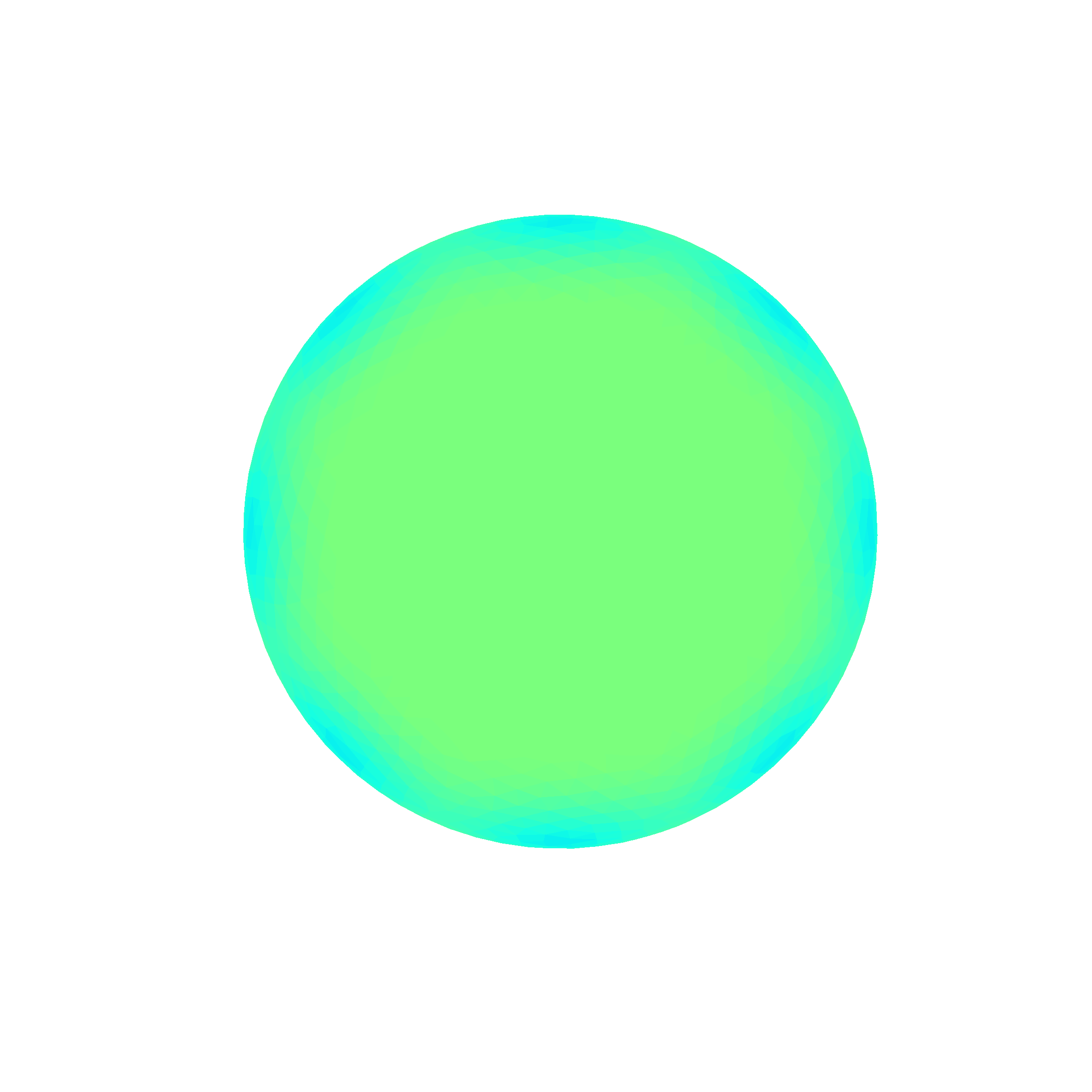}}
    \subfigure[$f(0.5, \boldsymbol{x})$]{
    \includegraphics[width=2.4cm]{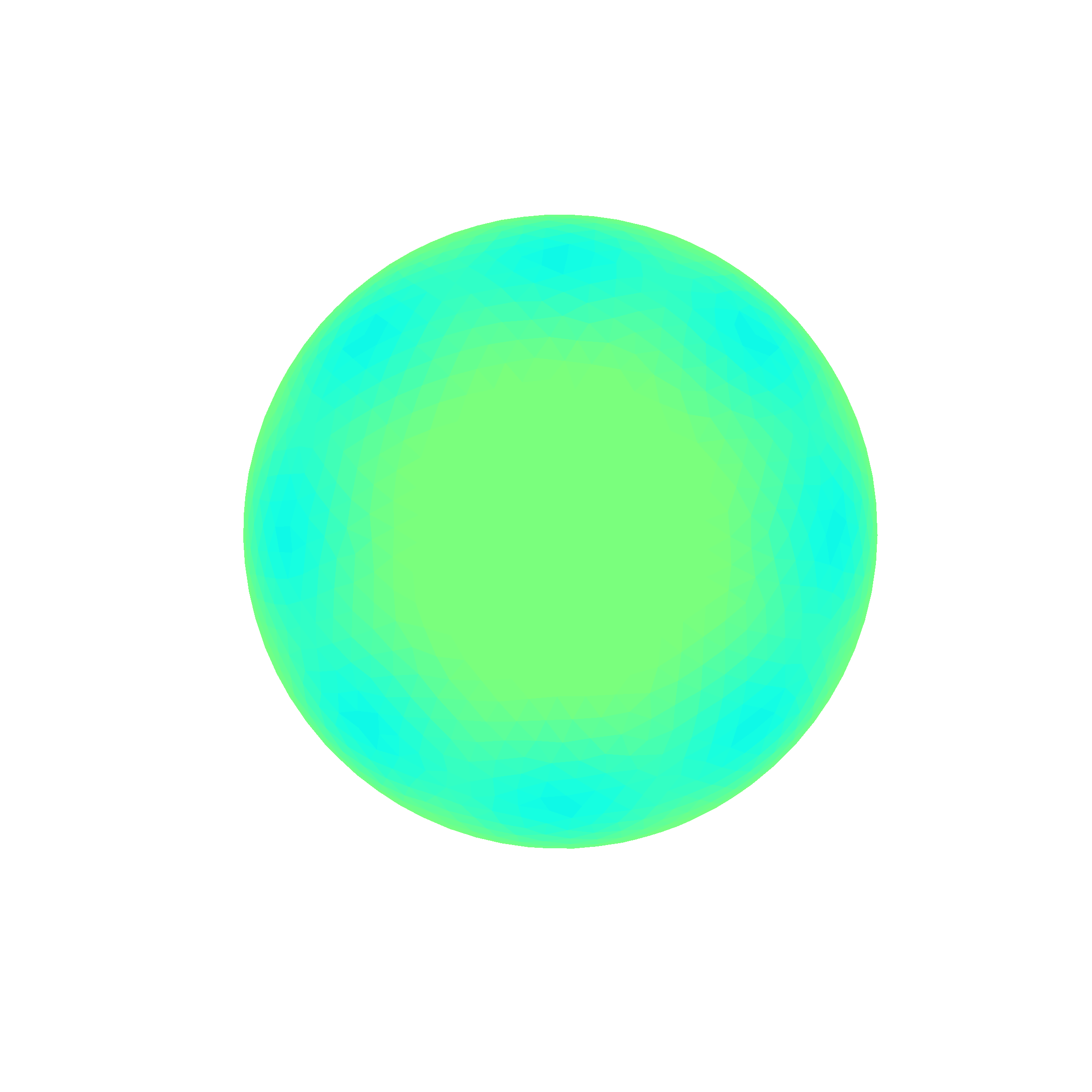}}
    \subfigure[$f(0.75, \boldsymbol{x})$]{
    \includegraphics[width=2.4cm]{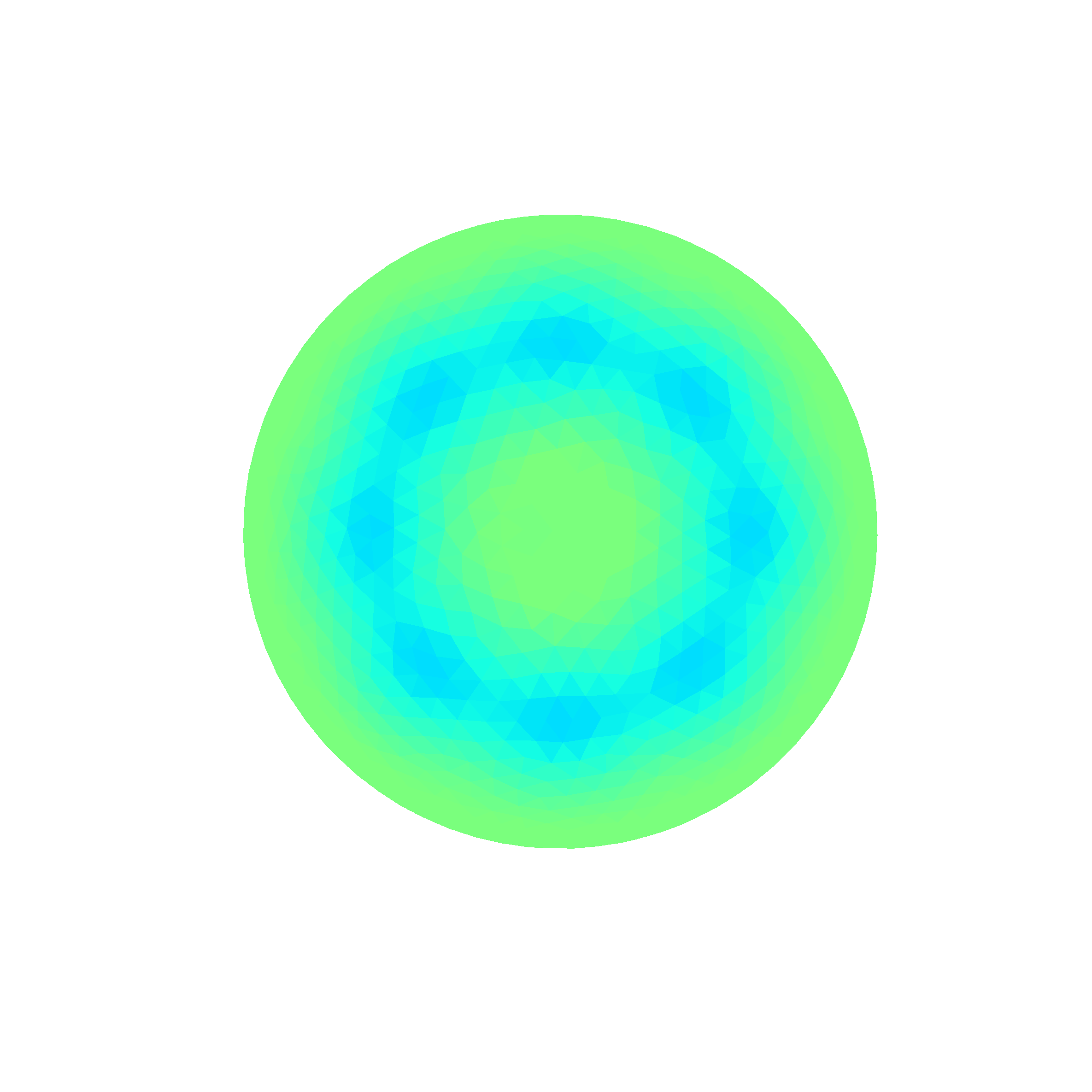}}
    \subfigure[$f(1, \boldsymbol{x})$]{
    \includegraphics[width=2.9cm]{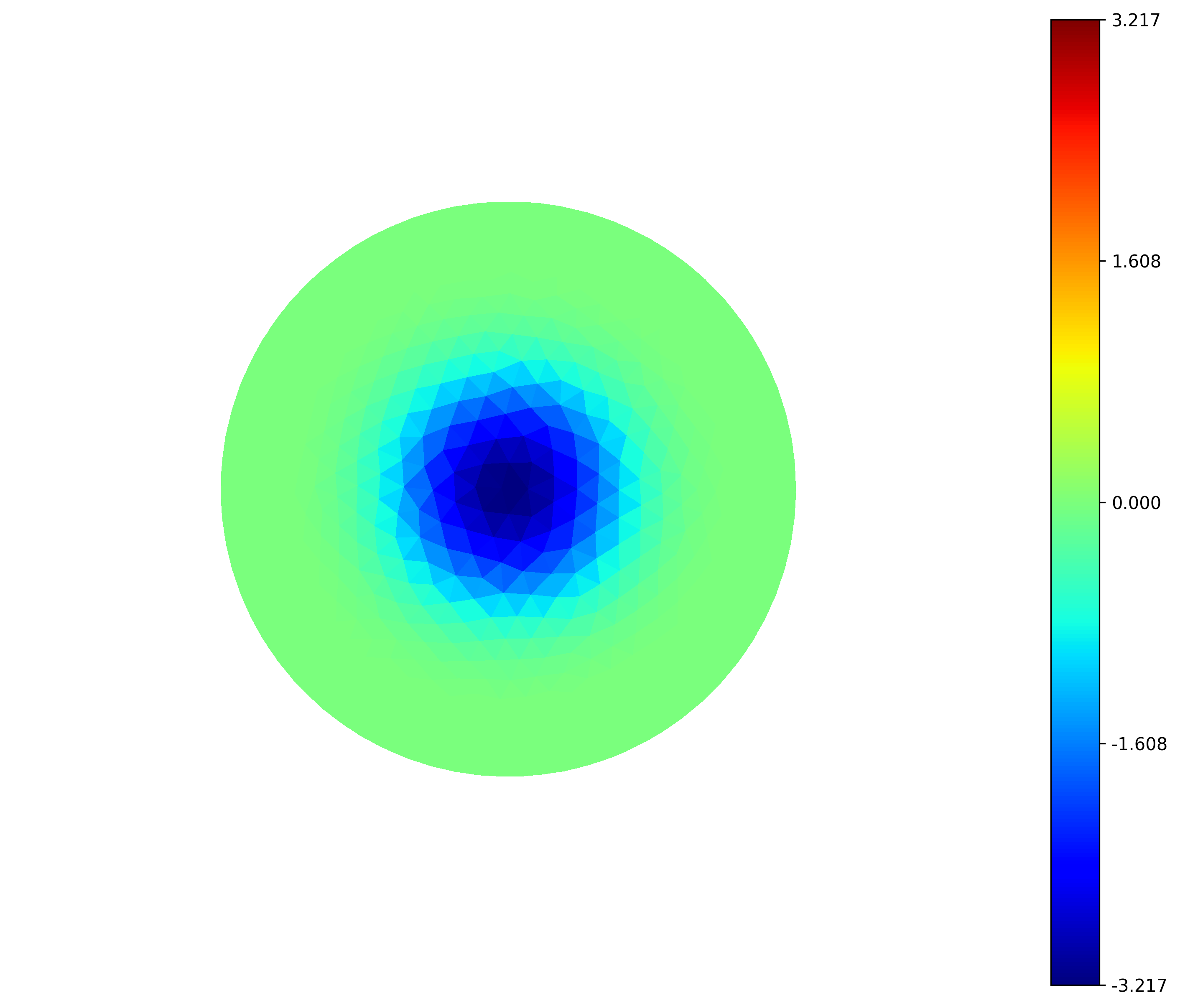}}\\
    \caption{SUOT test on point cloud for $S1$ and $S2$: source item $f$.}
    \label{SUOT-2-f}
    \end{center}
\end{figure}

As shown in Figure \ref{SUOT-2}, our method is capable to capture the phenomena of mass spliting and merging accurately. At the same time, symmetry is also preserved perfectly. In addition, the source term is shown in Figure \ref{SUOT-2-f}.

\begin{figure}[htbp]
    \begin{center}
    \rotatebox{90}{$~~~~~~\text{NN}$}
    \includegraphics[width=2.4cm]{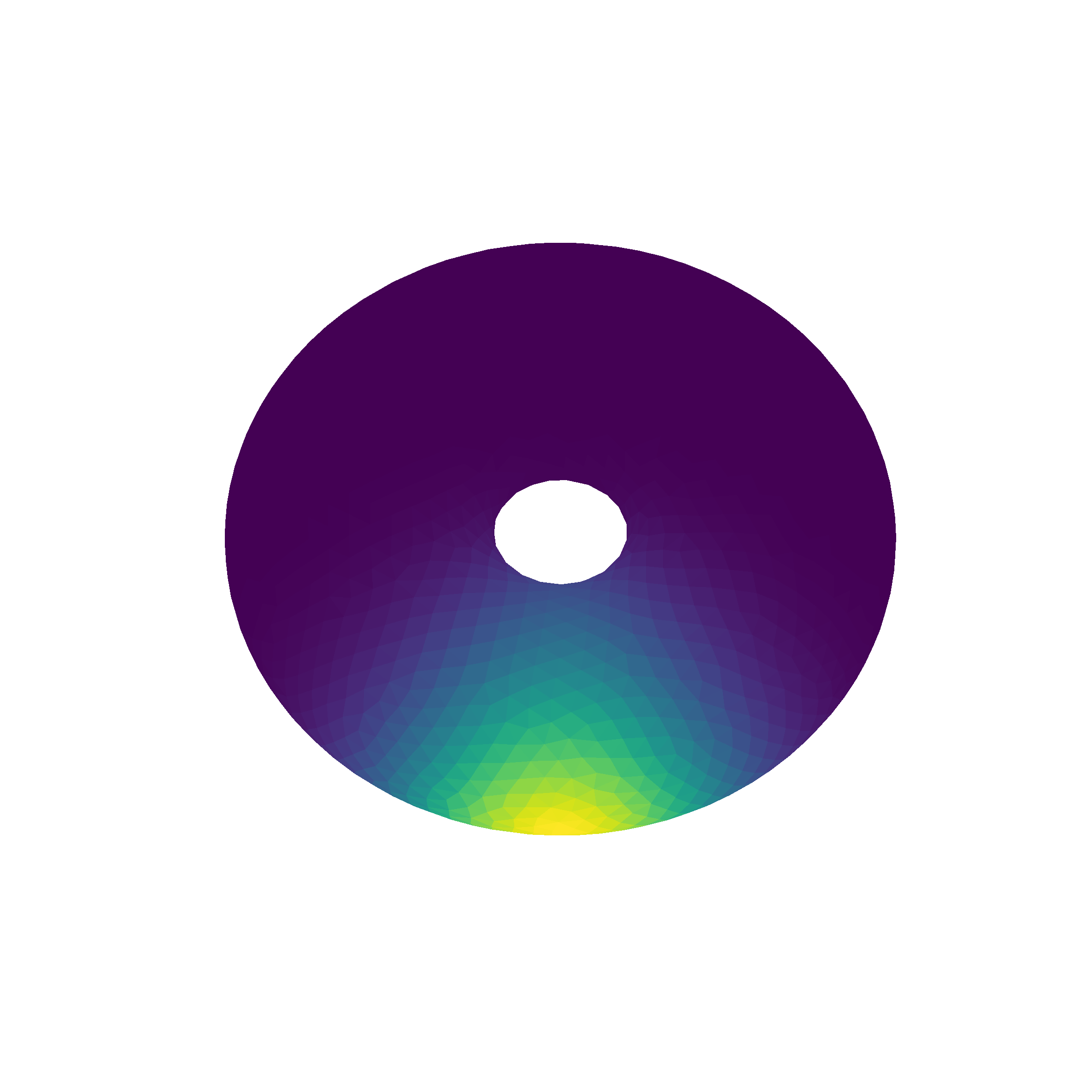}
    \includegraphics[width=2.4cm]{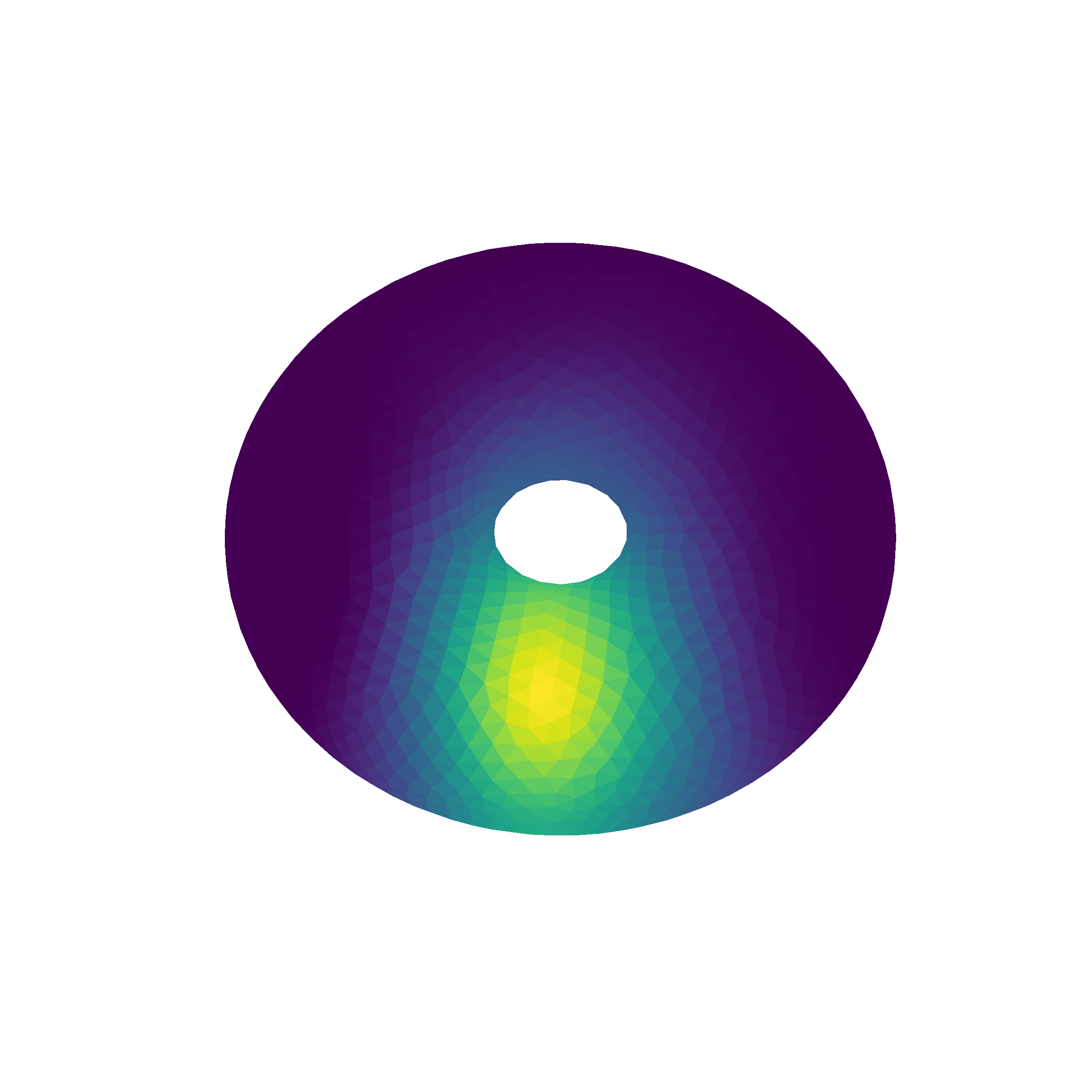}
    \includegraphics[width=2.4cm]{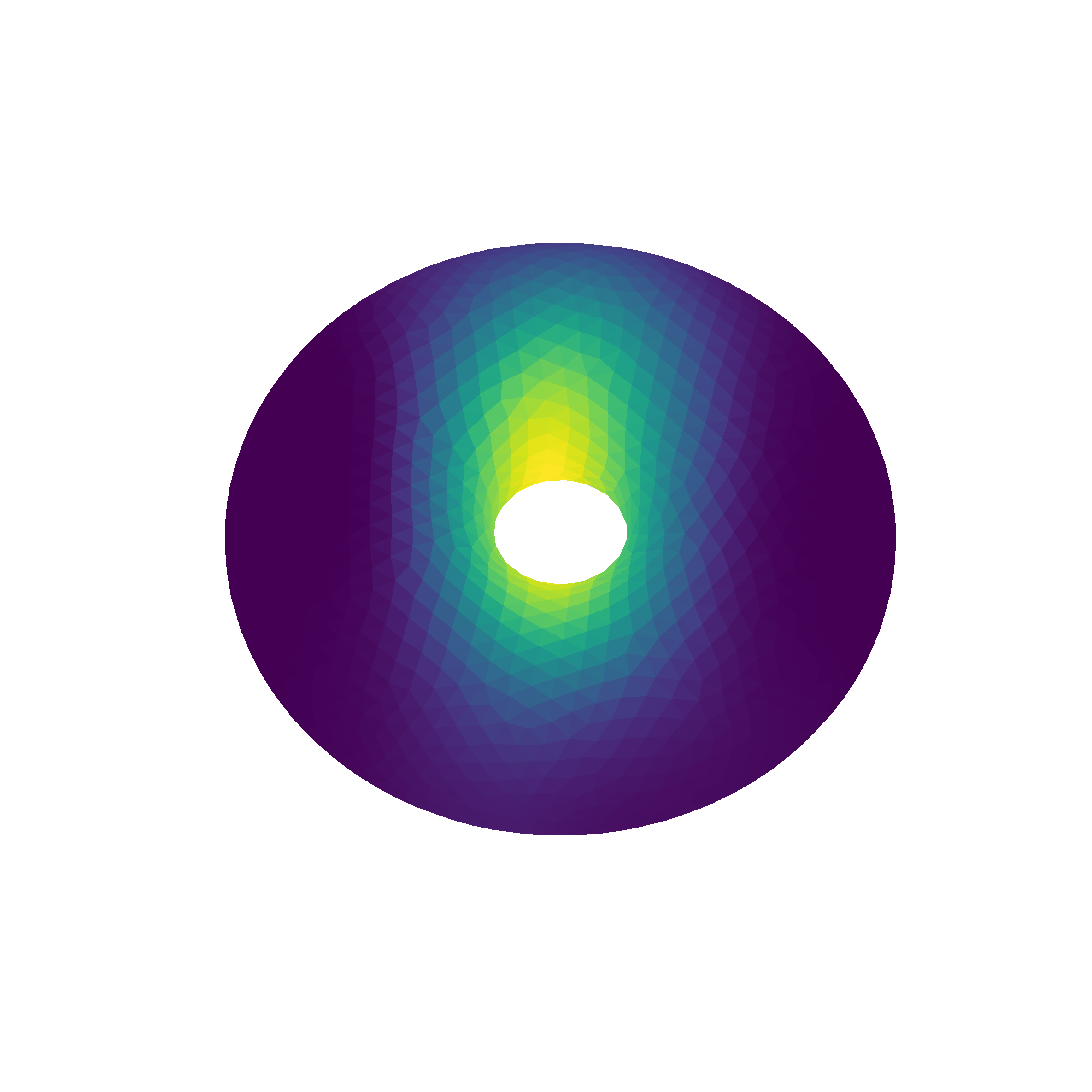}
    \includegraphics[width=2.4cm]{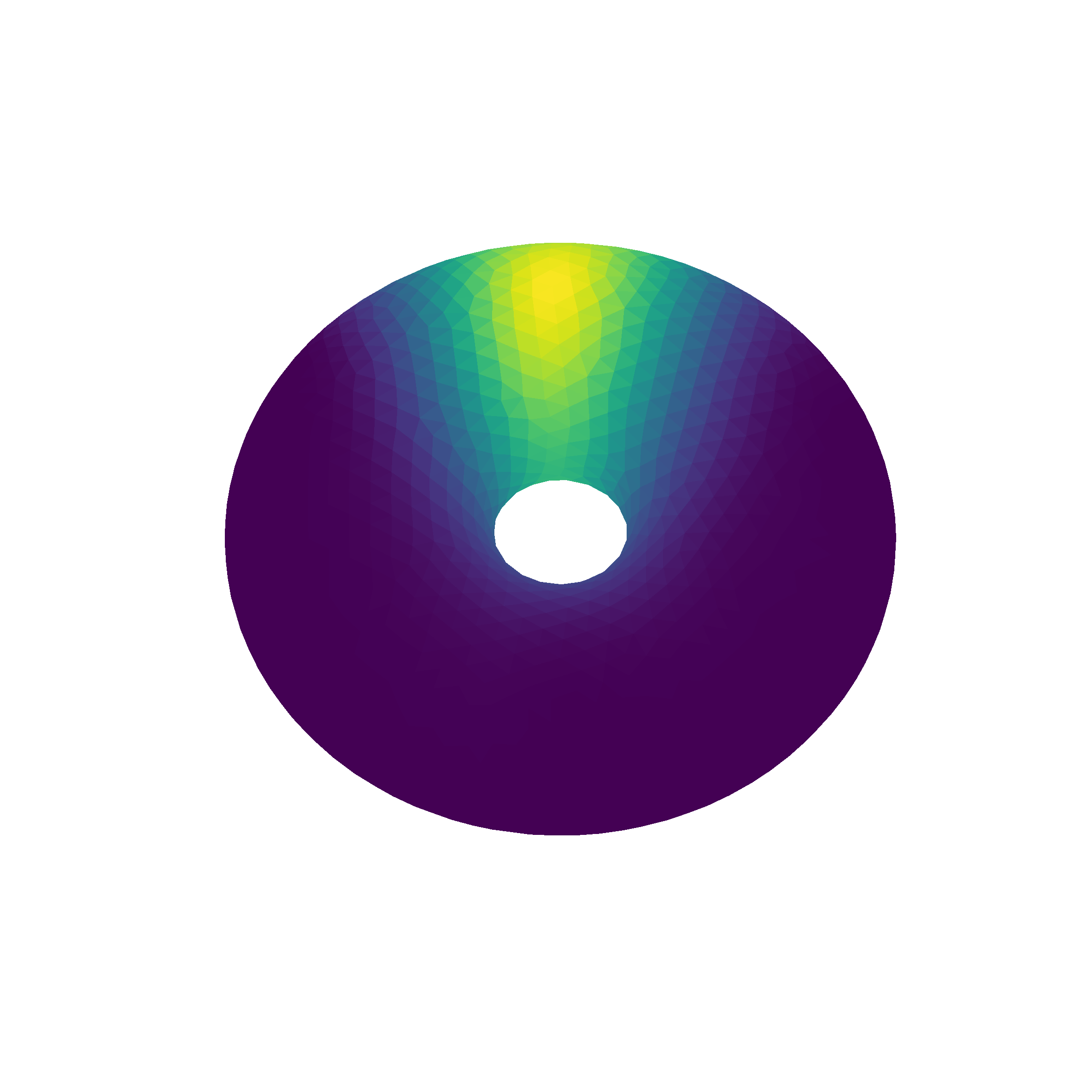}
    \includegraphics[width=2.4cm]{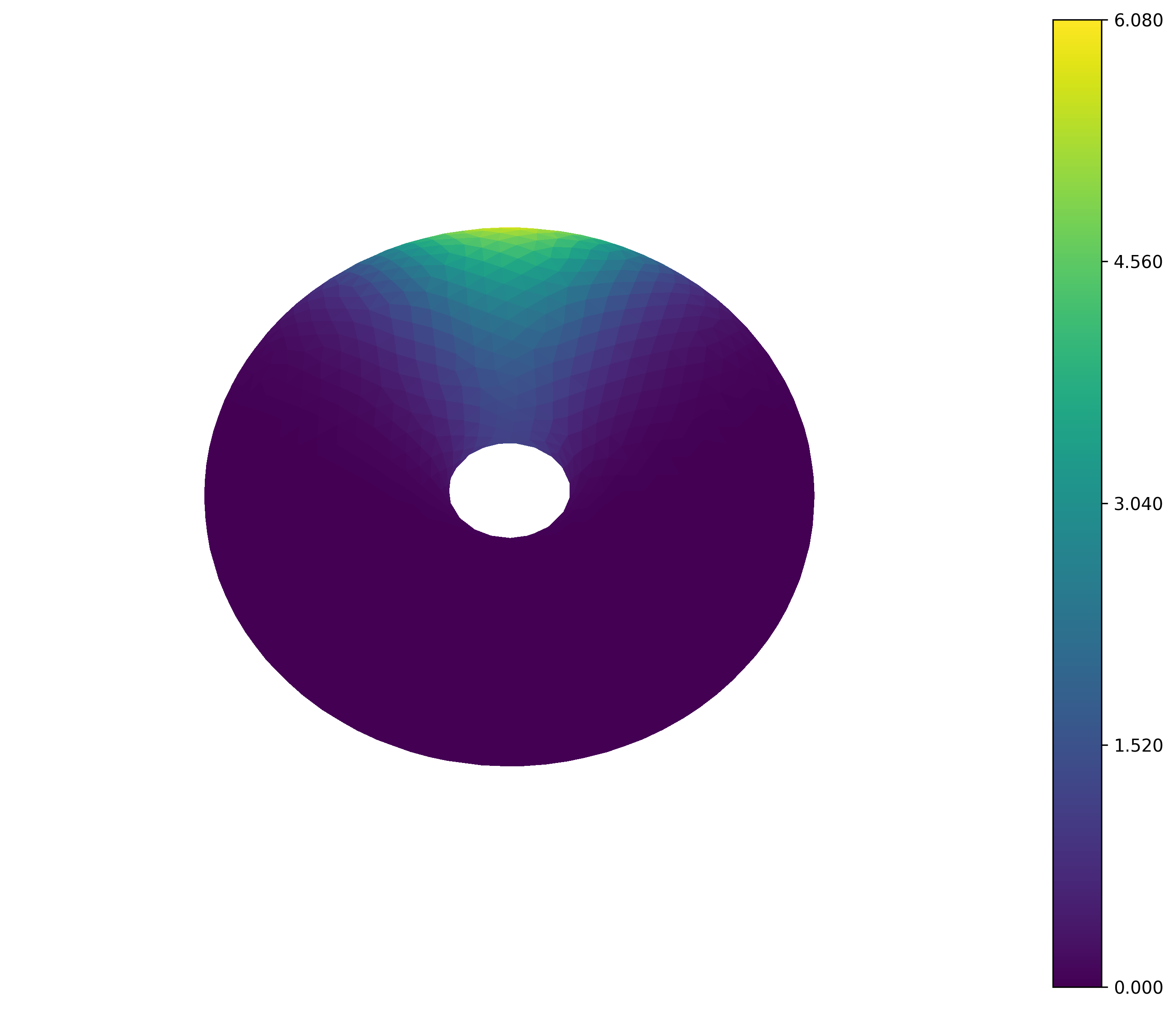}\\
    \vspace{5pt}
    
    \rotatebox{90}{$~~~~~~\text{GrSfem}$}
    \includegraphics[width=2.4cm]{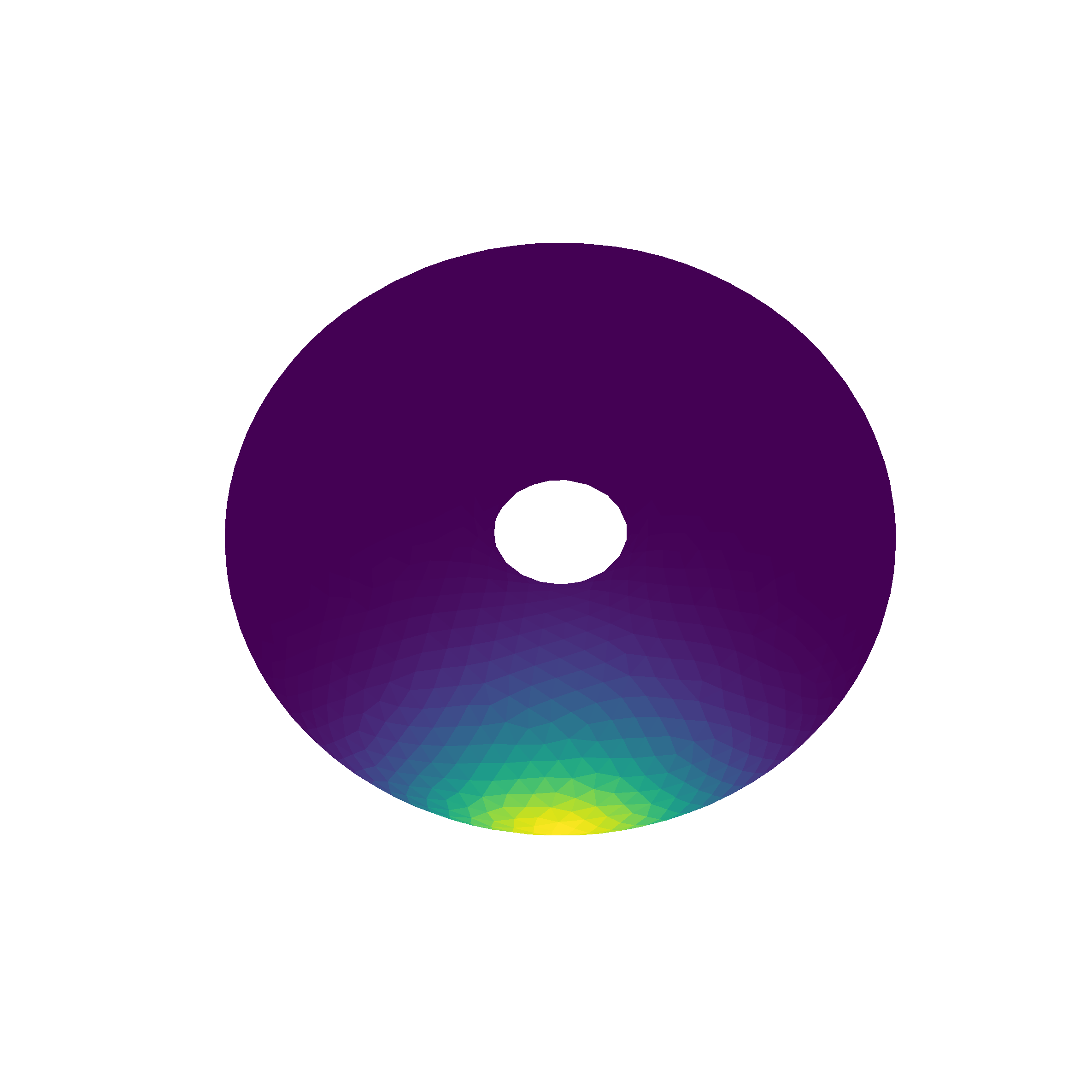}
    \includegraphics[width=2.4cm]{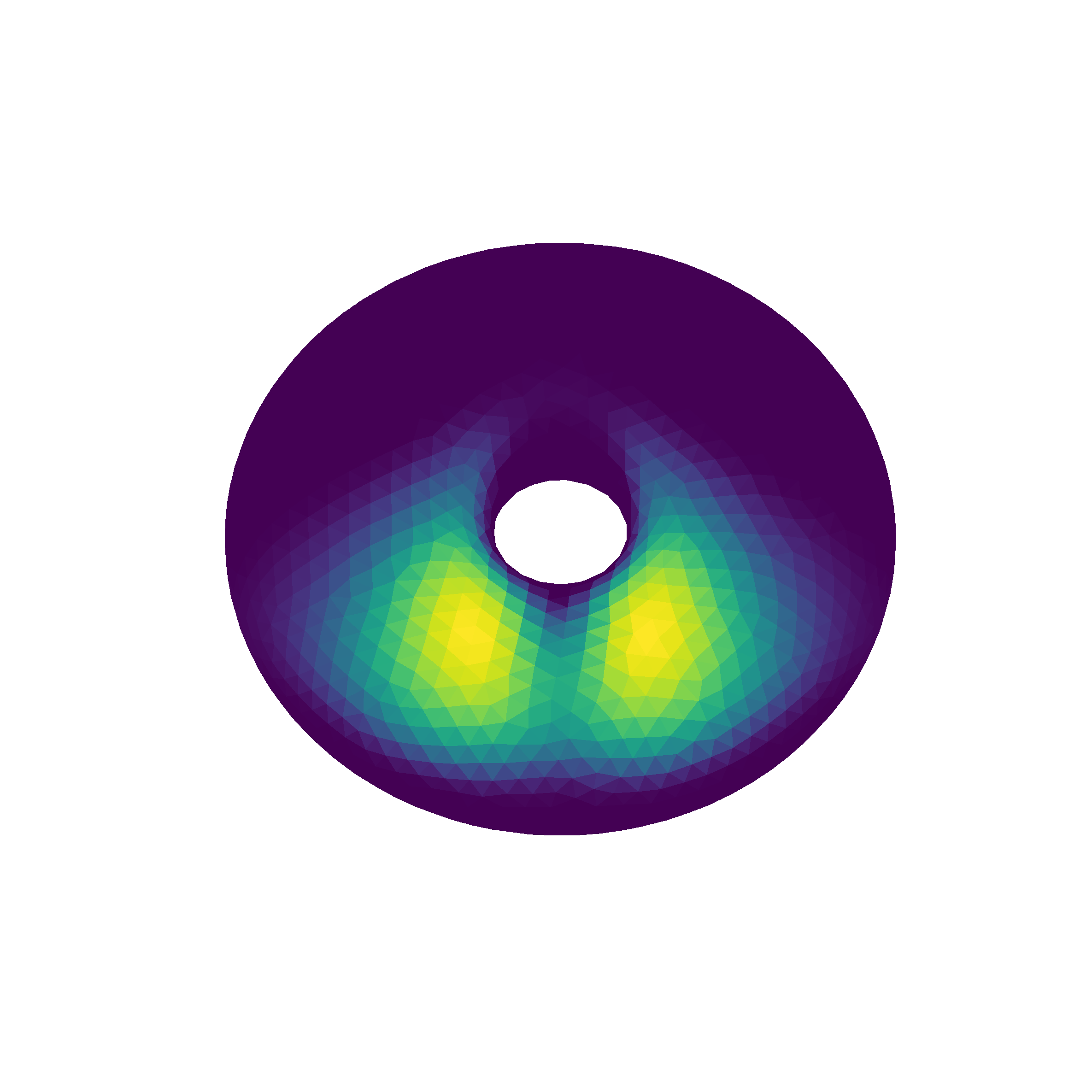}
    \includegraphics[width=2.4cm]{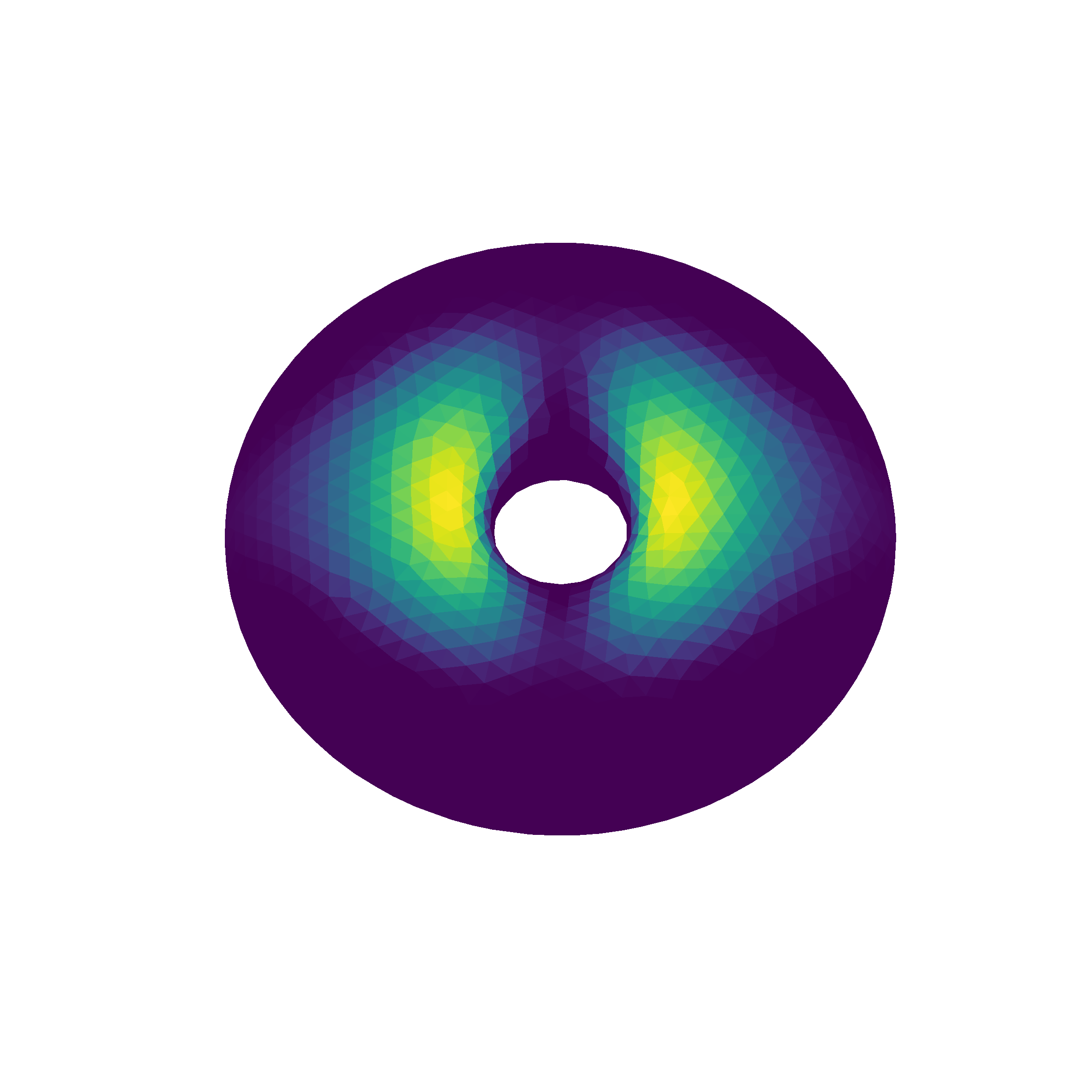}
    \includegraphics[width=2.4cm]{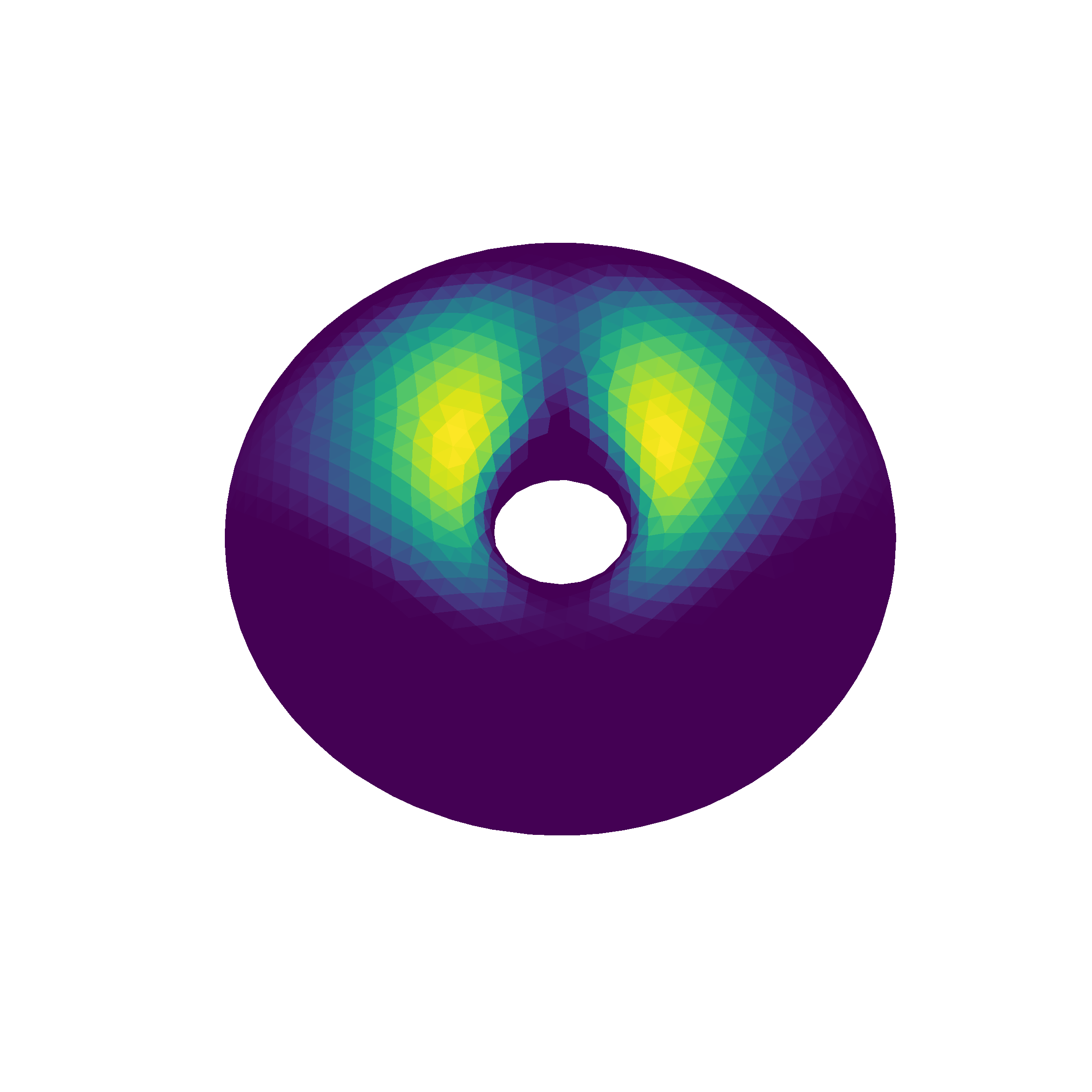}
    \includegraphics[width=2.4cm]{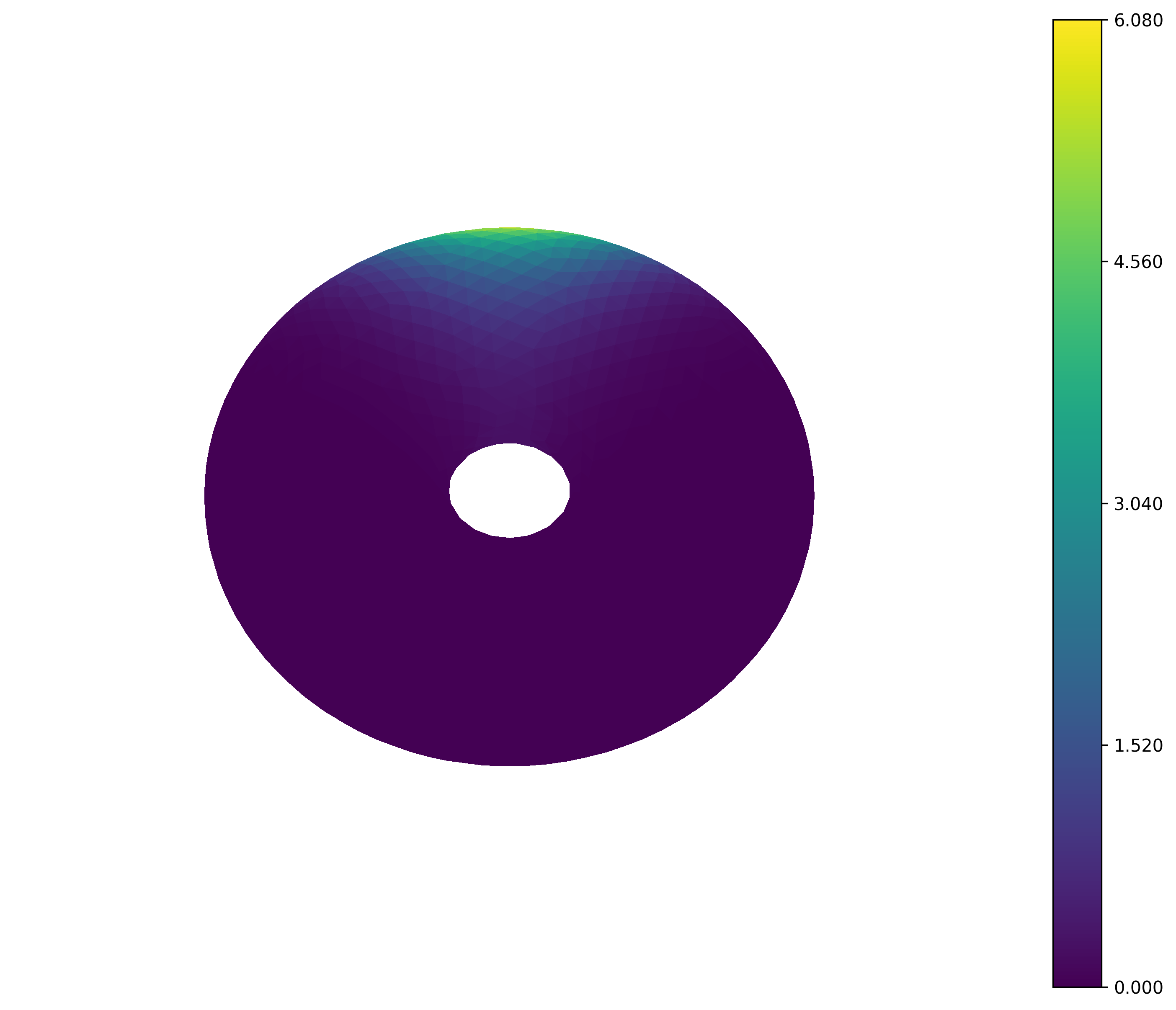}\\
    \vspace{5pt}
    
    \rotatebox{90}{$~~~~~~\text{TRBF}$}
    \subfigure[$\rho(0, \boldsymbol{x})$]{\includegraphics[width=2.4cm]{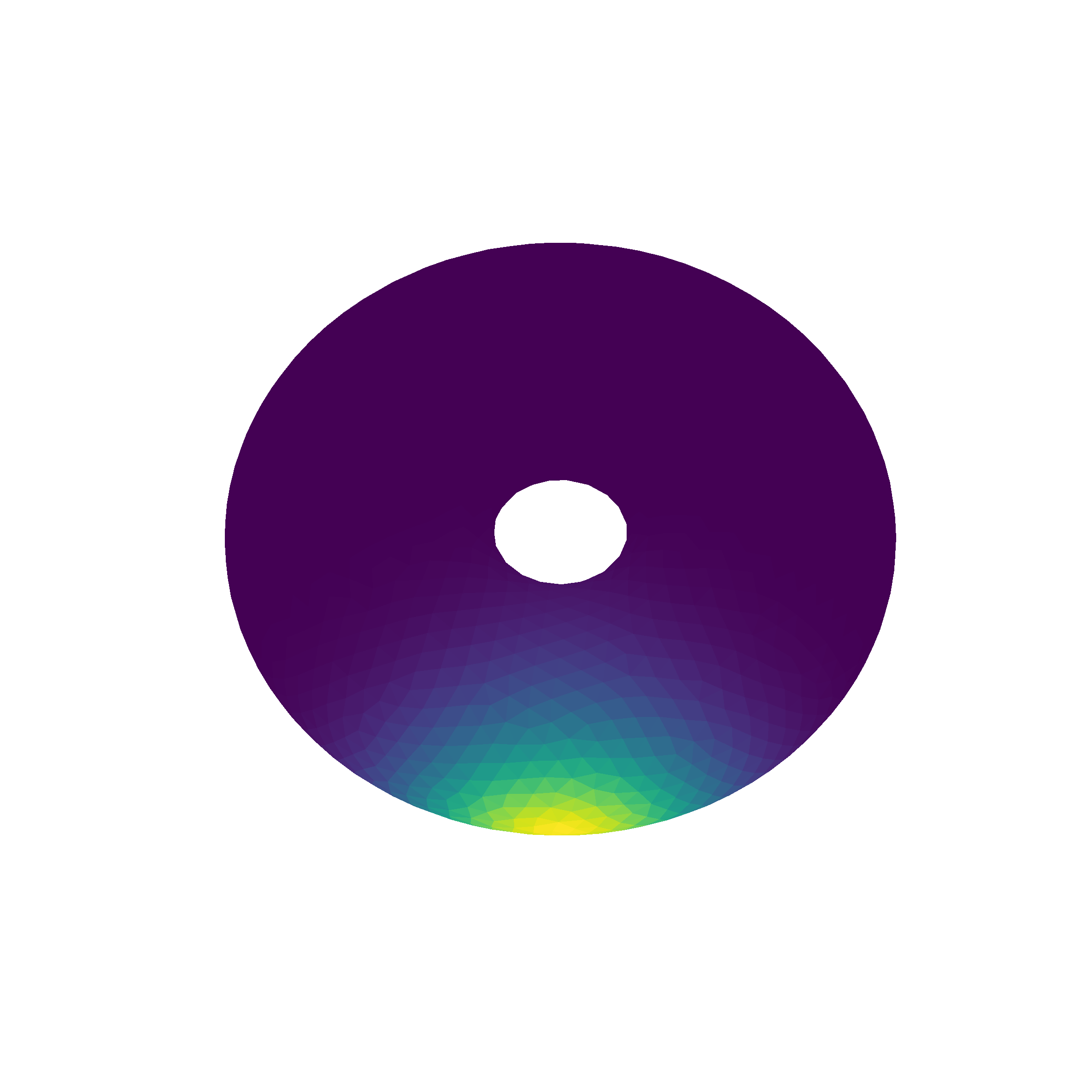}}
    \subfigure[$\rho(0.25, \boldsymbol{x})$]{\includegraphics[width=2.4cm]{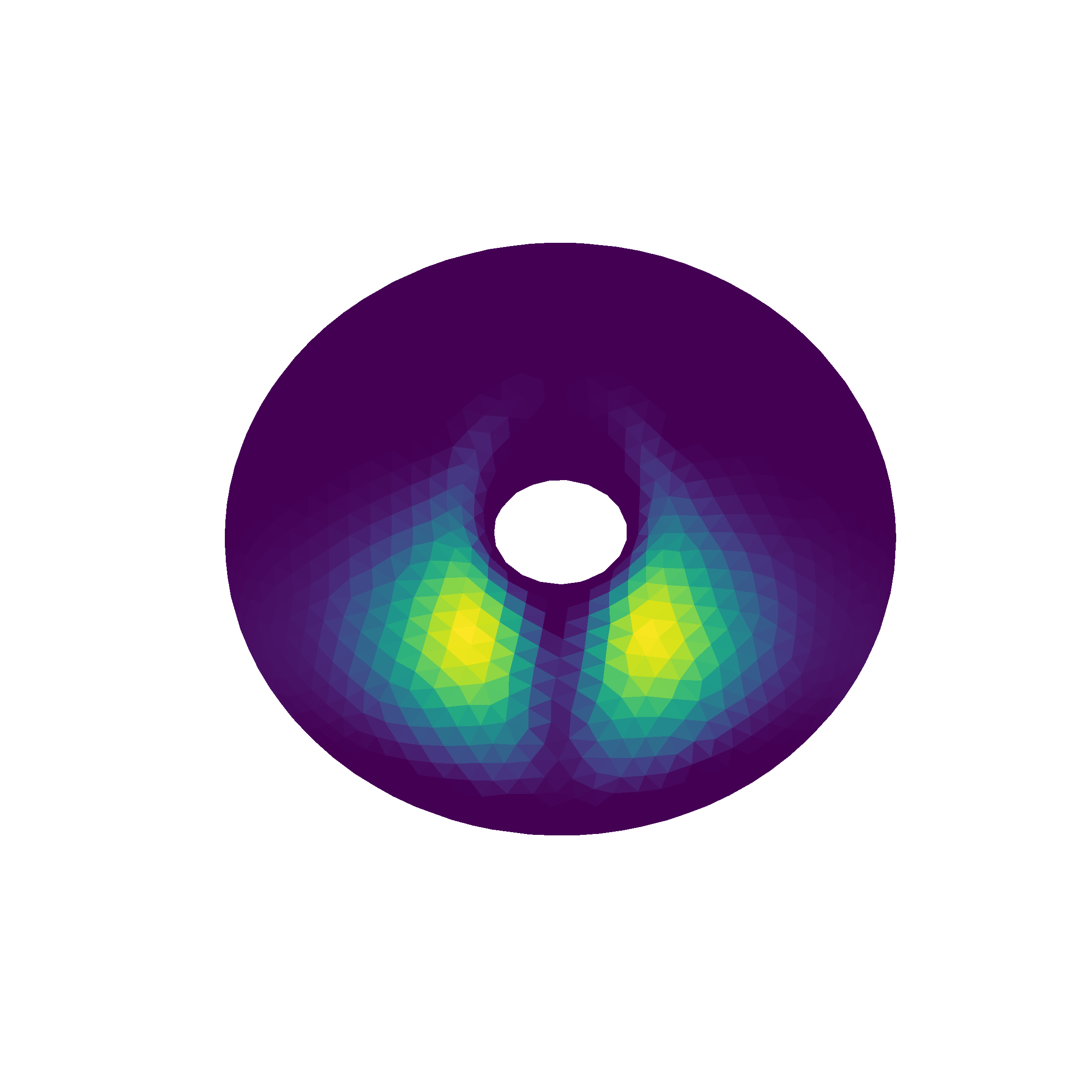}}
    \subfigure[$\rho(0.5, \boldsymbol{x})$]{\includegraphics[width=2.4cm]{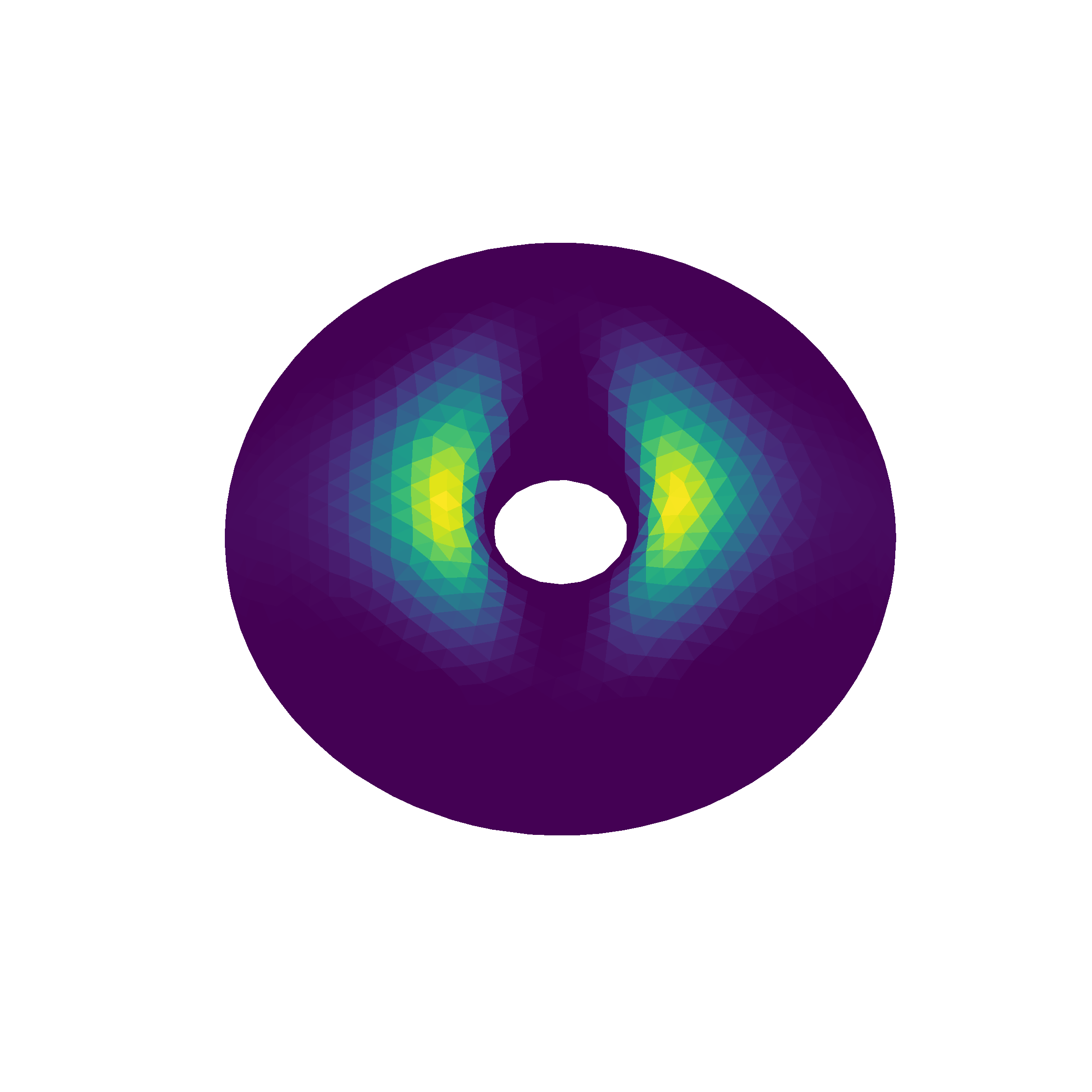}}
    \subfigure[$\rho(0.75, \boldsymbol{x})$]{\includegraphics[width=2.4cm]{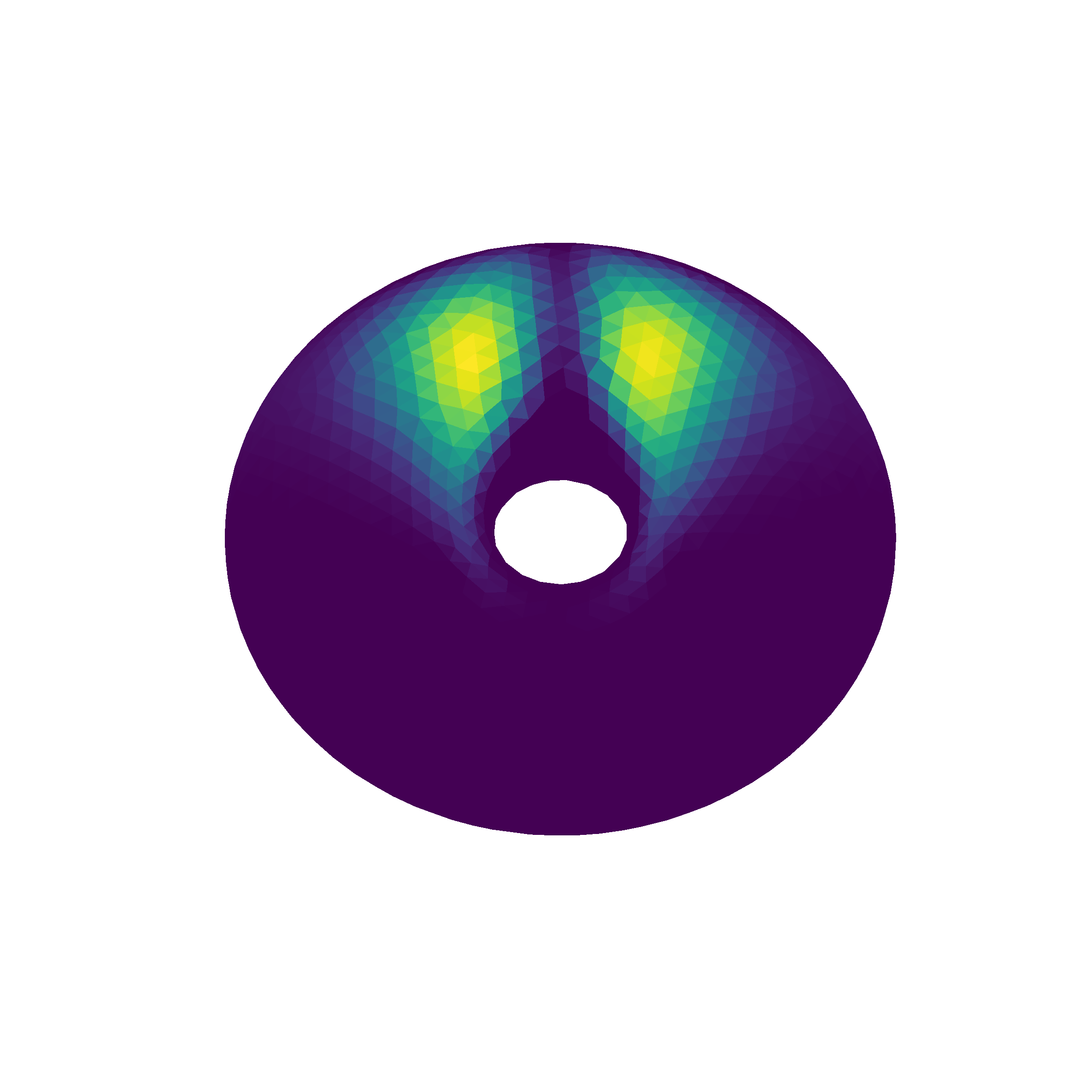}}
    \subfigure[$\rho(1, \boldsymbol{x})$]{\includegraphics[width=2.4cm]{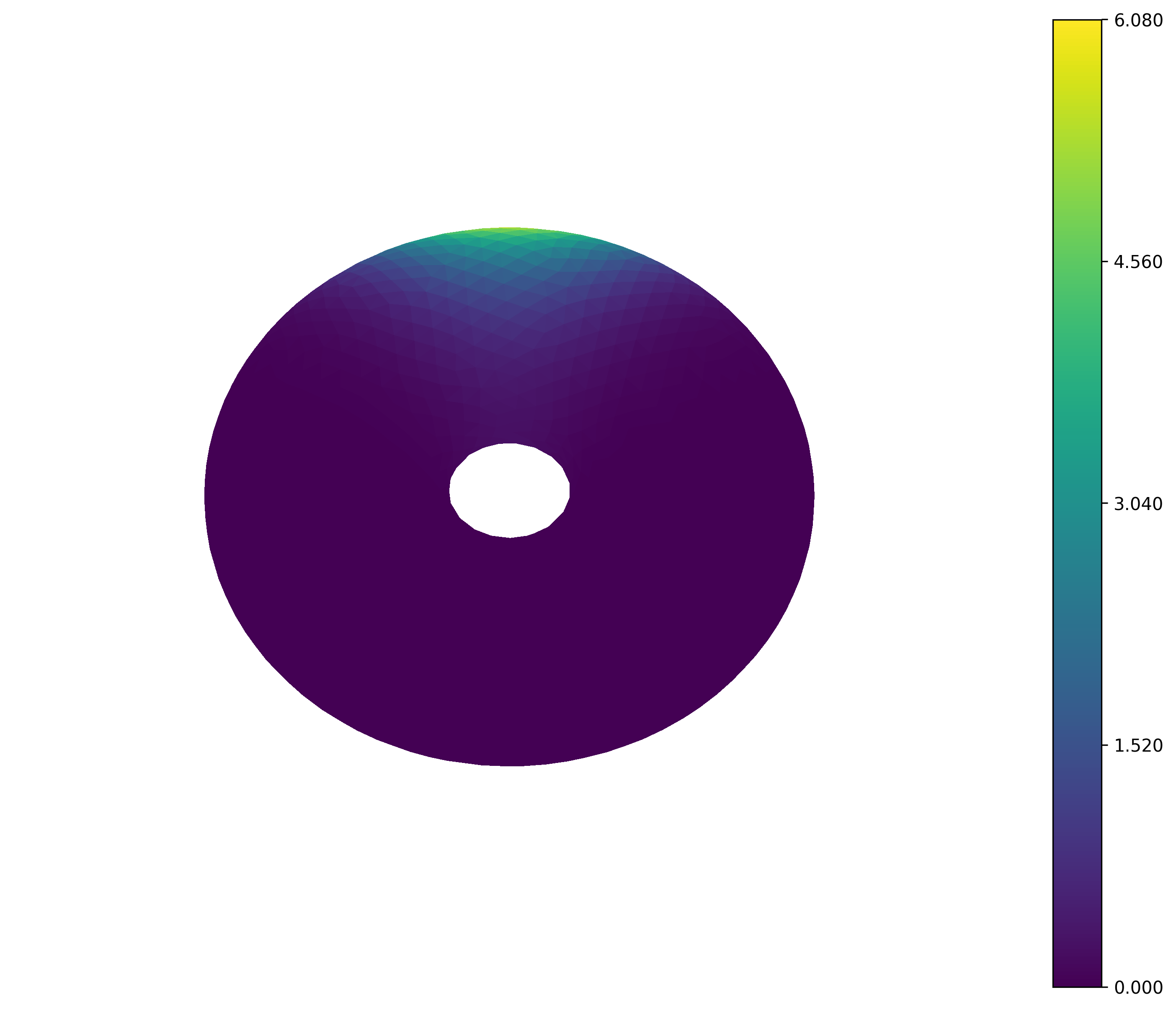}}
    
    \caption{Compare with different methods on torus. The first line: method based on neural network; The second line: surface finite element method based on gradient recovery; Last line: our tangent radial basis function method.}
    \label{compare-OT-Torus}
    \end{center}
\end{figure}

In addition, we compare the results on the torus with finite element based method (GrSfem) \cite{dong2024gradient} and a neural network based method (NN)\cite{pan2024network}. The results are shown in Fig.\ref{compare-OT-Torus}. 

It can be found that the results obtained by our method are almost consistent with those obtained by GrSfem. The accuracy of the neural network method is relatively low.

Finally, we compare the computational time and memory usage with GrSfem \cite{dong2024gradient} and NN \cite{pan2024network} on a sphere OT example. 

\begin{table}[htbp]
    \centering
    \caption{Memory usage (MB) with different size of point cloud.}
    \label{tab:compute-memory}
    \begin{tabular}{c|ccccc}
        \hline 
          $N_{n}$ & 390 & 1806 & 3934 & 5878 & 11406 \\
        \hline
        NN(cpu) & 51237 & 53331 & 98597 & 84969 & 144085 \\
        NN(gpu) & 1169 & 5228 & 10878 & 16202 & 32379 \\
        GrSfem(cpu) & 38 & 223 & 559 & 886 & 1905 \\
        TRBF(cpu) & 191 & 216 & 371 & 546 & 1400 \\
        \hline
    \end{tabular}
\end{table}

\begin{table}[htbp]
    \centering
    \caption{CPU time (s) with different size of of point cloud.}
    \label{tab:compute-time}
    \begin{tabular}{c|ccccc}
        \hline 
          $N_{n}$ & 390 & 1806 & 3934 & 5878 & 11406 \\
        \hline
       NN (GPU) & 10422 & 13578 & 18931 & 31746 & 48493 \\
        GrSfem & 14 & 27 & 48 & 65 & 119 \\
        TRBF& 93 & 442 & 1078 & 2146 & 3934 \\
        \hline
    \end{tabular}
\end{table}

The maximum memory usage and computational time are listed in Table \ref{tab:compute-memory} and Table \ref{tab:compute-time} respectively. Neural network based methods use much larger memory than finite element method or the proposed TRBF method. The computational time is also much longer. The memory usage of FEM and TRBF method are comparable. While the computational time of FEM method is much less than TRBF method. Notice that FEM requires mesh information and the mesh generation is not included in Table \ref{tab:compute-time}. Taking advantage of the mesh structure in FEM, The coefficient matrix is much sparser than that in TRBF method such that the assembling of coefficient matrix is much faster and the linear system is also much easier to solve. 

\subsection{Unbalanced optimal transport on general point cloud}\label{Subsec-OTGS}
Now, we test our method on general point cloud. 
The point clouds come from Keenan's 3D Model Repository\cite{lavenant2018dynamical}. 
The initial and terminal distributions are given in Table \ref{tab:SOT-general} of \ref{App-C}.
\begin{figure}[htbp]
    \begin{center}
    \includegraphics[width=2.4cm]{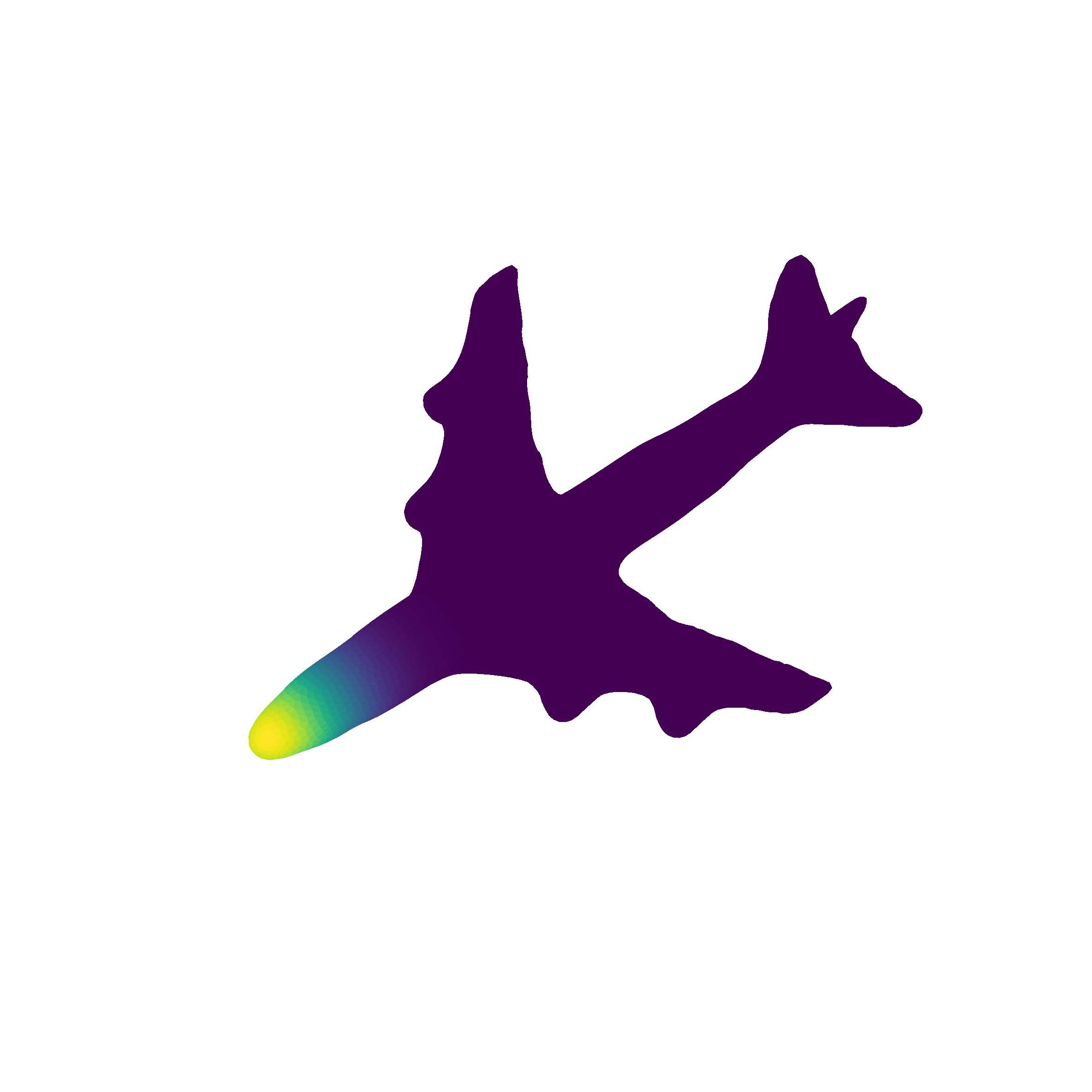}
    \includegraphics[width=2.4cm]{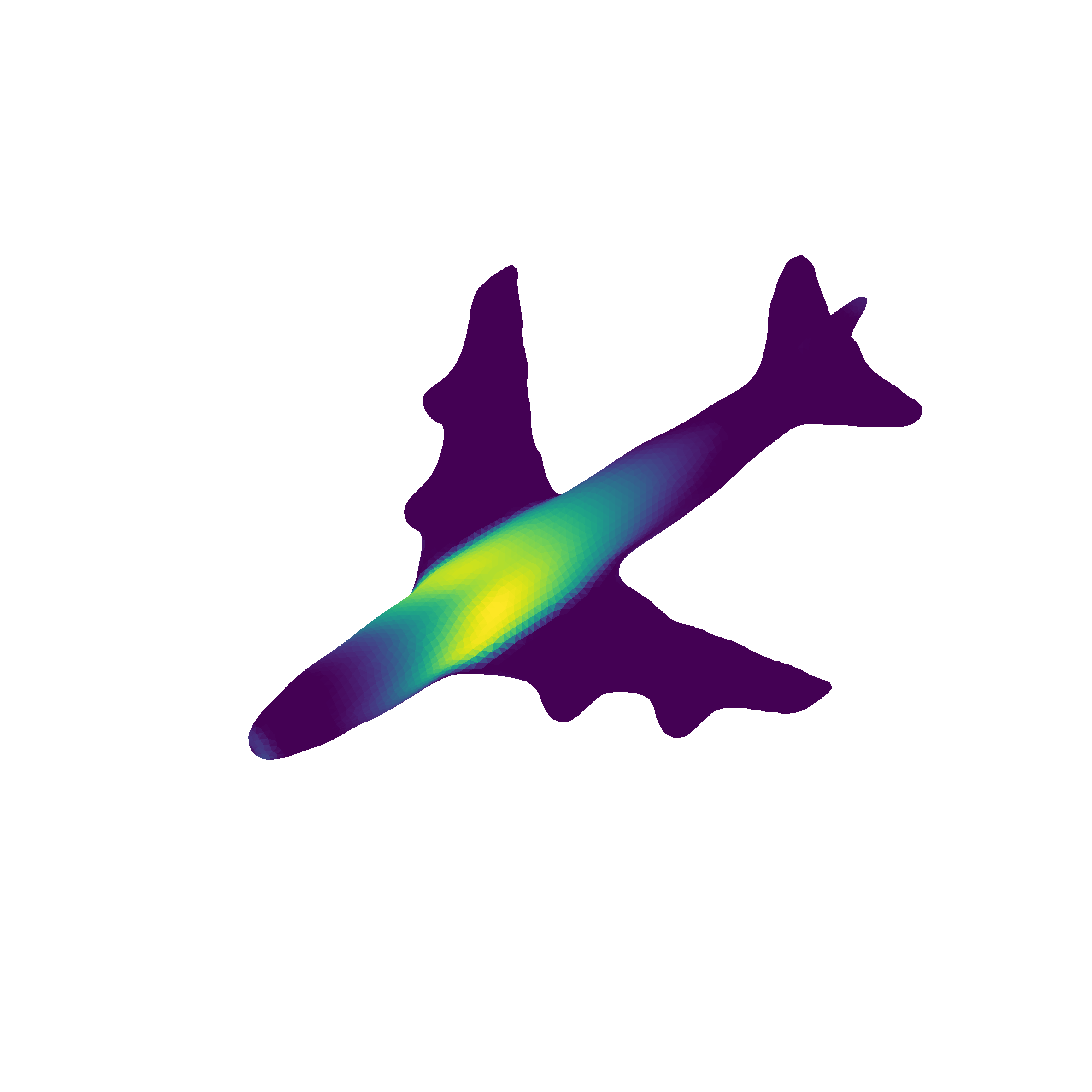}
    \includegraphics[width=2.4cm]{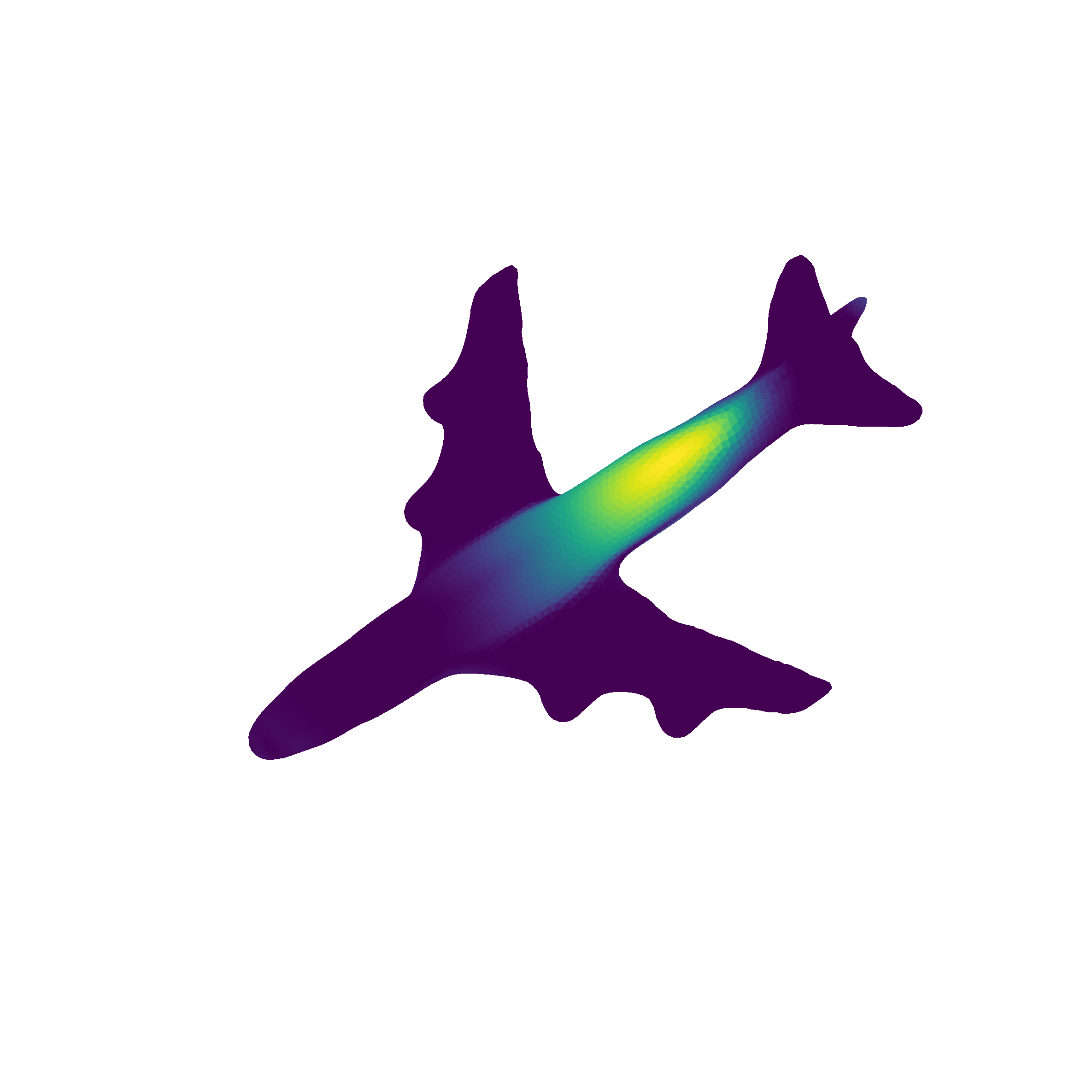}
    \includegraphics[width=2.4cm]{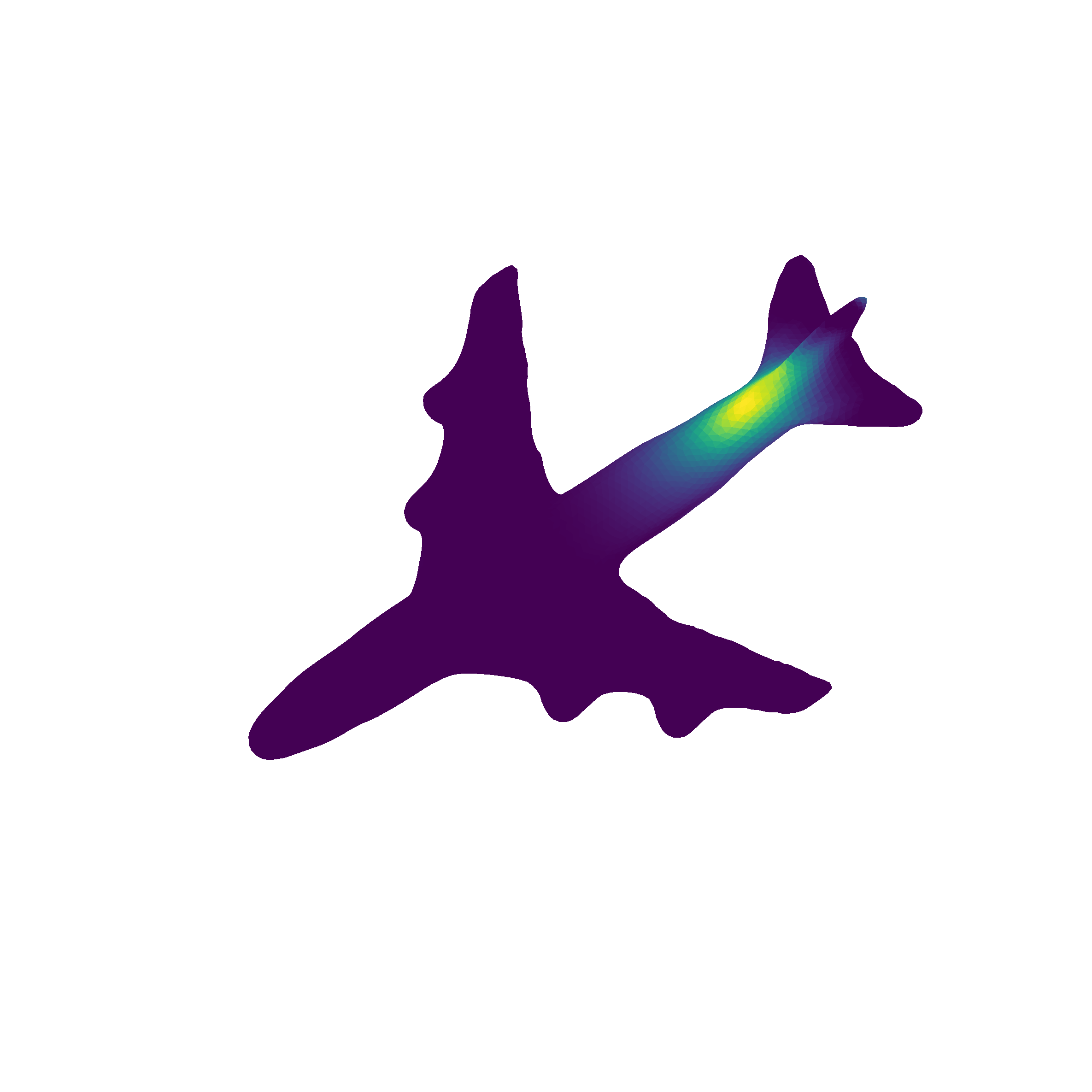}
    \includegraphics[width=2.9cm]{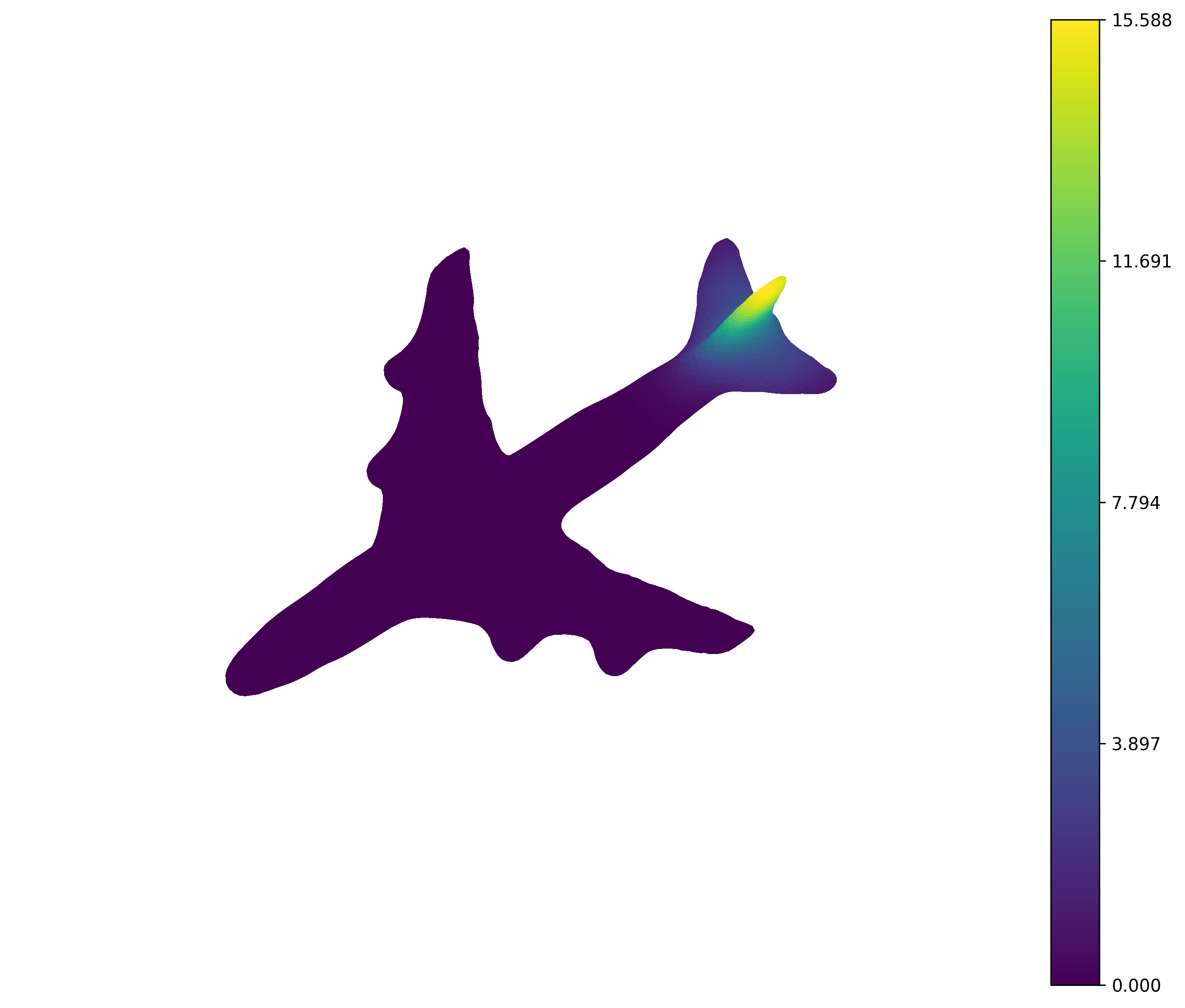}\\
    \vspace{5pt}

    \subfigure[$\rho(0, \boldsymbol{x})$]{\includegraphics[width=2.4cm]{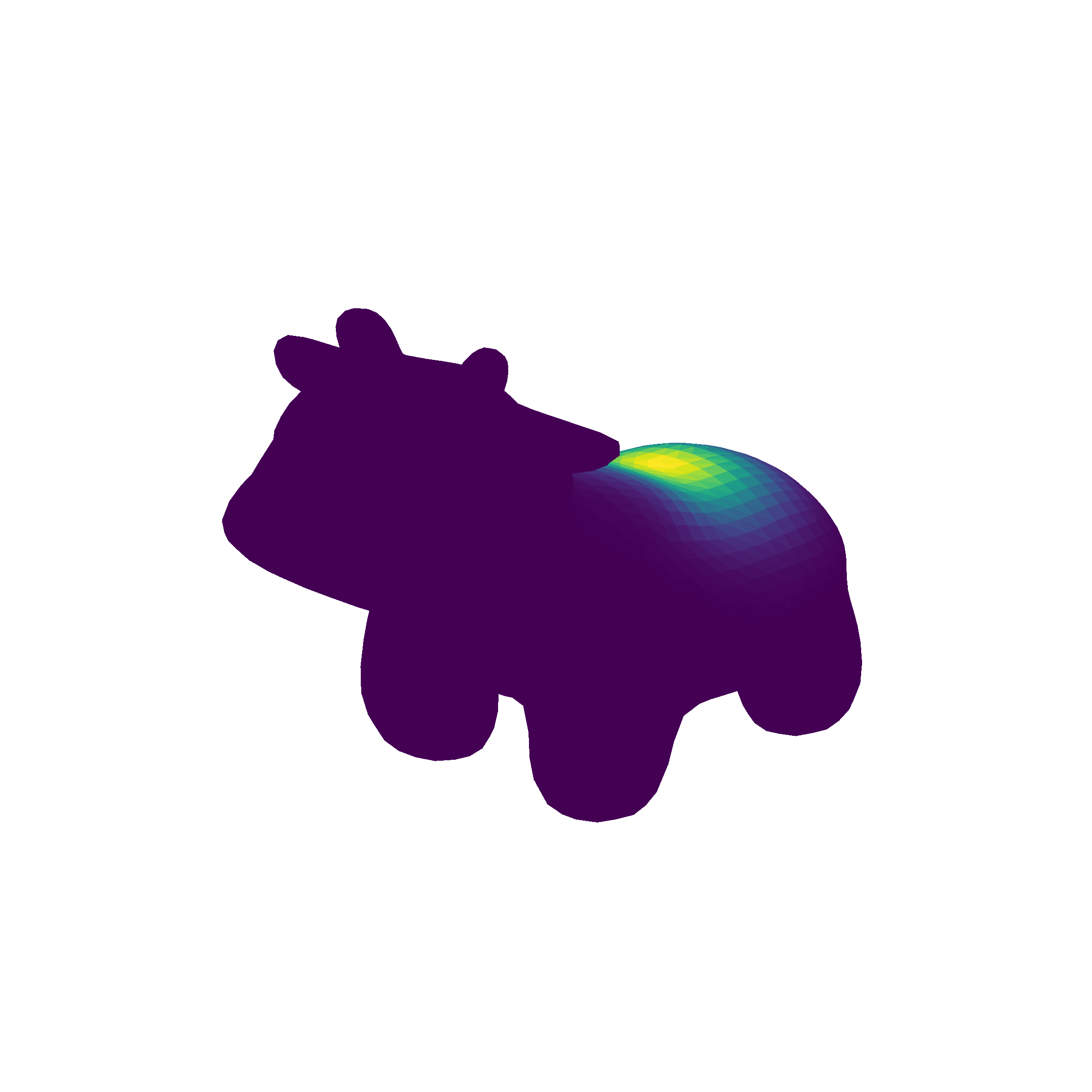}}
    \subfigure[$\rho(0.25, \boldsymbol{x})$]{\includegraphics[width=2.4cm]{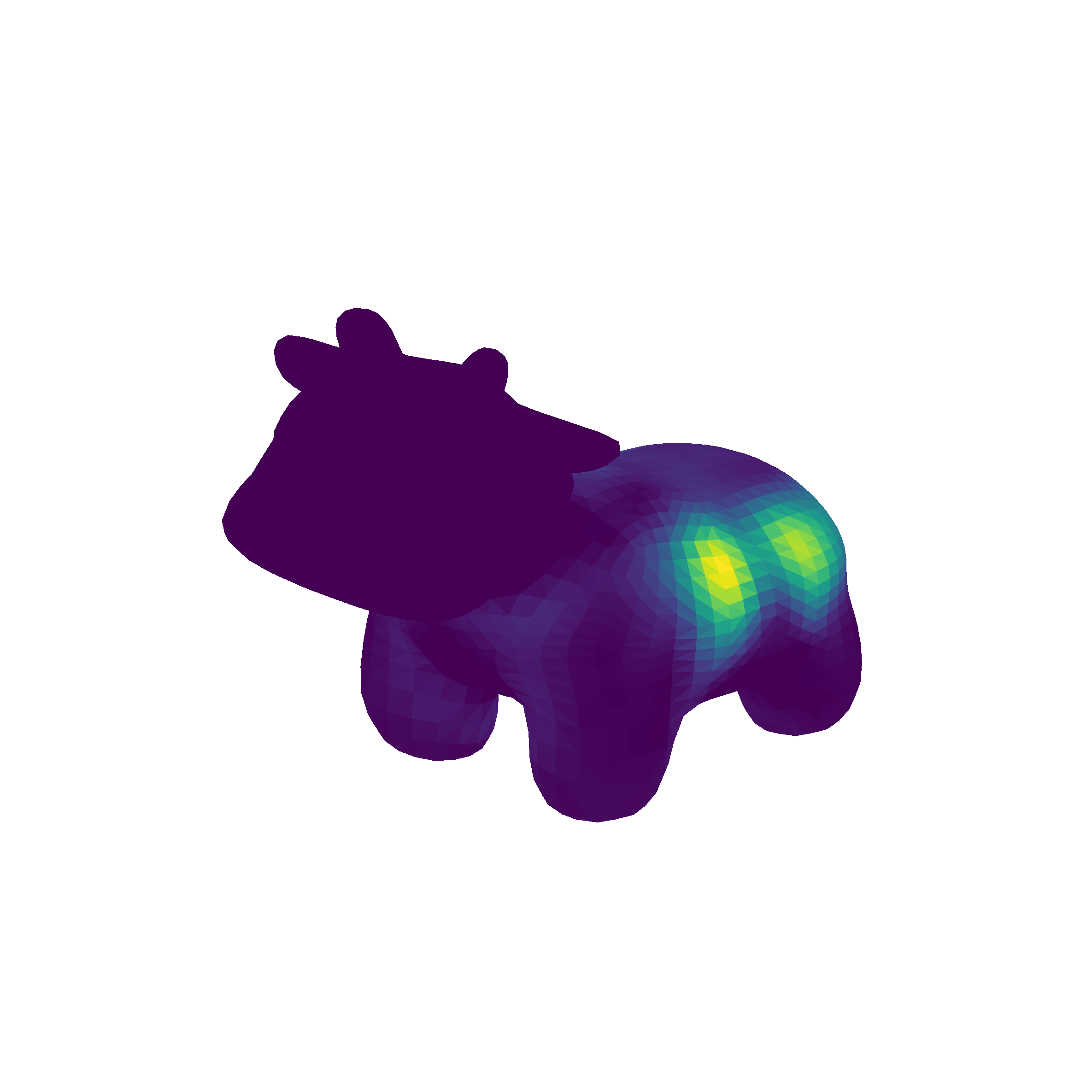}}
    \subfigure[$\rho(0.5, \boldsymbol{x})$]{\includegraphics[width=2.4cm]{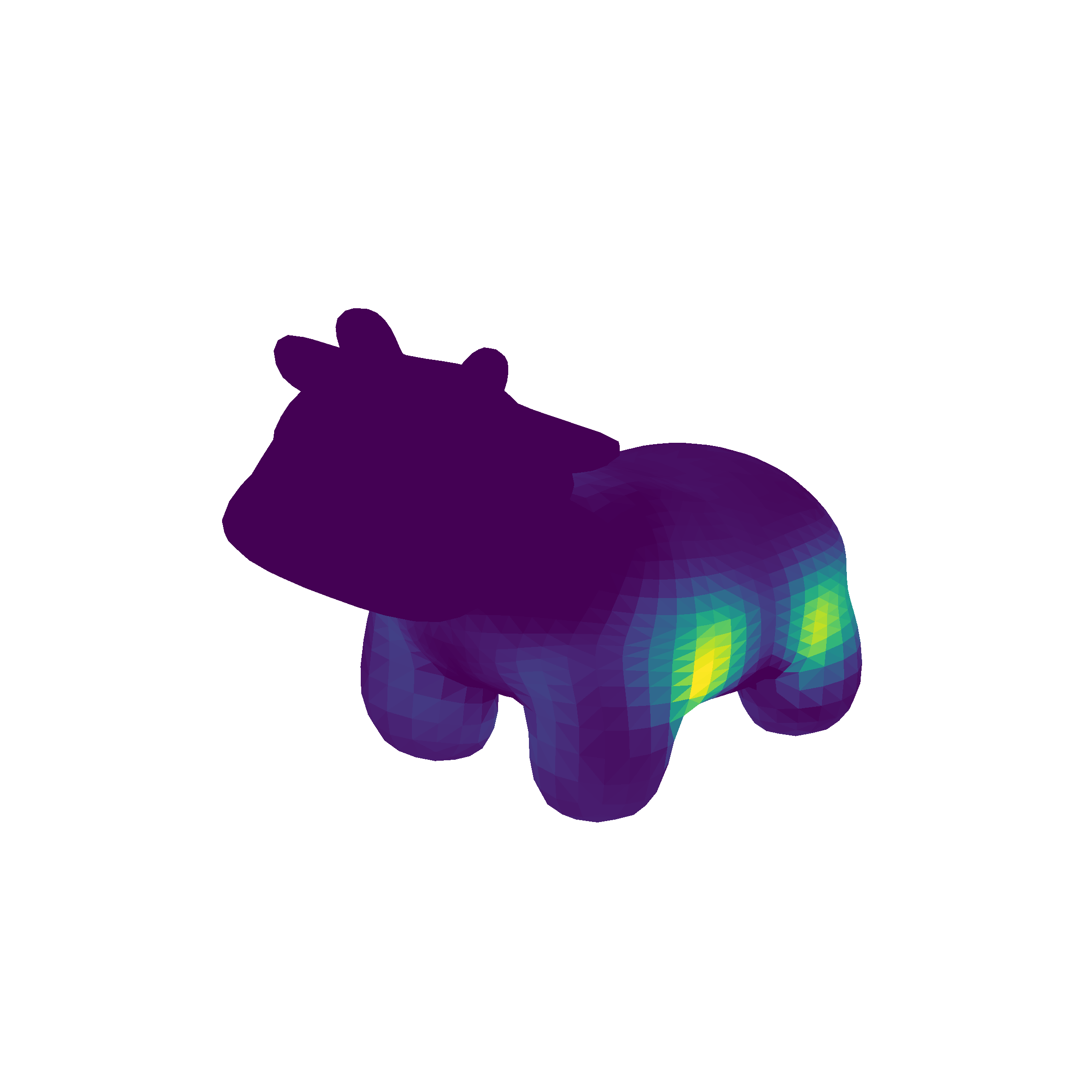}}
    \subfigure[$\rho(0.75, \boldsymbol{x})$]{\includegraphics[width=2.4cm]{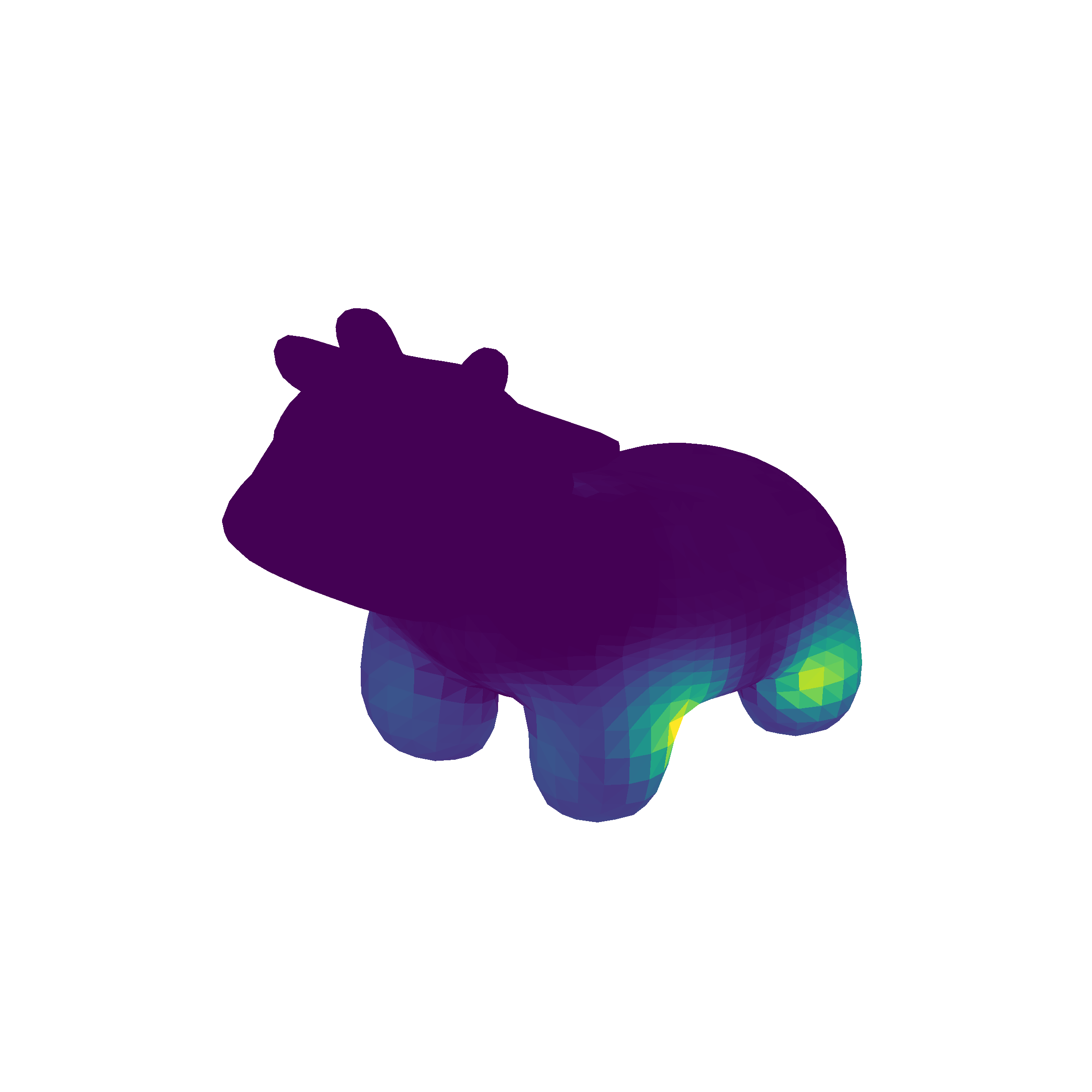}}
    \subfigure[$\rho(1, \boldsymbol{x})$]{\includegraphics[width=2.9cm]{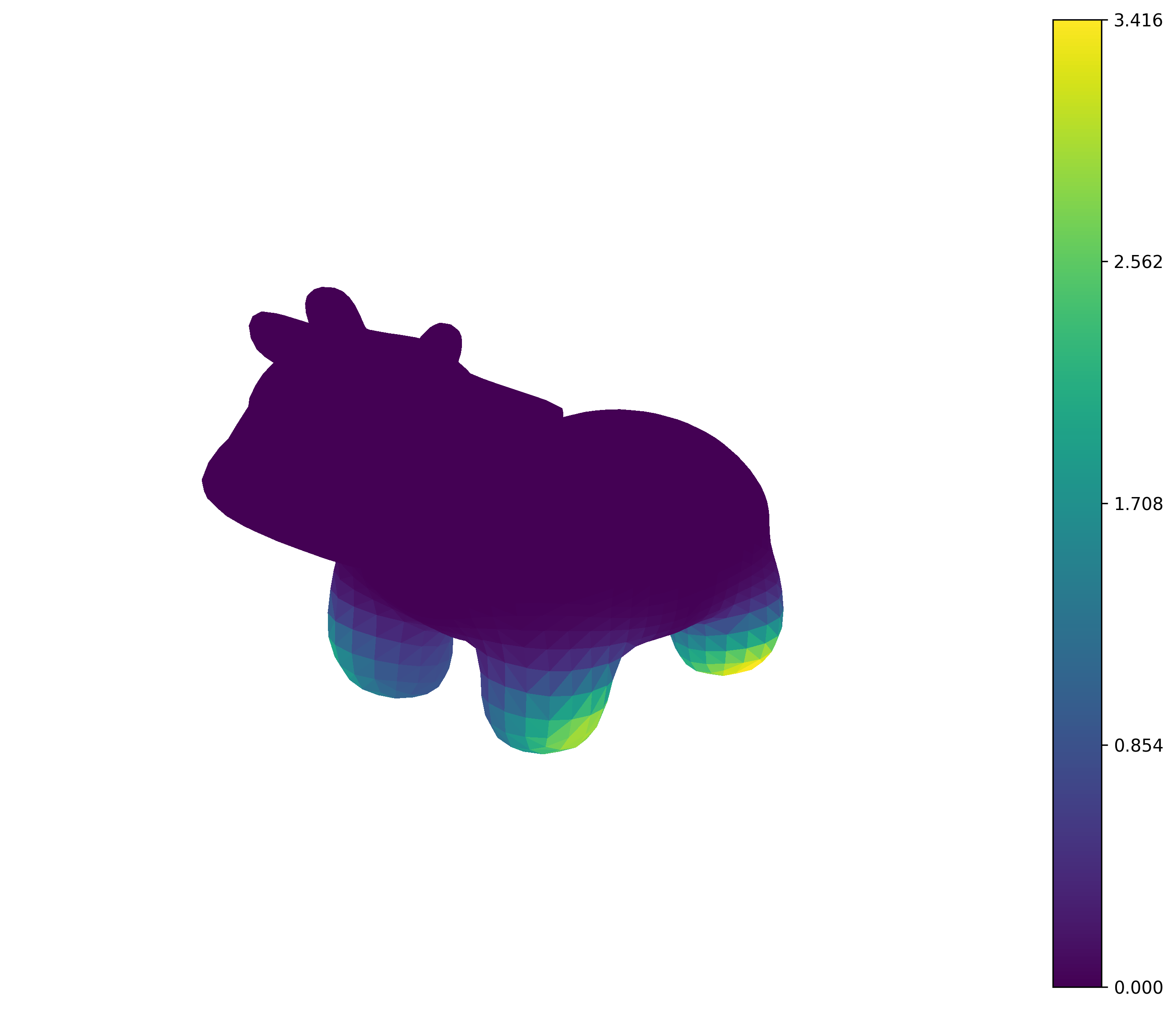}}\\
    \caption{SUOT test on general surfaces ($\beta=1.5$) and the calculation times are 1249s and 454s respectively.}
    \label{SUOT-general}
    \end{center}
\end{figure}

\begin{figure}[htbp]
    \begin{center}
    \includegraphics[width=2.4cm]{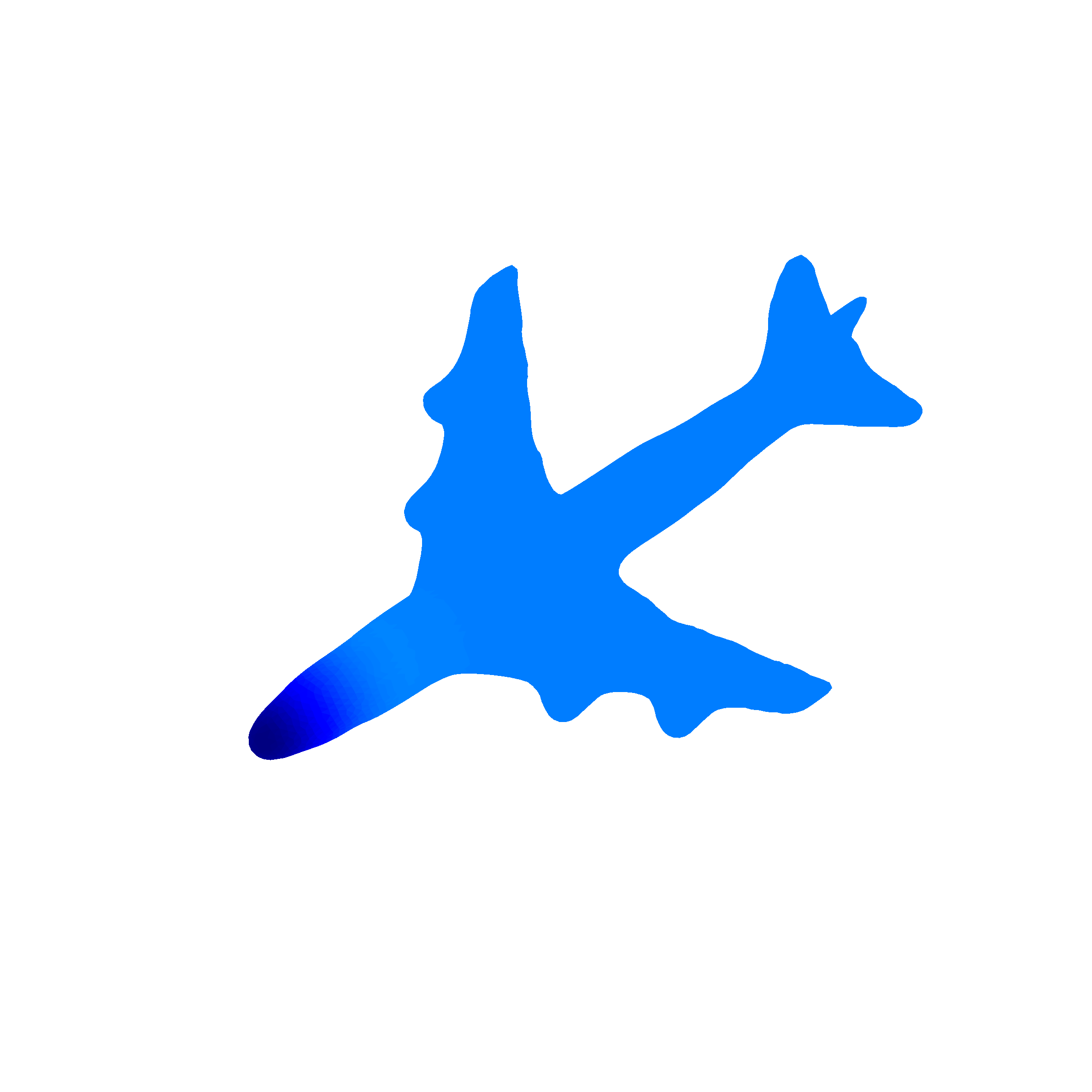}
    \includegraphics[width=2.4cm]{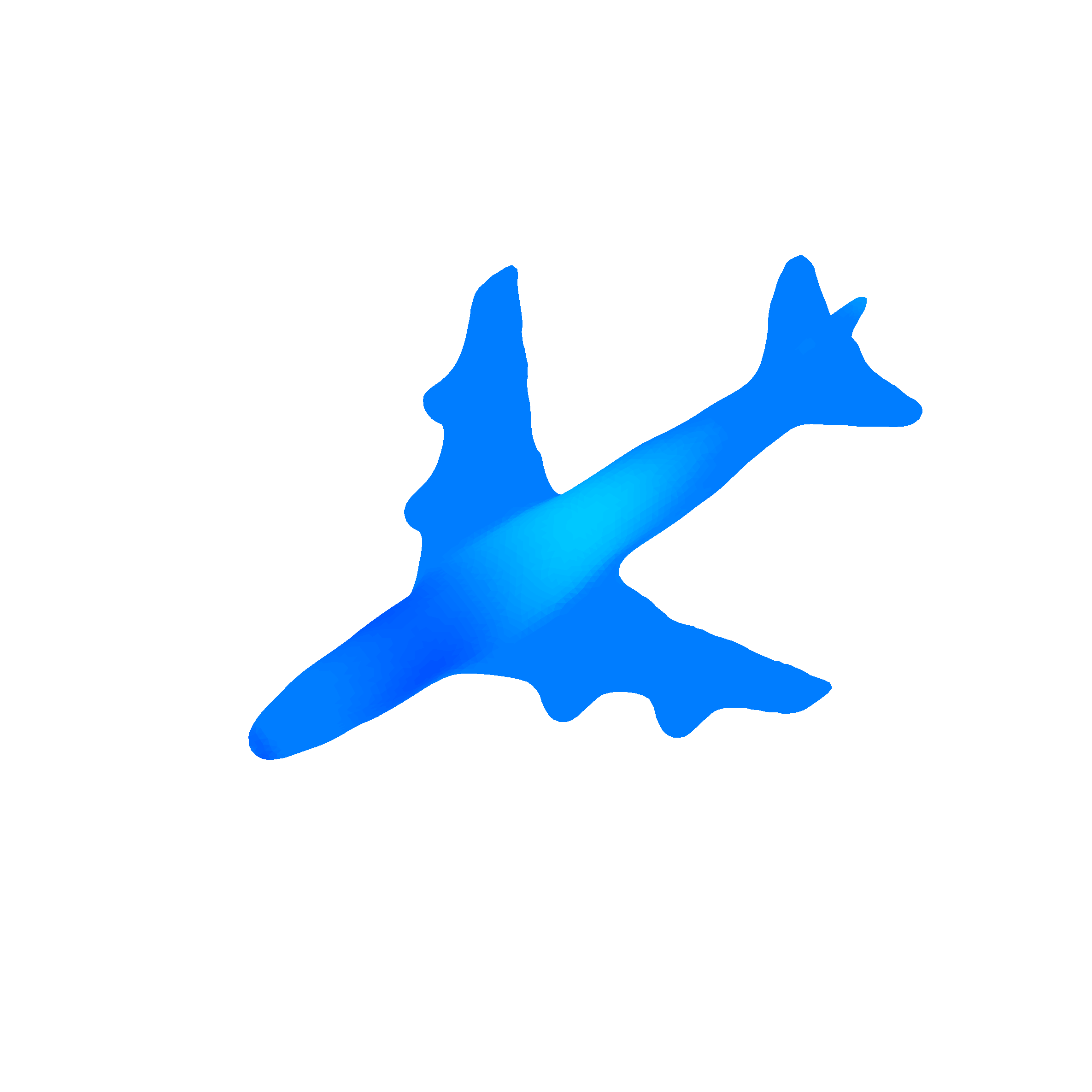}
    \includegraphics[width=2.4cm]{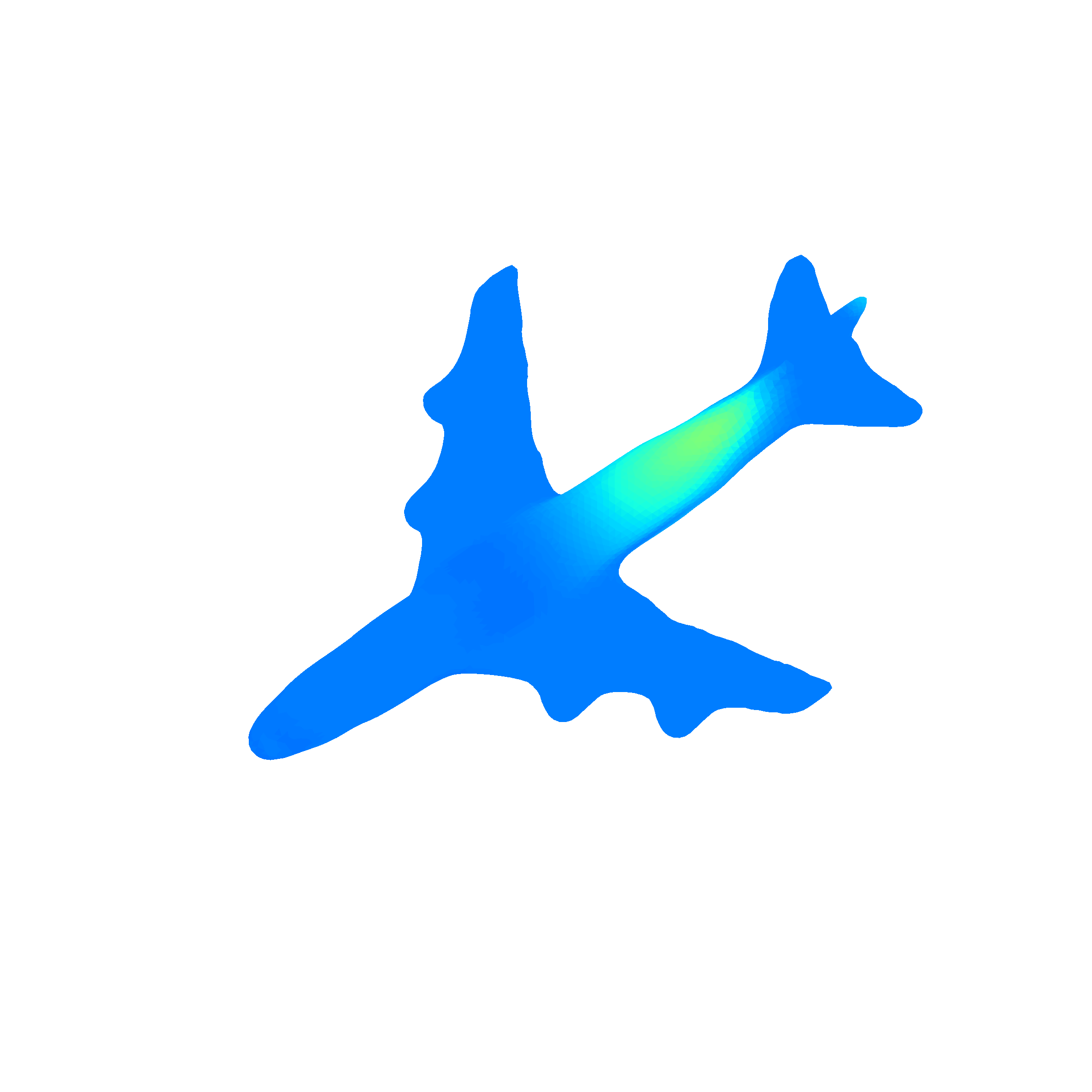}
    \includegraphics[width=2.4cm]{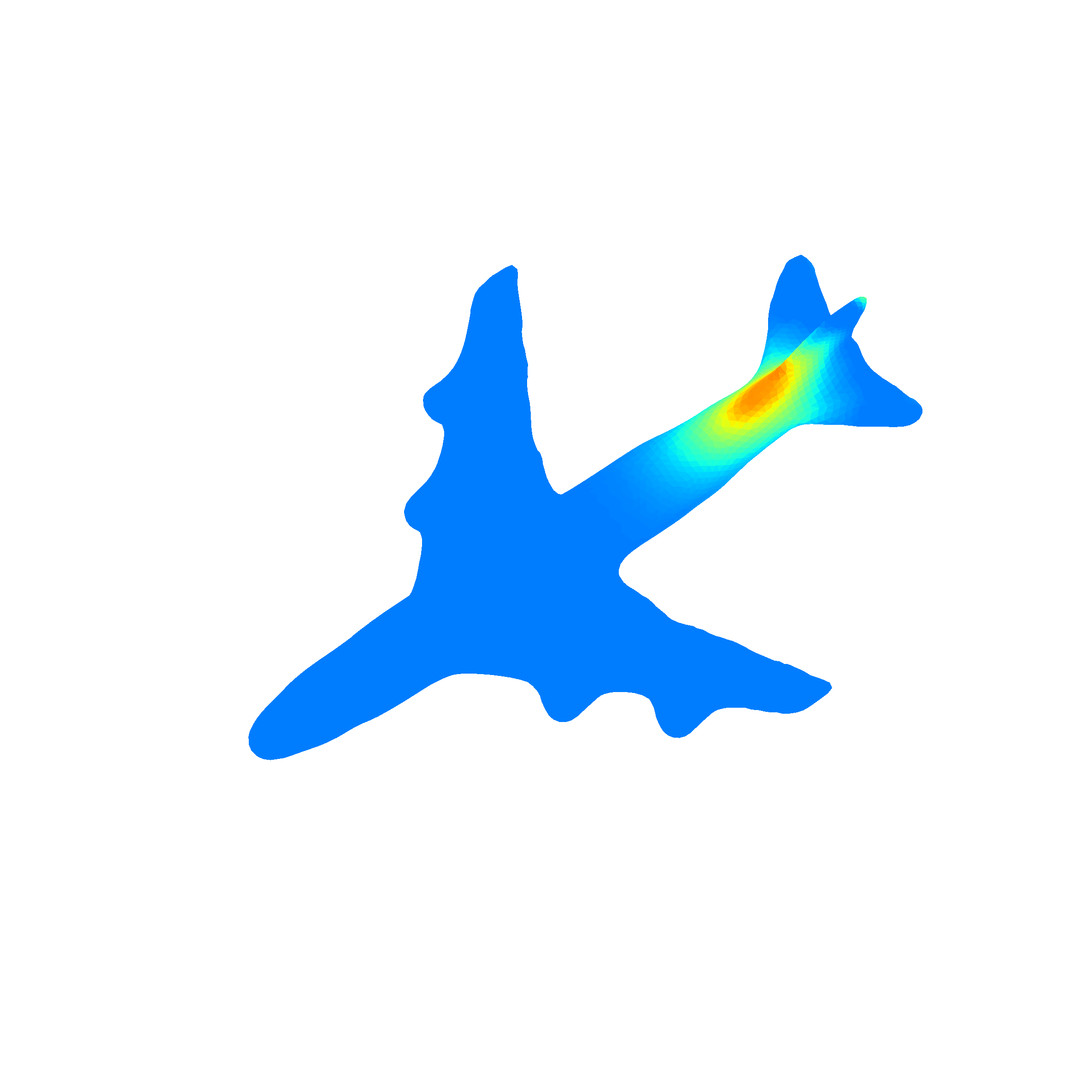}
    \includegraphics[width=2.9cm]{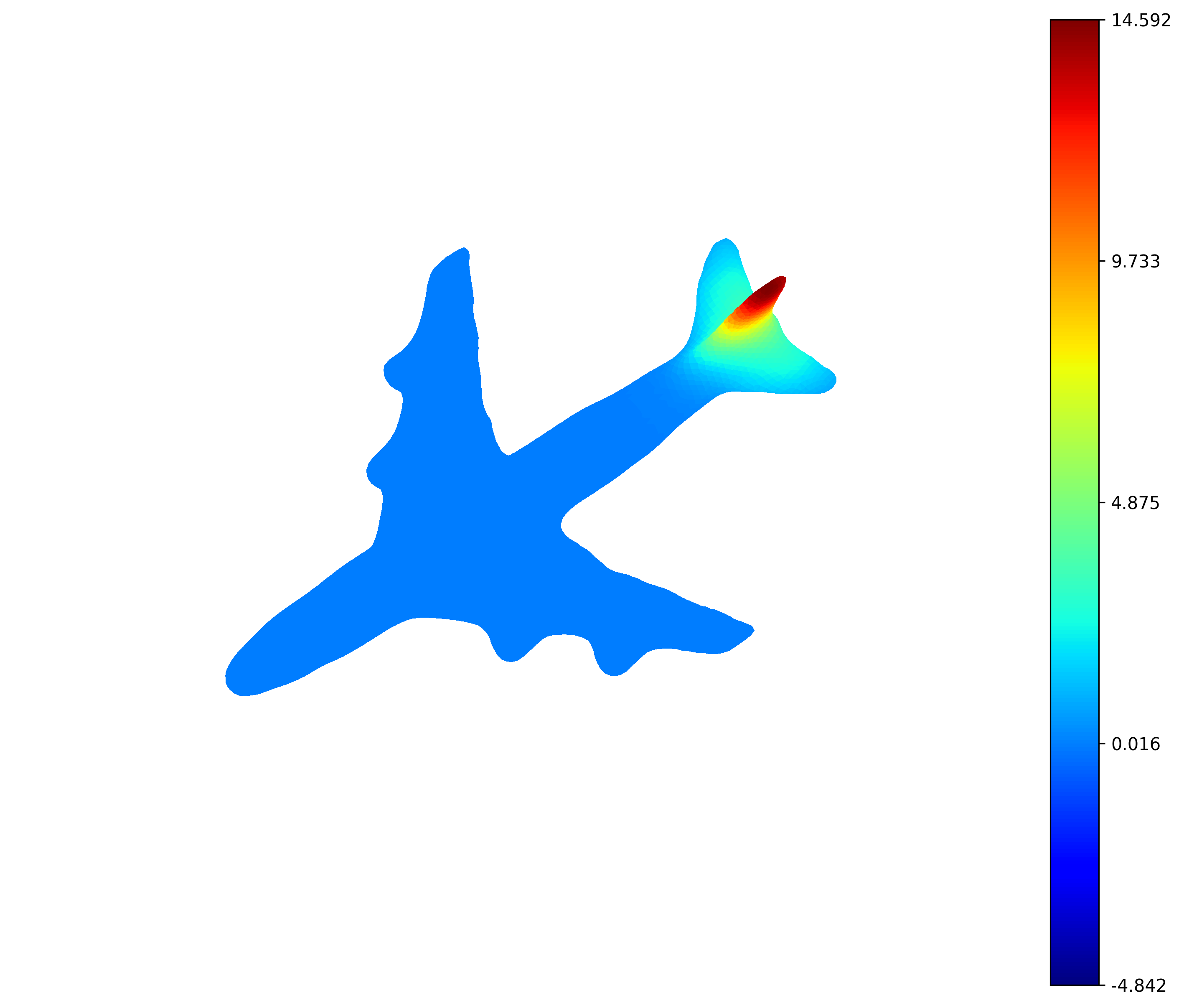}\\
    \vspace{5pt}

    \subfigure[$f(0, \boldsymbol{x})$]{\includegraphics[width=2.4cm]{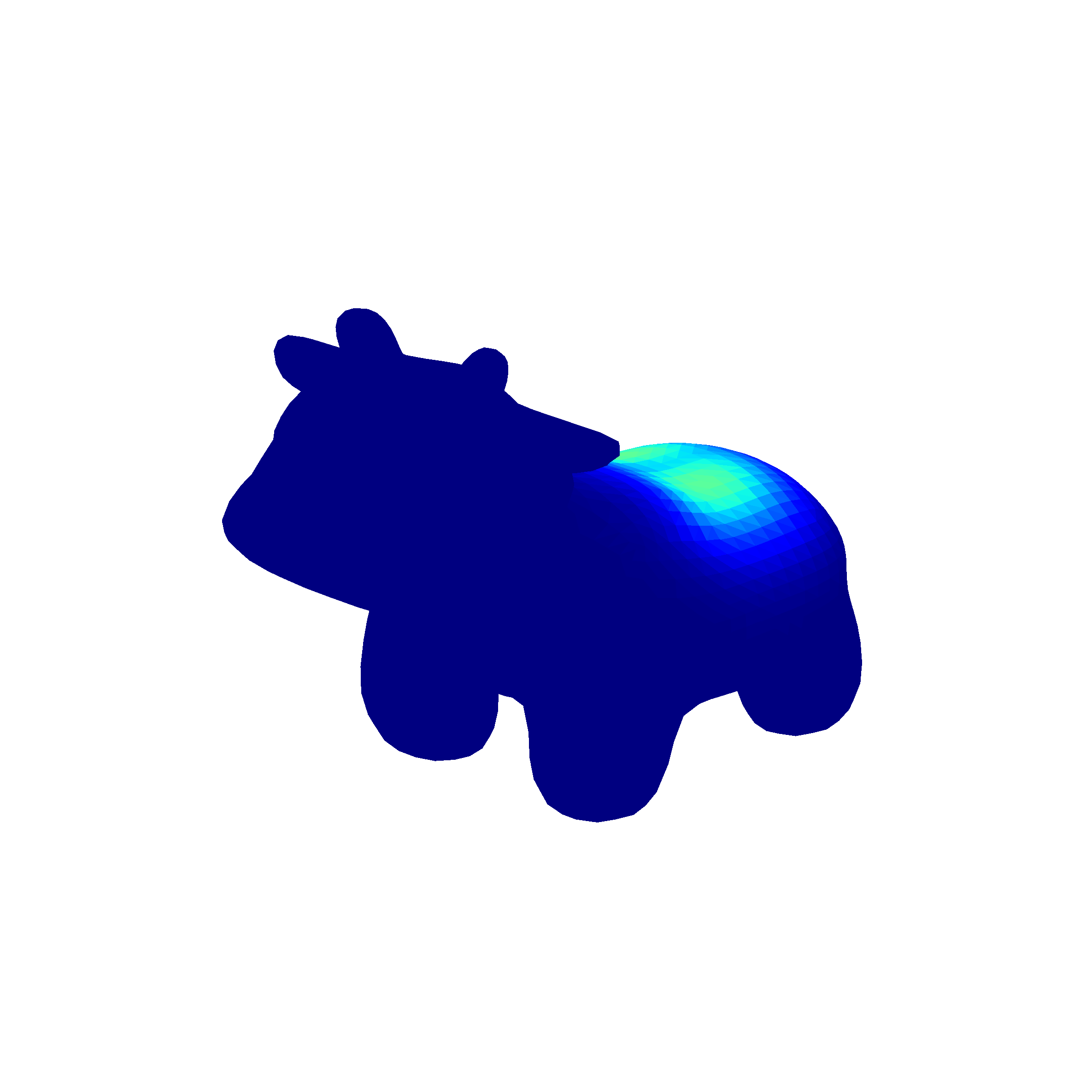}}
    \subfigure[$f(0.25, \boldsymbol{x})$]{\includegraphics[width=2.4cm]{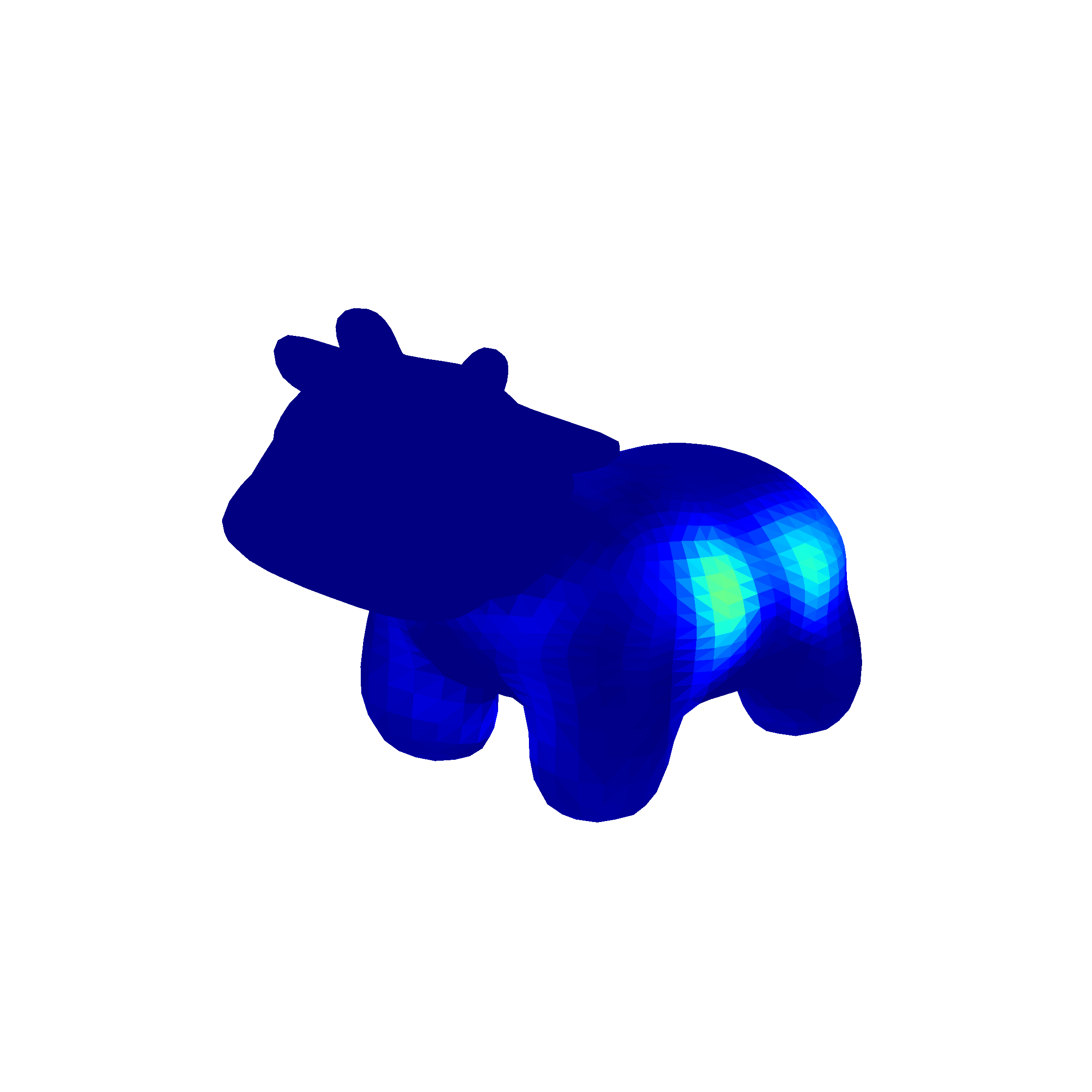}}
    \subfigure[$f(0.5, \boldsymbol{x})$]{\includegraphics[width=2.4cm]{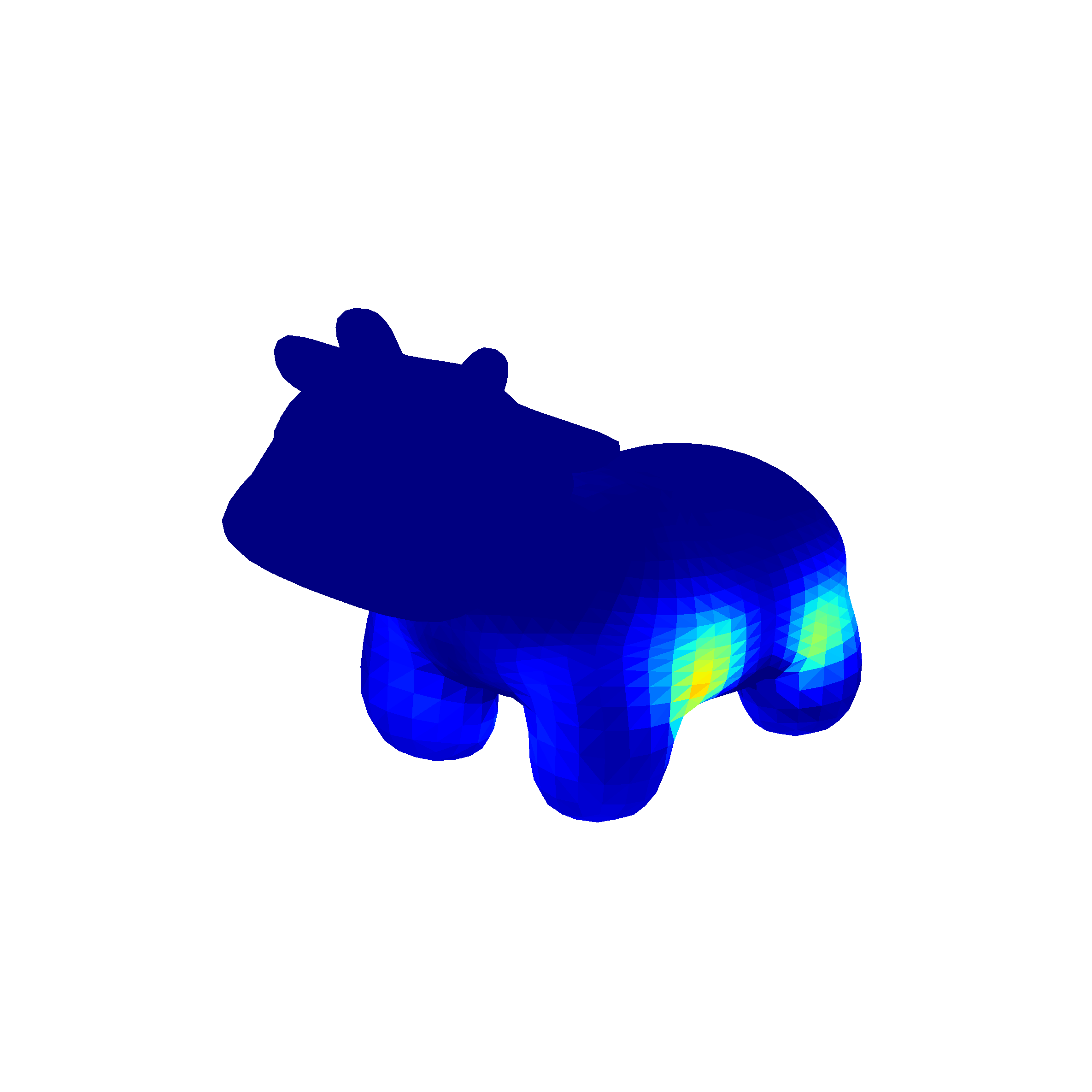}}
    \subfigure[$f(0.75, \boldsymbol{x})$]{\includegraphics[width=2.4cm]{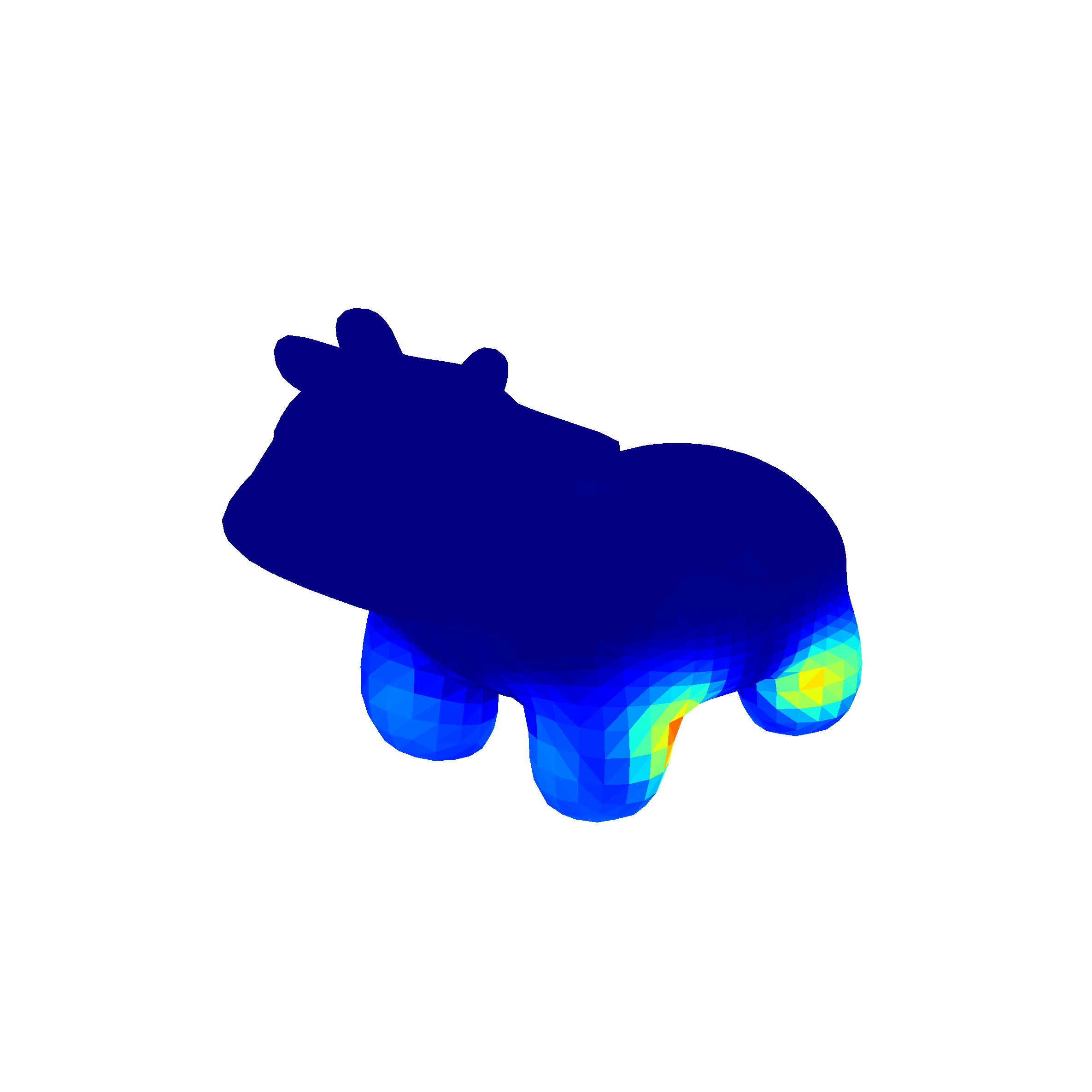}}
    \subfigure[$f(1, \boldsymbol{x})$]{\includegraphics[width=2.9cm]{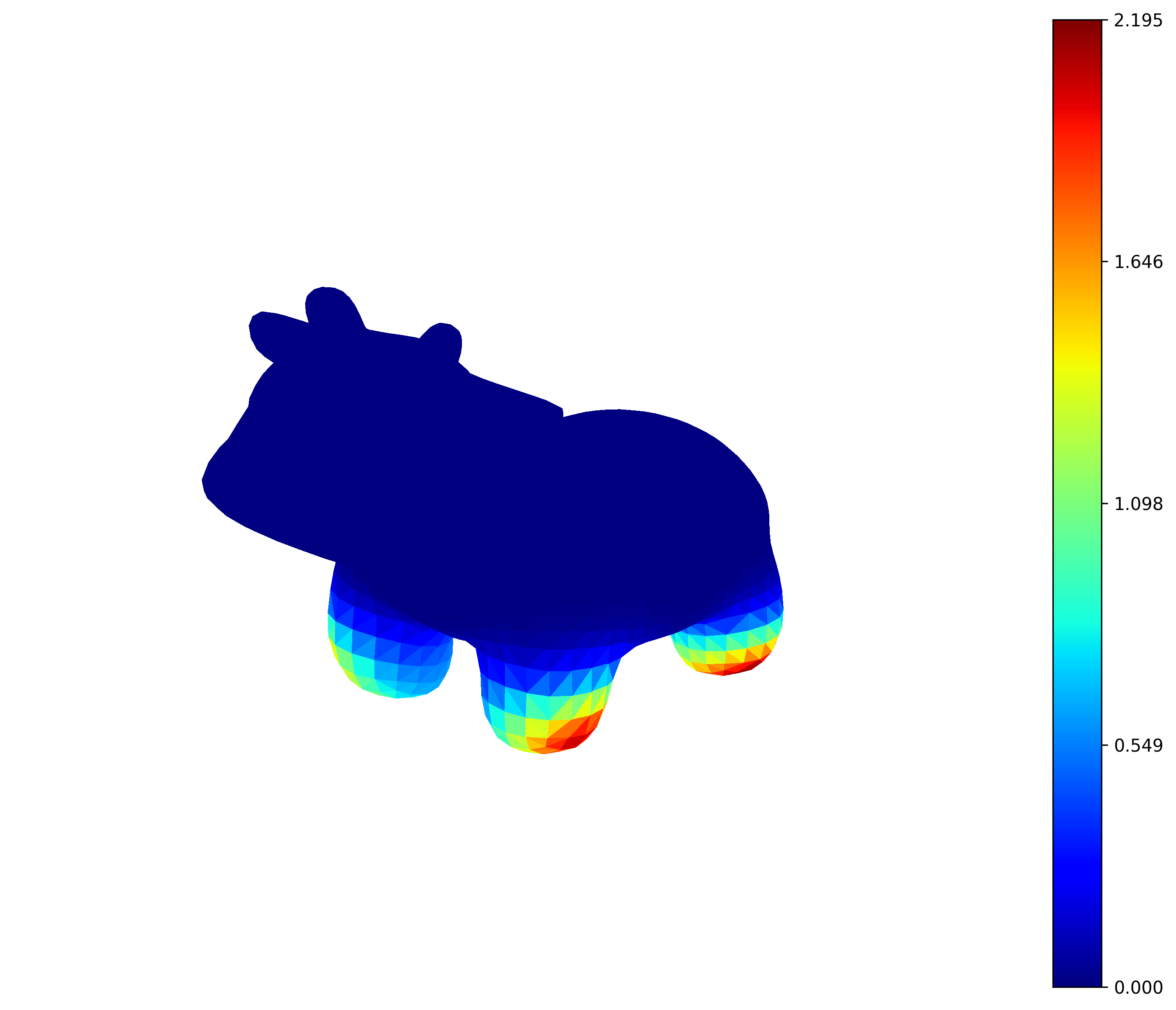}}\\
    \caption{SUOT test on general surfaces ($\beta=1.5$): source item $f$.}
    \label{SUOT-general-f}
    \end{center}
\end{figure}

As shown in Figure \ref{SUOT-general}, our method performs very well even in general point cloud. 
The source term is given in Figure \ref{SUOT-general-f}. 
In the example of airplane, mass initially decrease although the terminal mass is higher. 
This is also reasonable because it is expensive to move the mass from head to tail. 
So source term become more important than transport term.

\section{Conclusion}\label{Sec-Con}
We address the unbanlanced optimal transport problem on point clouds by combining the ADMM algorithm with the TRBF method.  
Building on Benamou and Brenier's dynamical formulation and the WFR metric, we introduce auxiliary variables that split the UOT problem into three subproblems such that the space-time Poisson equation becomes the key ingredient.  
To reduce the computational costs associated with mesh generation, especially for point cloud surfaces, we adopt the RBF method, a meshless approach that requires only points and normal vectors, avoiding the need to compute curvature.  
Additionally, we implement a fast algorithm to efficiently solve the algebraic equations involved in this subproblem.  
Numerical experiments on various point clouds demonstrate the effectiveness and efficiency of our proposed method. 
However, for general point cloud, the accuracy may not be very high. This is one direction we want to exploit further in the subsequent work.

\section*{Acknowledgement}\label{Acknow}
This work is supported by the National Natural Science Foundation of China (grants 92370125, 12301538) and the National Key R\&D Program of China (grant 2021YFA1001300).

\bibliographystyle{elsarticle-num} 
\bibliography{ref.bib}

\appendix
\section{Level set functions in Section 5.2}\label{App-A}
\begin{itemize}
    \item[-] \textit{Sphere}:
    \begin{equation*}\label{Sphere}
    \begin{aligned}
        (x-\frac{1}{2})^2 + (y-\frac{1}{2})^2 + (z-\frac{1}{2})^2 = \frac{1}{4}.
    \end{aligned}
    \end{equation*}
    \item[-] \textit{Ellipsoid}:
    \begin{equation*}\label{Ellipsoid}
    \begin{aligned}
        (x-\frac{1}{2})^2 + 3(y-\frac{1}{2})^2 + 6(z-\frac{1}{2})^2 = \frac{1}{4}.
    \end{aligned}
    \end{equation*}
    \item[-] \textit{Peanut}:
    \begin{equation*}\label{Peanut}
    \begin{aligned}
        \left((4(x-\frac{1}{2})-1)^2\right. & \left.+ 8(y-\frac{1}{2})^2+8(z-\frac{1}{2})^2\right)  \\
        & \left((4(x-\frac{1}{2})+1)^2 + 8(y-\frac{1}{2})^2 + 8(z-\frac{1}{2})^2\right) = \frac{6}{5}.
     \end{aligned}
    \end{equation*}
    \item[-] \textit{Torus}: 
    \begin{equation*}\label{Torus}
    \begin{aligned}
        \left(0.3 - \sqrt{(x-\frac{1}{2})^2 + (y-\frac{1}{2})^2}\right)^2 + (z-\frac{1}{2})^2 = \frac{1}{25}.
    \end{aligned}
    \end{equation*}
    \item[-] \textit{Opener}: 
    \begin{equation*}\label{Opener}
    \begin{aligned}
        \left(3(x-\frac{1}{2})^2(1-5(x-\frac{1}{2})^2) - 5(y-\frac{1}{2})^2\right)^2 + 5(z-\frac{1}{2})^2=\frac{1}{60}.
    \end{aligned}
    \end{equation*}
\end{itemize}

\section{Initial distribution \texorpdfstring{$\rho_{0}(\boldsymbol{x})$}{}, target distribution \texorpdfstring{$\rho_{1}(\boldsymbol{x})$}{} in Section 5.2}\label{App-B}
\begin{table}[htbp]
    \footnotesize
    \centering
    \caption{Initial distribution $\rho_{0}(\boldsymbol{x})$, target distribution $\rho_{1}(\boldsymbol{x})$ on different level set functions.}
    \label{tab:SOT-LS}
    \begin{tabular}{l|cc}
        \hline
         & $\rho_{0}(\boldsymbol{x})$ & $\rho_{1}(\boldsymbol{x})$\\
        \hline
        Sphere & $\hat{\rho}_{G}(\boldsymbol{x}, [0.5, 0.5, 0], 0.05\cdot\mathbf{I})$ & $\beta\hat{\rho}_{G}(\boldsymbol{x}, [0.5, 0.5, 1], 0.05\cdot\mathbf{I})$ \\
        Ellipsoid & $\hat{\rho}_{G}(\boldsymbol{x}, [0.5, 0.5, \frac{6+\sqrt{6}}{12}], 0.025\cdot\mathbf{I})$ & $\beta\hat{\rho}_{G}(\boldsymbol{x}, [0.5, 0.5, \frac{6-\sqrt{6}}{12}], 0.025\cdot\mathbf{I})$ \\
        Peanut & $\hat{\rho}_{G}(\boldsymbol{x}, [\frac{2+\sqrt{1+\sqrt{\frac{6}{5}}}}{4}, 0.5, 0.5], 0.025\cdot\mathbf{I})$ & $\beta\hat{\rho}_{G}(\boldsymbol{x}, [\frac{2+\sqrt{1-\sqrt{\frac{6}{5}}}}{4}, 0.5, 0.5], 0.025\cdot\mathbf{I})$ \\
        Torus & $\hat{\rho}_{G}(\boldsymbol{x}, [\frac{5+4\sqrt{\frac{1}{2}}}{10}, \frac{5+4\sqrt{\frac{1}{2}}}{10}, 0.5], 0.05\cdot\mathbf{I})$ & $\beta\hat{\rho}_{G}(\boldsymbol{x}, [\frac{5-4\sqrt{\frac{1}{2}}}{10}, \frac{5-4\sqrt{\frac{1}{2}}}{10}, 0.5], 0.05\cdot\mathbf{I})$ \\
        Opener & $\hat{\rho}_{G}(\boldsymbol{x}, [\frac{1+\sqrt{\frac{1+\sqrt{1+2\sqrt{\frac{1}{15}}}}{10}}}{2}, 0.5, 0.5], 0.025\cdot\mathbf{I})$ & $\beta\hat{\rho}_{G}(\boldsymbol{x}, [0.5, \frac{1-\sqrt{\frac{1+\sqrt{1+2\sqrt{\frac{1}{15}}}}{10}}}{2}, 0.5], 0.025\cdot\mathbf{I})$ \\
        \hline
    \end{tabular}
\end{table}
where
\begin{equation*}\label{Gaussian-manifold}
   \begin{aligned}
       \hat{\rho}_{G}(\boldsymbol{x}, \mu, \sigma) = 100e^{-\frac{\|\mu - \boldsymbol{x}\|^2}{\sigma}},
   \end{aligned}
\end{equation*}
and
\begin{equation*}\label{rho_mg}
    \footnotesize
    \begin{aligned}
        \rho_{mg1}&(\boldsymbol{x}) =  
        \frac{3}{8}\hat{\rho}_{G}(\boldsymbol{x}, [0.5, 0, 0.5], 0.025\cdot\mathbf{I}) + 
        \frac{3}{8}\hat{\rho}_{G}(\boldsymbol{x}, [0.5, 1, 0.5], 0.025\cdot\mathbf{I}) \\ 
        & + 
        \frac{3}{8}\hat{\rho}_{G}(\boldsymbol{x}, [0, 0.5, 0.5], 0.025\cdot\mathbf{I}) + 
        \frac{3}{8}\hat{\rho}_{G}(\boldsymbol{x}, [1, 0.5, 0.5], 0.025\cdot\mathbf{I}), \\
        \rho_{mg2}&(\boldsymbol{x}) =  
        \frac{3}{16}\hat{\rho}_{G}(\boldsymbol{x}, [0.5, 0, 0.5], 0.025\cdot\mathbf{I}) + 
        \frac{3}{16}\hat{\rho}_{G}(\boldsymbol{x}, [0.5, 1, 0.5], 0.025\cdot\mathbf{I}) \\ 
        & + 
        \frac{3}{16}\hat{\rho}_{G}(\boldsymbol{x}, [0, 0.5, 0.5], 0.025\cdot\mathbf{I}) + 
        \frac{3}{16}\hat{\rho}_{G}(\boldsymbol{x}, [1, 0.5, 0.5], 0.025\cdot\mathbf{I}) \\
        & + 
        \frac{3}{16}\hat{\rho}_{G}(\boldsymbol{x}, [\frac{2+\sqrt{2}}{4}, \frac{2+\sqrt{2}}{4}, 0.5], 0.025\cdot\mathbf{I}) + 
        \frac{3}{16}\hat{\rho}_{G}(\boldsymbol{x}, [\frac{2+\sqrt{2}}{4}, \frac{2-\sqrt{2}}{4}, 0.5], 0.025\cdot\mathbf{I}) \\
        & + 
        \frac{3}{16}\hat{\rho}_{G}(\boldsymbol{x}, [\frac{2-\sqrt{2}}{4}, \frac{2+\sqrt{2}}{4}, 0.5], 0.025\cdot\mathbf{I}) + 
        \frac{3}{16}\hat{\rho}_{G}(\boldsymbol{x}, [\frac{2-\sqrt{2}}{4}, \frac{2-\sqrt{2}}{4}, 0.5], 0.025\cdot\mathbf{I}), \\
    \end{aligned}
\end{equation*}

\begin{table}[htbp]
    \centering
    \caption{Initial distribution $\rho_{0}(\boldsymbol{x})$, target distribution $\rho_{1}(\boldsymbol{x})$ for SUOT.}
    \label{tab:SUOT}
    \begin{tabular}{l|cc}
        \hline
        & $\rho_{0}(\boldsymbol{x})$ & $\rho_{1}(\boldsymbol{x})$\\
        \hline
        S1 & $\hat{\rho}_{G}(\boldsymbol{x}, [0.5, 0.5, 1.0], 0.025\cdot\mathbf{I})$ & $\rho_{mg2}(\boldsymbol{x})$ \\
        S2 & $\rho_{mg2}(\boldsymbol{x})$ & $\hat{\rho}_{G}(\boldsymbol{x}, [0.5, 0.5, 1.0], 0.025\cdot\mathbf{I})$ \\
        \hline
    \end{tabular}
\end{table}
\section{Initial distribution \texorpdfstring{$\rho_{0}(\boldsymbol{x})$}{}, target distribution \texorpdfstring{$\rho_{1}(\boldsymbol{x})$}{} in Section 5.3}\label{App-C}

\begin{table}[htbp]
    \footnotesize
    \centering
    \caption{Initial distribution $\rho_{0}(\boldsymbol{x})$, target distribution $\rho_{1}(\boldsymbol{x})$ on different point clouds.}
    \label{tab:SOT-general}
    \begin{tabular}{l|cc}
        \hline
         & $\rho_{0}(\boldsymbol{x})$ & $\rho_{1}(\boldsymbol{x})$\\
        \hline
        Airplane & $\hat{\rho}_{G}(\boldsymbol{x}, [-0.015821, 0.957996, 0.055], 0.05\cdot\mathbf{I})$ & $\beta\hat{\rho}_{G}(\boldsymbol{x}, [-0.000874, -0.763727, 0.342374], 0.05\cdot\mathbf{I})$ \\
        Cow & $\hat{\rho}_{G}(\boldsymbol{x}, [0, 0.547798, 0.228164], 0.05\cdot\mathbf{I})$ & $\beta\rho_{c}(\boldsymbol{x})$ \\
        \hline
    \end{tabular}
\end{table}
where $\beta=1.5$ and
\begin{equation*}\label{rho_smg}
    \begin{aligned}
        \rho_{c}(\boldsymbol{x}) = & 
        \frac{1}{4}\hat{\rho}_{G}(\boldsymbol{x}, [0.291818,  0.788408, -0.690195], 0.025\cdot\mathbf{I})\\
        & + 
        \frac{1}{4}\hat{\rho}_{G}(\boldsymbol{x}, [0.368236 ,  0.0306267, -0.614976], 0.025\cdot\mathbf{I}) \\ 
        & + 
        \frac{1}{4}\hat{\rho}_{G}(\boldsymbol{x}, [-0.319732,  0.786428, -0.633454], 0.025\cdot\mathbf{I})\\ 
        & + 
        \frac{1}{4}\hat{\rho}_{G}(\boldsymbol{x}, [-0.368236 ,  0.0306267, -0.614976], 0.025\cdot\mathbf{I}), \\
    \end{aligned}
\end{equation*}
\end{document}